%% file: ReviewMain.tex
\documentclass[aps,twocolumn,bibnotes,amsmath,amssymb,notitlepage]{revtex4-1}
\usepackage{booktabs}
\usepackage{multirow}
\usepackage{lipsum}
\usepackage[table,xcdraw]{xcolor}
\usepackage{graphicx}
\usepackage{units}
\usepackage{mathrsfs}
\usepackage{setspace}
\usepackage[T1]{fontenc}

\usepackage[colorlinks,urlcolor=gray,citecolor=blue,linkcolor=red,linktocpage=true]{hyperref}

\usepackage[newparttoc,explicit, clearempty]{titlesec}%
\usepackage{titletoc}

\makeatletter
\renewcommand*\l@chapter[2]{%
  \ifnum \c@tocdepth >\m@ne
    \addpenalty{-\@highpenalty}%
    \vskip 1.0em \@plus\p@
    \setlength\@tempdima{1.5em}%
    \begingroup
      \parindent \z@ \rightskip \@pnumwidth
      \parfillskip -\@pnumwidth
      \leavevmode \bfseries
      \advance\leftskip\@tempdima
      \hskip -\leftskip
      #1\nobreak\hskip .5em \@plus 1fil 
      \nobreak\hb@xt@\@pnumwidth{\hss #2}\par
      \penalty\@highpenalty
    \endgroup
  \fi}
\makeatother

\expandafter\let\csname equation*\endcsname\relax
\expandafter\let\csname endequation*\endcsname\relax
\usepackage{amsmath}

\usepackage[]{lineno}

\usepackage{caption}
\DeclareCaptionLabelSeparator{vline}{ | }
\makeatletter
\def\justified{
  \let\\\@normalcr
  \@rightskip\z@skip \rightskip\@rightskip
  \leftskip\z@skip
  \parindent 0em\relax
  \setlength{\parfillskip}{0pt plus 1fil}}
\DeclareCaptionJustification{justified}{\justified}

\newcommand{\myparagraph}[1]{\paragraph{#1}\mbox{}\\}

\newcommand{\add}{\ensuremath{a_{\rm dd}}}

\newcommand{\pdd}{\Phi_{\rm{dd}}}
\newcommand{\edd}{\varepsilon_{\rm{dd}}}

\newcommand{\abg}{a_{\rm{bg}}}
\newcommand{\fdip}{f_{\rm{dip}}}

\newcommand{\be}{\begin{equation}}
\newcommand{\ee}{\end{equation}}
\newcommand{\bea}{\begin{eqnarray}}
\newcommand{\eea}{\end{eqnarray}}

\newcommand{\ket}[1]{|#1\rangle}

\newcommand{\bs}{\boldsymbol}
\newcommand{\Vlatt}{V_{\rm L}}
\newcommand{\vecr}{\ensuremath{\boldsymbol{r}}}
\newcommand{\vecu}{\ensuremath{\boldsymbol{u}}}
\newcommand{\vk}{\ensuremath{\boldsymbol{k}}}
\newcommand{\uK}{\ensuremath{\mu {\rm K}}}
\newcommand{\um}{\ensuremath{\mu {\rm m}}}
\newcommand{\as}{a}
\newcommand{\rd}{r_{\rm{d}}}
\newcommand{\Vdd}{V_{\rm{dd}}}
\newcommand{\muB}{\mu_{\rm{B}}}
\newcommand{\epsmin}{\Delta}
\newcommand{\kr}{k_{\rm{rot}}}

\newcommand{\rvdw}{r_{\rm vdW}}
\newcommand{\obdm}{\rho_{1}}

\begin{document}

\title{Dipolar physics: A review of experiments with magnetic quantum gases.}

\author{Lauriane Chomaz$^{1,2}$, Igor Ferrier-Barbut$^{3,4}$, Francesca Ferlaino$^{1,5}$, Bruno Laburthe-Tolra$^{6,7}$, Benjamin L.~Lev$^{8}$, Tilman Pfau$^{3}$. }

\vspace{5pt}
\address{$^1$ Institut f\"ur Experimentalphysik, Universit\"at Innsbruck, Technikerstrasse 25, 6020 Innsbruck, Austria}
\address{$^2$ Physikalisches Institut der Universit\"at Heidelberg, Im Neuenheimer Feld 226, 69120 Heidelberg, Germany}
\address{$^3$ Physikalisches Institut and Center for Integrated Quantum Science and Technology, Universit\"at Stuttgart, Pfaffenwaldring 57, 70550 Stuttgart, Germany}
\address{$^4$ Universit\'e Paris-Saclay, Institut d’Optique Graduate School, CNRS, Laboratoire Charles Fabry, 91127, Palaiseau, France}
\address{$^5$ Institut f\"{u}r Quantenoptik und Quanteninformation, \"Osterreichische Akademie der Wissenschaften, 6020 Innsbruck, Austria}
\address{$^6$ Universit\'e Sorbonne Paris Nord, Laboratoire de Physique des Lasers, F-93430 Villetaneuse, France}
\address{$^7$ CNRS, UMR 7538, LPL, F-93430, Villetaneuse, France}
\address{$^8$ Departments of Physics and Applied Physics and Ginzton Laboratory, Stanford University, Stanford CA 94305, USA}

\vspace{10pt}

\date{\today}

\begin{abstract}
Since the achievement of quantum degeneracy in gases of chromium atoms in 2004, the experimental investigation of ultracold gases made of highly magnetic atoms has blossomed. The field has yielded the observation of many unprecedented phenomena, in particular those in which long-range and anisotropic dipole-dipole interactions play a crucial role. In this review, we aim to present the aspects of the magnetic quantum-gas platform that make it unique for exploring ultracold and quantum physics as well as to give a thorough overview of experimental achievements.  

Highly magnetic atoms distinguish themselves by the fact that their electronic ground-state configuration possesses a large spin (as well as a large $g$ factor). This results in a large magnetic moment and a rich electronic transition spectrum.  Such transitions are useful for cooling, trapping, and manipulating these atoms. The complex atomic structure and large dipolar moments of these atoms also lead to a dense spectrum of resonances in their two-body scattering behaviour. These resonances can be used to control the interatomic interactions and, in particular, the relative importance of contact over dipolar interactions. These features provide exquisite control knobs for exploring the few- and many-body physics of dipolar quantum gases. 

The study of dipolar effects in magnetic quantum gases has covered various few-body phenomena that are based on elastic and inelastic anisotropic scattering.  Various many-body effects have also been demonstrated. These affect both the shape, stability, dynamics, and excitations of fully polarised repulsive Bose or Fermi gases. Beyond the mean-field instability, strong dipolar interactions competing with slightly weaker contact interactions between magnetic bosons yield new quantum-stabilised states, among which are self-bound droplets, droplet assemblies, and supersolids. Dipolar interactions also deeply affect the physics of atomic gases with an internal degree of freedom as these interactions intrinsically couple spin and atomic motion. Finally, long-range dipolar interactions can stabilise strongly correlated excited states of 1D gases and also impact the physics of lattice-confined systems, both at the spin-polarised level (Hubbard models with off-site interactions) and at the spinful level (XYZ models). In the present manuscript, we aim to provide an extensive overview of the various related experimental achievements up to the present.
\end{abstract}

\maketitle

  \tableofcontents

 \input{Section1/1_intro_v2.tex}

\input{Section2/2_Cr_to_Lanth.tex}

 \input{Section3/3_DipolarScattering.tex}

 \input{Section4/4_aa.tex}

 \input{Section5/5_instabilityanddroplets.tex}
 \input{Section6/6_Spinors.tex}

 \input{Section7/7_lattice_reorder.tex}

 \input{Section10/10_conclusions.tex}

\end{document}

%% file: Section1/1_intro_v2.tex
\section{Introduction to dipolar physics}
\label{sec:intro}
Ultracold gases have drawn considerable interest since the realisation of quantum degenerate Bose~\cite{Anderson1995oob,Davis1995bec,Bradley1997BEC} and Fermi~\cite{DeMarco1999oof, Schreck2001qbe,Truscott2001oof} gases in the mid-to-late 1990's. This interest stems from many quarters within the physics community, but especially from those interested in using ultracold gases as test-bed systems for  theoretical models, 
for exploring their properties as new---highly controllable---examples of strongly correlated matter, and for engineering them for quantum information processing~\cite{Bloch2008mbp,Bloch2012qsw,Pitaevskii2016bec}. 

Interparticle interactions fundamentally determine the  properties of a quantum gas. Even in the weakly interacting limit, they dictate its shape,  density, and the way it becomes excited. In the strongly interacting limit, even more drastic modifications of the system's properties can arise, such as the appearance of exotic phases or excitation modes not describable by effective single-particle models. On the other hand, interactions can lead to inelastic processes that cause population loss from a trap and limit the accessible range of, e.g., temperature and density.  

Quantum gases are typically dilute (compared to liquids and solids) and this allows their short-range interaction at low temperature to be  accounted for in a simple fashion by a two-body (isotropic) contact pseudo-potential~\cite{Pethick2002book,Stringari2016bec}. 
To go beyond the case of isotropic and short-range interactions---say using an ultracold system possessing strong dipolar interactions--- gives access to a wide variety of new physical phenomena~\cite{Goral2000bec,Marinescu1998caa,Baranov2012cmt,Baranov2002udg,Baranov2008tpi,Lahaye2009tpo,Stringari2016bec,Defenu2021lri}. 
This review focuses on the experimental achievements of the last fifteen years to study such physics using one particular example of a dipolar system, viz., ultracold quantum gases made of atoms possessing a large magnetic dipole moment.   

\subsection{Quantum gases with dipolar interactions}

Several platforms exist with which to study the effect of dipole-dipole interactions (DDIs) 
in the ultracold gas context.  For example, electric dipole moments may be induced using heteronuclear molecules~\cite{Carr2009cau,Bohn2017cmp,Moses2017nff} or Rydberg atoms~\cite{Saffman2010qiw,Loew2012aea,Labuhn2016ttd,Bernien2017pmb} in an electric field or through the use of light-induced dipoles~\cite{Lahaye2009tpo}.  We note that long-range interactions, beyond the dipolar $1/r^3$ scaling, can also be achieved in ultracold gas systems in other ways.  For example, one method uses optical cavity or waveguide-mediated interactions, which are fixed to be either global in range~\cite{Ritsch2013cai, Mivehvar2021cqe} or may be tuned between long and short range~\cite{Gopalakrishnan2009eca,Douglas2015qmb,Vaidya2018trp}.  Phonon-mediated interactions in trapped ion systems are another example of tunable-range interactions~\cite{Blatt2012qsw}. These systems often exhibit dipole strengths orders of magnitude larger than what is achievable with magnetic dipoles. However, other limitations can arise in these systems, e.g., short lifetimes, density limitations, and/or rapid dissipation.  We briefly discuss the case of electric dipolar systems before exclusively focusing on magnetic systems. 

\subsubsection{Electric dipoles}

There is no permanent electric dipole in an atom or in a molecule in its non-degenerate rotational ground state due to their rotational symmetry. Yet when an external electric field $E$  couples to the electric dipole moment operator, it mixes eigenstates of opposite parity. As the rotational symmetry is broken, an electric dipole moment is induced. The field to induce the electric dipole moment is lowest when the two states of opposite parity are closest in energy. 

Some systems  possess  degenerate states of opposite parity, which allows the induced electric moment to arise at vanishingly small electric field~\cite{wall2013sqm}. Rydberg states in a hydrogen atom are an example of such a system:  they possess electronically excited states of opposite parity that can be arbitrarily close. The electric dipole moment  scales as $n^{2}$, with $n$ the principal quantum number. Associating a Rydberg atom with a ground-state atom allows to form a Rydberg molecule with a permanent electric dipole moment~\cite{Li2011ahm}. Despite short lifetimes due to spontaneous emission, black-body radiation~\cite{Goldschmidt2016abi}, and collisions, Rydberg gases have, over the last 10 years, been the centre of much experimental and theoretical activity.  In particular,  experiments are ongoing that  investigate  strongly correlated dipolar gases, lattice spin models, and Rydberg molecules~\cite{Saffman2010qiw,Loew2012aea,Schauss2012oos,Schauss2015cii,Bernien2017pmb,Scholl2021qso,Ebadi2021qpo,Browaeys2020mbp}. They are even the basis for a competitive quantum computing platform, which has been pushed forward by several newly founded companies (https://pasqal.io/, https://coldquanta.com/, https://www.quera.com/, https://www.atom-computing.com/). 

A second, very productive field of research is the manipulation of heteronuclear molecules. In these, an electric field  mixes two rotational states (for example $N=0$ and $N=1$) within the electronic molecular ground state.  Ultracold molecular systems with a large electric dipole moment include: KRb~\cite{Ospelkaus2010cth, Moses2015coa,demarco2019afd,Valtolina2020deo}, NaK~\cite{Park2015udg,Will2016cmc,Voges2020ugo}, RbCs~\cite{takekoshi2012ttp, molony2014cou}, NaRb~\cite{Guo2016coa}, KCs~\cite{Grobner2016anq}, LiCs~\cite{Deiglmayr2008fou}, NaLi~\cite{Rvachov2017llu,Son2020cco}, SrF~\cite{Barry2014mot,Norrgard2016}, H$_2$CO~\cite{Prehn2016oco}, CaF~\cite{Anderegg2017rfm,Truppe2017mcb},
BaF~\cite{Albrecht2020bgc}and YO~\cite{Collopy2018mot,Ding2020sdc}, HO~\cite{Stuhl2012eco,Reens2017csf}. 
Due to their intrinsic complexity, cooling molecules has been an extremely challenging task.  
Recently, after many years of dedicated efforts~\cite{Moses2015coa,Park2015udg}, the first quantum degenerate gas of polar molecules has been achieved with KRb
~\cite{demarco2019afd, Valtolina2020deo}. 
Many of the molecular systems have been shown to experience strong and rapid losses, which unfortunately presents an additional challenge for creating dense and ultracold samples. 
For the case of KRb, it is believed that the exo-energetic reaction KRb  $+$ KRb $\longrightarrow$ K$_2 +$ Rb$_2$ drives the decay~\cite{Zuchowski2010rou}. For other molecular systems such as NaK and NaRb, for which the equivalent reactions are endo-energetic,  the lifetime also appears rather short at large densities for reasons that are yet to be fully understood~\cite{quemener2013sou}. The impact of losses could be reduced thanks to an ingenious control of their spatial dependence: Confining molecules in a quasi-two dimensional geometry enables one to take control off the stereodynamics of molecular reactions~\cite{deMiranda2011ctq}.  Producing molecules in three-dimensional optical lattices~\cite{Chotia2012lld,Moses2015coa} or optical tweezers~\cite{Liu2019mao, anderegg2019aot} prevents molecules from inelastically colliding due to their physical separation. Ultracold molecules now constitute a fast-expanding and promising field, especially for quantum simulation~\cite{Moses2017nff,Bohn2017cmp}. 

\subsubsection{Magnetic dipoles}\label{subsubsec:mag_dip}
In contrast to the situation with electric dipoles, elementary particles can have permanent magnetic dipoles even at zero field~\footnote{Searches for a permanent electric dipole moment (EDM) of elementary particles are ongoing~\cite{Chupp2019edm}; in particular, the search for the electron EDM is underway in several AMO systems.  See, e.g., Ref.~\cite{Andreev2018ilo}.}. As a consequence, the effect of magnetic DDIs on quantum gases can be studied under full rotational symmetry at arbitrarily small magnetic fields.  The magnetic dipole moment in atoms is primarily associated with the spin ($\hat {\bs S}$) and orbital ($\hat {\bs L}$) angular momentum of the electrons. The nucleus may also have a magnetic dipole moment, although it is three orders of magnitude smaller than the electron's. Nevertheless, the nuclear spin ($\hat {\bs I}$) couples to the electronic spin within the atom, giving rise to the hyperfine structure  ($\hat {\bs  F}=\hat {\bs  L}+\hat {\bs  S}+\hat {\bs  I}$). Therefore, the sensitivity of a given Zeeman sublevel to magnetic fields  indirectly depends on the nuclear spin. Only the fully stretched atomic state (i.e., maximal $F=L+S+I$ and $|m_F| = F$) reaches the full magnetic moment provided by the electrons. Here, and all along this review, $X$ and $m_X$ are the quantum numbers associated with the norm of the angular momentum operator $\hat {\bs X}$ and its projection along the quantization axis, respectively. Additionally,  we use the dimensionless version of the vectors and operators of angular momenta and spins such that the eigenvalues associated to the norm and projection of $\hat {\bs X}$ are simply $\sqrt{X(X-1)}$ and $m_X$. In the above, $X=\{F,L,S,I\}$. Throughout the remainder of this review, we will usually denote by $\hat{\bs  S}$ the total angular momentum of a magnetic particle. 

It is possible to study dipolar physics with alkali atoms~\footnote{For example, remarkable phenomena have indeed been observed in the case of a lattice interferometer of K~\cite{Fattori2008mdi} and in the context of spinor physics with Rb atoms~\cite{Vengalattore2008sms, Eto2014ood})}. However, the energy scale associated with DDIs is  rather small, typically in the Hz range. Therefore, significant focus has been on so-called highly magnetic atoms, such as chromium (Cr; with a dipole moment of 6 Bohr magnetons, $\mu_B$), erbium (Er; $7 \mu_B$) and dysprosium (Dy; $10 \mu_B$). 
In principle, other highly magnetic atoms can be studied, such as holmium (Ho; $9 \mu_B$), thulium (Tm; $4 \mu_B$) or europium (Eu; $9 \mu_B$)~\cite{Sukachev2010mot,Miao2014mot,Inoue2018mot,Davletov2020mlf}. Moreover, it was demonstrated that one can  use a Feshbach resonance to combine two Er atoms into a loosely bound molecule~\cite{Frisch2015udm}, which may possess up to twice the magnetic moment of the original atoms.  Likewise, a nearly 20$\mu_B$-large magnetic moment is accessible with Dy$_2$ molecules~\cite{Maier2015buf}.  These systems are described in detail in Sec.~\ref{sec:amgnetic_atoms}. 


\subsection{The dipole-dipole interaction}

Generally speaking, the DDI between two 
dipoles, $1$ and $2$, separated by $\vecr$ yields the following potential:
\begin{equation}
\label{vdd}
\Vdd(\vecr)=\frac{\gamma_{\rm d}}{4 \pi r^3} \left[\bs {d_1}\cdot\bs {d_2} -3 \frac{\left(\bs {d_1}\cdot\vecr\right) \left(\bs {d_2}\cdot\vecr\right)}{r^2} \right],
\end{equation}
where $\bs {d_i}$ is the dipole moment of particles $i=[1,2]$ and $\gamma_{\rm d}$ is the dipolar coupling constant and depends on the electric or magnetic nature of the dipoles.   This expression is valid at long distances, where electron orbitals do not overlap.
\begin{itemize}

\item \textbf{Magnetic dipoles:}\label{magdipoles}
Classically, the dipolar interaction between two magnetic particles corresponds to the interaction of the spin $\bs {S_1}$ of the particle 1 with the magnetic field created by the spin $\bs {S_2}$ of the particle 2, and vice-versa. Here, $\hat{\bs  S}$ is a generic angular momentum which in general is given by the total angular momentum $\hat {\bs  F}$, see~\ref{subsubsec:mag_dip}. For a magnetic particle of spin $\hat{\bs {S_i}}$, the dipole moment is given by $\hat {\bs {d_i}}=g_S \muB \hat {\bs {S_i}}$, where $g_S$ is the 
$g$-factor of the spin ${\bs S}$ and the dipolar coupling constant
$\gamma_{\rm d}=\mu_0$ is the vacuum magnetic permeability.
We denote $d^2=\frac{\mu_0 (g_S \muB)^2}{4 \pi}$  and $C_{\rm dd} = S^2 d^2$ so that $\Vdd(\vecr) \propto \frac{S^2 d^2}{r^3} = \frac{C_{\rm dd}}{r^3}$, see, e.g., Sec.~\ref{subsec:processes}. These constants set the DDI strength. 

\item\textbf{Electric dipoles:}
For electric dipoles, $\gamma_{\rm d}=1/\epsilon_0$, with $\epsilon_0$ the vacuum electric permittivity.

\end{itemize}

We now compare the magnetic and electric DDI strengths. Electric dipoles relate to charge displacement within a particle. Typical electric dipoles of molecules are of magnitude $e a_0$, given by the displacement of an elementary electric charge $e$ over the typical size of an atom, set by the Bohr radius $a_0$. In Rydberg atoms, the characteristic displacement of the electric charge is set by the Rydberg orbital radius, which scales as $n^2$, the square of the Rydberg principal quantum number $n$; typically $n$ is of order a few tens. The dipole moment of a Rydberg atom is thus typically $n^2$ times than that of an atom in the ground state. The atomic magnetic dipole scale is given by $\muB$. The typical ratio between the DDI strength of magnetic atoms and of polar molecules is therefore $\frac{\mu_0 \mu_B^2}{(e a_0)^2/\epsilon_0} = \alpha^2 /4$,
where $\alpha \approx 1/137 $ is the fine structure constant. The ratio is further reduced by a factor $n^4$ when comparing to Rydberg atoms. Thus, the typical magnetic DDI strength is orders of magnitude smaller than the typical electric DDI.

\subsection{Main characteristics of dipolar interactions}

In the absence of DDIs, ground state atoms interact through van der Waals interactions. These interactions are short ranged, $1/r^6$, and are typically isotropic because the electronic cloud of most atoms is spherically symmetric in the ground state~\footnote{Interestingly, however, lanthanide atoms exhibit anisotropic van der Waals interactions; see Sec.~\ref{sec:amgnetic_atoms}}. In contrast, the DDI introduced in Eq.~\eqref{vdd} has a long-range $1/r^3$ character. It is also anisotropic and can be either attractive or repulsive depending on the relative orientation of the dipoles;  in particular, its elastic part varies as $1-3\cos^2{\theta}$, where $\theta$ is the angle between the relative position of the particles and their direction of polarisation. 

\subsubsection{Definition of `long-range'}\label{longrange}

Whether an interaction, in particular the DDI, is long range depends on the exact system under study, its dimensionality, and  on the exact physical question addressed. We discuss below a number of physical questions that lead to slightly different definitions of the long-range character of the interaction at hand, with particular focus on power law potentials $U(r) \propto 1/r^n$. 

\begin{itemize}

\item \textbf{Collisional point of view (in 3D)}.  
Physically, for short-range interactions,  particles need to approach at small distances to interact. By decomposing the relative motion of the particles into the relative orbital angular momentum eigenstates (so-called partial waves), denoted by the quantum numbers $(l,m)$ for the momentum's norm and projection eigenvalues, one finds that at low collision energy, the centrifugal barrier  $\frac{2l(l+1) \hbar^2}{m r^2}$ prevents particles from  approaching  in higher partial waves $l>0$. That is,  the contributions from high partial waves vanish. For a $1/r^n$ interacting potential, the scattering phase shift  $\delta_l(k)$ at low collision momentum $k$ scales as $k^{2l+1}$ if $l<(n - 3)/2$ and as $k^{n-2}$ otherwise~\cite{Dalibard1999cdo,Landau1977book}. Therefore, for $n\geq 4$, the interaction is purely $s$-wave at low energy and short ranged.  In contrast, for $n=3$, $\delta_l(k) \propto k$ for all partial waves. Therefore, all partial waves contribute to the scattering process even at low collision energy. The interaction is then long range and can be felt beyond the centrifugal barrier. The long-range character of the DDI is spectacularly manifest  in the fact that polarised fermionic dipolar gases  thermalize despite the absence of $s$-wave interactions (due to the Pauli exclusion principle). This is in contrast to nondipolar polarised Fermi gases; see Sec.~\ref{sec:scatt}.

\emph{Note}: A thorough treatment of the above should account for the fact that the DDI is not a pure central potential $U(r) \propto 1/r^3$ due to its anisotropic character. This is of particular consequence for inelastic dipolar collisions, which necessarily involve the anisotropic character of the interaction---see Sec.~\ref{subsec:anisotropy}---and are actually \textit{short-range} processes at large magnetic field despite the same $1/r^3$ scaling as their elastic counterparts.  See Sec.~\ref{range_inel}.

We also remark that the scattering picture can be modified in the presence of strong confinement, in particular in reduced dimensions; see, e.g., Sec.~\ref{Sec1D}. 

\item  \textbf{Thermodynamic point of view}. Short-range interactions lead to an energy that is thermodynamically extensive. This is true when $\int^{\infty}_0 U(r) d^D r$ converges, which only happens when $n>D$, where $D$ is the spatial dimension. Thus, from this point of view, $1/r^3$ interactions are long range in 3D, but short range in 2D and 1D. Long-range-interacting systems possess peculiar thermodynamic properties, such as the non-equivalence of thermodynamic ensembles, the possibility for negative specific heat, and the spontaneous formation of structures. These arise because the hypothesis that the energy is additive in the thermodynamic limit---i.e., given the energy of two subsystems $A$ and $B$, $E(A \cup B) = E(A)+E(B)$---breaks down when the interaction between the subsystems cannot be neglected~\cite{Arimondo1977edo}.  The status of DDIs in 3D is therefore marginal since  while there are distant couplings between sub-systems $A$ and $B$, the integral $\int^{\infty} U(\vecr) d^3 \vecr$  \textit{does} converge due to the peculiar $d-$wave shape of DDIs.  

\item \textbf{Many-body physics perspective.} In the context of many-body physics, DDIs may lead to qualitatively new behaviour, even when $D<3$.    For example, the Mermin-Wagner theorem precludes the possibility of spontaneous breaking of a continuous symmetry and of long-range order in (homogeneous) low-D systems. However, in 2D, this applies only for short-range interacting systems with $n > 4$~\cite{hadzibabic2009tdb,Defenu2021lri}. Therefore, in this context, DDIs can be seen as long-ranged even in 2D.  Indeed, it has been predicted that ferromagnetic ordering should be stable in 2D for DDIs~\cite{Peter2012abo}.  The meaning of long range in 1D for the DDI will be addressed in Sec.~\ref{Sec1D}.
\item \textbf{Mathematical physics perspective.} Let us for completeness also briefly mention the mathematical physics point of view of the meaning of "long-range interaction." For an interaction potential  ${\sim}1/r^n$, the scattering wave is only described by an asymptotic outgoing spherical wave weighted by an angle-dependent scattering amplitude for $n>1$. For $n\leq1$, e.g., the Coulomb potential, there are logarithmic corrections to the general form of the outgoing spherical wave, which defines another border between long and short-range potentials.
\end{itemize}

\subsubsection{Consequences of anisotropy}
\label{subsec:anisotropy}
The anisotropic character of the DDI greatly impacts the properties of dipolar gases. It introduces profound differences from the point of view of two-body physics and scattering properties---see Sec.~\ref{sec:scatt}---and also on the collective many-body properties of quantum degenerate dipolar gases. In particular, the stability diagram of dipolar condensates is affected by an interplay between the anisotropy of the trap and the anisotropy of the interactions; this will be described in Secs.~\ref{Sec:RepulsiveGases} and~\ref{Sec:DCQD}. We now briefly describe a few basic consequences of this anisotropy.  

The DDI is attractive in one direction and repulsive in the other two directions.  The shape of the interaction  follows a $d$-wave form  mathematically described by the $j$ components of the spherical harmonics $Y_2^j$, in particular with $j=0$ in a fully polarized situation. This means that when integrated over all space in 3D, the DDI between polarised dipoles converges to zero for a 3D homogeneous gas. Consequently, the mean-field physics of dipolar gases is dominated by border and boundary effects: the average interaction between particles will strongly depend on the shape of the cloud. In particular, an elongated trap along the axis of the dipoles will favour the collapse of the gas due to the predominately attractive interaction. The stability of dipolar condensates as a function of geometry is described in Sec.~\ref{Sec:RepulsiveGases}.   

Another consequence of anisotropy is the existence of a special angle between the dipoles and  the interatomic axis, $\theta_m =\arccos \sqrt{1/3} \approx 54.74^\circ$, at which DDIs vanish.  More generally, controlling this angle can be used to tune the strength of DDIs, especially when performing experiments in reduced dimensions, as described in Sec.~\ref{Sec1D}. In a scheme inspired from NMR techniques, it has been suggested~\cite{Giovanazzi2002ttd} and demonstrated~\cite{Tang2018ttd} that by using time-varying magnetic fields, it is possible to time-average the DDI to reduce its amplitude or reverse its sign.  

Finally, the anisotropy of the interaction has fundamental consequences from the point of view of collisions. The interaction potential is not central, and therefore the orbital angular momentum does not need to be conserved during a collision. The expansion in spherical harmonics yields the selection rule for angular momentum transitions $\Delta l = (0, \pm 2)$.  Moreover, partial waves of differing $l$ that contribute to the scattering become coupled. Finally, the angular momentum of the atoms' internal state may also change during the collisions, opening the possibility for inelastic processes. 

\subsubsection{Physical processes associated with dipolar interactions}\label{subsec:processes}

In view of describing the physical processes at play when two dipolar particles collide, it is useful to rewrite the dipolar potential between atoms 1 and 2 in terms of quantum operators \cite{Hensler2003dri}: 
\begin{eqnarray}\label{ddispinform}
\Vdd(\vecr)= \frac{d^2}{r^3} \bigg[(S_1^z.S_2^z + \frac{1}{2} \left( S_1^+.S_2^- + S_1^-.S_2^+\right) \nonumber \\
-\frac{3}{4} \left( 2 z S_1^z + r^- S_1^+ + r^+ S_1^- \right) \times \nonumber \\
\left( 2 z S_2^z + r^- S_2^+ + r^+ S_2^- \right) \bigg],
\end{eqnarray}
where $(x,y,z)$ is the normalised unit vector connecting both atoms, $r^+ = (x + iy)$, $r^- =(x - iy)$, $S^+ = (S^x + iS^y)$, and $S^- = (S^x - iS^y)$.  Note that both here and throughout this review, we use the dimensionless version of the vectors and operators of angular momenta and spins.

We describe three physical processes that arise from this expression:
\renewcommand{\labelenumi}{\roman{enumi}.}
\begin{enumerate} 
\item \textbf{Elastic dipole-dipole interactions}, where the spin of each atom is conserved in time: 
\begin{equation}
\Vdd^{\rm el}(\vecr) = \frac{d^2}{r^3} S_1^z.S_2^z \left(1-3 z^2 \right). 
\label{eq:vddz}
\end{equation}
The experimental manifestation of this anisotropic Ising term on quantum degenerate dipolar gases has been extensively studied. It is the main process at play for most of the results presented in this review article: see Sec.~\ref{sec:scatt} on scattering physics, Secs.~\ref{Sec:RepulsiveGases} and \ref{Sec:DCQD} on the collective properties of dipolar gases, the stability diagram, the instability dynamics and the 
stabilisation of so-called dipolar droplets, Sec.~\ref{Sec1D} on integrability breaking in 1D gases, and Sec.~\ref{subsec:spinless_lattice} on the extended Bose-Hubbard model. 

\item \textbf{Exchange interactions}, where two atoms exchange one unit of spin (Zeeman) excitation,  while the total magnetisation and energy is conserved (in the absence of quadratic Zeeman effects):
\begin{equation}
\Vdd^{\rm ex}(\vecr) = -\frac{1}{4} \frac{d^2}{r^3} \left(S_1^+.S_2^- + S_1^-.S_2^+\right) \left(1-3 z^2 \right).
\label{eq:vddex}
\end{equation}
This exchange term can drive spin dynamics at constant magnetisation as described in Sec.~\ref{subsec:spindyn_bulk} and dictates the physics of spinor dipolar gases in deep lattices, which is the topic of Sec.~\ref{subsec:spinlattice}. We note that the elastic and the exchange terms result in an anisotropic Heisenberg-like term (the so-called XXZ model). 

\item \textbf{Relaxation terms} describe the modification of the longitudinal magnetisation of the pair of atoms during the collision. There are two possible processes:
\begin{eqnarray}\label{relaxationterms}
\Vdd^{{\rm rel}_1}(\vecr) &=& -\frac{3}{4} \frac{d^2}{r^3} (r^+)^2 S_1^-.S_2^-, \nonumber \\
\Vdd^{{\rm rel}_2}(\vecr) &=& -\frac{3}{2} \frac{d^2}{r^3} z r^+ (S_1^z.S_2^-+S_2^z.S_1^-), 
\end{eqnarray}
plus the conjugate processes.  Spin momentum and  angular orbital momentum exchange while the magnetic energy is transferred into kinetic energy.  The  second process  describes single spin flips, while the first describes double spin flips (i.e., both atoms flipping their spin). These terms underlie  most of the results presented in Secs.~\ref{Drelaxationsec} and~\ref{spinorsection}, the latter describing spinor physics with free magnetisation.
\end{enumerate}

\subsubsection{Two-body dipolar scattering}
\label{subsec:2Bdintro}
The cross section classically describes the area, transverse to the relative motion, within which two particles must meet to scatter. In other words, the scattering cross section is related to the typical distance at which the wavefunction of the relative motion is distorted by the interaction. Employing the Heisenberg uncertainty principle, this distance $\rd$ for the DDI is typically set by an interplay between the DDI strength $S^2 d^2/\rd^3$ and the energy cost to bend the wave function by an amount $\rd$. Setting $S^2 d^2/\rd^3 = \hbar ^2 /m \rd^2 $ defines the dipolar length~\footnote{The factor of 3 defining $\add$ comes from a normalisation imposed for the convenient  discussion of the many-body physics of dipolar gases; see Eq.~\eqref{eq:edd} below.}:
\begin{equation}\label{dipolarlength}
\add \equiv \frac{\rd}{3}= \frac{S^2 d^2 m}{3 \hbar^2}= \frac{C_{\rm dd} m}{3 \hbar^2},
\end{equation}
where $m$ is the atomic mass.  The order of magnitude of the scattering cross section is
\begin{equation}
\sigma \approx \rd^2 = \frac{S^4 d^4 m^2}{\hbar^4}.
\label{estimate}
\end{equation}

Likewise, one defines the range of the van der Waals potential $V_{\rm VdW}=-C_6/r^6$ as $\rvdw=(m C_6/\hbar^2)^{1/4}$, which sets the typical scattering cross section due to short-range interactions. The lengths $r_{VdW}$ and $\add$ are typically in the nm range; i.e., much larger than both the Bohr radius $a_0$ and the typical impact parameter at room temperature. 

In Sec.~\ref{sec:scatt}, we will describe the scattering theory for both dipolar and van der Waals interactions.  The DDI cross sections  are presented based on a first-order Born approximation, and the role of exchange statistics in these expressions is discussed. 

In contrast to the van der Waals case, the dipolar cross section depends on only the mass of the atoms and their dipole moment. Because it is independent of the details of the molecular potentials, dipolar scattering assumes a universal character. One remarkable aspect is that the dipolar cross section follows the same universal scaling of Eq.~\eqref{estimate} (up to numerical factors) regardless of particle exchange statistics. In particular, identical fermions have a finite dipolar cross section even at vanishingly small collision energy. This is a direct consequence of the long-range character of DDI, as discussed above. This topic will be discussed in Sec.~\ref{elasticdipolarscattering}. Inelastic dipolar scattering  will be discussed in Sec.~\ref{Drelaxationsec}.

Finally, by integrating over all directions of the collision, the scattering theory outlined above obscures one of the central features of dipolar scattering, the anisotropic dependence on the colliding angle, which has been observed in both Er and Dy. This is the topic of Sec.~\ref{subsec:anis_scatt}.

\subsubsection{Momentum-space DDI expression}

The form of the  Fourier transform of the interaction potential often provides insight regarding the  physics of interacting particles. For example, it facilitates the  description of  two-body scattering physics because scattering theory tends to formulate the wavefunction in terms of momentum states. It also proves convenient in discussing elementary excitations of a quantum gas, which, in a uniform system, are characterised by a well-defined momentum. 

The Fourier transform of the elastic part of the DDI, Eq.~\eqref{eq:vddz}, is
\begin{eqnarray}
\widetilde{\Vdd}(\vk) = \int e^{i \vk\vecr} \Vdd^{\rm el}(\vecr)d^3r 
&=& \frac{C_{\rm dd}}{3} \left[3 \cos(\theta_k)^2 -1 \right],
\label{vddtf}
\end{eqnarray}
where $\theta_k$ is the angle between $\vk$ and the polarisation of the dipoles. This form of the interaction is remarkable when compared to the contact interaction. While neither the Fourier transforms of the  contact interaction nor the DDI   depend on $k$, the Fourier transform of the DDI retains a nontrivial angular dependence. Such a feature can give rise to an anisotropic dispersion relation of excitations, as described in Sec.~\ref{Sec:RepulsiveGases}. 

That $\widetilde{\Vdd}(\vk)$ does not depend on the modulus of $\vk$ can be understood from dimensional analysis: $\int d^D r \exp(i k r) \frac{1}{r^3}$ is independent of $k$ for $D=3$. On the other hand, in a 2D system ($D=2$), we expect a different behaviour with a linear dependence on $k$ for small $k$. Quite generally, the fact that the Fourier transform of $\Vdd$ has a $\vk$ dependence  is an important feature for $D\neq 3$ systems, in that the DDI introduces a tendency in these systems to develop structured excitations (rotons, solitons) and exotic phases (supersolid, crystals of quantum droplets, etc.). These excitations are described in Sec.~\ref{Sec:RepulsiveGases}. The quantum droplets occur when the gas spontaneously forms stable spatial arrangements of liquid-like droplets in dipolar Bose--Einstein condensates (dBECs) driven to the instability point of mechanical collapse. This is related to the emergence of new phases stabilised by beyond-mean-field effects and is described in Sec.~\ref{Sec:DCQD}; see also Sec.~\ref{subsec:mf}.

\subsection{Many-body dipolar physics}

The two-body processes outlined in the previous paragraphs are the elementary phenomena behind the very rich phenomenology associated with many-body physics in dipolar quantum gases.  We introduce the various physical effects that will be further discussed in the Secs.~\ref{Sec:RepulsiveGases}--\ref{sec:lattice}.

\subsubsection{Dipolar Bose quantum gases}
\label{subsec:mf}

\myparagraph{{Spin-polarised dipolar Bose quantum gases in the mean-field regime}}

Many-body physics is often intractable. However, most of the first experiments on ultracold gases of magnetic atoms have been performed with weakly interacting Bose--Einstein condensates (BECs), and the associated theory is tractable because interatomic correlations are small and so mean-field theories apply.

Due to Bose stimulation, in which the population of bosonic atoms at low energy favours the occupation of a unique single-particle orbital, it is natural to propose a variational ansatz where the many-body wavefunction is assumed to be $
    \phi(r_1,...r_N;t)= \prod_{i=1,...,N} \psi(r_i,t)$.
The single-particle wave-function $\psi(r,t)$ is taken as a variational parameter to minimise the system's total energy. This approach leads to the well-known Gross--Pitaevskii equation (GPE), which has been found to describe most of the properties of dilute BECs~\cite{Bogoliubov1947ott,Gross1961soa,Pitaevskii1961vli,
Stringari2016bec}.

When all atoms are polarised (and the polarisation axis set to $z$), the GPE of a dBEC is
\bea
 i \hbar\frac{\partial}{\partial t} \psi =   \left[\frac{- \hbar^2 \nabla ^2}{2 m}+V_{\rm tr}(r) +g \left| \psi \right|^2+\Phi_{\rm dd}(r,t)\right] \psi, 
\label{GP}
\eea
where $V_{\rm tr}(r)$ is the trap potential, $g=\frac{4 \pi \hbar^2}{m} \as$ is the coupling constant describing contact interactions of $s$-wave scattering length $\as$; see also Secs.~\ref{subsec:fesbachsec} and \ref{sec:scatt}~\cite{Huang1957qmm,Pitaevskii2016bec}.  The mean field associated with the DDI is $\Phi_{\rm dd}(r,t)$~\cite{yi2000tac,Yi2001tco}:
\begin{eqnarray}
\label{meanfield}
\Phi_{\rm dd}(r,t) &=&  \int dr'  \left| \psi(\bs r',t) \right|^2 U_{\rm dd}(\bs r- \bs r'),\\
\label{Udd}
U_{\rm dd}(\bs r) &=&C_{\rm dd} \frac{1-3 \cos^2(\theta)}{\left| r \right|^3},
\end{eqnarray}
where $\theta$ is the angle between $\bs r$ and the polarisation axis $z$. Only the elastic part of the DDI, Eq.~\eqref{eq:vddz}, contributes due to polarisation in a fully stretched Zeeman substate. This term is non-linear and nonlocal. To quantify the strength of the DDI with respect to the contact interactions within a BEC, it is useful to introduce the dimensionless parameter
\begin{equation}
\edd=\frac{C_{\rm dd} m}{3 \hbar^2 \as}=\frac{\add}{\as}.
\label{eq:edd}
\end{equation}

We note that writing the GPE of a dipolar BEC in the form of Eq.~\eqref{GP} is in fact non-trivial. Its validity, which relies on describing the total inter-particle interactions via an effective pseudo-potential that is the simple sum of the contact pseudo-potential and the DDI potential, has been long debated. The efforts to prove the validity of this treatment as well as identify its limitation will be reviewed in Sec.~\ref{subsec:model_bose}.  
The applicability of the nonlocal GPE Eq.\,\eqref{GP} for the case of weakly interacting trapped BECs of magnetic atoms in the stable regime, e.g.,\,$\edd<1$ in a 3D isotropic trap, has been supported by numerous theory and experimental works. In this regime, the anisotropic and nonlocal character of Eq.~\eqref{meanfield} substantially modifies the static and dynamical properties of the BEC compared to contact-only BECs. This will be extensively discussed in Sec.~\ref{Sec:RepulsiveGases}.

\myparagraph{{Spinor dipolar Bose quantum gases}}

In the presence of spin degrees of freedom, the exact form of the GPE depends on the spin of the atoms and can be found in Ref.~\cite{Kawaguchi2012sbe}. Taking, for example, the case of a spin-1 atom -- i.e. the simplest example pertaining to bosonic physics -- the GPE takes the form 
\begin{eqnarray}
i \hbar \frac{d \psi_m}{dt}=&& \left[\frac{- \hbar^2 \nabla ^2}{2 m}+V_{\rm tr}(r) -pm+q m^2 \right] \psi_m \nonumber \\ 
&&+c_0 n \psi_m + c_1 \sum_{m'=-1}^1 {\bs S}.{\bs s}_{m,m'} \psi_{m'} \nonumber \\ 
&&+C_{\rm dd} \sum_{m'=-1}^1  \textbf{b}. {\bs s}_{m,m'} \psi_{m'},
\label{mfspinor}
\end{eqnarray}
where $\psi_m$ denotes the macroscopic wave function associated with the spin state of projection quantum number $m$. The terms in $p$ and $q$  describe the linear and quadratic Zeeman energy shifts of the spin states, respectively. The trap is assumed to be spin-independent. The terms proportional to $c_0$ and $c_1$ are spin-independent and spin-dependent contact interactions, respectively. The spin density vector is ${\bs S}$, and ${\bs s}=\{s^x,s^y,s^z\}$ are the spin matrices.  The DDIs are described by the term proportional to $C_{\rm dd}$, where the effective dipole field $\textbf{b}$ is defined by 
\begin{equation}
b_{\nu} = \int dr' \sum_{\nu'} Q_{\nu,\nu'} (\vecr-\vecr') S^{\nu'} (r'),
\label{meanfieldspinor}
\end{equation}
with
\begin{equation}
Q_{\nu,\nu'}(\vecr) = \frac{\delta_{\nu,\nu'}-3 r_{\nu} r_{\nu'}}{r^3},
\end{equation}
and $\nu,\nu'=\{x,y,z\}$.

Equation~\eqref{mfspinor} is central to the description of spinor dipolar physics, which is the subject of Sec.~\ref{spinorsection}. In general, the DDIs cannot be neglected when $C_{\rm dd}$ is comparable to either $c_0$ or $c_1$.  If $c_1\ll c_0$, then the DDIs can be significant even when $\edd\ll1$~\cite{StamperKurn2013sbg}.  Furthermore, magnetisation-changing processes effected by the term in Eq.~\eqref{meanfieldspinor} have no analogue in systems with only spherically-symmetric contact interactions.  Such processes start to play a role when $C_{\rm dd} n \simeq \{p,q\}$. 

Finally, it is useful to stress that in the mean-field regime, DDIs described by Eqs.~\eqref{meanfield} and~\eqref{meanfieldspinor} correspond to the average magnetic field produced by all atoms within the condensate. This is  due to the fact that  correlations between atoms have been neglected, which is the essence of the mean-field approximation. In the case of spinor gases, the effect of quantum fluctuations and correlations can, however, be significant, even in the weakly interacting regime.  This is due to entanglement or squeezing naturally arising in the spin degrees of freedom.   The consequences of DDIs on the properties and behaviours of gases with spin degrees of freedom will be detailed in Sec.~\ref{spinorsection}.

\myparagraph{{Elementary excitations of a spin-polarised dipolar Bose quantum gas}}
\label{excitations_intro}

The elementary excitations of a BEC are usually well described within a Bogoliubov treatment, which matches a linearisation of the GPE around the ground-state wavefunction
~\cite{Bogoliubov1947ott,Stringari2016bec}. The theory yields a simple dispersion relation for a uniform 3D gas ($V_{\rm tr} =0$): 
\begin{equation}
\epsilon(\vk) =\sqrt{\frac{\hbar^2k^2}{2m}\left(\frac{\hbar^2k^2}{2m}+2n\widetilde{V_{\rm int}}(\vk)\right)}.
\label{Bog}
\end{equation}
This describes the energy of the elementary excitation of momentum $\vk$ in a BEC of density $n$. Here, $\widetilde{V_{\rm int}}(\vk)$ is the Fourier transform of the total interaction potential. In contact-interacting gases, $\widetilde{V_{\rm int}}(\vk)=g$. For a dBEC, $\widetilde{V_{\rm int}}(\vk)=g+\widetilde{\Vdd}(\vk)$.  From the form of $\widetilde{\Vdd}(\vk)$ (see Eq.~\ref{vddtf}), one can infer that the energy of a collective excitation of a dipolar fluid depends not only on the magnitude of its wavevector but also on its propagation direction. The dispersion relation of elementary excitations of an isotropic, homogeneous dipolar fluid is anisotropic. For a dBEC, the dispersion retains the initial linear phonon character, but with an anisotropic speed of sound $c$; the dispersion relation remains monotonic.  
 
This picture is modified in a constrained geometry, where one externally lifts the spatial symmetry along at least one dimension by, e.g., imposing anisotropic trapping confinement~\cite{Ronen2006bmo}. The trap along the constrained dimension yields a new length scale. Because of the anisotropy and long-range character of the DDI, this length scale also becomes relevant to the description of the physics of the otherwise translationally invariant directions. In particular, it affects their elementary excitations.
That is, a BEC that is more tightly trapped along the dipoles' direction than transversely possesses  a favoured wavelength in its  dispersion relation at which the energy of the transverse excitations reaches a minimum~\cite{Goral2002gse,ODell2003rig,Santos2003rms,Giovanazzi2004iat,Ronen2006bmo}. This is referred to as a roton minimum in analogy to a similar minimum found in the dispersion relation of liquid helium~\cite{Landau1941tto,Landau1947ott,Henshaw1961mam}. These properties of dBECs are the topic of Sec.~\ref{subsec:excitationBEC}.

\myparagraph{{Mean-field instability and collapse}}

At the mean-field level, the mechanical stability of fluids may be understood from analysing the dispersion relation of their excitations~\cite{Stringari2016bec}. An instability occurs when the energy of an elementary excitation becomes zero, since then there is no cost for populating such a mode.  In the 3D homogeneous case, 
the lowest energy modes are the long-wavelength phonons. Furthermore, due to the DDI anisotropy, phonons propagating in the plane perpendicular to the dipoles cost the least amount of energy. The speed of sound $c$ reaches $0$ at $\edd=1$ in this direction, which identifies the threshold for mechanical collapse of a 3D homogeneous dipolar BEC. This is remarkable, because the instability, arising from the attractive part of the DDI, occurs in a gas with a finite (and positive) value of the short-range contact interaction.  Consequently, interactions are still present, even if cancelling at the mean field, and their beyond-mean-field contribution plays a crucial role in such a system; see Secs.~\ref{subsec:Stability} and~\ref{subsec:collapse}.  This collapse, corresponding to a phonon instability, is  called ``global collapse,''~\cite{Stringari2016bec,dodd1996roa,sackett1998gac,Roberts2001cco,Ticknor2008cto,Goral2002gse,Yi2001tco}. Generally speaking, in an ultracold quantum Bose gas, crossing the instability threshold leads to an implosion of the gas under the concomitant effects of two-body attraction and three-body inelastic collisions; these so-called Bose-Novas are well described by mean-field coherent dynamics~\cite{dodd1996roa,sackett1998gac,Donley2001doc,Gerton2000doo,Kagan1998ceb,Ueda1998mqt,Houbiers1996sob}. In a dipolar quantum Bose gas, at the mean-field level, the anisotropy of the DDI is expected to impact the geometry of the collapse and its dynamics.

Furthermore, anisotropic external trapping modifies the dispersion relation, yielding additional modifications of the stability criterion as well as of the subsequent collapse. In particular, the long-range character of the DDI brings the length scales of the trap into play. The instability may be induced by the softening of excitation modes of a nonphononic nature (e.g., modes of small wavelengths or with angular structures). In these cases, the instability threshold is expected to be shifted compared to the $\edd=1$ uniform value. 
In the collapse dynamics, structures at the corresponding length scale are then expected to be preferentially formed. The resultant ``local collapse'' corresponds to a ``modulational instability''~\cite{Ronen2007raa,Ronen2006bmo,Bohn2009hda,Parker2009sfd}. 
The collapse dynamics may reveal the properties of the underlying mode driving the instability.   
The different regimes of global and local instability, and the related collapse or collapsing dynamics of dipolar Cr, Er, and Dy dBECs, are described in Secs.~\ref{subsec:Stability} and~\ref{subsec:collapse}.

\myparagraph{{Quantum states stabilized by fluctuations: droplets and supersolids}}

Even if often well described by mean-field theory, dBECs are not classical fluids. 
As quantum fluids, they are liable to quantum fluctuations. Even at zero temperature, the vacuum population of its elementary excitations, yields interaction-induced modifications of the fluid's energy and ground state. 
The Bogoliubov treatment allows one to perturbatively take into account these effects~\cite{Bogoliubov1947ott,Stringari2016bec,Lee1957eae,Lee1957mbp}. The energy corrections are, in principle, negligible for weakly interacting gases; i.e., when $n a^3 \ll1$ and $n \add^3 \ll1$).   
However, the mean-field instability threshold described above occurs when the mean-field interactions are small, changing from repulsive to attractive on average. Importantly, while the overall interaction becomes negligible, the atoms still interact in a non-negligible way thanks to the competition of contact and dipolar interactions. In sufficiently dipolar gases, instead of a collapse, a remarkable phenomenon occurs at the instability threshold. Here, beyond mean-field effects provide sufficient repulsive interaction energy to stabilise the system.  This leads to exotic phases on the attractive side of the mean-field instability threshold, including liquid-like droplet states (a quantum state that is stabilised by the opposite effects of mean-field and beyond-mean-field interactions, and even in absence of trapping potential), droplet assemblies (a state formed of several independent quantum droplets, self-organised in a crystalline structure), and supersolids (a self-organised crystalline states with global superfluid properties. In a simplified picture, it can be viewed as a ground state consisting of an overlapping assembly of droplets where the droplets are allowed to maintain a common phase via particle exchange). These recently discovered states are discussed in Sec.~\ref{subsec:droplets} and \ref{subsec:supersolidity}.

\subsubsection{Dipolar Fermi quantum gases}
\label{subsec:fermi_intro}

Fermionic dipolar atoms are also of great interest for exploring new physics. A remarkable property of dipolar Fermi gases lies in the fact that polarised samples remain interacting even in the ultracold regime. This is unlike nondipolar Fermi gases; see Sec.~\ref{longrange}. Yet, the mean-field theory developed above does not appropriately describe fermionic ensembles because the ansatz used to write the many-body wavefunction is incompatible with the Pauli exclusion principle: it must be antisymmetrized due to fermionic exchange statistics. Therefore, it is generally not possible to neglect correlations in a Fermi gas at low temperature, even for small interactions. This makes a theoretical treatment of fermionic gases challenging. The simplest treatment of  mean-field theory that includes the antisymmetrization of the wavefunction replaces the product ansatz used in Sec.~\ref{subsec:mf} for $\psi(\vecr_1,\vecr_2,\vecr_3,...,\vecr_N)$ by a Slater determinant. This procedure is known to be sufficient for a pure state without interactions. With interactions, it may still be sufficient, but with the single-particle wave-functions modified compared to the noninteracting case. This approach constitutes the Hartree-Fock theory~\cite{Ashcroft1976ssp, Goral2001sto}.

The mean DDI energy for an ensemble of $N$ atoms in a state $\psi(\vecr_1,\vecr_2,\vecr_3,...,\vecr_N)$ is generally written as 
\begin{eqnarray}
E_{\rm dd} &= &\int dr_1...dr_N \psi^*(\vecr_1,\vecr_2,\vecr_3,...,\vecr_N) \\ \nonumber
&&\sum_{i,j} \Vdd(\vecr_i-\vecr_j) \psi(\vecr_1,\vecr_2,\vecr_3,...,\vecr_N) \\ \nonumber
&= &\frac{N(N-1)}{2}  \int dr dr' \int dr_3...dr_N \\ \nonumber
&&\psi^*(\vecr,\vecr',\vecr_3,...,\vecr_N) \Vdd(\vecr-\vecr') \psi(\vecr,\vecr',\vecr_3,...,\vecr_N). \\ \nonumber
\end{eqnarray}
Because of 
antisymmetrization, the integral $N(N-1)\times \int dr_3...dr_N \psi^*(\vecr,\vecr',\vecr_3,...,\vecr_N) \times \psi(\vecr,\vecr',r_3,...,r_N)$ does not reduce to $n(\vecr) n(\vecr')$ as in the bosonic case. Though by using the Slater determinant ansatz, it can be simplified to $ \obdm(\vecr,\vecr) \obdm(\vecr',\vecr') - \obdm(\vecr,\vecr') \obdm(\vecr',\vecr)$,
where 
\begin{eqnarray}
\obdm(\vecr,\vecr')= N(N-1)\times 
\int dr_2dr_3...dr_N \\ \nonumber
\psi^*(\vecr,\vecr_2,\vecr_3,...,\vecr_N) \times \psi(\vecr',\vecr_2,\vecr_3,...,\vecr_N)
\end{eqnarray}
is the one-body density matrix. Therefore, the DDI mean-field energy for the fermionic gas consists of two parts:
the usual term, also called the direct or Hartree term \begin{eqnarray}
E_{dd}^{\rm dir} = \frac{1}{2} \int dr dr' \Vdd (\vecr-\vecr') n(\vecr) n(\vecr'), \label{Edir}
\end{eqnarray}
and an unusual term, called the Fock or exchange term, resulting from the requirement for an antisymmetric wavefunction upon particle exchange: 
\begin{eqnarray}
E_{\rm dd}^{\rm exc} =- \frac{1}{2} \int dr dr' \Vdd (\vecr-\vecr') \obdm (\vecr',\vecr) \obdm (\vecr,\vecr').\label{Eexc}
\end{eqnarray}
This exchange term is zero in the case of a BEC. 
Based on Eqs.~(\ref{Edir}-\ref{Eexc}), and by performing the variational minimisation of the total energy with respect to $\psi$, one can derive semiclassical (Hartree-Fock) equations for the degenerate Fermi gas (DFG). 
To describe a trapped Fermi gas, one can use a local-density approximation, which assumes that the atoms feel a local DDI~\cite{Goral2001sto,Zhang2009fso, Zhang2010tpo,Baillie2010tac, Baillie2012mae,Wachtler2013lle}. 
This particular exchange interaction term, arising from the interplay of fermionic statistics and the nonlocal DDI, has several physical consequences---these will be the topic of Sec.~\ref{subsec:fermi}. 

\subsubsection{Dipolar gases in confined geometries}

As previously discussed in Sec.\,\ref{subsec:mf}, beyond-mean-field effects are typically weak for BECs in the weakly interacting regime (far from any instability). This is because the interaction energy is too small to create short-wavelength correlations in the gas. One way to reach strong correlations is to load the atoms into tight anisotropic traps or standing waves of light (so called optical lattice). Confined trapping geometries effectively reduce the atoms' kinetic energy  by restricting motion.  By doing so, they allow the interaction and kinetic energies to play competing roles in the determination of how the system organises~\cite{Mermin1966aof,Hohenberg1967eol,Giamarchi2003qpi,Bloch2008mbp,hadzibabic2009tdb,Bloch2008qca,Bloch2012qsw,Lewenstein2007uag}. In this review, we will discuss the experimental progress based on magnetic atoms in confined geometries; see Sec.~\ref{sec:lattice}. We note that important advances have been made with systems of polar molecules~\cite{Bohn2017cmp,Moses2017nff,Moses2015coa,Yan2013ood} as well as of Rydberg atoms~\cite{Saffman2010qiw,Barredo2016aab,Endres2016aba,Barredo2018std,Labuhn2016ttd,Bernien2017pmb}. Focusing on magnetic atoms, we discuss three main areas: (i) the physics in low-dimensional spaces and in particular 1D, where the motion of the particles arise only in some directions of space and is frozen transversely; (ii) the physics of spinless particles whose motion occurs, this time, along the specially confined directions in space, and in particular, along directions of a periodic external potential formed by an optical lattice. This realises extended Hubbard models for spinless dipolar particles; (iii) the case of spinful dipolar particles in such periodic external potentials, leading to quantum magnetism and XYZ models.

\myparagraph{Dipolar gases in lower dimensions}

We now discuss a special case of lattice-confined geometries wherein atoms remain free to move in one or two directions of space while being tightly trapped (frozen) in the other(s).  Such gases effectively realise lower-dimensional systems.  Quantum physics in lower dimensions is fundamentally different from that in our usual 3D world. For instance, in both 1D and 2D, quantum fluctuations preclude long-range order, and, in 1D, bosons can act like fermions and vice-versa~\cite{Tonks1936tce,Girardeau1960rbs,Mermin1966aof,Hohenberg1967eol,Giamarchi2003qpi,Bloch2008mbp,hadzibabic2009tdb}. Exotic strongly correlated states may arise and interactions play a crucial role.

In 1D, many aspects of quantum physics for particles interacting via short-range potentials are understandable at an analytic level.  In particular, solvable models, such as the Lieb-Liniger model, can often be evoked to describe such systems~\cite{Lieb1963eao,Giamarchi2003qpi}.  When such models break down, e.g. by introduction long-range interactions, these systems can serve as testbeds for exotic strongly correlated many-body physics~\cite{Sinha2007cdg,Deuretzbacher2010gsp,Deuretzbacher2013egs}.  In the particular case of 1D dipolar gases, the DDI lifts integrability~\cite{Yurovsky2008cca}, 
thereby introducing chaotic dynamics that allows the gas to thermalize. Furthermore, control of the dipole orientation provides a knob with which to control the integrability-breaking mechanism and the induced thermalisation rates; see ~\cite{Tang2017tni} and Sec.~\ref{Sec1D}.  

Excited states of 1D Bose gases can possess correlations stronger than ideal Fermi gases~\cite{Astrakharchik2005btt}. These so-called super-Tonks-Girardeau states have been observed in a narrow range of attractive contact interaction strengths in nondipolar gases~\cite{Haller2009roa}. Repulsive dipolar interactions have been shown to completely stabilise these highly excited states regardless of contact interaction strength~\cite{Kao2020tpo}. Dipolar stabilisation has provided access to a quantum holonomy of the underlying Hamiltonian that allows the gas to be topologically pumped to higher energy states.  These prethermal states realise a form of quantum many-body scar state wherein a strongly correlated excited state evades thermalisation in an otherwise chaotic system~\cite{Kao2020tpo,Serbyn2020qmb}. This physics will be discussed in Sec.~\ref{Sec1D}. Though initially explored in Refs.~\cite{Tang2017tni,DePalo2020vba}, future work could aim to provide a more general understanding of 1D collisional physics in the presence of both the van der Waals interaction and the DDI, especially near a Feshbach resonance.

\myparagraph{Extended Hubbard model}

Another regime of interest arises when the motion of the particles takes place in a periodic external potential. This case, easily achieved by confining atoms in light standing waves, has been considered for a long time, in the ultracold community, and raised wide interest due to its similarity to the physics of electrons gases in crystals, and the possibility to realise very clean Hubbard models. The introduction of DDI within such lattice systems, yield novel physics even for spinless particles, by bringing new terms into the standard Hubbard Hamiltonian  which standardly comprises a contact on-site interaction and tunnelling. Due to the DDI's character, the new terms are anisotropic on-site and off-site interaction terms. 
The new interaction terms in dipolar lattices introduce competition between numerous energy scales. 
This yields exotic dynamical behaviours, excitations, as well as novel phases~\cite{Dutta2015nsh,Baranov2012cmt}. In experiments, the relevance of the extended Hubbard model for dipolar bosons has been demonstrated, and the additional interactions terms quantified. The most exotic phases predicted based on the extended Hamiltonian have for now remained elusive. Current achievement and prospects are discussed in Sec.~\ref{subsec:spinless_lattice}.


\myparagraph{Spin physics in optical lattices}

Finally, we will discuss lattice systems with spin degrees of freedom. By bringing into play dipolar exchange and relaxations terms---see Sec.~\ref{subsec:processes}---such systems realise models with a rich range of exotic dynamics and phases~\cite{Dutta2015nsh,Baranov2012cmt,Bloch2008mbp}. In particular, the off-site term induced by dipolar spin-exchange processes yield generic XYZ Heisenberg models. Of particular interest, a growth of quantum correlations is expected in such systems under the effect of inter-site spin-exchange interactions. Lattice spin models realised with magnetic atoms are discussed in Sec.~\ref{subsec:spinlattice}. 

\subsubsection{Light-induced coupling of spins in magnetic atoms}

The engineering of synthetic coupling involving the particle's spin, such as spin-spin or spin-orbit coupling,  opens the door to the realisation of exotic states such as, on the many-body level, topological superfluids~\cite{Galitski2013soc,Sato2009nat,Jiang2011mfi,Zhu2011pna,Alicea2012ndi,Ruhman2015tsi,Nascimbene2013rod,Celi2014sgf} as well as highly nonclassical or topological spin states~\cite{Kitagawa1993sss,Lian2012sfn,Cui2013sgf,Ho2015lss}, that can even be produced at the single-atom level for large-spin atoms. Spin-dependent light shifts or optical Raman dressing of internal spin states may be used to effect such coupling for the introduction of abelian (magnetic field-like) or non-abelian (spin-orbit coupling-like) gauge fields~\cite{Goldman2014lig,Dalibard2016itt}.  Unfortunately, however, light fields also heat the atoms due to spontaneous emission, limiting the lifetime of systems. This may be circumvented by exploiting the  level structure of magnetic atoms such as  Dy and Er, while  allowing dipolar interactions to play a role in the physics.  Dipolar relaxation then sets new limits on the lifetime of these gases.  Fortunately, there exist platforms in which the rate of dipolar relaxation is significantly reduced.  One method uses a tightly confining potential in one or more spatial dimensions to suppress relaxation via phase-space restriction; see Refs.~\cite{Pasquiou2010cod,Pasquiou2011sra,depaz2013rdo} and Sec.~\ref{subsec:inel_confin}. Another method employs fermions in a large magnetic field so that dipolar relaxation may be suppressed due to Fermi statistics; see Refs.~\cite{Pasquiou2010cod,Burdick2015fso} and Sec.~\ref{fsuppress}. Exploiting such possibilities has yielded the realisation of long-lived SOC Fermi gases~\cite{Burdick2016aco}. Related achievements and prospects are discussed in Sec.~\ref{SecSOC}.

%% file: Section2/2_Cr_to_Lanth.tex
\section{The magnetic atoms}
\label{sec:amgnetic_atoms}

The experimental research on dipolar quantum gases in the degenerate regime began in Stuttgart with the first production of a Bose-Einstein condensate made of chromium atoms in 2004~\cite{Griesmaier2005bec}. This achievement has since then attracted great interest, 
both from theorists and experimentalists. A second Cr BEC machine soon became available at Villetaneuse~\cite{Beaufils2008aop}. The interest in dipolar quantum degenerate gases was further sparked when it was shown that atoms with even larger magnetic moments than Cr, such as erbium~\cite{McClelland2006lcw} and dysprosium~\cite{Lu2010tud}, could be efficiently laser cooled. Soon after, Bose and Fermi degenerate gases of both these lanthanide (Ln) atoms were produced, first at Urbana-Champaign (the group moved to Stanford in 2011)~\cite{Lu2011sdb, Lu2012qdd}, then at Innsbruck \cite{Aikawa2012bec, Aikawa2014rfd}.  Degenerate Fermi gases of Cr have also been produced in Villetaneuse~\cite{Naylor2014cdf}. These achievements, and the subsequent experiments, have stimulated much theoretical interest and activity in these systems. 
In response to these achievements, the field of dipolar gases made of magnetic atoms is now rapidly expanding. Experiments worldwide are being constructed to explore the fascinating properties of  Ln atomic gases and many additional groups have realised gases in the ultracold~\cite{Dreon2017oca,Ravensbergen2018ado,Ilzhofer2018tsf,Sukachev2010mot,Miao2014mot,Cojocaru2017lac,Inoue2018mot,Tsyganok2018pcc,Chalopin2018qes,Evrard2019ems,Seo2020epo,Lunden2020etc} and quantum degenerate~\cite{Kadau2016otr,Ulitzsch2017bec,Lucioni2018ddb,Trautmann2018dqm,Ravensbergen2018poa,Tanzi2019ooa,Tanzi2019ssb,Davletov2020mlf,Phelps2020ssp} regimes. 

In this section, we discuss the properties and the special features of the highly magnetic atomic species currently available in the quantum-degenerate regime, and in particular, their electronic structure and energy spectrum in comparison to the alkali atoms. We first recall a few features of Cr  (Sec.~\ref{subsec:Cr}; see also Ref.~\cite{Lahaye2009tpo}) before presenting the magnetic Ln atoms (Secs.~\ref{subsec:Ln}). We describe the basic method for cooling and trapping such species. Ultracold gases are typically created and confined in vacuum chambers using the techniques of Zeeman atomic beam slowers (ZSs), magneto-optical traps (MOTs), magnetic traps (MTs) and/or optical dipole traps (ODTs)~\cite{Cohentannoudji2011aia}.  In the case of magnetic atoms, pure MTs are of limited efficacy due to dipolar-relaxation-induced atom loss~\cite{Newman2011mri,Connolly2010lsr,Hensler2003dri}; see also Sec.~\ref{Drelaxationsec}.  The ZS, MOT and ODT techniques are more effective for magnetic atoms. The application of these slowing, cooling and trapping methods must take into account the special electronic structure of these magnetic atoms, which we will discuss.   

In addition, we discuss the interactions of light with these atoms, specifically in regard to their large total orbital momentum; see Sec.~\ref{subsec:atomlightLn}. In Sec.~\ref{subsec:fesbachsec}, we discuss short-range scattering properties of the magnetic atoms including their scattering length $\as$ and Feshbach resonances (FRs).  Feshbach resonances  enable  the wide tunability, in both sign and  amplitude, of $\as$. This allows one to control the dipolar character of magnetic gases too, since such properties often depend on the relative strength of the DDI to the contact interaction, $\edd=\add/\as$; see Sec.~\ref{subsec:mf}.  Moreover, FRs enable the production of more magnetic particles via the association of two atoms into a molecule. We discuss the related possibilities in both Cr and Ln atoms. We finally compare the overall tunability of the collisional properties of Cr and Ln in Sec.~\ref{subsec:compareSystems}.



 
\subsection{Chromium}
\label{subsec:Cr}
When the first ultracold-gas experiment with Cr atoms was started by the Stuttgart group, which began in Konstanz, prior to their move to Stuttgart, the aim was in fact to create a new instance of degenerate Fermi gas (due to the large abundance of its fermionic isotope). They realised only later, during a visit of K.~Rzazewski in 1998, that it would be a great candidate for dipolar physics~\cite{Goral2000bec}. Focusing on the latter physics, using bosons, they achieved the first BECs of
$^{52}$Cr~\cite{Griesmaier2005bec}. Chromium-52 atoms have a purely electronic spin of $S=3$ and a $g$-factor of $g_S=2$ in the ground state (denoted $^7$S$_3$ using the standard notation $^{2S+1}$L$_J$).  The DDIs between these atoms are 36 times stronger than in the most magnetic case of alkalis because its magnetic dipole moment is $\unit[6]{\mu_B}$, whereas alkalis' moments are at most $\unit[1]{\mu_B}$.  The most abundant isotope of Cr is $^{52}$Cr; see table~\ref{table_expgases}. The atom has two other bosonic isotopes, $^{50}$Cr and $^{54}$Cr, which have never been Bose-condensed. The fermionic isotope $^{53}$Cr was cooled to degeneracy via a sympathetic cooling technique in 2014~\cite{Naylor2014cdf}. Besides their dipolar character, ground state Cr atoms realise large spin systems (see Sec.~\ref{subsec:large_spin}) whose magnetic properties are driven by both the DDI and their relatively strong, spin-dependent, isotropic contact interactions.   The relevant scattering lengths are $a_{S=6}$=102.5 $a_0$~\cite{Pasquiou2010cod}, $a_{S=4}$=56 $a_0$, $a_{S=2}=$-7 $a_0$~\cite{Werner2005oof}, $a_{S=0}=$ 13.5 $a_0$~\cite{depaz2014das}; see Sec.~\ref{subsec:fesbachsec} for details. 

Laser cooling of Cr  was pioneered by J.~McClelland for the purpose of creating collimated atomic beams for atom lithography~\cite{Bradley2000mot}.  A motivation for this work was based on the fact that the Cr ion is a colour centre in various host materials; e.g., it gives the red colour to a ruby crystal. This, controlling and positioning individual single colour centres could be possible through the three-dimensional laser cooling of Cr.  Additionally,  Cr sticks to most surfaces, and as such, Cr serves as an excellent material for etch masks:  Using transversely cooled Cr atomic beams, one and two-dimensional structures  at a resolution of a few tens of nanometers, as well as three-dimensional structured doping, were demonstrated~\cite{Oberthaler2003ota}.

In the relatively high-density regime relevant for laser cooling, a large light-assisted inelastic cross-section was observed \cite{chicireanu2006smo,Volchkov2014edc} close to the Langevin limit \cite{Julienne1991cco}. This loss mechanism creates an intrinsic limitation to the density and number of atoms that can be efficiently captured in a Cr MOT. However, as the cooling transition  $^7$S$_3\rightarrow^7$P$_4$ is not perfectly closed, metastable states ($^5$D$_{3,4}$) are populated during the cooling process and are trapped in the quadrupole field of the MOT due to their large magnetic dipole moment. This provides continuous loading into a dark, but trappable, state that is immune to light-assisted collisions.  Typically a few $10^8$ atoms accumulate in this trap, with their number limited in density by inelastic
collisions between these metastable atoms~\cite{Schmidt2003clo}. Using a repumper on the intercombination line $^5$D$_{3,4} \rightarrow^7$P$_{3,4}$ produces a magnetically trapped sample in the $^7$S$_3$ ground state~\cite{Stuhler2001clo}.  The sample is sufficiently dense that evaporative cooling may proceed.  However, such cooling is then limited by dipolar relaxation collisions, flipping spins into untrapped states and inducing heating from the released Zeeman energy~\cite{Hensler2003dri, Pasquiou2010cod}. This may be avoided by loading into a crossed ODT an evaporatively precooled and spin-polarised sample, i.e., one in the strong-field seeking, lowest-energy Zeeman state.  The gas could then be evaporatively cooled  all the way to degeneracy~\cite{Griesmaier2005bec}.  Subsequent production schemes of Cr degenerate gases start by accumulating atoms in the metastable $^5$D$_{3,4}$ and $^5$S$_2$ states directly from the MOT into an ODT.  The atoms were then repumped to the ground state where all-optical evaporative cooling in the ODT produced the BEC~\cite{Beaufils2008aop}.  Additional details regarding the key techniques and strategies to produce Cr BECs are described in an earlier review~\cite{Lahaye2009tpo}.

In contrast to the Lns, the ground state in Cr is an ``$S$-state," which means that the mutual van der Waals interactions are isotropic. However, there are still sufficient Zeeman substates to provide a rich structure in the asymptotic molecular states. The DDI additionally provides a coupling between these states. Therefore, even without hyperfine structure, as it is the case for bosonic Cr ($I=0$), a number of narrow FRs exist, as were first found and characterised in Ref.~\cite{Werner2005oof}. See section~\ref{subsec:fesbachsec} for details. 

\subsection{Lanthanides} 
\label{subsec:Ln}
Atoms with multiple valence electrons and non-$S$ electronic ground state, such as the magnetic Lns, are of increasing interest to the study of strongly dipolar phenomena in atomic quantum gases. Among the magnetic Lns, Dy~\cite{Lu2011sdb, Lu2012qdd} was the first to be brought to quantum degeneracy, shortly followed by Er~\cite{Aikawa2012bec, Aikawa2014rfd}; see Sec.~\ref{subsec:BECLn}. We note that Yb has been the first Ln to be Bose-condensed~\cite{Takasu2003ssb}, but because of its closed-shell character, Yb has zero magnetic moment and is more similar to the alkaline earth atoms---it will not be discussed here. In addition, BECs of Tm were recently achieved~\cite{Davletov2020mlf}. We will now mostly focus on Dy and Er because they are the most widely employed for the quantum gas research reviewed here. 

As summarised in Table~\ref{table_expgases}, Er and Dy have a number of special features that make them particularly appealing for quantum-gas experiments. Both Er and Dy possess many  naturally abundant isotopes with a wide variability of properties, including  several bosonic and fermionic isotopes. Besides different quantum statistics (bosonic or fermionic), the isotope variety also offers a useful diversity and tunability of scattering properties:  Each isotope has a different background value of $\as$ and a distinct Feshbach spectrum, the latter of which can be used to further tune $\as$; see Sec.\,\ref{subsec:fesbachsec}. Besides binary collisions, the rates of multibody collisional processes are also expected to change from one isotope to the other. This includes, in particular, the three-body inelastic collisions, which typically induce detrimental losses and heating in cold gases.   The wide variability in collisional properties offer more opportunities for finding isotopes that can be efficiently cooled to quantum degeneracy. We note that this isotope variety is special to Er and Dy among the magnetic Lns: Eu, Tm and Ho all have only one stable (bosonic) isotope. This was one of the reasons for first focusing on Er and Dy within the Ln series.  

Magnetic Lns have a large magnetic dipole moment in the electronic ground state (e.\,g.\,, $\unit[7]{\mu_B}$ for Er and $\unit[10]{\mu_B}$ for both Dy and Tb), and, 
because their mass appears in the DDI strength, the corresponding dipolar lengths are several times larger than Cr's ($\add=15\,a_0$), with $\add=65.5\,a_0$ and $\add=131\,a_0$ in Er and Dy, respectively. In dipolar BECs (dBECs), the strength of the dipolar character of a quantum gas is proportional to the ratio $\edd=\add/\as$; see Sec.~\ref{subsec:mf}. Magnetic Lns are sufficiently dipolar that their $\edd$ are typically of the order of $1$ at background level (i.e., away from FRs).

In addition to their strong dipolar character, magnetic Lns feature a large orbital momentum quantum number, $L$, in their ground state ($L=5$ for Er and $L=6$ for Dy). This is a major difference from Cr, where the large angular momentum arises purely from the electronic spin $S$ while $L=0$. The large $L$ value in the Ln case induces an orbital anisotropy, which causes the  van der Waals interactions to be anisotropic, in addition to the DDI~\cite{Petrov2012aif,Kotochigova2014cib,Ming2018oqm}.  As we will discuss in Sec.\,\ref{subsec:fesbachsec}, this orbital anisotropy has important consequences for the scattering properties of Lns \cite{Krems2004eia,Connolly2010lsr,Kotochigova2011ait,Petrov2012aif,Kotochigova2014cib}, in particular for interspin interactions, as well as for their atomic polarizability~\cite{Chu2007dpo,Dzuba2011dpa,Lepers2014aot,Li2017oto,Kao2017ado,Becher2017apo,Ravensbergen2018ado}; see Secs.~\ref{subsec:atomlightLn},~\ref{subsec:fesbachsec}, and~\ref{subsec:spin_int}. 

In addition, and similar to Cr, magnetic Lns realise large-spin systems; see Sec.~\ref{subsec:large_spin}. Finally, the large masses leads to low recoil energies, which is beneficial for optical trapping and laser cooling.

\subsubsection{Atomic energy spectrum of magnetic Lns}
\label{subsec:Lnspectrum}

The electronic configuration of magnetic Lns is $[Xe]4f^{m}6s^{2}$.  It is  characterised by a xenon-like core, an inner open $4f$ shell with $m$ valence electrons, and an outer closed $6s$ shell. Due to the electron vacancies in the inner shell, magnetic Ln are often called {\em submerged-shell} atoms. The unfilled $4f$ shell plays a particularly important role in their high magnetism and orbital anisotropy. 

\begin{figure*}[ht!]
\centering
\includegraphics[width=1.\textwidth]{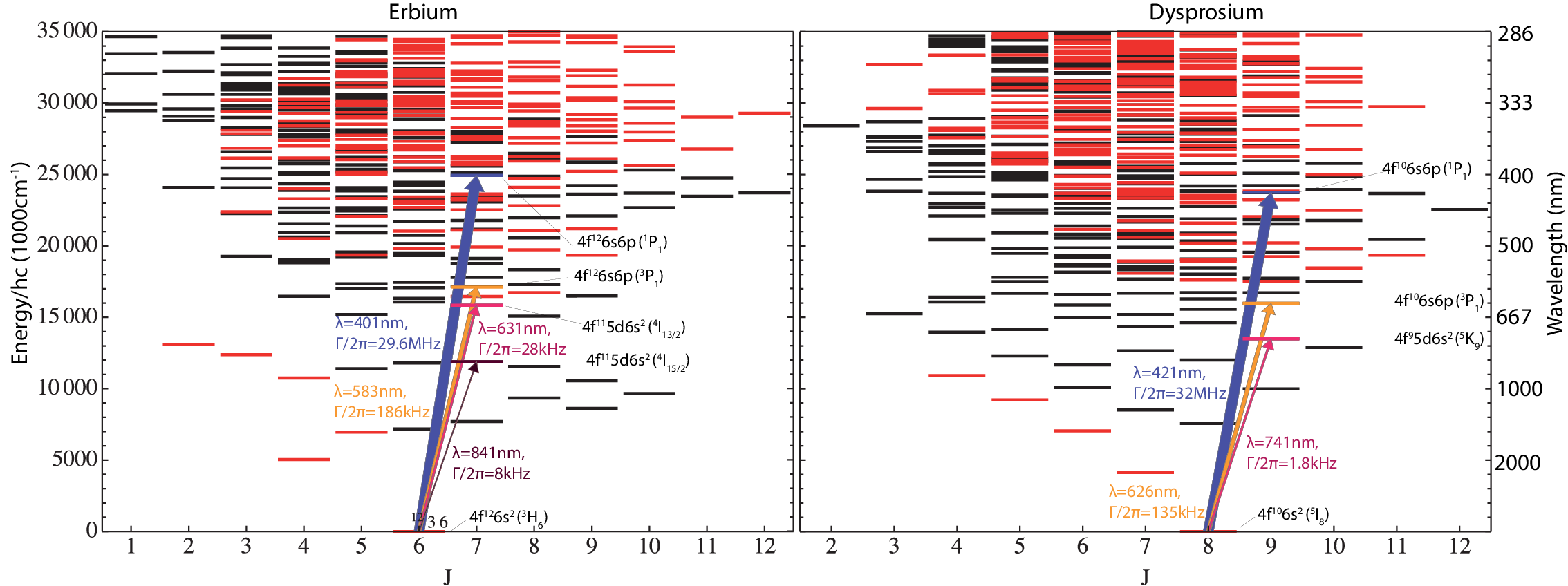}
\caption{Energy spectrum of Er and Dy. 
The levels shown in red (black) have an even (odd) parity. Electric dipole transitions couple the ground state, which has an even parity, to energy levels of odd parity and with a total angular momentum within the interval $J=[5,7]$ for Er and $J=[7,9]$ for Dy. Both species feature broad transitions in the blue ($401{\rm nm}$ for Er and $421 {\rm nm}$ for Dy), intercombination lines at $583{\rm nm}$ for Er and $626{\rm nm}$, and narrow cooling transitions at $631{\rm nm}$ and $841{\rm nm}$ for Er and $741{\rm nm}$ for Dy. These transitions  have been used in experiments to laser cool the atomic samples. 
} 
\label{fig:levelscheme}
\end{figure*}

Figure~\ref{fig:levelscheme} shows the atomic energy spectra for the cases of Er and Dy, up to a wavenumber of $\unit[25{,}000]{cm^{-1}}$. Both species have an even-parity ground state with a large 
$J$ and many excited states of odd and even parity of various quantum numbers. A comprehensive set of spectroscopic data of all Ln elements can be found in Refs.\,\cite{Martin1978ael,Ban2005lct,Ralchenko2011uru}. Note that theory results that are based on the Cowan suite of codes~\cite{cowan1981tto,wyart2011oti} and used to estimate the polarizability of Lns predict atomic transitions that have not yet been observed~\cite{Lepers2014aot,Li2017oto}; see also Sec.~\ref{subsec:atomlightLn}. This reveals the still incomplete knowledge of Ln atomic spectra and their properties. 

Unusually for those accustomed to alkali atoms, many of the energy levels in the electronic spectrum of Ln atoms do not conform to the usual LS-coupling scheme. In the LS scheme, the total electron spin $S$ and angular orbital momentum $L$ couple to form $\bs{J}=\bs{L}+\bs{S}$, but the large spin-orbit coupling of the electrons in Lns renders this scheme sub-optimal for some of the electronic levels. 
In this case, a $J_{1}J_{2}$-coupling scheme is more appropriate~\cite{Wybourne2007oso}. In this $J_{1}J_{2}$ scheme, the electrons in each shell couple independently in a LS-coupling scheme, and then the total angular momentum quantum numbers from the different shells sum together. For instance, $4f$ electrons give rise to  $J_{1}$ and the $6s$ ones to $J_{2}$, and $\bs{J}=\bs{J_{1}}+\bs{J_{2}}$ with quantum number $J$, which is denoted as $(J_{1},J_{2})_{J}$. For Lns, the LS scheme remains relevant for the ground state while the excited states (where spin-orbit coupling is stronger) typically follow a $J_{1}J_{2}$-coupling scheme.
Finally, the bosonic isotopes of Dy and Er have nuclei with an even number of protons and neutrons. This results in a zero nuclear spin $\mathbf{I}$ and no hyperfine structure. In contrast, the fermionic isotopes have an even number of protons and an odd number of neutrons, resulting in a nuclear spin $\mathbf{I} = 5/2$ and $\mathbf{I} = 7/2$ for Dy and Er, respectively. We note that other magnetic Lns, such as Tm, Eu or Ho, have only bosonic isotopes and those isotopes have a hyperfine structure.

Lanthanides' atomic spectra offer a rich collection of $J\longrightarrow J+1$ optical lines, including broad-, narrow- and ultra-narrow-linewidth transitions. Many of such transitions can be used for optical manipulation and laser cooling, and they are readily accessible with common lasers. As a rule of thumb, the linewidth of the transitions is larger for short-wavelength (i.\,e.\, high energy) transitions.  The strongest line in Er (Dy) at 401 (421)~nm, the intercombination line at $583$ ($626$)~nm, and the narrow line at $841$ ($741$)~nm are used in experiments for laser cooling, as discussed in the next section. The narrower resonances have also been used for spin manipulation or coupling, see Sec.~\ref{spinorsection}. Furthermore, the spectra of the $J\longrightarrow J$ and $J\longrightarrow J-1$ transitions are equally rich and relevant for optical manipulation schemes~\cite{Schmitt2013soa}. Among the rich spectra of Lns, an interest is also growing for the ultranarrow resonances of Hz-level linewidth, see e.g. refs~\cite{Studer2018lso,Golovizin2019isc,Petersen2020sot,Patscheider2021ooa}. 
Finally, the orbital anisotropy of Lns also yields a dependence of the matter-light interaction on the atomic spin. This presents both challenges (how to obtain equal trapping of all spin states?) and advantages (realisation of spin-dependent potentials, spin-orbit coupling, etc.) for ultracold gas experiments; see also latter discussions.

\subsubsection{Optical cooling, trapping, and evaporative cooling of open-shell Ln}
\label{sec:coolingLn}
Trapping and cooling of open-shell Ln atoms have been achieved using  ZSs, MOTs, and ODTs, but without the use of magnetic trapping. The overall efficiency is similar to that of alkali-metals~\cite{Cohentannoudji2011aia}, though there are some key differences that we now describe.
 
\myparagraph{Optical cooling and MOTs}\label{subsec:Ln_cool}
The broadest laser cooling cycling transitions of Lns are open, meaning that there are a multitude of metastable states to which the excited state can decay through spontaneous emission.  Repumping the population back to the cooling transition is not practically feasible.  Fortunately, two solutions exist: a repumperless MOT on these broad transitions and a MOT using a closed transition with small or intermediate linewidth. We detail below the working principle of these two schemes.

$\bullet$\textbf{Broad-line MOTs}

In 2006, McClelland \textit{et al.} presented a repumperless Er MOT~\cite{Ban2005lct,McClelland2006lcw}.  Despite the open nature of the $\sim$30-MHz linewidth, 401-nm transition, neither the MOT nor the ZS required a repumper, and $10^6$ atoms were confined.    Subsequently, transverse cooling~\cite{Leefer2010tlc,Lu2010tud,Leefer2011vot} and a MOT~\cite{Lu2010tud,Youn2010dmo} containing several $10^8$ atoms were reported for Dy, both using a similarly wide transition at 421~nm.  Likewise, MOTs of Ho~\cite{Miao2014mot}, Tm~\cite{Sukachev2010mot,Sukachev2011lco,Vishnyakova2014tsl} and Eu~\cite{Inoue2018mot} have  been formed.  The surprising success of the repumperless MOT derives from two key properties of magnetic Lns atoms:  i) They possess a surprisingly small branching ratio  (${<}10^{-5}$) of decay to the metastable states; and ii) the  lifetime of these metastable states in the quadruple MT  of the MOT is long.  The former property allows a sufficient number of atoms to go through the ZS without decaying to metastable states.   The latter property means that atoms in metastable states are not lost, but remain trapped in the MT of the MOT (due to the atoms' large magnetic moment) 
until they eventually decay to the ground state and undergo cooling cycles again.

An unusual, anisotropic sub-Doppler cooling effect was observed inside these Er, Dy, and Tm MOTs~\cite{Berglund2007sdl,Lu2010tud,Youn2010dmo,Youn2010asd,Sukachev2010sdl}.  The effect is a consequence of the fact that the Land\'{e} $g$ factors of  the ground and  excited states are nearly the same, yielding nearly zero differential Zeeman shift on the cooling transition.  This allows $\sigma^+\sigma^-$ polarization-gradient cooling~\cite{Cohentannoudji2011aia,Dalibard1989lcb} to exist even in a large magnetic field.  The atoms are therefore exposed to both Doppler and sub-Doppler cooling mechanisms inside the MOT. The larger-population Dy blue-line MOTs exhibit a small core of sub-Doppler cooled atoms, as in the Er MOT, but this core is surrounded by a larger-population shell of hotter, Doppler-cooled atoms.  Atoms beyond a certain distance from the MOT's quadrupole MT center feel a large-enough magnetic field to disrupt the sub-Doppler cooling mechanism beyond this radius. The temperature of this core of colder, sub-Doppler cooled atoms is highly anisotropic, with the temperature of atoms along the quadrupole MT axis hotter than those in the quadrupole plane of symmetry, or vise-versa depending on the ratio of cooling laser intensity along these directions~\cite{Youn2010asd}.  This unusual anisotropic sub-Doppler cooling effect is likely due to the countervailing tendency of atomic polarisation to  lock its orientation to a direction favoured by the laser optical pumping versus its tendency, due to the large magnetic dipole moment of Dy, to align according to the local magnetic field of the MT~\cite{Youn2010asd}.

$\bullet$\textbf{Multi-stage MOTs}

While ZSs and MOTs on broad transitions can perform the initial stages of laser cooling and trapping of open-shell Lns, they cannot cool the atoms low enough to load ODTs.  Further cooling may be provided by using so-called intercombination-line MOTs.  These are formed using laser cooling lines on electric dipole semi-allowed (intercombination) transitions.  
Fortunately, all of these narrow transitions are closed, obviating the need for repumping lasers.  

In one of the realised schemes, the atoms are cooled using, first, a broad-line MOT, and then, in a second stage, a colocated MOT on a very narrow transition $<$10~kHz. A similar scheme was also used for cooling strontium atoms~\cite{Katori1999mot}. The first narrow-line open-shell Ln MOT was demonstrated in  Ref.~\cite{Berglund2008nlm} with Er, using such a scheme on the 8-kHz wide, 841-nm transition.  This very narrow line, however, leads to an unusual requirement for stable MOT operation:  The cooling lasers must be tuned to the \textit{blue} (i.e., positive frequency detuning) side of the transition, rather than the \textit{red}.  This is because Er's large magnetic moment causes the magnetic Zeeman force to dominate the optical radiation forces,  even in the small magnetic fields encountered near the MT centre.  Blue detuning also optically pumps the atoms to the magnetically trappable weak-field-seeking states. Under conditions of blue detuning, the MOT forms below the quadrupole MT where  gravitational, radiation, and magnetic Zeeman forces mutually balance.  The 8-kHz-wide, 841-nm Er MOT provided $\sim$2~$\mu$K gases~\cite{Berglund2008nlm}.  A similar blue-detuned MOT for Dy on its 1.8-kHz-wide, 741-nm transition was able to cool 10$^7$ Dy atoms loaded from the broad MOT to $\sim$2~$\mu$K~\cite{Lu2011soa,Lu2011sdb} and works for all high-abundance isotopes of Dy. More recently, a 841-nm Er MOT operating with red-detuning have been demonstrated~\cite{Phelps2020ssp}. This was made possible by its loading from an intermediate-linewidth MOT operating from the 583-nm transition, as in Ref.~\cite{Frisch2012nlm}; see below. This additional cooling stage enabled temperatures as low as $400\,$nK to be reached and phase-space densities as high as 0.05. 

$\bullet$\textbf{Single-stage intermediate-linewidth MOTs}

An alternative approach replaces the double-MOT scheme with a single MOT whose transition linewidth is intermediate, i.e., in the 100's of kHz range.  This is similar to the scheme employed for ytterbium atoms~\cite{Kuwamoto1999mot}. The narrow linewidth provides Doppler-limited cooling below 10's of $\mu$K, sufficiently low to directly load an ODT, yet is broad enough to allow the capture of atoms from a Zeeman-slowed atomic beam.  This intermediate-linewidth scheme was first developed for an open-shell Ln MOT in  Ref.~\cite{Frisch2012nlm} using Er.  This scheme has then become the most widely employed, and MOTs of all high-abundance Er~\cite{Frisch2012nlm, Ilzhofer2018tsf,Seo2020epo} and Dy~\cite{Maier2014nlm,Dreon2017oca,Ravensbergen2018poa,Ilzhofer2018tsf,Muhlbauer2018soo,Lucioni2018ddb,Phelps2020ssp,Lunden2020etc} isotopes, both fermionic and bosonic, have been created in various labs. Double-species MOTs of Er and Dy have also been achieved using this scheme~\cite{Ilzhofer2018tsf}. It has recently been employed in Tm~\cite{Cojocaru2017lac,Tsyganok2018pcc}. Typically, MOTs containing several $10^7$ to $10^9$ atoms with final temperatures of $\sim$6--13~$\mu$K are achieved. These low temperatures allow direct loading of the atoms into relatively low-power ODTs. The intermediate-line MOT can also load narrow-line MOTs operated with red detuning, which enables rapid  cooling to lower temperatures before loading an ODT~\cite{Phelps2020ssp}. While intermediate-line MOTs have been successfully  loaded directly from Zeeman-slowed atomic beams~\cite{Frisch2012nlm, Ilzhofer2018tsf,Maier2014nlm,Dreon2017oca,Ravensbergen2018poa,Ilzhofer2018tsf,Muhlbauer2018soo,Lucioni2018ddb,Phelps2020ssp}, recent schemes have enhanced the capture efficiency of the narrow-line MOT by using angled Zeeman slower beams on the broad 421-nm transition in between the output of the ZS and the position of the MOT~\cite{Lunden2020etc,Seo2020epo}. This provides a factor 20 gain in population in the final MOT.  

The intercombination lines in Er, Dy and Tm are at wavelengths of 583\,nm, 626\,nm and 530.7\,nm  with linewidths of 190~kHz, 136~kHz, and 320~kHz, respectively. 
These MOTs are operated at very large \textit{red} detunings, intensities, and  magnetic gradients~\cite{Frisch2012nlm,Maier2014nlm,Ulitzsch2017bec,Dreon2017oca,Ravensbergen2018poa,Ilzhofer2018tsf}. In this configuration, the atoms feel a radiative force only over a portion of the trap volume. The force is negligible at the centre, while the magnetic gradient brings the atoms back into resonance on an ellipsoidal shell whose radius is set by the competition of light detuning and Zeeman energy shift~\cite{Loftus2004nlc2}.  This yields a MOT capture volume that increases with the light detuning.  
Similar to the MOTs achieved on extremely narrow transition described above, the atoms are displaced below the trap centre and are in an unconventional bowl-shape due to gravitational effects.  This  enabled a  single-beam narrow-line MOT in Er~\cite{Berglund2008nlm} due to magnetic trapping. A more practical scheme using only five MOT beams was demonstrated~\cite{Ilzhofer2018tsf,Phelps2020ssp}, where the sixth beam coming from the top was unnecessary due to the MOT located below the zero of the magnetic field. Conveniently, in the intermediate-line MOTs, the atoms are spin-polarised into the lowest Zeeman sublevel, rather than the highest, as is the case for the blue-detuned narrow-line MOTs~\cite{Berglund2008nlm,Lu2011sdb}.  The latter require an RF adiabatic rapid fast passage step to transfer atoms from the weakest to the strongest field-seeking states. References.~\cite{Dreon2017oca,Cojocaru2017lac} additionally demonstrate that light-assisted collisions are the limiting factor to the performance of the intermediate-line MOT scheme for the bosonic isotopes. Such losses are minimised at large detuning.  In all schemes, the fermionic MOTs have the lowest trap populations relative to natural abundance.  This is thought to arise from complications  due to inefficient optical pumping and loss channels arising from the existence of hyperfine structure in the fermionic isotopes~\cite{Lu2010tud,Youn2010dmo,Frisch2014dqg,Maier2014nlm}.

\myparagraph{Optical dipole trapping and cooling to quantum degeneracy}\label{subsec:trapingLn}

Despite the complex electronic structure of open-shell Lns, optical dipole trapping is in practice very similar to that of alkali-metal atoms; see Ref.~\cite{Grimm2000odt}.  Long-lived optically trapped quantum gases of Lns have  been made at a variety of wavelengths. These in include:  1560~nm, 1070~nm, 1064~nm, 741~nm, and 532~nm for Dy~\cite{Lu2011sdb,Lu2012qdd,schmitt2016sbd,Kao2017ado,Chalopin2018als}; 1570~nm, 1070~nm, 1064~nm, and 532~nm for Er~\cite{Aikawa2012bec,Aikawa2014rfd,Becher2017apo};  and 532~nm for Tm~\cite{Tsyganok2018pcc,Vishnyakova2014tsl,Davletov2020mlf}.  A quasielectrostatic optical dipole trap of Er near 10.6~$\mu$m has also been reported~\cite{Ulitzsch2017bec}. We will review the special features of optical trapping with Ln in Sec.\ref{subsec:atomlightLn}. In the remainder of the present section, we will describe how the resulting conservative traps have been used to achieve quantum degenerate gases of both Dy and Er via standard evaporative cooling.

$\bullet$\textbf{BECs}\label{subsec:BECLn}

Forced evaporative cooling in a crossed ODT---overlapping ODTs that create a ``dimple'' trap at their intersection---is quite efficient for open-shell Ln due to the contributions to the elastic cross-section provided by dipolar collisions; see Sec.~\ref{elasticdipolarscattering}.  However, compared to the ODT evaporation of weakly dipolar species, care must be taken to ensure that the dipolar gas does not become mechanically unstable and collapse during the evaporation process; see Sec.~\ref{Sec:DCQD}. 

The first degenerate gas of an open-shell Ln was created in 2011~\cite{Lu2011sdb}:  Nearly pure BECs of $^{164}$Dy were observed with a population of ${\sim}10^4$ atoms at a density of $10^{14}$~cm$^{-3}$.  Soon thereafter, BECs of $^{168}$Er were made, with $2\times 10^5$ atoms~\cite{Aikawa2012bec,Frisch2014dqg}.  BECs of other isotopes of Dy, specifically $^{162}$Dy with $10^5$ atoms and $^{160}$Dy with $10^3$ atoms~\cite{Tang2015bec} and  $^{166}$Er with $10^5$ atoms~\cite{Chomaz2016qfd} were produced; see Sec.~\ref{subsec:compareSystems} and table~\ref{table_expgases} for a summary.  
More recently, various quantum degenerate mixtures of Er and Dy with typically few $10^4$ atoms in each component~\cite{Trautmann2018dqm}  were also achieved using the isotopes $^{166}$Er, $^{168}$Er and $^{170}$Er (which was previously uncondensed), and $^{162}$Dy and $^{164}$Dy. In 2020, the first BEC of Tm was achieved~\cite{Davletov2020mlf}.


The evaporation efficiency of Lns greatly depend on the choice of bias magnetic field applied $B$.  This is due to the  complex scattering behaviour of the Lns tuned by the extremely dense spectra of FRs; see  Secs.~\ref{subsec:feshbachLn} and~\ref{subsec:compareSystems}. Both two-body (elastic) and three-body (inelastic) scattering rates strongly depend on $B$. We note that, up to now, efficient evaporative cooling has been reported only at relatively low $B$, from a few hundreds of mG to a few tens of G. 
Moreover, the orientation of the bias field can also play a role in determining the efficiency of evaporative cooling, in particular because it impacts the stability of the interacting system. To reduce the contribution of the attractive DDI, ODTs are typically cigar- or pancake-shaped during evaporation; see also Sec.~\ref{subsec:Stability}.  We note that a particularly efficient evaporation scheme consists in changing the trap geometry toward more pancake shapes in the final stage of the scheme~\cite{Frisch2014dqg,Maier2016iic,Chomaz2016qfd,Trautmann2018dqm}. 

$\bullet$\textbf{DFGs}
\label{subsec:dFGLn}

The first degenerate dipolar Fermi gas was created using a sympathetic cooling scheme.  Fermionic $^{161}$Dy was co-trapped with bosonic $^{164}$Dy to provide sympathetic cooling using this dipolar mixture~\cite{Lu2012qdd}.  The resulting deeply degenerate $^{161}$Dy quantum gas had a population of $10^4$ atoms at $T/T_F = 0.2$~\cite{Lu2012qdd}.   As Sec.~\ref{elasticdipolarscattering} describes in more detail, the direct evaporative cooling of a spin-polarised gas of magnetic fermions is also possible thanks to long-range universal elastic dipolar scattering. In Ref.~\cite{Lu2012qdd}, such a direct evaporative cooling was also performed, yielding an assembly of a few thousand $^{161}$Dy atoms cooled down to $T/T_F = 0.7$.  The first deeply degenerate DFG of Er used this direct approach to reach degeneracy with a spin-polarised gas of $^{167}$Er of ${\sim}7\times10^4$  atoms at $T/T_F \lesssim 0.2$ and densities exceeding $4\times 10^{14}$~cm$^{-3}$~\cite{Aikawa2014rfd}. This work confirmed the efficacy of this cooling mechanism arising from universal elastic dipolar scattering.  Similarly large dipolar DFGs of $^{161}$Dy were created through the optimisation of the crossed ODT shape---i.e., a tighter trap should be used to evaporatively cool fermions---and evaporation procedure~\cite{Burdick2016lls}. This direct-evaporation scheme is very attractive, as it involves a far simpler experimental procedure than sympathetic cooling. Moreover, the evaporative cooling efficiency was found to be as high as that of bosonic isotopes. More recently, quantum degenerate mixtures of the fermionic $^{161}$Dy isotope with bosonic $^{168}$Er~\cite{Trautmann2018dqm} as well as with fermionic $^{40}$K~\cite{Ravensbergen2018poa} were produced, combining direct evaporative cooling and sympathetic cooling. Note that $^{167}$Er DFGs could not be achieved in a 1064-nm ODT and a distinct setup must be implemented that uses a 1570-nm crossed-ODT. The $^{167}$Er is rapidly lost from  1064-nm ODTs perhaps because of light-induced collisions  related to the isotope's hyperfine structure adding complexity to its electronic spectrum~\cite{Aikawa2014rfd}. Finally, we highlight that the $B$ value plays a similarly important role in setting the evaporative cooling efficacy fermionic open-shell Lns as well. Indeed, for such polarised fermions, a large number of FRs also exist, with a far greater density than even for bosons due to the fermions' hyperfine structure; see also Sec.~\ref{subsec:feshbachLn}. 
 
 \subsection{Atom-light interactions in magnetic atoms} 
 \label{subsec:atomlightLn}
Atom-light interactions are at the heart of cold-atom experiments. They provide exquisite control and diagnostic tools, and all the observations discussed in this review rely on them.  Indeed, these interactions  enable, e.g., atomic cooling (see e.g., Sec.\,\ref{subsec:Ln_cool}), imaging of density distributions, trapping with  conservative potentials (see e.g. Sec.\,\ref{subsec:trapingLn}), controlling internal atomic degrees--of--freedom (i.e., their spin) (see e.g. Secs.\ref{spinorsection},\ref{subsec:spinlattice}), and coupling of this spin to their external motion (see e.g. Sec.\ref{SecSOC}). The interaction between atoms and light is generally described 
by the Hamiltonian term~\cite{Steck2017}
 \begin{equation}
  \label{eq:atomlight}
 \hat{V}_{\rm AL}= \hat{\bs D}.\hat{\bs E}.   
 \end{equation} 
The electric-dipole approximation has been applied in deriving this expression and is justified by the fact that the light wavelength is much larger than the size of the atom. Here, $\hat{\bs D}$ is the atomic electric dipole operator, whose origin arises from the displacement operators of the electrons times the electric charge $-e$, and $\hat{\bs E}$ is the electric field operator that accounts for all possible photon modes with wavevector $k$ and polarisation $s$. 
The operator $\hat{\bs D}$ of an atom is determined by its electronic ground state and electronic excited state spectrum. Magnetic atoms, whose electronic structure has been shown to be remarkable, are expected to present unusual light-matter interaction properties compared to alkali atoms. In this section, we will partly review these particularities, focusing on the aspects most relevant for the manipulation of quantum gases, i.e., under the typical condition of a coherent laser field sufficiently far detuned from  electronic excited states. 
 
 \subsubsection{Atomic polarisability tensor}
 
In presence of a (monochromatic) laser field, a semiclassical approximation may be used for $\hat{\bs E}$:  $\hat{\bs E}={\bs E}(\hat{\bs r},t)= {\mathcal E}_0(\hat{\bs r})\left({\hat \varepsilon}\exp(-i\omega_0 t+\phi_0(\hat{\bs r}))+c.c.\right)$ with ${\mathcal E}_0$,  $\omega_0$, ${\phi}_0$ and $\hat \varepsilon$ the amplitude, frequency, phase and polarisation unit vector of the laser field. Within the semi-classical approximation, 
the atomic dipole operator is non-zero  thanks to the charges' displacement induced by the electric field.  This results in the simplified form
 \begin{equation}
  \label{eq:alpha}
 \hat{\bs D}(\bs r,t)=\alpha\left(\omega_0 
 \right){\bs E}(\hat{\bs r},t).   
 \end{equation}
Here, $\alpha$ is the complex dynamical atomic polarisability at frequency $\omega_0$. It is a sum of all electronic transition contributions, which typically scale linearly with the transition linewidth (setting the electronic transition strength) and vary inversely to the detuning of $\omega_0$ to the transition frequency~\cite{Steck2017,Grimm2000odt}. The real part of $\alpha$ relates to the (reactive) dipole force while its imaginary part relates to the  (dissipative) radiation pressure force and to the photon-scattering rate. Practically, the former determines the AC Stark shifts induced by the light on the atomic levels' energies and, thus, to the induced dipole trap depth. The latter limits the trap lifetime.
 
Because of the internal structure of the atoms, the atom-light coupling depends on the  total angular momentum of the electronic state $\bs F$ and on the light polarisation $\bs \epsilon$. This yields a (rank-2) tensor structure for the polarisability $\alpha$. It is again a sum over all electronic transitions, but now accounting for the various electric dipole transition elements between the ground and excited states' sublevels as well as their possible spatial anisotropy.   The polarisability tensor $\alpha$ can be decomposed into a scalar $\alpha_s$, vector $\alpha_v$, and tensor $\alpha_t$ contribution such that~\cite{wyart2011oti,Deutsch2010qca,lekien2013dpo}
  \begin{eqnarray}
\nonumber  
&&\alpha(\omega_0) =  
\alpha_s(\omega_0)\bs{\bar{\bar{1}}}+
\alpha_v(\omega_0)\frac{({\bs \epsilon}^* \times {\bs \epsilon}).{\bs F}}{2F}+ \\
\label{eq:alphatensor}
&&\alpha_t(\omega_0)\left[\frac{3({\bs \epsilon}.{\bs F})({\bs \epsilon^*}.{\bs F})+3({\bs \epsilon^*}.{\bs F})({\bs \epsilon}.{\bs F})-2{\bs F}^2}{2F(2F-1)}\right]. 
 \end{eqnarray}
The scalar part is independent of the angular momentum operator, while the higher-rank contributions yield spin-dependent terms. In this respect, vector and tensor polarisabilities set the strength of Raman coupling between atomic Zeeman sublevels. They also provide tools for optical spin manipulation and spin-dependent trapping. Their contribution's dependence on the light polarisation additionally induces an anisotropy of the atom-light coupling, varying with the relative angle between the quantisation axis (set by the magnetic field) and the light-beam polarisation vector.

\subsubsection{Tensor polarisability in magnetic atoms}
 
A major difference in the atom-light interactions involving magnetic atoms versus, e.g., alkali atoms, lies in their large vectorial and tensorial polarisabilities. This arises from the large electronic spin of the magnetic atoms, which yields strong spin-orbit coupling in electronic states with $L\neq 0$, both in ground and excited states. This strong spin-orbit coupling results in differences between the electric dipole matrix elements coupling between the different fine-structure levels of the ground and the electronically excited states, which ultimately leads to spin-dependent and anisotropic behaviour of the atom-light coupling. In the Cr case, the anisotropy comes from spin-orbit coupling in the excited states~\cite{Chicireanu2007aoc, depaz2013nqm, lepoutre2017sma}, while in the Ln case, it arises from the combined effect of spin-orbit coupling in both the ground and excited states~\cite{Dzuba2011dpa,Lepers2014aot,Li2017oto,Li2017aot,Kao2017ado,Li2017aot,Becher2017apo,Ravensbergen2018ado,Chalopin2018als}.
The existence of fine structure in both the excited and the ground state electronic configurations of Lns---and the large energy splitting within these fine structure manifolds---results in a different scaling of the vector and tensor polarisabilities versus that of the alkalies~\cite{Geremia2006tpa,Cui2013sgf}.  That is, both the vector and tensor polarisabilities scale as $\propto \Delta_a^{-1}$, while in alkali metals the scaling is  $\propto\Delta_a^{-2}$ for all detunings $\Delta_a$ greater than the hyperfine splitting, where $\Delta_a$ is the (average) detuning of the light from the excited states. 

The large vector and tensor polarisabilities of magnetic atoms have been investigated via different means. In Cr, spin-dependent optical trapping was indirectly revealed in the study of many-body spinor dynamics, both in bulk~\cite{lepoutre2017sma} and in lattices~\cite{Lepoutre2018csm}. Spin-dependent quadratic light shifts, resulting from the tensor polarizability close to the 427.6 nm-transition (laser at 427.85 nm), were also used to produce controllable spin mixtures~\cite{depaz2013nqm}.
 
A detailed experimental investigation of the characteristics of atom-light coupling in Lns using the newly available ultracold samples was  crucial for a proper understanding of these systems.  Indeed, up to now, theoretical predictions for Ln atom-light interactions have been difficult and remain incomplete. This is due to the complexity of the many-body electronic structure of the atoms and the limited knowledge available. Much progress has recently been made in developing sophisticated numerical tools to analyse the electronic level structure of Dy~\cite{Dzuba2011dpa,Li2017oto}, Er~\cite{Lepers2014aot}, Ho~\cite{Li2017aot}, and Tm~\cite{Golovizin2017mfd}, see also Sec.~\ref{subsec:Lnspectrum}. Recent measurements of Er, Dy, and Tm dynamic (total, scalar) polarisabilities based on trap frequencies measurements have resulted in close agreement between this theory and the experiments~\cite{Becher2017apo,Ravensbergen2018ado,Golovizin2017mfd,Tsyganok2019stv}.  In addition, thanks to a direct comparison with an alkali atom with a precisely known polarizability (potassium), unprecedented accuracy and precision on the Dy polarizability have been obtained in experiment, allowing theory test~\cite{Ravensbergen2018ado}.

The unusually large vector and tensor polarisabilities of Lns (as compared to alkali atoms like Rb) have also been experimentally investigated. The large vector and tensor polarisability of Dy was first studied using a laser field close to the 741-nm transition~\cite{Kao2017ado}.  Near this transition, a so-called \textit{``tune-out''} wavelength was found, at which the total light shift vanishes. Kao \textit{et al.} observed a strong dependence of the tune-out wavelength on the light polarisation angle, which is a consequence of the large tensor polarisability of Dy. This  introduces a novel experimental knob with which to tune the wavelength where the AC polarisability vanishes. Becher \textit{et al.}~\cite{Becher2017apo}  probed the large tensor part of the polarisability of Er grund-state, by testing the dependence of the optical trap's frequencies (thus its depth) on the light's polarisation axis. A sizeable anisotropy effect (few percent) was observed at conventional wavelengths (1570~nm, 1064~nm, and 532~nm) far away from atomic transitions. The spin-dependent light shifts were later used to rapidly control the spin dynamics in an assembly of lattice-confined fermionic Er~\cite{Patscheider2019cde}, see also Sec.~\ref{subsec:spinless_lattice}. A similar scheme  employing the  ellipticity of  light polarisation was applied to gases of Tm atoms and enabled the measurement of both the tensor and vector parts of its ground-state polarisability at 532~nm~\cite{Tsyganok2019stv}.

The locations of other special wavelengths at which the ground and excited state polarizibilities match have been predicted in Dy~\cite{Dzuba2011dpa}.  These so-called ``magic" wavelengths are of great utility to research involving atomic clocks and optical lattices for quantum simulation~\cite{Ye2008qse}.  By making  use of Dy's large tensor polarisability at the conventional trapping wavelength of 1070~nm, the authors Ref.\,\cite{Chalopin2018als} developed an analogous \textit{``magic polarisation''} scheme wherein  the AC Stark shifts of the ground and excited states of a transition (here at 626~nm) are tuned to be identical. The authors then apply their findings to develop a Doppler cooling scheme on a trapped sample with improved efficiency. Large tensor light shifts close to the 626\,nm transition of Dy were also used to generate effective spin-coupling terms~\cite{Chalopin2018qes,Evrard2019ems}; see also Sec.~\ref{SecSC}. Other important consequences of these large vector and tensor polarisabilities in Lns relate to increasing the strength of Raman coupling schemes and to the possibility of efficient spin-orbit coupling; see Sec.~\ref{SecSOC}.   

The unconventional features of atom-light coupling in these magnetic atoms come into play in various ways in this review; see in particular Secs.\,\ref{spinorsection} and~\ref{subsec:spinless_lattice}. Experiments have only begun to explore the possibilities that are offered by the magnetic atoms' electronic structure for the purpose of manipulating the properties of dipolar quantum gases.

\subsection{Feshbach resonances in magnetic atoms}\label{subsec:fesbachsec}

Collisions in ultracold dilute gases are known to be well described within two-body scattering theory~\cite{Landau1965book,Dalibard1999cdo,Pethick2002book}. 
When the interparticle interaction potential is short ranged, as is the case for  the van der Waals force, the scattering at ultralow temperatures is characterised by a single parameter, the scattering length $\as$~\footnote{We point out that within first order-Born approximation, the DDI does not modify the van der Waals scattering length but simply appears in the effective Hamiltonian as distinct potential.}. In brief, $\as$ describes the phase shift $\delta_0$ between the incoming and outgoing waves describing the relative motion of two atoms colliding under the influence of the total interparticle interaction $V(r)$ at vanishing energy, where $r$ is the  atomic separation; see Sec.~\ref{scatttheory} for details.  
This simple picture is modified when accounting for the internal spin degree of freedom of colliding atoms.  This can cause $\as$ to be spin-dependant.  Moreover, magnetic fields can shift the relative energy of different spin collisional channels. In the presence of coupling between different channels, this induces resonant collisions at particular field values.  These are called Fano-Feshbach resonances---Feshbach resonances (FRs) for short---and we  now review their key properties and primary applications.  We will then provide  details particular to the case of magnetic atoms.  See Refs~\cite{Tiesinga1993tar,Pethick2002book,Chin2010fri,Bloch2008mbp} for more information.

A \emph{collisional channel} is defined by a set of bare internal states of the two \emph{free} atoms associated with a given partial wave component of their relative motion.  These partial waves are represented as spherical harmonics with quantum numbers $\ell$ and  $m_{\ell}$ for the norm orbital angular momentum and its projection, respectively; see Sec.~\ref{scatttheory} for details. 
Each collisional channel corresponds, by projection of $V(r)$, to a distinct \emph{molecular} (\emph{scattering}) potential $V_{\rm ch}(r)$. In the context of FRs, the different channels can be coupled by off-diagonal terms of the potential at finite $r$. 
In addition, the asymptotic values of $V_{\rm ch}(r \rightarrow \infty)$, so-called \emph{dissociation thresholds}, may be tuned  relative to one another, e.g., via the Zeeman effect if the two channels have different magnetic moments. Note that for bosons (fermions) in the same spin states, only channels with even (odd) values of $\ell$ are allowed.

Two atoms collide in the \emph{entrance} channel, defined by the internal states in which they are prepared, and typically  $\ell=0$~\footnote{For short-range interacting gases in the ultracold regime, this is the only possibility. Note that when the DDI comes into play, other entrance channels of $\ell\neq 0$ may need to be considered.}. The initial state of the pair, also called \emph{scattering state}, lies in the continuum of the associated scattering potential, with a small (kinetic) energy $E$ above the dissociation threshold $E_{\rm o}$. The entrance channel is thus an \emph{open} channel for the collision, as the atoms can be infinitely far away in this channel. A FR in the collision of the two atoms occurs when a bound state of a different \emph{(closed) channel} couples to the entrance channel.  The closed channel has an energy $E_{\rm cl}$ that can be tuned around $E$~\cite{Feshbach1958uto,Fano1961eoc}. This second channel is \textit{closed} because its dissociation threshold is higher than $E$ and the atoms can not reemerge from the collision in this channel; i.e., $|r|\rightarrow \infty$. The coupling term between the two channels induces a mixing of the scattering and bound states, and the atoms, during their approach, can be temporarily captured in a quasi-bound state. 
This behaviour resonantly alters the scattering properties of the pair, resulting in a change in $\as$. 

The FRs are a particularly convenient tool in ultracold atomic systems for changing the scattering length thanks to the ease with which one may  tune the relative energy of the channels $E_{\rm cl}-E_{\rm o}$, in particular using magnetic fields.   These \emph{magnetic} FR allow $\as$  to effectively vary with the magnetic field $B$ around the resonance centre $B_0$ as
\begin{equation}
    \as(B)=\abg\left(1-\frac{\Delta}{B-B_0}\right),
\end{equation}
where $\abg$ is the background value away from any resonance.  That is,  $\abg$ is the scattering length of the single open channel. $\Delta$ is the resonance width and relates to strength of the coupling between the bound and scattering states~\cite{Moerdijk1995riu}. 

We note that, besides magnetic FRs, optically induced FRs are also possible. These utilise an open channel involving the electronically excited state. Optical control of FRs open new prospects beyond magnetic tuning, as it allows for ultrafast and local control of the interparticle interactions. This idea has been theoretically proposed~\cite{Fedichev1996ion,Bohn1997pfi} and experimentally realised in alkali~\cite{Fatemi2000ooo,Theis2004tts,Thalhammer2005iao}, alkali-earth~\cite{Zelevinsky2006nlp,Blatt2011moo} and non-magnetic Ln atoms~\cite{Enomoto2008ofr,Yamazaki2010ssm}. It has not yet been applied to magnetic atoms. Yet, the rich atomic spectra of magnetic Lns (see e.g.\,Sec.~\ref{subsec:atomlightLn}) make optical FRs promising candidates for controlling $\as$ while minimising heating and atom loss associated with photo-association and light-induced inelastic collisions; see e.g.~\cite{Ciurylo2005oto}.  Finally, we note that magnetic FRs could also be controlled optically, using state-dependent light shifts. Such a tuning was demonstrated with alkali atoms~\cite{Bauer2009coa,Fu2013oco,Williams2013rii,Cetina2015doi,Clark2015qdw,Jagannathan2016oco}. This technique may also bear fruit in magnetic atomic systems thanks to the large spin-dependence of their light coupling. 

In addition to $\as$ tuning, the FR provides access to a weakly-bound dimer (i.e., molecule) state.  This is the closed-channel bound state  dressed by the open-channel scattering state. Close  to the resonance, the dimer becomes extremely weakly bound and its binding energy and size conform to universal scaling laws, which are solely dictated by $\as$ and are insensitive to short-range details. It is called a \emph{halo dimer} because of its large spatial extent. This weakly-bound dimer state can be populated through a dynamical tuning of $E_{\rm cl}-E_{\rm o}$ via, e.g., a ramp of the $B$-field.  This  leads to the formation of so-called \emph{Feshbach molecules}~\cite{Donley2002amc,Chin2010fri, Koehler2006poc,Ferlaino2009ufm}. We note that such molecules can be further transferred to a more deeply bound state via coherent  optical adiabatic transfer schemes~\cite{Bergmann1998cpt}. 

In the case of alkali-metal atoms in their absolute ground state, the coupling between the channels involved in magnetic FRs  is of hyperfine origin. 
In magnetic atoms,  the coupling can be of a distinct origin due to the anisotropy of the interparticle interaction potential. This anisotropy couples channels of different $\ell$ and $m_{\ell}$ and thus may yield FRs; see also Secs.\,\ref{subsec:anisotropy},\,\ref{subsec:processes}. We note that the bosonic isotopes of Cr, Er, and Dy 
do not possess hyperfine structure, unlike alkali atoms.  Thus, in these atoms, the hyperfine FR mechanism is absent, and FRs arise  from only interaction anisotropy. In contrast, their fermionic isotopes possess hyperfine structure, and thus both coupling mechanisms are present. Furthermore, we highlight that magnetic atoms have large total $F$ in their ground state. This means that there are many more collision channels that can be coupled than in alkali-metal atoms.  Finally, we note that the long-range character of the DDI also enables entrance channels of $\ell \neq 0$, even for ultracold temperature, increasing once more the number of channels that must be accounted for in the collision problem. Next, we review the collisional properties of Cr and Ln atoms, respectively.

\subsubsection{Feshbach resonances in Chromium}

\begin{figure}[ht!]
\includegraphics[width=\columnwidth]{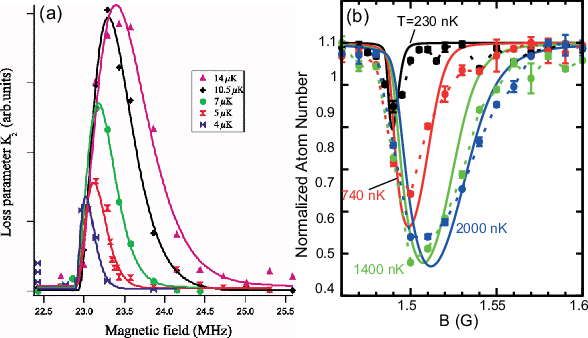}
\caption{\textit{Temperature variations of two FR line-shapes (a) around $B=8.3\,$G in $^{52}$Cr, adapted from Ref.~\cite{Beaufils2009fri} and (b) around $B=1.5\,$G in $^{168}Er$, adapted from Ref.~\cite{Maier2015eoc} (Temperature values given in legend.) In (a), the line shape has been reinterpreted in terms of loss parameter, and the solid line results from an analytical modelling of the loss process with a $d$-wave entrance channel. In (b), the solid lines result from coupled-channel calculations on the three-body process in such a channel. }} \label{dwave}
\end{figure}

The bosonic isotopes of Cr 
have a spherical orbital wave function in its ground state ($^7S_3$) and  no nuclear spin. Thus, the short-range van der Waals scattering of two ground-state bosonic Cr atoms is isotropic and no interaction of hyperfine origin between collision channels arises. Consequently, a single anisotropic DDI potential explains the emergence of FRs in this case. The DDI couples channels with orbital angular momentum $\Delta \ell=0,2$, and the total spin can vary by $\Delta S =0,2$; see also Secs.\,\ref{subsec:anisotropy},\,\ref{subsec:processes}. Thanks to its relative simplicity, the scattering features in this case are very well accounted for by multichannel scattering calculations.

The first observation of FRs between spin-polarised $^{52}$Cr atoms in their absolute ground state (leading to a total spin of the entrance channel $S=6$, $m_S=-6$) revealed the existence of 14 FRs between 0 and 600\,G~\cite{Werner2005oof,Stuhler2007uca}. Coupled-channel calculations show that the relevant closed channels are those with $\ell=2,4$ and with $S=2,4$ and 6.  This  means that  second and fourth-order mixing are relevant. In addition, it was shown that essentially a single closed-channel bound state contributes to each FR. In this way, all except one of the FRs could be assigned. The DDI provides the necessary coupling terms, 
but only slightly affects the FR positions through the molecular potential $V_{\rm ch}(r)$ themselves. Finally, the background scattering length can also be theoretically estimated from the comprehensive coupled-channel calculations, giving $\abg=112(14)a_0$. 

Remarkably, two FRs, the ones with smallest $B_0$, are explained by a $d$-wave entrance channel resonant with a $s$-wave closed channel~\cite{Werner2005oof}. Such a  $d$-wave entrance channel is intrinsic to the DDI coupling mechanism. One of these two FRs was later  experimentally characterised~\cite{Beaufils2009fri}. The corresponding atom loss feature is asymmetric in $B$ and shows an unconventional dependence with decreasing temperature, shifting to lower $B$ and with decreasing width and amplitude; see Fig.\,\ref{dwave}a. The atom losses are shown to occur via three-body recombination, but with a recombination coefficient depending linearly on the density $n$, in contrast to the usual $n^2$ dependence of three-body processes. This unusual behaviour is attributed to the slow tunnelling through the centrifugal barrier in the collision channel, and the adiabatic elimination of the fast collision with a third colliding atom. A precise analysis of this resonance yields a more accurate estimate of the scattering length: $\abg=102.5(4)a_0$~\cite{Beaufils2009fri}.

Besides the scattering between atoms in their lowest spin states, the analysis of Feshbach spectra and their comparison to coupled-channel calculations 
also provides information about the strengths of the spin-dependent scattering. Generally speaking, scattering strengths depend on the total spin of the pair of atoms colliding, which is preserved during the collision in absence of the DDI. The total spin $\mathcal{S}$ of the pair defines a spin channel, corresponding to a given molecular potential and thus to a distinct scattering length $a_{\mathcal{S}}$. The spin-polarised case of Cr corresponds to ${S=6}$ and the background scattering length mentioned above is actually that of $a_{S=6}$. In Ref.~\cite{Werner2005oof}, additional spin channels were characterised, with $a_{S=4}=56\,a_0$, $a_{S=2}=-7\,a_0$~\cite{Werner2005oof}.  Additional analysis of bulk spin relaxation dynamics resulted in a determination of $a_{S=0}= 13.5\,a_0$~\cite{depaz2014das};  see also Sec.~\ref{sec:scatt}. The large differences between these values provide evidence for the importance of spin-dependent scattering in Cr. Knowing this is important for investigating spinor physics with Cr; see Secs.~\ref{spinorsection}--\ref{sec:lattice}.

\subsubsection{Feshbach resonances in magnetic Lns}\label{subsec:feshbachLn}

\begin{figure*}[ht!]
\includegraphics[width=0.98\textwidth,trim={0cm 1cm 0cm 0cm},clip]{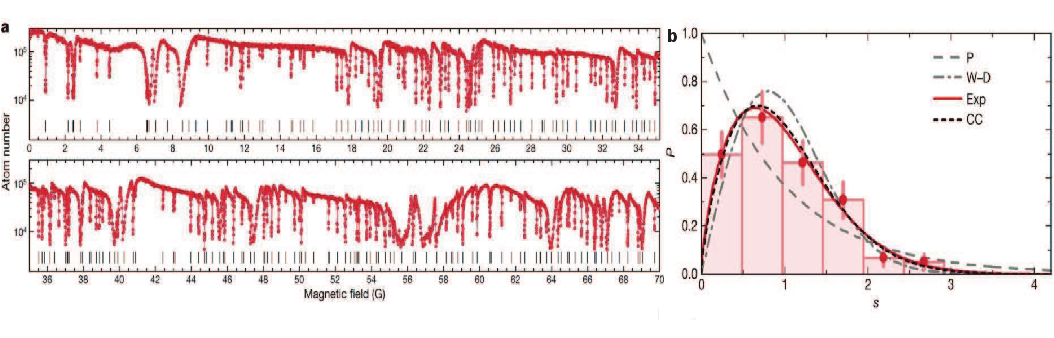}
\caption{\textit{(a) Trap loss spectroscopy measurements performed on a cold ($T=330\,$nK) gas of $^{168}$Er polarised in its absolute ground state, measured after holding 400-ms at $B$. (b) Nearest-neighbour spacing distribution from spectrum (a) (dots). $s$ is the spacing in $B$, renormalised by the mean spacing (inverse of the resonance density). It is compared to the uncorrelated Poisson distribution (dashed line) and to the fully-correlated Wigner Dyson distribution (dashed dotted line) via a fit to the Brody distribution (solid line), measuring the degree of correlation. The short dashed line shows results from coupled channel calculation with $\ell$ up to 20. Figure adapted from Figs.~1a and 4a from Ref.~\cite{Frisch2014qci}}} \label{FREr}
\end{figure*}

The case of magnetic Ln atoms is more intricate than that of Cr because of their  electronic structure is not spherically symmetric. This results in an anisotropy of the van der Waals interaction, which changes the short-range physics~\cite{Kotochigova2011ait,Petrov2012aif,Kotochigova2014cib}. Because both short- and long-range interaction potentials are anisotropic in Lns, they both induce coupling between molecular channels, which leads to FRs. 
Both anisotropic potentials were predicted to substantially contribute to the  character, distribution, and prevalence of FRs in collisions among bosonic atoms~\cite{Petrov2012aif}.  A large number of scattering channels contribute to each resonance, making it hard for coupled-channel calculations based on partial-wave decomposition to converge. Indeed, in Ref.~\cite{Petrov2012aif}, $\ell$ up to 10 were  considered, and later, this was extended up to 20~\cite{Frisch2014qci}.  An analytical model estimates that $\ell$'s up to $\gtrsim 40$ must be considered to reproduce experimental observations~\cite{Frisch2014qci}. Needless to say, no perturbative treatment can be safely applied. 

First experimental observations of FRs in magnetic Lns were performed on $^{168}$Er and revealed an unusually large number of FRs within a narrow magnetic field range~\cite{Aikawa2012bec}.  Soon after, theory work of Ref.~\cite{Petrov2012aif} predicted that the high Feshbach spectral density is a general feature of magnetic Ln atoms. Systematic high-resolution trap-loss spectroscopy later extended the observations of dense FRs in two bosonic Er isotopes from 0 to 70~G~\cite{Frisch2014qci}; see Fig. \ref{FREr}a. A statistical analysis of the Feshbach spectra revealed correlations between the resonance locations, with very similar characteristics for the two bosonic isotopes. Based on the formalism of random matrix theory, these correlations were quantified and found to be consistent with chaotic behaviour in the scattering of these atoms; see Fig. \ref{FREr}b. 
The fermionic case was also studied and shows a ten-fold  larger density of FRs, increasing from 2.7 resonances per Gauss for the bosons to 25.7 resonances per Gauss for the fermions. This was attributed to the additional role of the hyperfine structure. A similarly high density of FRs was concurrently reported  for four Dy isotopes, three bosonic and one fermionic, up to 6 G~\cite{Baumann2014ool}. These measurements also revealed a much higher density for the fermionic isotopes as well as an intriguing temperature dependence of the Feshbach spectrum.

The chaotic behaviour of both Er and Dy was thoroughly analysed and compared in the collaborative work of Ref.~\cite{Maier2015eoc}. Interestingly, even though Dy shows a higher density of FRs (4.3 resonances per Gauss), the degree of correlation in the resonance locations is similar for the two species, and in fact, slightly larger for Er.  This work also introduced a new scheme for coupled-channel calculations, employing a basis comprised of $B=0\,$G Hamiltonian eigenstates.  Such a basis is more amenable to non-central scattering potentials, allowing more rapid computational convergence. This theoretical analysis established the leading role played by the van der Waals anisotropy in the chaotic scattering behaviour: the DDI alone cannot account for the correlations. The slightly larger degree of correlation observed in Er FRs may be attributed to a larger short-range anisotropy.  

The temperature dependence was  further analysed in Ref.~\cite{Maier2015eoc}. At a statistical level, spectra at higher temperature exhibit a larger density of resonances with the degree of correlation that is nearly unchanged. The temperature dependence of individual resonances reveals behaviour compatible with $d$-wave entrance channel predictions, similar to the case in Cr~\cite{Beaufils2009fri}; see Fig.\,\ref{dwave}b. The weakening and disappearance  of higher partial-wave entrance channels in the ultracold regime were proposed as an explanation for the variation of the FR density with temperature. A subsequent work using Ho atoms also revealed a dense FR spectrum as well as a similar temperature dependence ~\cite{Khlebnikov2019rtc}. In contrast to the Er system~\cite{Maier2015eoc}, the chaotic character of the Ho FR spectrum was observed to increase with $T$, ranging from intermediate at $T~2\mu$K to almost fully chaotic at $T~12\mu$K. In the high-temperature regime, the shifts of Ho $d$-wave resonances deviate from a linear dependence on $T$, which is expected at low temperatures, as observed in Er~\cite{Maier2015eoc}.
The authors speculate that this behaviour is responsible for the observed change in FR statistics.

Because of the complexity and non-perturbative character of the coupling between channels, the FRs cannot in general  be assigned to a particular bound state. A few assignments were,  however, assigned in the low magnetic field region of Er~\cite{Frisch2015udm}.  This was accomplished by comparing measurements of the molecular state binding energy with coupled channel calculations.

Broader FR resonance can be observed on top of the forest of narrow FR features in the Ln atoms.  Such features were already observable in the early FR scans of Ref.~\cite{Frisch2014qci}, where, e.g., a  3.5\,G-wide FR is found at 57\,G for $^{168}$Er; see Fig.~\ref{FREr}a. A detailed study of such features was performed in $^{164}$Dy~\cite{Maier2015buf} up to 600\,G and revealed several more broad features. Two of these broad features were extensively characterised via spectroscopy of the molecular state binding energy. Universal properties were found that are characteristic of open-channel-dominated $s$-wave FRs. Particularly noticeable is the decoupling of the broad FR from the chaotic background of the narrow FRs, allowing one to observe the universal behaviour of the halo dimer binding energies over several Gauss to the low-$B$ side of the FRs; see Fig.~\ref{Fig:Dy76G}. A similar study was later reported  in $^{162}$Dy~\cite{Lucioni2018ddb}: two broad FRs with $s$-wave character are found at magnetic fields in the 20--30~G range.  Broad FRs are also observed in Ho~\cite{Khlebnikov2019rtc}.

\paragraph{Scattering length measurements}
\label{scatlengthmeas}

As is the case for other multi-electron atoms~\cite{Pethick2002book}, ab initio calculations fail to predict the scattering lengths of Ln atoms.  Instead, these must be experimentally measured. A common technique involves the observation of cross-rethermalisation:  one observes the rethermalisation dynamics of a thermal sample after exciting it along one of its the three spatial directions~\cite{Monroe1993moc}; the thermalization rate is proportional to the elastic cross section, which in turn is proportional to the square of the scattering length. The proportionality relies on a theory prediction for the average number of collisions necessary for thermalisation and depends on the particular partial waves and interaction involved in the collision~\cite{DeMarco1999mop}.  Bohn \textit{et al.}\cite{Bohn2014dsa} calculated this for the anisotropic DDI, which was verified experimentally in Ref.~\cite{Aikawa2014ard} using spin-polarised fermionic Er. This theory allowed Tang \textit{et al.}~\cite{Tang2015sws,Tang2016esw} to extract scattering lengths from cross-rethermalisation measurements of two Dy isotopes~\cite{Tang2015sws,Tang2016esw}.  The theory-experiment comparison resulted in $\as=122(10)\,a_0$ for $^{162}$Dy and $\as=92(8)\,a_0$ for $^{164}$Dy at $B=1.58\,$G. This field is at least $0.1\,$G away from any FR in either  species.
These $\as$-values, as well as their $B$-dependence, were also measured in  Ref.~\cite{Tang2016aeo} using a different method involving a detailed analysis of the anisotropic expansion of thermal gases after  release from their confining potentials., The analysis relied on the competition of the DDI and the contact interaction; see Sec.~\ref{subsec:magnetostriction} for details. The results are consistent with the cross-rethermalisation experiments: $\as=154(22)\,a_0$ for $^{162}$Dy and $\as=96(22)\,a_0$ for $^{164}$Dy, but with larger error.  A similar measurement near a 5~G FR in $^{162}$Dy yielded a background scattering length of $\as=157(24)\,a_0$, which is  consistent with the lower-field measurement, but larger than that obtained with cross-rethermalization~\cite{Tang2015sws,Tang2016esw}.
A third scheme relies on the universal scaling of the molecular binding energy close to broad FRs to extract scattering lengths in Dy. Such an analysis used a relatively high-field resonance in $^{164}$Dy to obtain  $\abg=91(16)\,a_0$~\cite{Maier2015buf}, which is in good agreement with the low-$B$ values of Refs.\,\cite{Tang2015sws,Tang2016aeo}. A similar study in $^{162}$Dy yielded $\abg=220(50)\,a_0$~\cite{Lucioni2018ddb}. This is consistent with the $\as=154(22)\,a_0$ measurement in Ref.\cite{Tang2016aeo}, but larger than both the 5-G measurement in Ref.\cite{Tang2016aeo} and the $\as=122(10)\,a_0$ measurement of Ref.\,\cite{Tang2015sws}.
  
For Er, preliminary cross-rethermalisation measurements were first reported in Ref.~\cite{Frisch2014dqg} for the four bosonic isotopes in the low-field regime. A simpler analysis of the rethermalisation rates than used in Ref.\,~\cite{Tang2015sws} yielded $\abg=81(10)\,a_0$ for $^{164}$Er, $\abg=72(13)\,a_0$ for $^{166}$Er, $\abg=200(23)\,a_0$ for $^{168}$Er, and  $\abg=221(22)\,a_0$ for $^{170}$Er. In $^{170}$Er, the background scattering length was first thought to be negative, but more recent measurements have questioned this~\cite{Trautmann2018dqm}. These Er $\abg$ measurements were taken by averaging data obtained at different $B$-values between 0.2\,G and 1\,G, away from FRs. A more precise estimate was later developed to extract $\as$ and its $B$ dependence by loading a quantum gas into a deep 3D lattice and performing lattice-modulation spectroscopy~\cite{Baier2016ebh}. This probes the Mott-insulator excitation gap, which is related to the scattering length through the on-site interaction energy; see also Sec.~\ref{subsec:spinless_lattice}.  This technique yielded  $\as=137(1)a_0$ at 0.4\,G for $^{168}$Er~\cite{Baier2016ebh}, and for  $^{166}$Er, it was used to make a sequence of measurements of  $\as$ versus $B$ from  0-to-2.5\,G~\cite{Chomaz2016qfd}. In this range, the data are well-described by overlapping FRs with a $B$-dependent background scattering length $\abg(B) =62(4)+k B$, with $k=5.8(1.2)\,a_0/{\rm G}$. 

In the case of Ho, recent experiments reported scattering length values using cross-rethermalisation.  The first such measurement yielded a background value of $\abg=144(38)\,a_0$~\cite{Khlebnikov2019rtc}, while a subsequent experiment reported $\abg=90(11)\,a_0$ via a more extensive set of measurements~\cite{Davletov2020mlf}. 

We note that a debate regarding the exact values of the scattering lengths in bosonic Er and Dy in the quantum degenerate regime has arisen in light of measurements of these parameters extracted from the many-body physics of the droplet and supersolid states; see also Sec.~\ref{Sec:DCQD}.  These differ from the two-body collision-based methods described above.  Central to the debate is the  appropriateness of the description of the many-body physics based on perturbative mean-field and local-density approximations, on the one hand, and on the other, regarding the relevance, in the regime of interest, of corrections beyond the Born approximation and the role of the momentum dependence in the two-body scattering description. That is, the  many-body theory used to extract scattering lengths in such finite-size quantum systems could be as problematic as the use of ultracold thermal gases in the former collision-based measurements.   Those questions are beyond the scope of the current section, but additional discussions may be found in Secs.~\ref{subsec:theory_droplet} and~\ref{subsec:droplet_props}. 

Concerning the scattering length itself, the most extended discussions focus on the $^{164}$Dy case. Studies of single quantum droplets provide estimates of the background scattering length via a theory-experiment comparison of different many-body properties, namely the critical atom number for the existence of a self-bound droplet state and the frequencies of droplet's elementary excitations~\cite{schmitt2016sbd,FerrierBarbut2018smo,Boettcher2019qci}. These investigations yield values of $\abg=62.5\,a_0$ and $\abg=69(4)\,a_0$ at $B\approx 6.6\,$G and $B=0.8\,$G, respectively. While they are mutually compatible, they are consistently lower than the measurement results in Refs.~\cite{Tang2015sws,Maier2015buf,Tang2016aeo} obtained with thermal gases. Ferrier-Barbut \textit{et al.}~\cite{FerrierBarbut2018smo}  speculate that the discrepancy could come from the different temperatures of the samples used in those two sets of experiments and related momentum dependence of the scattering~\footnote{We note that a similar down-shift of the scattering length can be observed in  Er  based on lattice modulation spectroscopy of a Mott insulator  versus that from cross-rethermalisation of a thermal sample. This might also be interpreted as arising from the momentum dependence of the scattering.}. 
Reference~\cite{Boettcher2019qci} extends the measurements of the critical atom number of a self-bound quantum droplet (see Sec.\,\ref{subsec:droplet_selfbound}) and was able to extract the ratio between the background scattering lengths of the two dysprosium isotopes to be $a_{\rm bg,162} / a_{\rm bg,164} = 2.03(6)$. Using $a_{\rm bg} =  69(4) a_0$ for \textsuperscript{164}Dy~\cite{FerrierBarbut2018smo} yields a value of $a_{\rm bg} =140(7) a_0$ for the background scattering length of \textsuperscript{162}Dy. This agrees with most of the values obtained from  thermal samples within their uncertainty~\cite{Tang2015sws,Lucioni2018ddb,Tang2016aeo}.
In the Er case, we note that sensitive probes based on collective excitations of single macro-droplets~\cite{Chomaz2016qfd},  supersolids~\cite{Natale2019eso}, and roton excitations of dBECs~\cite{Chomaz2017oot,Petter2019ptr} show a good agreement between experiment and theory using the lattice-calibrated scattering lengths values.

Finally, we note that little information is available on Ln spin-dependent scattering, as opposed to the well-characterised Cr system. One exception is the case of the two lowest energy weak-field seeking spin states of fermionic $^{167}$Er, whose scattering have been experimentally investigated in Ref.~\cite{Baier2018roa}; see also Sec.~\ref{spinorsection}. Extensive loss spectroscopy measurements were performed in a limited $B$-field range (up to $2$\,G) and for varying mixture compositions.  This  enabled the identification of both intra-spin FRs and inter-spin FRs; spin-polarised loss spectra have also been observed in $^{161}$Dy for its lowest energy weak-field seeking spin states~\cite{Burdick2016lls}. Lattice modulation spectroscopy also characterised the $B$-to-$\as$ conversion close to a relatively broad inter-spin FR. The background scattering for the interaction of the two lowest spin-states is  $\abg=91(8)\,a_0$.  Theoretical investigations have also investigated spin-dependant resonances ~\cite{Ming2018oqm}. 

An important difference between Lns and Cr lies in the origin of spin-dependent scattering. 
In contrast to Cr, the dominant effect in Lns arises from their orbital anisotropy (i.e., $L\neq 0$; see also Sec.\,\ref{subsec:Ln}).  This induces an interaction term that couples the angular momentum of each individual atom to the orbital momentum of the pair and results in sets of molecular potentials with very different character versus Cr atoms.
More systematic studies, especially experimental, are needed to fully elucidate the spin-dependent collisional physics. A relevant open question relates to the density of the Feshbach spectra in Ln spin mixtures and in fermionic Ln isotopes.  Regarding the latter, the density of FR spectra at high field in fermionic Dy seem to be spin-dependant~\cite{Burdick2016lls}. Ericson fluctuations, known from resonances in nuclear physics, may provide an appropriate framework for explicating the field dependence of the overlapping loss spectra~\cite{Mayle2012sao} observed in Ref.~\cite{Burdick2016lls}. Interspecies scattering involving Ln atoms is also of recent interest~\cite{Trautmann2018dqm,Ravensbergen2018poa,Ravensbergen2020rif,Zaremba2018mtf}.

\subsubsection{Feshbach molecules of magnetic atoms}

Besides tuning of $\as$, FRs provide the ability to associate a pairs of free atoms into a loosely bound molecule~\cite{Donley2002amc,Chin2010fri, Koehler2006poc,Ferlaino2009ufm,Moses2017nff}. A FR 
couples two states, the state of the two free particles scattering in the entrance channel and the bound state of the closed channel. The relative energy of the two states is tunable via the $B$-field thanks to their different magnetic moments.  The coupling of the two states yields an avoided crossing at $B=B_0$ where their bare energies coincide. By ramping $B$ across the resonance (from $B>B_0$ to $B<B_0$), one can then drive the population of the molecular state by (adiabatically) following the low-energy state~\cite{Donley2002amc,Chin2003sdo,Regal2003mop,Cubizolles2003pol,Herbig2003poa,Hodby2005peo}. Alternative approaches use small-amplitude modulation of $B$ at a frequency resonant with the bound-state energy for $B \lesssim B_0$~\cite{Hanna2007aom,Thompson2005ump}, or even further away from the FR via radiofrequency coupling~\cite{Weber2008aou,Klempt2008rfa}.  Such protocols are commonly used in alkali metal experiments, with typical conversion efficiency of few tens--of--percents in the bosonic 
case and close to unity in the fermionic case. Note that these molecules are not in their internal ground-state, but in a highly excited (rovibrationnal) state. Close to the FR, the properties of the weakly-bound molecule, referred to as a \emph{halo-dimer}, follow a universal behaviour.

By producing Feshbach molecules of magnetic atoms, one creates ultracold gases of dipolar particles with even larger permanent dipole moments.  Indeed, compared to the bare atoms, the mass $m$ of the molecule is doubled while the magnetic moment $\mu$ is also nearly doubled, resulting in a dipolar length $\add \propto m\mu^2$ eight times longer. 


An experiment using Er  has demonstrated and studied the production of an ultracold sample of Er$_2$ Feshbach molecules using $B$-field ramps~\cite{Frisch2015udm}. Pure molecular samples were  produced by the resonant removal of the remaining free atoms. The gas contained about $2\times 10^4$ Er$_2$ at 300\,nK with typical densities of $\sim8\times 10^{11}$\,cm$^{-3}$ and lifetimes of $\sim 10\,$ms. Ultracold samples of molecules in four different  molecular states, associated with four different low-field FRs of $^{168}$Er, were produced. The properties of the Feshbach molecules have been further measured, in particular the magnetic moment, which were found to be  $\mu/\muB>11$ for three of the four states. By contrast, the bare atomic value is $\mu/\muB\approx 7$.
The impact of the confinement geometry and relative dipole orientation on the scattering properties of the Feshbach molecules has also been investigated. In particular, a reduction of the relaxation rate in quasi-two dimensional geometry with  out-of-plane dipole orientation was observed. 

In the Cr case, the study of Feshbach molecules has focused on a  single peculiar resonance with an $d$-wave entrance channel~\cite{Beaufils2009fri,Beaufils2010rfa}. However,  no ultracold sample has yet been produced. In this case, the molecular state is very short lived and studies have focused on the remarkable atom-loss behaviour associated with the special character of the resonance. 
\begin{figure}[htbp]
  \includegraphics[width=\columnwidth]{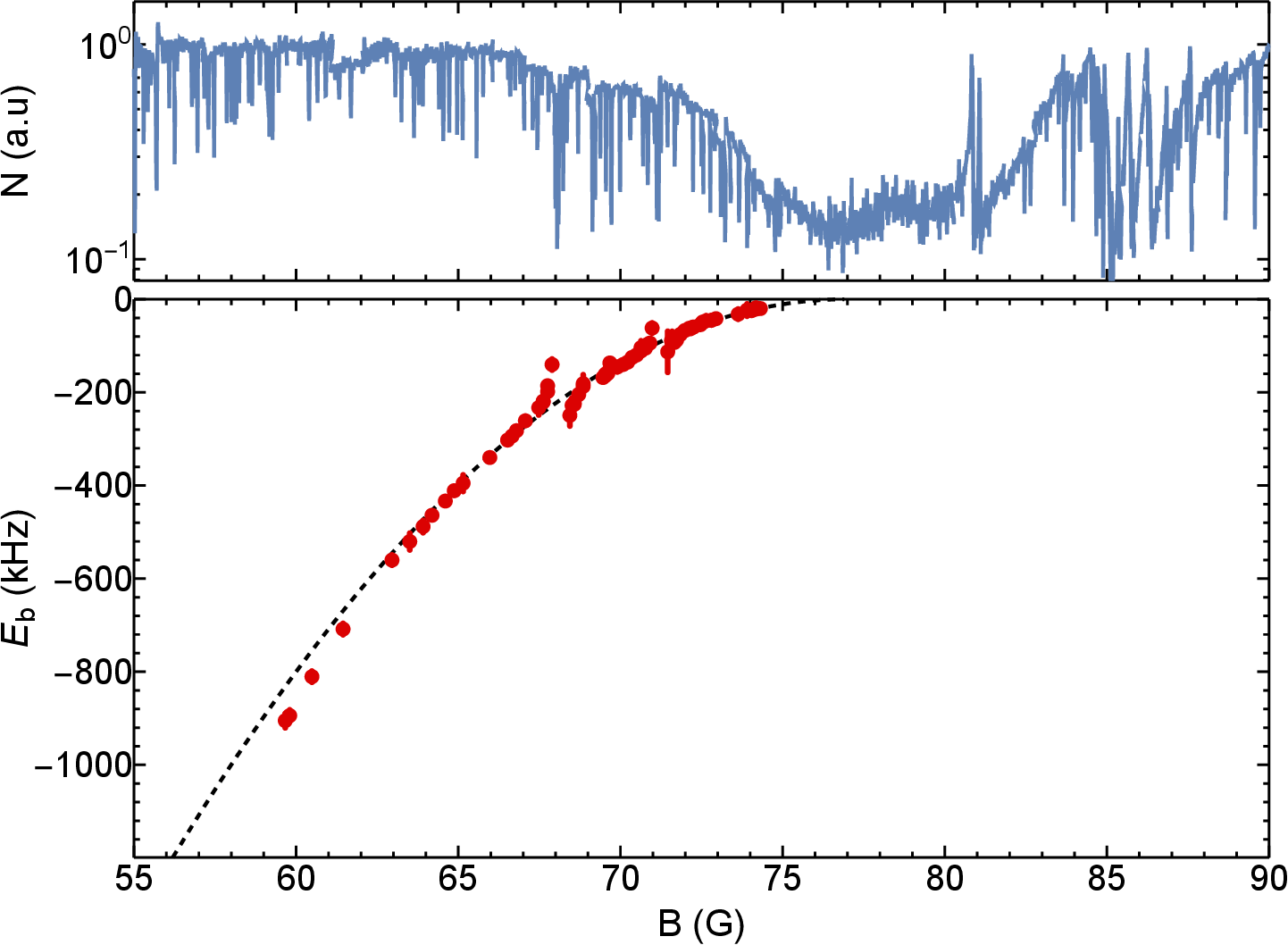}
  \caption{Atom-loss spectrum (top) and measured molecular binding energy (bottom) of $^{164}$Dy near a broad FR. The atom loss spectrum exhibits a very high density of narrow resonances and one broad feature. The binding energy of the weakly bound dimer displays a universal quadratic behaviour (dashed line). The very weak variation of the binding energy as a function of magnetic field shows that the dimer has a magnetic moment very close to that of two free atoms $\mu\approx20\muB$. Data originally published in~\cite{Maier2015buf}. \label{Fig:Dy76G}}
\end{figure}

In the Dy case, studies have focused on the properties of the halo dimers close to the broad FRs found at relatively large $B$ in two bosonic isotopes\,\cite{Maier2015buf,Lucioni2018ddb}, see also Sec.~\ref{subsec:feshbachLn}. The corresponding magnetic moments range from $\mu \approx 18\muB$ to $\mu \approx 20\muB$; see Fig.~\ref{Fig:Dy76G}. A striking observation is that these halo dimers decouple from the many bound states coming from the overlaid forest of narrow FRs. This makes the broad FRs very promising for the production of ultracold gases of highly magnetic molecules. They have not yet been investigated, but this constitutes an interesting prospect for future work.

\begin{table*}[t!]
\centering
\caption{Summary of the main atomic properties, scattering properties, and available quantum gases of highly magnetic atoms made of Cr, Er or Dy.  The electronic ground state is denoted ``g.s." The exchange statistic---fermionic (f) or bosonic (b)---is listed under ``stat." The natural abundance is under ``abound. denotes.: $I$ is the nuclear spin quantum number. $\edd$ is  the background value $\add/\abg$. ``Degenerate gas" lists the  properties  of the  most populous gas of that isotope produced, as reported in journal articles. For all these, the BEC's are almost pure; i.e., the thermal fraction is small. }
\label{table_expgases}
\begin{tabular}{|c|c|c||c|c|c|c|p{2.5cm}|p{1cm}|p{6cm}|}
  \hline
  atom & g.s. & $\mu_m(\muB)$ &  stat.:isotope & abund. & $I$& $\add (a_0)$& $\abg(a_0)$ & $\edd$ & degenerate gas\\
  \hline
  
  \hline
 Cr & $^{7}S_3$& 6 &  b:$^{52}$Cr & 83.79\% &0 & 15.3 & 102.5(4)~\cite{Pasquiou2010cod} & 0.150(6) & BEC, $5\times 10^4$~\cite{Griesmaier2005bec,Beaufils2008aop}\\
  
  \cline{4-10}
 &&& f:$^{53}$Cr & 9.5\% &$3/2$& 15.6 & -- &--& DFG, $10^3$, $\frac{T}{T_{\rm F}}\sim 0.6$~\cite{Naylor2014cdf}\\
  
  \cline{4-10}
 &&& b:$^{50}$Cr & 4.35\% &0& 14.7 &40(15)~\cite{Schmidt2003dot} & 0.37(14) & none \\
  
  \cline{4-10}
 &&& b:$^{54}$Cr & 2.36\% &0& 15.9& unknown &  & none \\

   \hline 
    Er & $^{3}H_6$& 7 & b:$^{168}$Er &26.8\%&0 & 66.3 & 137(1)~\cite{Baier2016ebh} & 0.484(3) & BEC, $1.5\times 10^5$~\cite{Frisch2014dqg,Aikawa2012bec,Phelps2020ssp}\\
    
\cline{4-10}
 &&& b:$^{166}$Er & 33.6\% &0 & 65.5 & 68(5) at B=1G~\cite{Chomaz2016qfd} & 0.96(7) & BEC, $10^5$~\cite{Chomaz2016qfd}\\
 
 \cline{4-10}
 &&& f:$^{167}$Er &23\%& 7/2& 65.9 & -- & --& DFG: 
 $4\times10^4$, $\frac{T}{T_{\rm F}}=0.1$~\cite{Frisch2014dqg,Aikawa2014rfd,Phelps2020ssp}\\
 
 \cline{4-10}
 &&& b:$^{170}$Er & 15\% &0 & 67 & preliminary estimate 221(22)~\cite{Frisch2014dqg} & 0.30(3) & BEC, $~10^4$~\cite{Trautmann2018dqm}\\
 
  \cline{4-10}
 &&& b:$^{164}$Er & 1.6\% &0 & 64.7 & preliminary estimate 81(10)~\cite{Frisch2014dqg} & 0.8(1) & none\\
 
  \cline{4-10}
 &&& b:$^{162}$Er & 0.14\% &0 & 63.9 &  unknown &  & none \\
 
    \hline 
    Dy & $^{5}I_8$& 10 &  b:$^{164}$Dy &28.3\%& 0& 130.7 & disagreeing estimates: 69(4)~\cite{Boettcher2019qci,FerrierBarbut2018smo} or 92(8)~\cite{Tang2015sws,Maier2015buf,Tang2016aeo} & 1.9(1) or 1.4(1) & BEC, $1.5\times 10^4$~\cite{Lu2011sdb,Maier2016iic, Kadau2016otr}, $3.5\times 10^4$~\cite{Chomaz2019lla}, with typically sizeable thermal fraction remaining.\\
    
    \cline{4-10}
 &&& b:$^{162}$Dy & 25.5\%& 0&129.2& most precise estimates are 122(10)~\cite{Tang2015sws}--157(4)~\cite{Tang2016aeo} & 0.92(5)-1.06(9)& BEC, $10^5$~\cite{Tang2015bec},$5\times10^4$\cite{Tanzi2019ooa,Boettcher2019tsp}\\
 
 \cline{4-10}
 &&& f:$^{163}$Dy &24.9\%&5/2&130& -- & -- &  none\\ 
 \cline{4-10}
 &&& f:$^{161}$Dy &18.9\%&5/2&128.4& -- & -- &  DFG: 
$4\times 10^4$, $\frac{T}{T_{\rm F}}=0.1$~\cite{Burdick2016lls}, $8\times 10^4$, $\frac{T}{T_{\rm F}}=0.085(10)$\cite{Ravensbergen2018poa}\\
 \cline{4-10} 
&&& b:$^{160}$Dy & 2.3\%& 0&127.6& unknown & & BEC, $10^3$~\cite{Tang2015bec}\\

\hline
  
\end{tabular}
\end{table*}

\subsection{Comparison between the different systems 
}\label{subsec:compareSystems}

Table~\ref{table_expgases} summarises the electronic, isotopic, scattering-related properties as well as the characteristics of all the available degenerate quantum gases for all isotopes, both bosonic and fermionic, of the highly magnetic elements, Cr, Er and Dy.  The background $\edd$ varies among the elements and isotopes; e.g., it is $\edd\sim 0.16$ for Cr, varies  between $0.9\lesssim \edd\lesssim1.9$ for Dy isotopes, and is between $0.3\lesssim \edd\lesssim1$ for Er isotopes. 
All isotopes exhibit FRs, which provide opportunities for tuning $\edd$; see Secs.\,\ref{subsec:fesbachsec},~\ref{Sec:RepulsiveGases}, and \ref{Sec:DCQD}. 

With respect to degenerate gases, all three species offer relatively large BECs of roughly similar sizes. 
The maximum population of the BEC depends on the isotope, not only because of its natural abundance, but also because of its scattering properties. In particular, isotopes with a background value of $\edd\gtrsim 1$, or $\edd<0$ (i.e., with a negative scattering length), are more challenging to condense.  They typically require a tuning of $\as$ during the evaporation process by setting  $B$  close to a FR to stabilise the quantum gas at the mean-field level. This is in particular the case for the most abundant isotope of Dy, $^{164}$Dy, for which the BEC numbers are otherwise  smaller than for the $^{162}$Dy isotope and Er's most populous BEC. In fact, we point out that, due to the extreme density of FRs in the spectra of open-shell Lns, the choice of the $B$ value at which the evaporative cooling scheme is performed is a crucial  parameter to adjust for all isotopes of Er and Dy, and even more so for their fermionic isotopes. The population of the achieved quantum gases greatly depend on it. 
Large and cold degenerate Fermi gases of Dy and Er are also possible due to the ability to perform direct evaporative cooling. The quantum degeneracy of fermionic Cr has been achieved, but so far, is smaller and not as cold with respect to the Fermi temperature. 

%% file: Section3/3_DipolarScattering.tex
\section{Ultracold dipolar scattering}
\label{sec:scatt}

The notion of scattering in the ultracold regime has been briefly introduced in Secs.~\ref{sec:intro} and~\ref{sec:amgnetic_atoms}. We will now extend our overview on ultracold collisional physics before describing the special features of the scattering induced by the DDI.  Topics include the universal properties of elastic dipolar scattering (i.e., coming from spin-conserving collisions; see Sec.~\ref{elasticdipolarscattering}); the special features of inelastic dipolar scattering (i.e., dipolar relaxation), including its local character and suppression via exploiting quantum statistics and  confinement (Sec.~\ref{Drelaxationsec}); and finally the anisotropic character of (elastic) dipolar scattering (Sec.~\ref{subsec:anis_scatt}).

\subsection{Relevant aspects of scattering theory}
\label{scatttheory}

Collisions among atoms in dilute ultracold gases are described within quantum scattering theory, which accounts for the long-distance behaviour of the wavefunction encompassing the relative motion of two colliding atoms, the so-called \emph{collisional wavefunction}~\cite{Landau1965book,Pethick2002book,Dalibard1999cdo}. In brief, any incident wave is decomposed in plane waves of momentum $\vk$, whose scattering yields outgoing spherical waves of amplitude $f(\vk,{\boldsymbol n})$ in the direction ${\boldsymbol n}=\vecr/r$. The scattering amplitude $f$ directly relates to the interaction potential $V(\vecr)$ and, within the first-order 'Born' approximation, $f$  simply reduces to the Fourier transform of $V(\vecr)$. The scattering cross section $\sigma$ then corresponds to the spherical integration of the square norm of $f$ over the scattering direction  $\sigma(\vk)= \int |f(\vk,{\boldsymbol n})|^2d^2n$.

\subsubsection{Partial wave expansion}

In the case of an isotropic potential $V(\vecr)=V(r)$, expanding the collisional wavefunction in spherical harmonics is a very powerful tool as each of the resulting radial waves are decoupled from  one another. These are indexed by $l$ (and $m_l$), corresponding to the  quantum numbers for the orbital angular momentum norm (and projection). For an isotropic potential, only spherical harmonics with projection $m_l=0$ contribute, while in the case of anisotropic interactions, $m_l\neq 0$ harmonics may also play a role.  Each spherical harmonic component of the expansion is an independent solution of the 1D Schr\"{o}dinger equation, where the potential $V(r)$ is augmented by a centrifugal term $U_{{\rm c},l}(r) = \hbar^2l(l+1)/m r^2 $. For the isotropic case, asymptotically, the outgoing  wave  differs by only a phase shift $\delta_l$ from the incoming wave. Therefore, the general scattering amplitude $f$ decomposes into 
\begin{eqnarray}
\nonumber
f(k,\theta) &=& \frac{1}{2ik}\sum_{l=0}^{\infty}(2l+1)(e^{2i\delta_l(k)}-1)P_l(\cos\theta)\\
\label{fexpression}
&=&\sum_{l=0}^{\infty}f_l(k,\theta),
\end{eqnarray}
where $\theta$ describes the angle between $\vk$ and $\vecr$ and varies between 0 and $\pi$. Here, $P_l$ denote the ordinary Legendre polynomials. The scattering cross section then reads
\bea\label{sigmaexpression}
\sigma(k) &= & \sum_{l=0}^{\infty}\sigma_l(k), \\ \sigma_l(k) &= &\frac{4\pi}{k^2}(2l+1)\sin^2[\delta_l(k)].
\eea

\subsubsection{Ultracold limit and short-range interactions}
\label{ultracoldshortrange}
For an interaction of short range $b$, we can define the temperature below which the system is in the ultracold collisional regime  as  $k_BT\ll \hbar^2/m b^2$.  In this regime, the centrifugal barrier at $r=b$ is much larger than the typical kinetic energy of the colliding particles such that they do not feel the interaction potential in $l>0$ harmonics (they are reflected at $r>b$). Therefore, only $l=0$ contributes to the scattering.
  
\paragraph{Scattering length:}
In the ultracold limit, the scattering amplitude and the cross section tend to finite values given by the $k \rightarrow 0$ limit of their $l=0$ contributions: $f \approx f_0 \rightarrow -a_{\rm s}$, and $\sigma  \approx \sigma_0\rightarrow 4\pi a_{\rm s}^2$.  This is for the case of distinguishable particles given by the sum over all $l$ in Eq.~\eqref{sigmaexpression}; see  Sec.~\ref{statisticshortrange} for the role of quantum statistics in scattering. The $s$-wave scattering length $\as$ is then the only (non-universal) parameter describing the ultracold scattering physics and it relates to the $s$-wave phase shift via $\as = \lim_{k \rightarrow 0} \delta_0(k)/k$. Hence, in the ultracold regime, the details of the short-range interaction are all represented by the value of a single parameter.  Moreover, potentials yielding the same value of $\as$ are interchangeable. It is therefore useful to employ a particularly simple form of the potential, the contact pseudo-potential $U_s(r) = g\delta(r)$, with $g=\frac{4\pi\hbar^2 \as}{m}$~\cite{Pitaevskii2016bec,Pethick2002book,Dalibard1999cdo,Fermi1934,Huang1957qmm}~\footnote{This simple formula is not regular everywhere. As described in Refs.~\cite{Dalibard1999cdo,Fermi1934,Huang1957qmm}, a more rigorous regular version exists.}.

\paragraph{Power-law potentials and the case of van der Waals interactions:}

The scaling of $\delta_l(k)$ can be deduced from the scattering potential form by solving the associated 1D Schr\"{o}dinger equation.
As introduced in Sec.\,\ref{sec:intro}, in the case of a power-law interacting potential $V(r)=V_n(r) = -C_n/r^n$, one finds that  $\delta_l(k)$ scales as $k^{2l+1}$ if $l<(n - 3)/2$ and as $k^{n-2}$ otherwise~\cite{Landau1977book,Dalibard1999cdo}. Therefore, the ultracold regime described above only holds for potentials decreasing  sufficiently fast, as $n>3$. 

Typically, this condition is satisfied by the  dominant interaction between two atoms in a dilute gas, the van der Waals interaction, which is an $n=6$ power law. This scaling arises from the virtual exchange of photons between two fluctuating electric dipoles:  while on average the electric dipole of each atom is zero (in the absence of an electric field), a dipolar fluctuation of the electronic distribution of one atom can distort the other's, leading to the product $r^{-3}{\cdot} r^{-3}$ for the interaction dependence. The ultracold limit is  defined (in the absence of a direct DDI) by $k_BT\sim \hbar^2/m b^2$.  Associating $b$ with the van der Waals length $\rvdw=(m C_6/\hbar^2)^{1/4}$, we find that the ultracold limit is typically reached for temperatures of hundreds of $\uK$.

In the ultracold limit, the scattering cross section  for partial waves $l={0,1}$ at low collision energy $E=\hbar^2k^2/m$ follows the so-called Wigner threshold law~\cite{Wigner1948otb}: 
\be \label{WignervdW}
\sigma_l(E) \propto k^{4l}\propto E^{2l},
\ee
while for $l>2$, $\sigma_l(E) \propto k^{6}\propto E^{3}$. Only the $l=0$ cross section does not vanish in the $E\rightarrow 0$ limit. This is in contrast to the  case of interacting \textit{polarised} dipoles in which $n=3$. In this case, the above scaling laws---based on the assumption of a short-range potential---are not valid, and we will see (Sec.\,\ref{elasticdipolarscattering}) that all partial waves contribute.

\subsubsection{Role of quantum statistics}
\label{statisticshortrange}
Quantum statistics plays a prominent role in the ultracold regime, as the wavefunction should be appropriately symmetrised for collisions among identical particles, i.e., atoms of the same isotope in the same spin state.  Indeed, the total collisional wavefunction of any pair of atoms---which is comprised of the tensor product of wavefunctions describing both their spin and orbital degrees of freedom---must be symmetric (antisymmetric) under particle exchange for bosons (fermions). For example, if their spin state is identical, then  their motional state must be symmetric (antisymmetric), and thus only even (odd) $l$ partial waves contribute to the partial wave decomposition of Eqs.~\eqref{fexpression} and~\eqref{sigmaexpression}. 
These expressions may be rewritten as:
\be\label{fexpression2}
f(k,\theta) = \frac{1}{ik}\sum_{l=0}^{\infty}\epsilon(l)(2l+1)(e^{2i\delta_l(k)}-1)P_l(\cos\theta),
\ee
where $\theta$ now varies between 0 and $\pi/2$ and $\epsilon(l)$ depends on the statistics of the particles. For bosons (fermions), $\epsilon(l)=0$ for $l$ odd (even) and 1 in the diametric case $l$ even (odd). Similarly:
\bea\label{sigmaexpression2}
\sigma(k) &=& \sum_{l=0}^{\infty}\sigma_l(k),\\ \sigma_l(k) &=& \frac{8\pi}{k^2}\epsilon(l)(2l+1)\sin^2[\delta_l(k)].
\eea 

For van der Waals interactions, the Wigner threshold law, Eq.~\eqref{WignervdW}, implies that the cross section for the lowest partial wave accessible to fermions, $l=1$, vanishes as $E^2$, while for bosons it tends toward a constant $8\pi\as^2$. One key consequence lies in the fact that while ultracold bosonic gases can be evaporatively cooled to degeneracy via short-range elastic thermalizing collisions, identical fermions cannot.

\subsection{Universal elastic dipolar scattering}\label{elasticdipolarscattering}

The above cross sections for  van der Waals interactions may be said to be non-universal, in that they depend on a parameter $\as$ that cannot be calculated from fundamental constants and fixed parameters of the system alone (like mass or quantum numbers).  Indeed, $\as$ is calculable ab initio within state-of-the-art methods only for light systems consisting of few electrons, such as hydrogen and helium~\cite{Przybytek2010raq,Piszczatowski2009tdo}. Accurately predicting the scattering length for more complex systems is very challenging. In contrast, we will see that it is possible to derive the dipolar cross section from first principles. This is due to the long-range nature of $V_{\rm dd}$, which often  allows one to neglect the short-range part of the interaction potential. The longer-range character of the DDI  ($n=3$) also leads to a different Wigner threshold law; namely, one that exhibits a constant, non-zero cross section even for identical fermions.  

In contrast to the form of the $\delta_l(k)$ scaling presented above, for $n=3$, $\delta_l(k) \sim k$ for all partial waves such that they all contribute to the scattering cross section even in the low-energy limit. Specifically, one can write $\delta_l(k) \approx Ak^{2l+1} + Bk$, where $A$ is the strength of the short-range part of the potential and $B$ is the strength of the long-range dipolar portion beyond the centrifugal barrier.   $A$ and $B$ may be calculated under the first-order Born approximation~\footnote{This approximation assumes that the interaction is sufficiently weak that there is no rescattering during the collision.  This means that the incident field can be taken as the total field at the scatterer.}~\cite{Lagendijk1986sea, Shlyapnikov1994dka,Fedichev1996idp,Bohn2009qud,Bohn2014dsa}.  This yields the total cross sections $\sigma_{l\textrm{ even}}$ ($\sigma_{l\textrm{ odd}}$) for collision in even (odd) partial waves:
\bea
\sigma_{l\textrm{ even}} &=& 4\pi \as^2 + \frac{16\pi}{45}D^2, \label{eq:sigmaeven}\\
\sigma_{l\textrm{ odd}} &=& \frac{16\pi}{15}D^2,
\label{eq:sigmaodd}
\eea  
with $D = S^2 d^2 m/\hbar^2 = 3 a_{dd}$ and $a_{dd}$ as  defined in Eq.~\eqref{dipolarlength}. For indistinguishable bosons (fermions) $\sigma=2\cdot\sigma_{l\textrm{ even}}$ ($\sigma=2\sigma_{l\textrm{ odd}}$), while for distinguishable particles $\sigma=\sigma_{l\textrm{ even}}+\sigma_{l\textrm{ odd}}$.

Regardless of the exchange statistics of the atoms or the value of $\as$, the cross sections of dipolar atoms are finite and energy independent as $k\rightarrow 0$. Since $D$  contains only fundamental constants and known quantum numbers, we may consider its contribution to the cross section universal---i.e., not depending on the details of short-range physics~\cite{Bohn2009qud}.  However, this statement ignores the effects of molecular potentials at certain values of magnetic field, as pointed out in Refs.~\cite{Lagendijk1986sea,Pasquiou2010cod} and later discussed in Sec.~\ref{subsec:inel_crosssec}. In addition, as the dipole strength is increased, corrections beyond the first-order Born approximation can cause $D$ to differ from its universal value~\cite{Oldziejewski2016pos}. 

An important consequence of the dipolar Wigner threshold law is the fact that identical dipolar fermions can evaporatively cool themselves even at low temperatures. This effect was used to cool spin-polarised fermionic polar molecules of KRb, using forced evaporation in an optical dipole trap, to $T \approx 2T_F$~\cite{Ni2008ahp}---most recently $T\approx 0.6T_F$ was achieved in a similar system~\cite{Valtolina2020deo}.  Subsequently, a Fermi gas of identical, spin-polarised $^{161}$Dy atoms was evaporated to 
$T/T_F \sim 0.7$ with ${\sim}$3$\times$10$^3$ atoms remaining~\cite{Lu2012qdd}.  The deeply degenerate regime was then reached  with identical fermions of $^{167}$Er~\cite{Aikawa2014rfd}, achieving $T/T_F \approx 0.1$   with several times $10^4$ atoms remaining.  Similar temperatures and trap populations have also been reached with $^{161}$Dy~\cite{Burdick2016lls}. 

In Ref.~\cite{Aikawa2014rfd}, the spin purity of the sample was demonstrated and the elastic cross section was measured via cross-dimensional rethermalisation measurements~\cite{Monroe1993moc}; see also Sec.~\ref{Anisotropicrether}. The extracted elastic cross-section at small $T/T_{\rm F}$ was in good agreement with the universal value of Eq.~\eqref{eq:sigmaodd} without adjustable parameters.  Moreover, this study showed that the evaporation efficiency is among the highest achieved in an ultracold gas experiment. This stems from the Pauli suppression of inelastic processes for single-spin fermionic assemblies.  Together, these studies demonstrated the efficacy of universal elastic dipolar scattering for degenerate Fermi gas production.  This is a far simpler technique than sympathetic cooling with spin or species mixtures~\cite{Ketterle2008mpa}.

\subsection{Universal inelastic dipolar scattering:  Dipolar relaxation}\label{Drelaxationsec}

A large dipole moment enhances not only elastic dipolar scattering but also the \textit{inelastic} rate; see also Sec.~\ref{subsec:processes}.  While rapid elastic scattering is useful for efficient evaporative cooling, inelastic scattering---dipolar relaxation in this case---usually leads to gas heating and population loss.  Inelastic dipolar scattering arises from the anisotropy of the interaction and causes atoms to change their spin state.  In the presence of a finite external magnetic field, the degeneracy between the sublevels of the ground state hyperfine manifold is lifted via the Zeeman effect. In magnetic fields inducing Zeeman splittings larger than $k_BT$, atoms in metastable spin states relax to lower-energy, stronger-field-seeking Zeeman substates.  The Zeeman energy is released into the orbital motion of the collision pair, conserving total orbital momentum by increasing the atoms' kinetic energy. Atoms in magnetic traps whose spins have flipped to  strong-field seeking states are lost.  Lost too are atoms confined in  traps---magnetic or otherwise---that are shallower than the released Zeeman energy. Otherwise, the relaxed atoms may remain confined but the change in their motional state leads to an overall heating after rethermalisation via elastic collisions.    We note that for Zeeman splittings similar to or smaller than $k_B T$, the reverse process called dipolar \textit{promotion} is allowed.  This was  used for demagnetisation cooling of a chromium gas; see Sec.~\ref{demagcooling} for details. For Zeeman splittings similar to the harmonic trap excitation energy $\hbar\omega$, the change in the motional state of the colliding atoms is spectroscopically resolvable. This constitutes an atomic equivalent of the Einstein-de-Haas effect; see Sec.~\ref{freemag} for details and Ref.~\cite{Lahaye2009tpo}. 

Only atoms in their lowest-energy spin state---i.e., the maximally stretched strong-field-seeking state---are immune to inelastic dipolar scattering (and spin exchange)~\cite{Pethick2002book}. In this section, we will show that dipolar relaxation can be seen as a short-range process for a large-enough magnetic field. As a direct consequence, we will describe how confinement can be used to control dipolar relaxation. We will also see  that exchange statistics plays a role in these inelastic scattering processes. In particular, in Fermi gases,  dipolar relaxation is suppressed for certain spin configurations.

\subsubsection{Inelastic cross sections}
\label{subsec:inel_crosssec}
The DDI in Eq.~\eqref{ddispinform} admits non-zero matrix elements between pairs of atoms with differing spin and orbital angular momentum.  The relevant selection rules are  $\Delta m^{i}_F = 0, \pm 1$ for each of the two atoms $i={1,2}$ and  $\Delta l = 0,\pm2$ for their orbital motion.  To conserve total angular momentum, $\Delta m_l + \sum_{i=1,2}\Delta m^{i}_F = 0$  must be satisfied.  The cross sections may be calculated in a first-order Born approximation treatment~\cite{Hensler2003dri,Pasquiou2010cod}\footnote{Note that previous theoretical treatments appeared  in Refs.~\cite{Lagendijk1986sea,Shlyapnikov1994dka,Fedichev1996idp}.}. The appropriate expression, neglecting all other interatomic potentials~\cite{Pasquiou2010cod}, but including direct and exchange terms, is
\bea\label{sigmaborn}
&&\sigma = \left(\frac{m}{4\pi\hbar^2}\right)^2\frac{1}{k_{\rm i}k_{\rm f}} \left[ \int|\widetilde{\Vdd}(\boldsymbol{k}_{\rm i}-\boldsymbol{k'})|^2\delta(|\boldsymbol{k'}|-k_{\rm f})\textrm{d}\boldsymbol{k'} \right. \nonumber\\
 &&+  \left. \tilde{\epsilon}\int  \widetilde{\Vdd}(\boldsymbol{k}_{\rm i}-\boldsymbol{k'})  \widetilde{\Vdd}^{*}(-\boldsymbol{k}_{\rm i}-\boldsymbol{k'})\delta(|\boldsymbol{k'}|-k_{\rm f})\textrm{d}\boldsymbol{k'}\right],
\eea
where $\widetilde{\Vdd}(\boldsymbol{k})$ is the Fourier-transformed  matrix elements of Eq.~\eqref{ddispinform} and  $\tilde{\epsilon} = [1,0,-1]$ for identical bosons, distinguishable particles, and identical fermions, respectively~\cite{Hensler2003dri,Lahaye2009tpo}.  The $\boldsymbol{k_{\rm i}}$ and $\boldsymbol{k_{\rm f}}$ are the initial and final relative wavevectors, respectively.  

The cross sections for a collision channel starting with a two-body  stretched spin-state $|F,m_{F}=+F;F,m_{F}=+F\rangle$ and ending in either the same state ($\Delta m_l=0$) or a single ($\Delta m_l=1$) or double ($\Delta m_l=2$) spin-flipped state form a hierarchy in powers of the spin $F$~\cite{Hensler2003dri,Pasquiou2010cod}:
\bea
\sigma_{0} &=& \frac{16\pi}{45}F^{4}\left(\frac{\mu_{0}(g_{F}\mu_{B})^{2}m}{4\pi\hbar^{2}}\right)^{2}[1 + \tilde{\epsilon} h\left(1\right)], \label{elastic}  \\ \label{oneflip}
\sigma_{1} &=&  \frac{8\pi}{15}F^{3}\left(\frac{\mu_{0}(g_{F}\mu_{B})^{2}m}{4\pi\hbar^{2}}\right)^{2}[1 + \tilde{\epsilon} h\left(\frac{k_{\rm f}}{k_{\rm i}}\right)]\frac{k_{\rm f}}{k_{\rm i}} ,\\ \label{twoflip}
\sigma_{2} &=&  \frac{8\pi}{15}F^{2}\left(\frac{\mu_{0}(g_{F}\mu_{B})^{2}m}{4\pi\hbar^{2}}\right)^{2}[1+ \tilde{\epsilon} h\left(\frac{k_{\rm f}}{k_{\rm i}}\right)]\frac{k_{\rm f}}{k_{\rm i}}.
\eea
In these expressions, $\sigma_{0}$ is the elastic cross section discussed in the previous section; $\sigma_{1}$ and $\sigma_2$ are the cross sections for single spin-flip and double spin-flip dipolar relaxation processes, respectively. The kinematic terms $\frac{k_{\rm f}}{k_{\rm i}}= \sqrt{1+\frac{m\Delta E_{\Delta m_l}}{\hbar^{2}k^{2}_{i}}}$ are dictated by the energy conservation condition and depend on the number $\Delta m_l$ of spin flips involved in the collision.  This is because $\Delta E_{\Delta m_l}$ depends on the change in kinetic energy of the atomic pair. Typically $\Delta E_{\Delta m_l}=\Delta m_l\Delta E$, where the Zeeman energy splitting $\Delta E \propto B$. The value of $h\left(x\right)$, defined on $(1,\infty)$, monotonically increases from $h\left(1\right) = -1/2$ to $h(x\rightarrow\infty) = 1-4/x^{2}$~\cite{Hensler2003dri,Bohn2009qud,Pasquiou2010cod}. 
 
The inelastic cross sections in Eqs.~\ref{oneflip} and~\ref{twoflip}  contain only fundamental constants, atomic quantum numbers, and kinematic variables.  Therefore, just like the elastic dipolar collisions, one may consider inelastic dipolar scattering a universal process in the first-order Born approximation limit. We note that Ref.~\cite{Pasquiou2010cod} extended the cross section calculation to consider the effect of molecular potentials, and in particular, that of the short-range van der Waals interaction. The authors demonstrate theoretically and experimentally the effect of a finite scattering length on the dipolar relaxation cross section. While Eqs.~(\ref{oneflip}-\ref{twoflip}) are good approximations at low magnetic field for which $k_{\rm f}\as\ll1$, the effect of the molecular potential becomes important when $B$ is such that $k_{\rm f}\as\gtrsim 1$. This leads to  deviations from  universal behaviour (see below).
 
\subsubsection{Inelastic cross section measurements}

Experiments typically do not give direct access to the cross section, but rather to collisional rates. In a gas at thermal equilibrium, the dipolar relaxation collisional rate is deduced from the above cross section via $\beta_{dr} = 2\langle (\sigma_{1}+\sigma_{2}) v_{\rm rel} \rangle_{\textrm{\small{th}}}$, where $v_{\rm rel}=2\hbar k_{i}/m$ describes the initial relative velocity between a pair of particle and $\langle . \rangle_{\textrm{\small{th}}}$ defines a thermal average over the gas. This collisional rate has two contributions, one from collisions involving single spin-flips and one from double spin-flips respectively.  The latter is $\beta_{dr}^{(1,2)} = 2\langle \sigma_{1,2} v_{\rm rel} \rangle_{\textrm{\small{th}}}$. 

Following earlier studies of weakly magnetic atoms~\cite{Stoof1988sea,Moerdijk1996cta,Gerton1999drc}, dipolar relaxation was quantitatively studied in bosonic Cr~\cite{Hensler2003dri,Weinstein2002eco}.  In these studies, $^{52}$Cr was magnetically trapped in the highest energy (i.e., low-field-seeking) Zeeman sublevel  $|S,m_s\rangle = |3,3\rangle$.  Dipolar relaxation in Cr takes the atoms initially in the state $\ket{S,m_s;S,m_s} = \ket{3,3;3,3}$  to  $\ket{3,3;3,2}_\mathcal{S}$ ($\ket{3,2;3,2}$) via a single (double) spin flip, where $\mathcal{S}$ denotes the symmetrised total wave function. The spin-flipped atoms may be lost from the trap or, if not, heat the gas due to their increased kinetic energy. The single and double spin-flip processes were characterised by studying both the rate of population loss from $|S,m_s\rangle = |3,3\rangle$ and the rate of temperature increase in the  gas.  The overall dipolar relaxation rate  was found to be four orders--of--magnitude larger than in weakly magnetic alkali-metal atoms~\cite{Hensler2003dri}.  Moreover, the data were in reasonable agreement with cross section predictions based on the first-order Born approximation given above, i.e., Eqs.\eqref{oneflip}~and~\eqref{twoflip}. 

Dipolar relaxation processes limit the temperature and density achievable in  magnetically trapped gases of highly magnetic atoms. In contrast, optical traps enable the trapping of all Zeeman sublevels, and in particular of the lowest of these states, thus providing immunity to dipolar relaxation and opening the way to reaching higher gas phase-space densities. Additional dipolar relaxation data for Cr~\cite{Pasquiou2010cod} and Dy~\cite{Burdick2015fso} at  lower temperatures (100's of nK) were performed using optical trapping~\footnote{We note that the dipolar relaxation rates have also been measured in  bosonic Dy at a few hundred $\mu$K~\cite{Lu2010tud} and Er and Tm at hundreds of mK~\cite{Connolly2010lsr}, as well as Ho and Dy at similar temperatures~\cite{Newman2011mri}.}. In optical traps, because of limited trap depth, the particles involved in dipolar relaxation are typically lost (except at very low magnetic field, see Sec.~\ref{spinorsection} for details) and loss spectroscopy enables the characterisation of the collision rate. These additional experimental results also agreed well with predictions using the first-order Born approximation. Because the first-order Born approximation is valid for only weakly interacting systems (stronger interactions cause higher-order terms to be non-negligible), these findings imply that the atoms' DDI are not so strong as to invalidate this convenient approximation.  To conclude, the first-order Born approximation should hold for dipolar collisions among any element (Dy being the most magnetic stable element of the periodic table) as long as the atomic density remains on a similar scale as investigated in these works.  Such densities are  typical of repulsive three-dimensional gases.  

We note that the magnitude of $\sigma_1$ for Dy at $B=0$ can be up to 100$\times$-larger than that of the alkali metal caesium. In addition, Ref.~\cite{Burdick2015fso} pointed out that, while the ratio of $\sigma_1/\sigma_2 = F^{-1}\ll1$ for the stretched states of large-spin atoms, i.e., those with $|m_F|\approx F$, these same atoms exhibit large ratios of $\sigma_2/\sigma_1$ for states near $m_F = 0$.  Indeed, $\sigma_2$ for these states can be \textit{larger} than $\sigma_1$ for the stretched states:  One cannot avoid dipolar relaxation simply by employing spin states near $m_F = 0$.

\subsubsection{Interaction range for dipolar relaxation}
\label{range_inel}
While the elastic dipolar  interaction is long-ranged, Ref.~\cite{Pasquiou2010cod} shows that the dipolar relaxation processes are intrinsically short-ranged, despite the same $1/r^3$ scaling. This effect stems from the reduced overlap between the incoming and outgoing waves due to the increase in kinetic energy.  Specifically, while the integral overlap of Eq.~\eqref{sigmaborn} for $\sigma_0$ involves incoming and outgoing waves oscillating at the same frequency ($k_{\rm i}$), for inelastic collisions, the  incoming and outgoing wavefunctions oscillate at different frequencies and become spatially mismatched at long distance, averaging to zero. Assuming a vanishingly small collision energy (low $T$), the range for this effective cancellation of the overlap, $R_{\rm dr}$,  scales as $1/k_{\rm f}$ .  This is set by the release of Zeeman energy in the spin relaxation. Hence, the range of dipolar relaxation decreases with $B$ following: 
\begin{equation}
R_{\rm dr}(B) \approx \frac{\hbar}{\sqrt{m g_F \muB B}}.
\label{rdr}
\end{equation}
In terms of partial-wave decomposition, dipolar relaxation occurs at the interparticle distance $R_{\rm dr}$ that matches the distance at which the energy released in dipolar relaxation $\Delta E= g_{F}\mu_{B}B$ is compensated for by the centrifugal energy of the output channel. This simple statement assumes that the  input channel is $l=0$, which is usually the case at low $T$.  (To satisfy the $\delta l = 2$ selection rule, the output channel is $l=2$.) 

The localised character of dipolar relaxation has been revealed in the interplay between dipolar relaxation process and other interatomic (molecular) potentials in Ref.~\cite{Pasquiou2010cod}.  This results in  variations of the relaxation rate with $B$ that deviate from that expected from the first-order Born approximation when neglecting the effect of molecular potentials; specifically, the rate can dip far below this expectation at finite $B$. This strong reduction of dipolar relaxation is shown to arise from a node in the initial wavefunction, preventing the presence of pairs of atoms at an inter-particle distance corresponding to the distance $R_{\rm dr}$ at which dipolar relaxation occurs. This provides a factor of twenty reduction in the relaxation rate versus the rate predicted by the first-order Born approximation.  This reduction relies on a non-universal molecular potential configuration of $^{52}$Cr and it cannot be directly extended to the other highly magnetic atoms due in particular to the complexity of their Feshbach resonance spectrum.
 
 \subsubsection{Quantum statistics and dipolar relaxation: Bose enhancement and Fermi suppression. }\label{fsuppress}
 
Now that we have discussed the short-range nature of dipolar relaxation, we pivot to a discussion of how this fact combines with quantum statistics to  affect relaxation rates.   As with the short-ranged van der Waals interaction (see Sec.~\ref{statisticshortrange}), one might expect a suppression of collisions between identical fermions  due to quantum  statistics. Indeed, this is the case, because identical fermions cannot surmount the centrifugal barrier of the lowest quantum statistically allowed partial wave $l=1$:  since identical fermions cannot closely approach, the short-ranged inelastic DDI does not lead to dipolar relaxation of their spin.  Conversely, we will show below that the relaxation rate is enhanced for identical bosons, since they can collide on the barrierless $l=0$ partial wave.

The dependence of the relaxation rate on quantum statistics is manifest in the form of the terms containing the  ratios of incoming and outgoing momenta in Eqs.~\eqref{oneflip} and~\eqref{twoflip}. At high $B$ and low $T$, the kinematic terms diverge as $\langle \frac{k_{\rm f}}{k_{\rm i}} \rangle_{\rm th} \propto \sqrt{B/T}$, where the $\rm th$ denotes a thermal average.  In this limit, $\sigma_1$ vanishes as $4\sqrt{T/B}$ for indistinguishable fermions ($\tilde{\epsilon}=-1$), while it increases as  $2\sqrt{B/T}$ for indistinguishable bosons ($\tilde{\epsilon}=+1$) and is $\sqrt{B/T}$  for distinguishable particles ($\tilde{\epsilon} = 0$).  The relative suppression ratio in this limit becomes $\sigma_{1}^{\textrm{fermions}}/\sigma^{\textrm{bosons}}_{1}\propto 2T/B$.  This scaling exhibits  two important facts:  i) Dipolar relaxation rates grow  worse for identical bosons versus $B$, e.g.,  $\sigma_1$ for bosonic Dy is $10^3$ times larger than Cs's  at only a 1-G field; ii) $\sigma_{1,2}$ can be \textit{suppressed} for identical fermions, e.g., dipolar relaxation in fermionic Dy is only $10{\times}$-worse than Cs's at a few G, as experimentally demonstrated in Ref.~\cite{Burdick2015fso}.  It is remarkable, and counterintuitive, that the more exothermic the fermionic dipolar relaxation would be, the less likely they are to occur.  This constitutes another striking example of the role quantum statistics plays in ultracold collisions.

\begin{figure}[t!]
   \includegraphics[width=0.95\columnwidth]{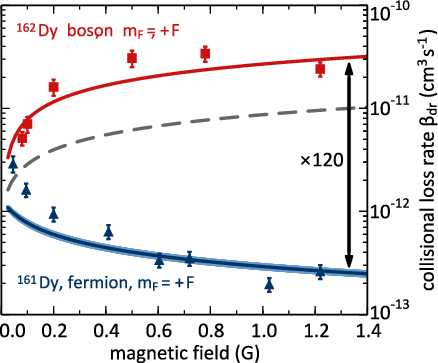}
\caption{Dipolar relaxation for bosonic $^{162}$Dy ($m_F=+8$) and fermionic $^{161}$Dy ($m_F=+21/2$). Two-body collisional loss rates for $^{162}$Dy (squares) and $^{161}$Dy  (triangles).  Curves are calculated collisional loss rates using Eq.~\eqref{oneflip} at $T \approx 400$~nK with no free parameters. Adapted from Ref.~\cite{Burdick2015fso}.} \label{suppDR}
\end{figure}

Figure~\ref{suppDR} shows measured dipolar relaxation rates for spin polarised bosonic and fermionic Dy.  Reference~\cite{Burdick2015fso} also shows the fermionic suppression of dipolar relaxation of spin mixtures for processes in which the output state of the fermions consists of identical spins~\cite{Pasquiou2010cod}. This corresponds to the time-reversed, state-changing process of indistinguishable fermions colliding. Importantly, the stability of a spin mixture in the lowest-two Zeeman states opens interesting prospects for the study of many-body spin physics.  In particular, one could explore how the DDI impacts the BEC--to--BCS crossover (see also Sec.~\ref{subsec:fermi})~\cite{Baranov2012cmt}. The stability of these states was also exploited for the long-lived implementation of artificial spin-orbit coupling in Dy gases; see Sec.~\ref{SecSOC}. 

\subsubsection{Control of dipolar relaxation via confinement}\label{subsec:inel_confin}

\begin{figure}[t]
\centering
\includegraphics[width= 8 cm]{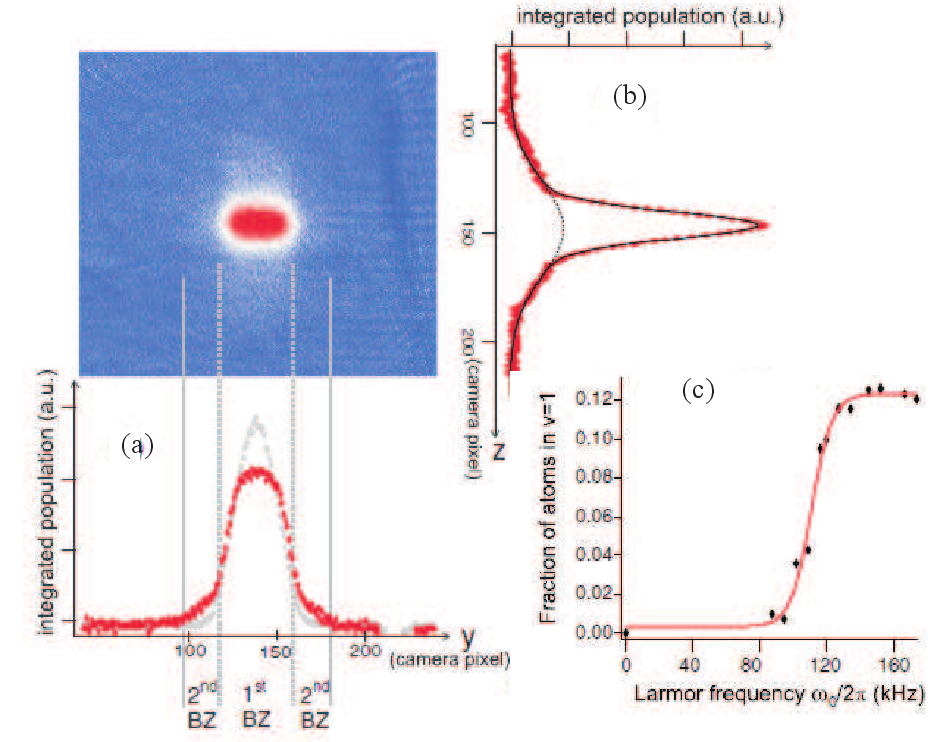}
\caption{Control of dipolar relaxation in reduced dimensions. The false colour absorption image is a band mapping of the atom population trapped in a 2D optical lattice. (a) Integrated population profile along $z$ yields populations in the different Brillouin zones (BZ). Below the threshold magnetic field for dipolar relaxation, only the first BZ is populated (grey profile with the sharper peak). Above the threshold, the first transverse excited band (second BZ) is populated (red profile with the wider peak). (b) Integrated population profile along $y$ shows the (1D) velocity distribution in the sites of the 2D lattice. Above threshold, vibrational de-excitation of atoms from excited bands create a non-Gaussian distribution. (c) Fraction of atoms detected in the first excited band as a function of the Larmor frequency after a 25-ms duration allowed for dipolar relaxation. The lattice depth is $25 E_r$ which sets a threshold Larmor frequency of 120~kHz. Below threshold, dipolar relaxation is reduced by three orders of magnitude. Adapted from~\cite{Pasquiou2011sra}.}
\label{relaxation1D}
\end{figure}

Confinement of atoms in lower dimensions can also suppress dipolar relaxation rates through the phase-space reduction of outgoing scattering channels.  That is, the relaxation rate can be lowered via the interplay between the range of the dipolar relaxation process and the length scales of the confinement potential.  Following Eq.~\eqref{rdr}, when the magnetic field $B$ is small, dipolar relaxation takes place at large inter-particle distances. Dipolar relaxation thus strongly depends on the trapping geometry, provided the size of the cloud is comparable to $R_{DR}$. This phenomenon was demonstrated in~\cite{Pasquiou2010cod,Pasquiou2011sra,depaz2013rdo} for the case of 1D, 2D, and 3D optical lattices. In Ref.~\cite{Pasquiou2010cod}, a 1D optical lattice was shown to reduce dipolar relaxation in Cr by a factor of seven compared to a 3D trap when $\Delta E$ was smaller than the gap to first transverse excited state. The dipolar relaxation rate was further suppressed---by three orders--of--magnitude---in a 2D optical lattice of $^{52}$Cr~\cite{Pasquiou2011sra}. This was achieved below the threshold $B$-field determined by this gap condition for $\Delta E$.  Interband transitions mediated by dipolar relaxation were observed above this threshold, as illustrated in Fig.~\ref{relaxation1D}. In the specific case of a deep 3D optical lattice~\cite{depaz2013rdo}, it was observed that dipolar relaxation becomes a resonant process as a function of $B$ due to the quantisation of kinetic energy in the 3D-band structure. We see that by adjusting confinement and the magnetic field, dipolar relaxation can be either eliminated (e.g., to study spinor physics at constant magnetisation), or be used to couple different lattice bands. This then realises an intrinsic nonlinear spin-orbit coupling in the lattice~\cite{depaz2013rdo,sun2007ote}.

Modifications to dipolar relaxation rates were also demonstrated in Ref.~\cite{Pasquiou2010cod} using RF fields.  While only an increase in relaxation rates was experimentally demonstrated, the work showed that RF-dressing might serve as a valuable tool to control dipolar relaxation.

\subsubsection{Dipolar relaxation and spinor physics}

Section~\ref{spinorsection} will describe in more detail why the large spin of the highly magnetic atoms presents attractive prospects for the construction of exotic spinor states and many-body phases.  Dipolar relaxation can play a positive role in this regard:  magnetisation-changing collisions open the door to the exploration of spinor physics under conditions of free magnetisation. For example, controlling the field at the mG level makes the Zeeman energy $\Delta E$ comparable to thermal excitations of $\sim k_{\rm B}$100~nK.  This induces spontaneous, incoherent demagnetisation from the absolute ground state.  Further control of  the field down to the 100-$\mu$G level allows $\Delta E$ to be  of the order of  spin-dependent interactions.  This would enable the observation of spinor phases driven by contact interactions under free magnetisation.  This degree of control has already been achieved in current ultracold atoms experiments~\cite{Pasquiou2011sdo}. With even better control, possible experimentally,  one can enter the regime where dipolar relaxation is relevant from the many-body point of view.   For example, if $R_{\rm dr} > \bar{d}$, where $\bar{d}=n^{-1/3}$ is the average particle distance, then many-body effects may arise. Interesting vortex structures may appear due to the Einstein-de-Haas effect since the orbital angular momentum is increased from spin relaxation~\cite{kawaguchi2006edh,Santos2006scb,gawryluk2007red}. Strongly rotating and interacting Fermi gases approaching a Laughlin state may appear~\cite{Peter2013ddf}.

In general, however, dipolar relaxation has unfortunate consequences for the study of many-body spinor physics in fields larger than those considered above or outside of 2D or 3D lattices where it can be strongly suppressed.  High densities are needed to enhance the interaction energy, but fast dipolar relaxation can render systems with metastable spin states too fragile to observe in (near-)equilibrium situations.  Indeed, dipolar relaxation for all metastable spin states in bosons is particularly severe, and only a  few metastable spin configurations for fermions are sufficiently long-lived~\cite{Burdick2015fso}. For example, long-lived mixtures of $\ket{F,-(F-1);F,-F}$ were achieved in fermionic Dy and Er~\cite{Burdick2015fso,Baier2018roa} (see Sec.~\ref{subsec:fermispin}) and the same set of states was used to generate spin-orbit coupling in fermionic Dy~\cite{Burdick2016lls}; see Sec.~\ref{SecSOC}. The study of out-of-equilibrium dynamics is also possible for time-scales up to the dipolar relaxation time scale~\cite{depaz2016psd,lepoutre2017sma}; see Sec.~\ref{subsec:spindyn_bulk}.

\subsection{Anisotropic scattering}
\label{subsec:anis_scatt}

\begin{figure}[t!]
\includegraphics[width=0.95\columnwidth]{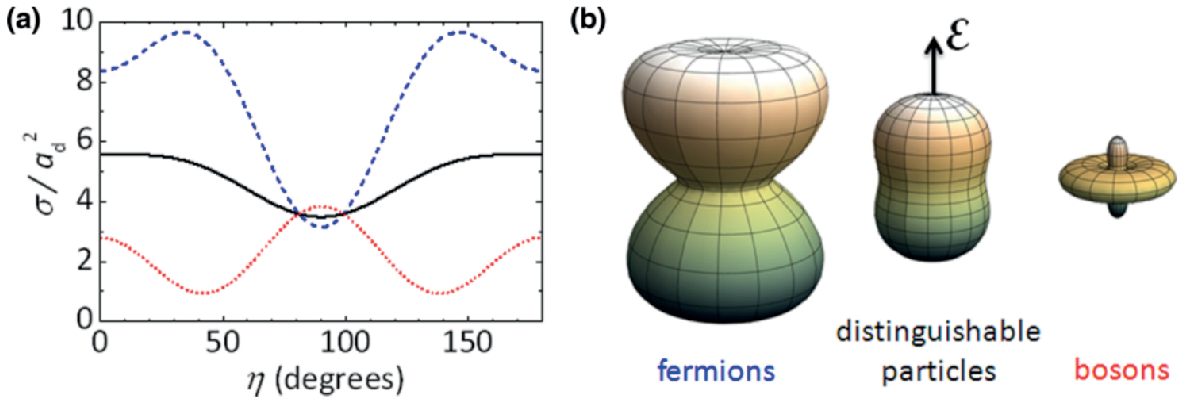}
\caption{a) Total dipolar cross section $\sigma=\int  \frac{d\sigma}{d\Omega}(\vk,\vk')d\Omega_{\vk'}$ in units of the dipolar length $a_{\rm d} = \add/3$ versus the angle $\eta$ between the incident scattering direction $\hat{\vk}$ and the polarisation direction $\hat\epsilon$.  The curves are for distinguishable particles (solid black), identical fermions (dashed blue), and identical bosons (dotted red). b) 3D plots of $\sigma$ as a function of $\hat{\vk}$ with $\hat\epsilon$ set to the vertical axis.  Reproduced from Ref.~\cite{Bohn2014dsa}.} \label{totalxsec}
\end{figure}

We now return to the discussion of elastic dipolar scattering and explore manifestations of the anisotropic nature of the DDI on scattering into different directions in space.  This 3D scattering is characterised by the dipolar differential cross section  $\frac{d\sigma}{d\Omega}$. From the scattering theory reviewed in Sec.~\ref{scatttheory}, we see that the differential cross section  directly relates to the scattering amplitude $\frac{d\sigma}{d\Omega} (\vk,{\boldsymbol n}) =|f(\vk,{\boldsymbol n})|^2$. The expressions $\frac{d\sigma_{\rm B/F}}{d\Omega}$, including both the contact interaction and DDI, for identical bosons and fermions in the first-order Born approximation are, respectively:
 \bea
 \nonumber
&&\frac{d\sigma_{\rm B}}{d\Omega}(\vk,\vk^{\prime})=\frac{D^2}{8}\left[\frac{4}{3}-\frac{2\as}{3\add}\right.\\ 
&&\left.-\frac{2(\hat{\vk}.\hat{\boldsymbol\varepsilon})^2+2(\hat{\vk}^{\prime}.\hat{\boldsymbol\varepsilon})^2-
4(\hat{\vk}.\hat{\boldsymbol\varepsilon})(\hat{\vk}^{\prime}.\hat{\boldsymbol\varepsilon})(\hat{\vk}.\hat{\vk}^{\prime})}{1-(\hat{\vk}.\hat{\vk}^{\prime})^2}\right]^2,\label{DifferentialScatteringB}
\eea
 \bea
 \nonumber
&&\frac{d\sigma_{\rm F}}{d\Omega}(\vk,\vk^{\prime})=\\ &&\frac{D^2}{8}\left[\frac{4(\hat{\vk}.\hat{\boldsymbol\varepsilon})(\hat{\vk}'.\hat{\boldsymbol\varepsilon})-2\left[(\hat{\vk}.\hat{\boldsymbol\varepsilon})^2+(\hat{\vk}'.\hat{\boldsymbol\varepsilon})^2\right](\hat{\vk}.\hat{\vk}^{\prime})}{1-(\hat{\vk}.\hat{\vk}^{\prime})^2}\right]^2,\label{DifferentialScatteringF}
\eea
where $\vk$ and $\vk^{\prime}$ denote the relative momenta before and after the collision, with directions $\hat{\vk}=\frac{\vk}{|\vk|}$ and $\hat{\vk}'=\frac{\vk'}{|\vk'|}$~\cite{Bohn2014dsa,Sykes2015ndo}.  The dipole moments are aligned along the $B$-field direction  $\hat{\boldsymbol\varepsilon}$.  These are nontrivial functions of the relative orientations of the three directions in the problem, the incident and scattered wavevectors and the polarisation direction of the dipoles. The total dipolar cross sections are found by integrating these expressions over  $d\Omega_{\hat\vk'}$ and are shown in Fig.~\ref{totalxsec}. For a general angle $\eta$ between $\hat{\vk}$ and $\hat\epsilon$, these are
\bea
\nonumber
\sigma_B(\hat{\bf p}_{\rm rel}) &=& D^2\frac{\pi}{9}[72\as^2-24\as(1-3\cos^2\eta)\\
&&\,\, +11 - 30\cos^2\eta + 27\cos^4\eta ]\\
\sigma_F(\hat{\bf p}_{\rm rel}) &=& D^2\frac{\pi}{3}\left[3+18\cos^2\eta -13\cos^4\eta\right].
\eea
Note that the results of Eqs.~\ref{eq:sigmaeven} and~\ref{eq:sigmaodd} correspond to angle-averaged versions of the above expressions.

\subsubsection{Anisotropy of rethermalisation}  
\label{Anisotropicrether}

\begin{figure}[t!]
\includegraphics[width=0.95\columnwidth]{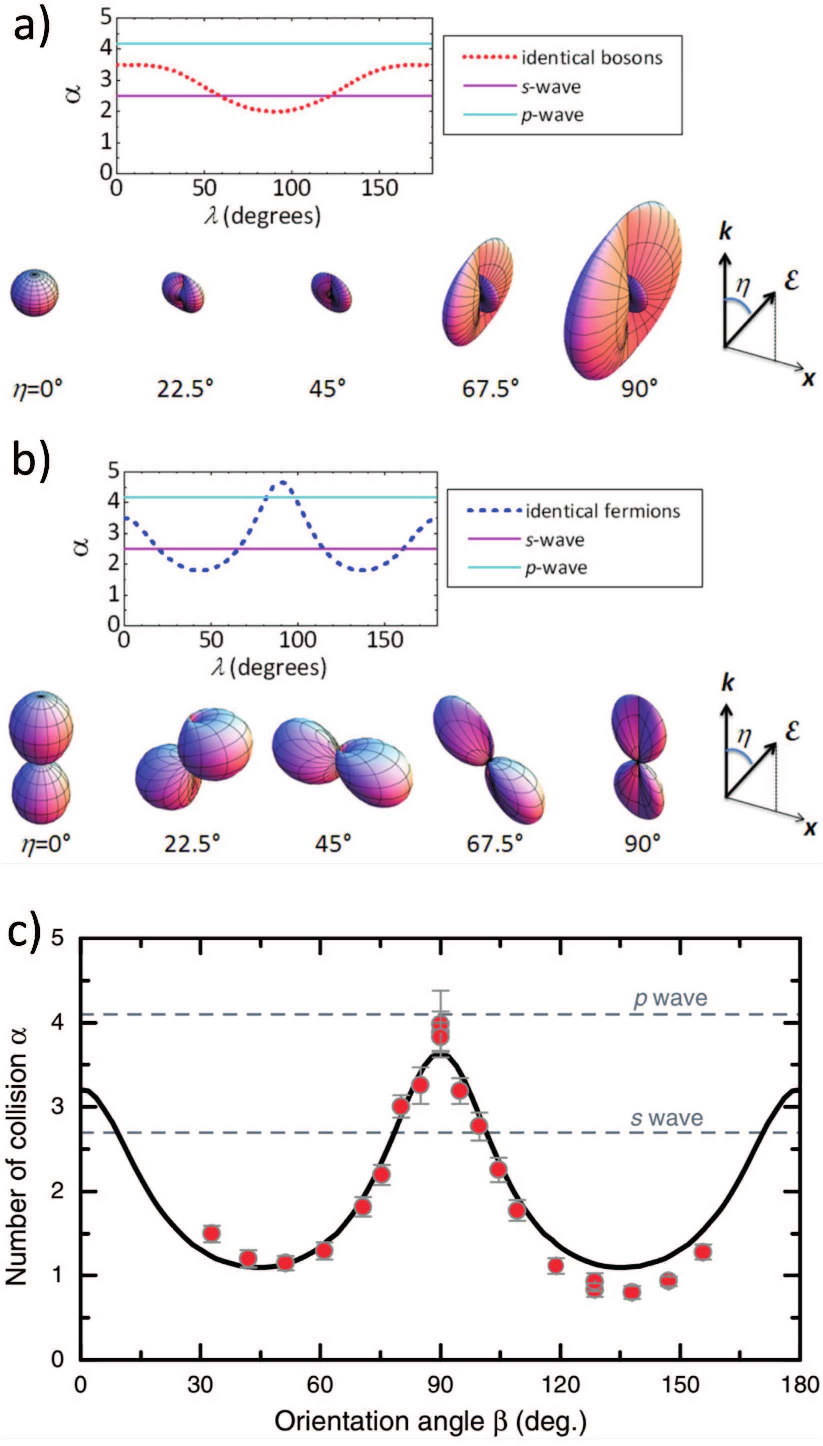}
\caption{ (a,b) Upper panels: Number of collisions per rethermalisation, $\alpha$, versus the angle between the quantization axis $\hat\epsilon$ and the trap axis along which rethermalisation occurs, $\lambda$. Lower panels:  3D plots of $d\sigma/d\Omega$ as a function of the scattering direction $\hat\vk'$, for different values of $\eta$.  $\eta$ is the angle between $\hat\vk$, set to the vertical $z$ axis, and the polarisation direction $\epsilon$ in the $xz$-plane. Plots are for a) identical bosons and b) identical fermions.   Reproduced from Ref.~\cite{Bohn2014dsa}. c) Comparison between theory and measurement of $\alpha$ for ultracold dipolar fermions of $^{167}$Er.  Reproduced from Ref.~\cite{Aikawa2014ard}.} \label{diffxsec}
\end{figure}

The anisotropy of the differential cross section may reveal itself in the thermalisation dynamics, and in particular, via the thermalisation rate  as a function of the angle $\lambda$ between $\hat{\boldsymbol\varepsilon}$ and the axis along which energy is imparted. This can be studied in cross-dimensional rethermalisation experiments. Cross-dimensional rethermalisation (relaxation) is a method for measuring the total elastic cross section wherein the gas is brought out of thermal equilibrium by heating it along one or two of three directions~\cite{Monroe1993moc}.  One then observes the time for the temperature in the third direction to equilibrate.  The time constant $\tau$ for this typically exponential relaxation process may be related to the cross section via $\tau = \alpha/n v \sigma$, where $n$ is the average number density, $v$ is the mean relative speed, and $\sigma$ is the elastic cross section.  The anisotropy of the differential cross section appears through the parameter $\alpha$, which is the number of collisions, on average, for rethermalisation.   For $s$-wave ($p$-wave) collisions, $\alpha = 2.5$ (25/6 $\approx$ 4.17)~\cite{DeMarco1999mop}.   However, for dipolar particles, $\alpha$ becomes angle dependent due to the anisotropic differential cross section~\cite{Bohn2014dsa}.  Its value is plotted in Fig.~\ref{diffxsec}(a,b) versus the angle $\lambda$ between the axis along which energy is imparted and $\hat{\boldsymbol\varepsilon}$.  We see that $\alpha$ can change by more than a factor two versus angle for fermions, while sightly less than two for bosons.  The control of $\lambda$ is therefore an important tool for optimising, e.g., evaporative cooling efficiency:  rethermalisation of fermions (bosons) is far more efficient at 45$^\circ$ (90$^\circ$) than at 0$^\circ$ or 90$^\circ$ (0$^\circ$).

The large anisotropy in the rethermalisation of dipolar gases was experimentally demonstrated in ultracold gases of fermionic $^{167}$Er~\cite{Aikawa2014ard}. For fermionic atoms, $\sigma$ is simply given by the universal formula Eq.~\eqref{eq:sigmaodd} so that a measurement of  $\tau$ directly provides $\alpha$. Figure~\ref{diffxsec}(c) shows measurements of $\alpha$ in agreement with the dependence predicted in Ref.~\cite{Bohn2014dsa}.  The authors also reported the reduction in rethermalisation rate due to Pauli-suppression of scattering once $T$ was  sufficiently lower than $T_F$ (see Sec.~\ref{elasticdipolarscattering} and Eqs.\ref{eq:sigmaeven} and~\ref{eq:sigmaodd}).  Unlike $\alpha$, the suppression factor was observed to be  independent of $\lambda$, implying that the occupation of the final density of states leading to Pauli-blocking is mostly unaffected by the anisotropy of the DDI.

Cross-dimensional rethermalisation experiments in ultracold gases where also performed in bosonic isotopes of $^{164}$Dy, $^{162}$Dy~\cite{Tang2015sws} and of $^{164}$Er, $^{166}$Er, $^{168}$Er and $^{170}$Er~\cite{Frisch2015udm}. These studies measured the unknown values of the scattering lengths $\as$ of these atoms by using either the theoretically predicted $\alpha$ or by using more extensive dynamics simulations to convert the  measured $\tau$ into values of $\as$; see Eq.~\eqref{eq:sigmaeven} and Sec.~\ref{scatlengthmeas} for more details. In Ref.~\cite{Tang2015sws}, the authors used a thorough theoretical analysis informed by data from different values of the angle $\lambda$ to extract $\as$. This study confirmed the anisotropic behaviour as well as provided evidence of hydrodynamical effects in the rethermalisation dynamics~\cite{Sykes2015ndo}.  

These hydrodynamic effects were manifest as modifications of the mechanical oscillations of the gas after the gas was kicked out of thermal equilibrium. It was observed that the large magnetic dipole moment of Dy provides a sufficiently large elastic collision cross section for the gas to lie near the hydrodynamic collisional regime.  These effects were predicted in Ref.~\cite{Sykes2015ndo}, where it was shown that the relaxation no longer follows a simple exponential in the hydrodynamic regime when the collisional rate is similar to the trapping frequencies, as observed in Ref.~\cite{Tang2015sws}.  Nevertheless, cross sections were able to be extracted from the data through close comparison to Monte Carlo simulations.  Section~\ref{scatlengthmeas} describes the scattering length measurements obtained as a result.  Hydrodynamic behaviour arising from the DDI has also been observed in the expansion aspect ratio of Dy thermal gases~\cite{Tang2016aeo}; see also Sec.~\ref{subsec:thermal_anis}.

\subsubsection{Anisotropic scattering halos}

\begin{figure}[t!]
\includegraphics[width=0.95\columnwidth]{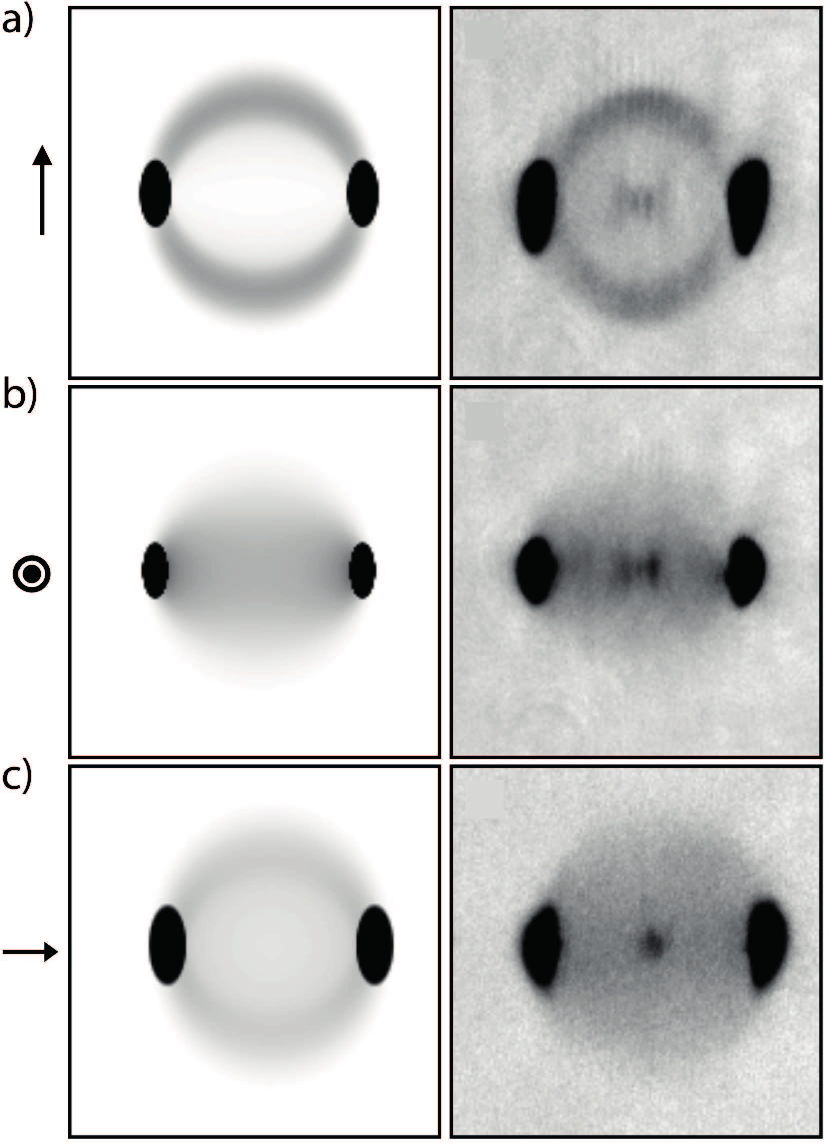}
\caption{2D images of the scattering halos from the collision of two BECs of $^{162}$Dy that are counter-propagating along the horizontal direction of the images. Left column are simulations, while the right column are experimental data.  Dipoles are aligned using a $B$-field oriented: a) along the vertical direction, b) out of the page, and c) along the horizontal direction, as indicated by the arrows. Adapted from Ref.~\cite{Burdick2016aco}.} \label{halos}
\end{figure}

The anisotropy of the differential cross sections may be directly visualised by observing the shape of scattering ``halos'' of atoms from two gases that have undergone a head-on collision.  Reference~\cite{Burdick2016aco} describes such an experiment: One creates two dipolar BECs of $^{162}$Dy counter-propagating at momentum $\pm2\hbar k_R$ using Bragg diffraction.  This energy is much larger than the internal momentum distribution width of the individual BEC. When the BECs propagate through one another, atoms scatter away from the forward and backward directions of the BECs' motion.  A halo rapidly expands from the centre-of-mass position, as shown in the data of Fig.~\ref{halos}.   The BECs are sufficiently dilute that atoms scatter at most one time.  The shape of the halo therefore reveals information about the two-body differential cross section.  

The data visually reveal a novel regime of quantum scattering.  Rather than exhibiting halos with shapes dominated by a single partial wave---e.g., the spherically symmetric halos from the $s$-wave scattering of ultracold identical bosons~\cite{Legere1998qsi,Chikkatur2000sae,Thomas2004ios,Buggle2004ido}---anisotropic shapes are manifest.  This is a consequence of the superposition of a large number of partial waves due to high-order angular momentum coupling through the elastic DDI.  Moreover, the particular halo shape strongly depends on $\eta$, as expected from Eq.~\eqref{DifferentialScatteringB}.  The shapes of the halos in the data agree with Monte Carlo simulations based on Eq.~\eqref{DifferentialScatteringB} and described in Ref.~\cite{Sykes2015ndo}.

%% file: Section4/4_aa.tex
\section{Fully polarised repulsive bulk dipolar gases}\label{Sec:RepulsiveGases}

In this chapter, we will consider dipolar quantum gases of atoms fully polarised in the lowest Zeeman state. We will cover only harmonically trapped gases in the absence of a three dimensional lattice; see Sec.~\ref{sec:lattice} for dipolar lattice systems. Most of the experimental activity has been focused on dipolar bosonic gases and in particular dBECs (Sec.~\ref{subsec:bose}).  Nevertheless, we will also include the relevant results on dipolar DFGs (Sec.~\ref{subsec:fermi}) as well as results on dipolar gases in lower dimensions, in particular in 1D (Sec.~\ref{Sec1D}).
 
 \subsection{Weakly interacting three-dimensional dipolar Bose gases}
 \label{subsec:bose}
 
A majority of the phenomena arising from the DDI that have been observed in bulk BECs, in particular in the regime where the overall interactions are effectively repulsive, are well accounted for by a mean-field (MF) description of the interactions. After a short description of the MF theory of dipolar Bose gases in Sec.~\ref{subsec:model_bose}, this section will focus on experimental observations of related phenomena. These observations can be understood as hydrodynamic measurements, as they concern the linear response from the equilibrium state of the quantum fluid; e.g., its time evolution in a trap or in free space as well as its elementary collective excitations. These aspects will be reviewed in Secs.~\ref{subsec:magnetostriction} and \ref{subsec:excitationBEC}. Beside modifying the speed of sound and giving it a directional dependence, the DDI profoundly affects the structure of the elementary excitation spectra of trapped BECs by the interplay of long-range anisotropic interactions with the trap geometry. In particular, a dispersion relation showing a local minimum in energy, i.e., a roton mode, was predicted and observed, as discussed in Sec.~\ref{subsec:excitationBEC}. Furthermore, the DDI influences the  stability of BECs in traps. Due to the anisotropy of the DDI, the effective sign of the MF interaction depends on the density distribution, which, in turn, is not only determined by the trap geometry but also by the DDI. This interplay results in a non-trivial stability diagram as a function of $\as$ and of the anisotropy of the trapping potential, as well as instability behaviours which also connects to the underlying modifications of the excitations spectrum of the BEC mentioned above. This will be reviewed in Sec.~\ref{subsec:Stability}. The stability analysis and related stabilised states of dipolar Bose gases beyond a MF treatment is the subject of Sec.~\ref{Sec:DCQD}.
 
 \subsubsection{The mean-field description}\label{subsec:model_bose}
 
 \myparagraph{Interaction potential in dilute gases}\label{subsec:interaction_potential_bose}
As introduced in Sec.~\ref{subsec:mf}, polarised dipolar bosons in an ultracold dilute ensemble can be considered to be interacting via a potential of the form: 
\begin{eqnarray}\label{interaction}
	V_{\rm {int}}(\boldsymbol r)=\frac{4\pi\hbar^2\,\as}{m}\delta(\boldsymbol r)+\frac{3\hbar^2\,\add}{m}\frac{1-3\cos^2\theta}{r^3}.\label{Eq:Pot}
\end{eqnarray}
Again, $\as$ is the $s$-wave scattering, while $\add=\mu_0m\mu^2/12\pi\hbar^2$  is the dipolar scattering length.   The interparticle separation vector is $\bs r$, with norm $r$ and angle $\theta$ with respect to the dipole orientation. In the following, we will assume a polarising magnetic field aligned along the $z$-axis, and a harmonic trap  with frequencies $\omega_{x,y,z}$. Unless otherwise specified, we also consider that the trap has a cylindrical symmetry along $z$ such that $\omega_{x}=\omega_{y}=\omega_{\rho}$, and
\begin{eqnarray}\label{Eq:trap}
	V_{\rm{tr}}=\frac m2\left(\omega_\rho^2\,\rho^2+\omega_z^2\,z^2\right).
\end{eqnarray}
The trap aspect ratio is 
\begin{eqnarray}
	\lambda=\omega_z/\omega_\rho.
\end{eqnarray}

The  validity of Eq.~\ref{interaction} as an effective pseudo-potential written as a simple sum of a contact pseudo-potential $g\delta(\bs r)$ (or its more rigorous, regularised version~\cite{Huang1957qmm}) and the DDI potential Eq.\,\eqref{Udd} has long been debated~\cite{Lahaye2009tpo,Pitaevskii2016bec}. This expression for the potential results from the first-order Born approximation applied to a molecular potential of the dipole-dipole type dominating at long distance plus a van der Waals potential dominating at short range~\cite{yi2000tac,Yi2001tco,Derevianko2003apf,Derevianko2005eap,Derevianko2005rhy,Bohn2009qud}. While it is a good approximation for weak dipoles and away from scattering resonances, modifications may be necessary for systems with large $\edd$. In particular, the $s$-wave scattering length may depend on $d$~\cite{yi2000tac,Yi2001tco,Mazets2007mos}, requiring a renormalisation of the dipolar length in Eq.~\ref{interaction}~\cite{Bortolotti2006sli,Oldziejewski2016pos}. In the case  of the most magnetic of the atoms, Dy, with $\edd \sim 0.5-2$, corrections to $\add$ were predicted to be of 10\% at temperature of 100\,nK and only 2\% at 10\,nK~\cite{Oldziejewski2017epo}.   

 \myparagraph{Mean-field approximation}\label{subsec:mf_bose}

The MF approximation, which underlies common approaches to describe the properties of bulk BECs, consists of replacing the field operator by its mean value, $\psi(\boldsymbol r,t)$. The approximation assumes that the BEC mode is macroscopically occupied~\cite{Bogoliubov1947ott,Gross1961soa,Pitaevskii1961vli,Stringari2016bec}. This effectively ignores fluctuations of the bosonic field around its mean value. In the absence of the DDI, the corrections to this approximation scale as $\sqrt{n\as^3}$. In the presence of the DDI, one  naively expects the MF approximation to be valid when $\sqrt{n\as^3},\,\sqrt{n\add\,^3}\,\ll1$. A proper treatment of fluctuations around MF values, outlined in Sec.\,\ref{Sec:DCQD}, leads to a more restrictive condition: $\sqrt{n\as^3}\ll1,\edd\lesssim 1$. Note that the length $\add$ is defined such that $\as=\add$ ($\edd= 1$) marks the limit of the mechanical stability of a homogeneous isotropic dBEC, and other authors sometimes call $3\add$ the dipolar length (which arises from a consideration of the two-body problem~\cite{Bohn2009qud}); see Sec.\ref{subsec:2Bdintro}.

Considering a gas of $N$ atoms interacting via Eq.~\eqref{Eq:Pot}, 
the MF approximation results (at zero temperature) in a nonlocal Gross-Pitaevskii equation (GPE) given in Eq.~\eqref{GP} and reprinted here:  
\begin{eqnarray}
	 i\hbar\frac{\partial\psi}{\partial t}(\boldsymbol r,t)=\left[-\frac{\hbar^2}{2m}\nabla^2+V_{\rm{tr}}(\boldsymbol{r})+g|\psi|^2+\pdd(\boldsymbol r)\right]\psi,
\end{eqnarray}
with $g=4\pi\hbar^2\as/m$ and  $\pdd(\boldsymbol r)=\int d\boldsymbol{r'}U_{\rm{dd}}(\boldsymbol{r'}-\boldsymbol{r})|\psi(\boldsymbol{r'})|^2$; see also Eq.~\eqref{meanfield}.
The corresponding energy functional is 
\begin{eqnarray}
	 \nonumber
	 E[\psi]=&&\int \left[-\frac{\hbar^2}{2m}|\nabla\psi|^2+V_{\rm{tr}}(\boldsymbol{r})|\psi|^2\right.\\
	 &&\left.+\frac g2|\psi|^4+\frac12|\psi|^2\pdd(\boldsymbol r)\right]dr,\label{Eq:Energy}
\end{eqnarray}
which can be minimised to obtain the equilibrium ground state of the BEC~\cite{Lahaye2009tpo,Baranov2002udg,Baranov2008spb,Baranov2012cmt}. In these equations, the DDI yields a nonlocal contribution $\pdd(\boldsymbol r)$, whose sign depends on the shape of the density distribution. We note that various analyses demonstrated the applicability of the nonlocal GPE in accurately describing the properties of trapped dipolar gases~\cite{Ronen2006dbe,Bortolotti2006sli}. 

\subsubsection{Magnetostriction}\label{subsec:magnetostriction}

\myparagraph{Magnetostriction of dBECs}

When placed in a cylindrical harmonic trap, a nondipolar gas ($\edd=0$) assumes the geometry of the trap. A dipolar system behaves differently because of the presence of $\pdd$ in Eq.~\eqref{GP}. Specifically, the dipole term $\pdd(\boldsymbol r)$ has a saddle-shape, tending to elongate the distribution along the magnetic field. That is, to minimise its energy, the dBEC elongates along the dipolar axis due to the attractive part of the DDI. This effect is called magnetostriction and is of course not unique to quantum gases as it has been known since Joule discovered the effect in iron~\cite{Joule1847ote}. Note that the magneto- or electro-strictive effects deform a system solely due to the presence of a homogeneous field which breaks rotational symmetry. The forces responsible are due only to the interactions between the particles because the external field has no gradient. 

$\bullet$ \textbf{Application of the Thomas-Fermi approximation:}

Rather than numerically solving Eq.~\eqref{GP}, magnetostriction can be analysed by the use of the Thomas-Fermi (TF) approximation, valid for high atom numbers and sufficiently large total interaction strength~\cite{Stringari2016bec}.  This allows one to  neglect the kinetic energy term in Eq.~\eqref{GP}. For $\edd=0$ (i.e., a nondipolar BEC), one can easily show that the density of a BEC acquires an inverted parabolic shape with a so-called TF radius that scales as $R_i \propto 1/\omega_i$, $i=x,y,z$~\cite{Pitaevskii2016bec}. In the dipolar case,  Refs.~\cite{ODell2004eho,Eberlein2005eso} showed that a dBEC retains the parabolic shape in this (TF) limit.  However, its aspect ratio 
\begin{eqnarray}
	\kappa=R_\rho/R_z,\label{Eq:Kappa}
\end{eqnarray}
defined by the ratio of the TF radii perpendicular ($R_\rho$) to the radius along ($R_z$) the dipole direction, no longer reflects the aspect ratio of the trap. This is due to the contribution of $\pdd$ to the mean energy in Eq.~\eqref{Eq:Energy}, which one can calculate analytically:
\begin{eqnarray}
	E_{\rm{dd}}=-\frac47\edd\,g\,\frac{n_0\,N}{2}\fdip(\kappa),\label{Eq:Edd}
\end{eqnarray}
where $n_0$ is the density at the trap centre. The function $\fdip(\kappa)$ encompasses the anisotropy of the interaction energy, it is a decreasing function of $\kappa$~\footnote{Expression (\ref{Eq:Edd}) is valid for an inverted parabola solution of the TF approximation. If one uses a Gaussian variational ansatz for the wavefunction of the BEC, it becomes $E_{\rm{dd}}=-\edd\,g\,\frac{n_0\,N}{2}\fdip(\kappa)$.}. The complete expression is~\cite{Giovanazzi2003beo}
\begin{equation}
	\fdip(\kappa)=\frac{1+2\kappa^2}{1-\kappa^2}-\frac{3\kappa^2\rm{arctanh}\sqrt{1-\kappa^2}}{(1-\kappa^2)^{3/2}}.\label{Eq:FOfKappa}
\end{equation}
Note that an angular integration of the DDI over an isotropic distribution is zero, leading to $\fdip(1)=0$. $\fdip$ is bounded by two limits: fully collinear dipoles (attractive DDI) $\lim_{\kappa\to 0}\fdip=1$, while side-by-side dipoles (repulsive DDI) yield $\lim_{\kappa\to\infty}\fdip=-2$. Thus, one readily sees that $E_{\rm{dd}}$ is reduced by lowering $\kappa$. In the TF approximation,  minimising $E$ leads to a transcendental equation for $\kappa$~\cite{Yi2001tco,Giovanazzi2003beo,Eberlein2005eso}: 
\begin{eqnarray}
	\lambda=\kappa\left(\frac{1+2\edd-\frac{3\edd\fdip(\kappa)}{1-\kappa^2}}{1-\edd+\frac{\kappa^2}{2}\frac{3\edd\fdip(\kappa)}{1-\kappa^2}}\right)^{1/2}.\label{Eq:magnetostriction}
\end{eqnarray}
The solution of this equation for a given $\lambda$ thus gives the degree of magnetostriction. 
A solution exists for any $\lambda$ provided that $\edd\leqslant1$. The absence of a MF solution will be discussed at the end of this section, and for now we assume there exists one. 

We note that assuming a 3D Gaussian shape for the dBEC is another common approximation~\cite{Eberlein2005eso,yi2000tac,Yi2001tco,Santos2000BEC,Goral2002gse}. This approximation will be later reviewed in Sec.~\ref{subsec:global_stability}. Interestingly, and despite the difference in the approximations, such a Gaussian variational ansatz for $\psi$ yields the same equation  as Eq.\,\eqref{Eq:magnetostriction} for the aspect ratio.  We note that both the TF and the Gaussian approximations can describe a regular BEC wavefunction with only maximal density at the trap centre. They fail in describing more complex phenomena that can occur in dBECs, e.g., when operating close to the MF instability threshold; see  Secs.\,\ref{subsec:local_instab} and~\ref{subsec:supersolidity}. In this paragraph, we focus on the shape of stable dBECs.     

$\bullet$ \textbf{In-situ measurements:}

\begin{figure}[hbt]
  \includegraphics[width=0.98\columnwidth]{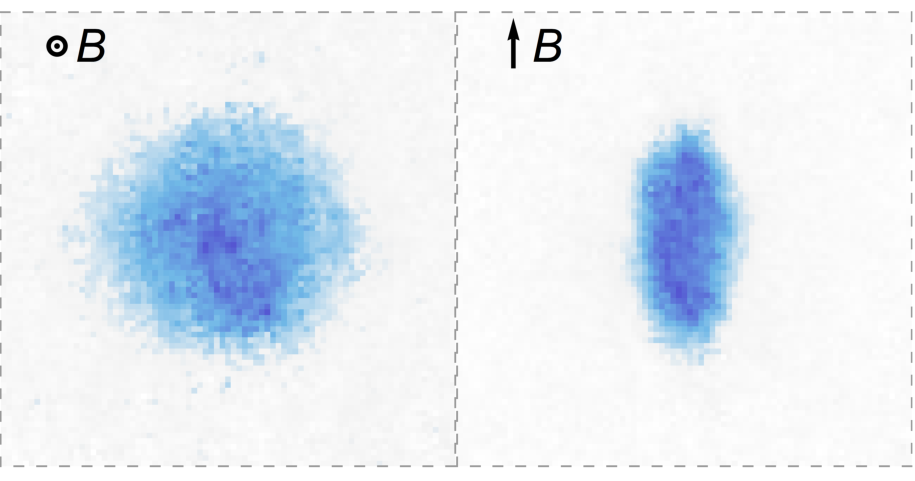}
 \caption{In situ imaging of magnetostriction of a \textsuperscript{162}Dy dBEC. The atoms are held in a pancake-shaped trap, nearly isotropic in the plane. The field is out of plane in the left image, and in-plane for the right image, while the trap is kept unchanged. Adapted from Ref.\,\cite{Wenzel2018acv}. }\label{Fig:magnetostriction}
\end{figure}

We now describe the distortion of the gas shape inside the trap. Quantitative in-situ measurements of magnetostriction have only recently been possible:  The first in-situ images of dipolar BECs were reported in Ref.\cite{Kadau2016otr} and in-situ magnetostriction images first presented in Ref.\,\cite{Wenzel2018acv}. An example is shown in Fig.~\ref{Fig:magnetostriction}.

$\bullet$ \textbf{Magnetostriction in the TOF dynamics of BECs:}

Observing the (hydrodynamic) expansion of the gas trap release is a common way to reveal BEC physics~\cite{Holland1997eoi,OHara2002ooa}, and in particular the impact of interactions on quantum gases. 
The hydrodynamic equations for the expansion of a (nondipolar) BEC were first presented in Ref.~\cite{Castin1996bec}, showing inversion of ellipticity during TOF. Solutions to the hydrodynamic equations for dipolar gases can  also be found~ \cite{ODell2004eho,Eberlein2005eso,Giovannazi2006edo}. The effect of the DDI is the same as in trap: energetics favour a distortion in the expansion that aligns the dipoles head-to-tail.  

\begin{figure}[htbp]
  \includegraphics[width=\columnwidth]{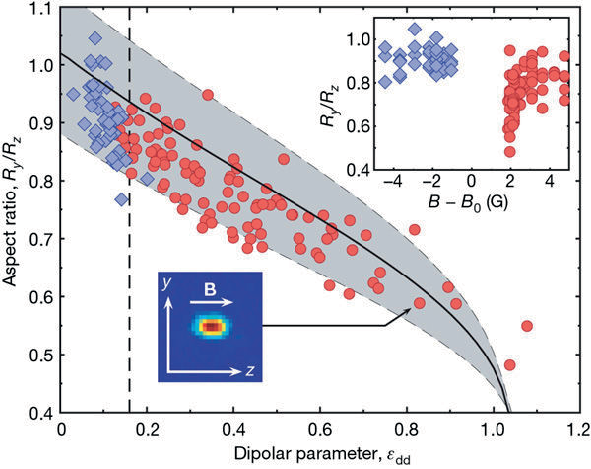}
  \caption{Aspect ratio of an expanding \textsuperscript{52}Cr BEC, measured 2\,ms after its release from a trap, as a function of $\edd$.  $\edd$ was controlled using a  a FR centred about $B_0$. Blue (red) dots: data with $B<B_0$ ($B>B_0$). Dashed line: background $\edd$. Solid line: prediction from Eq.~\eqref{Eq:Edd} without adjustable parameters. Shaded area: uncertainties in $\omega_{\rho,z}$. Inset, same data as a function of $B-B_0$. From \cite{Lahaye2007sde}.\label{Fig:Lahaye2007}}
\end{figure}

The first observation of magnetostriction in a gas was observed in the hydrodynamic expansion of a \textsuperscript{52}Cr BEC~\cite{Stuhler2005ood}. This was manifest as a dependence of the inversion of ellipticity on the magnetic field direction. Comparing the experimental dynamics with solutions to the hydrodynamic equations, a value for $\edd$ for \textsuperscript{52}Cr away from FRs was extracted \cite{Griesmaier2006cca}. The value agrees with most precise values of the background scattering length $\abg$ obtained more recently~\cite{Pasquiou2010cod}; see Sec.~\ref{subsec:fesbachsec}. Using a FR to lower $\as$---and thus to enhance dipolar effects into the $\edd>1$ regime---magnetostriction in situ and in TOF was strong enough so that a complete suppression of ellipticity inversion was observed~\cite{Lahaye2007sde}. The variation of the aspect ratio of an expanding cloud with $\edd$ can be seen in Fig.~\ref{Fig:Lahaye2007}. 

\myparagraph{Magnetostriction and TOF dynamics of ultracold, non-condensed gases}\label{subsec:thermal_anis}

Observation of magnetostriction in gases requires low temperatures and high densities. Otherwise,  thermal energies are orders-of-magnitude larger than the DDI.  As mentioned above, the use of dBECs enabled the observation of magnetostriction both in TOF and in situ. 
Magnetostriction effects could also be observed in ultracold, but non-degenerate thermal gases of sufficiently strong dipolar atoms.

The dimensionless parameter relevant for DDI corrections to the ideal gas law is obtained is obtained by comparing the mean dipolar energy to the mean kinetic energy: $\eta= E_{\rm dip}/E_{\rm k}$. For a non-degenerate gas, $E_{\rm dip}\sim S^2d^2n$ and $E_{\rm k}\sim k_{\rm B}T$, such that $\eta\propto(n\lambda_{\rm th}^3)\times(k_{\rm th}\add)$, with $k_{\rm th}=2\pi/\lambda_{\rm th} =\sqrt{mk_{\rm B}T}/\hbar$ the thermal wavenumber, and $\lambda_{\rm th}$ the thermal wavelength. To maximise $\eta$ in a thermal gas, we may choose Dy, which provides a $\add = 131\,a_0$; see Sec.~\ref{sec:amgnetic_atoms}). At typical experimental densities for a gas with a temperature just above its BEC $T_c$, $\eta$ remains $\ll1$.  Thus, at best, the DDI only weakly modifies the TOF expansion dynamics. Nevertheless, DDI effects on time-of flight dynamics have been observed, as we now discuss.

\begin{figure}[t!]
  \includegraphics[width=\columnwidth]{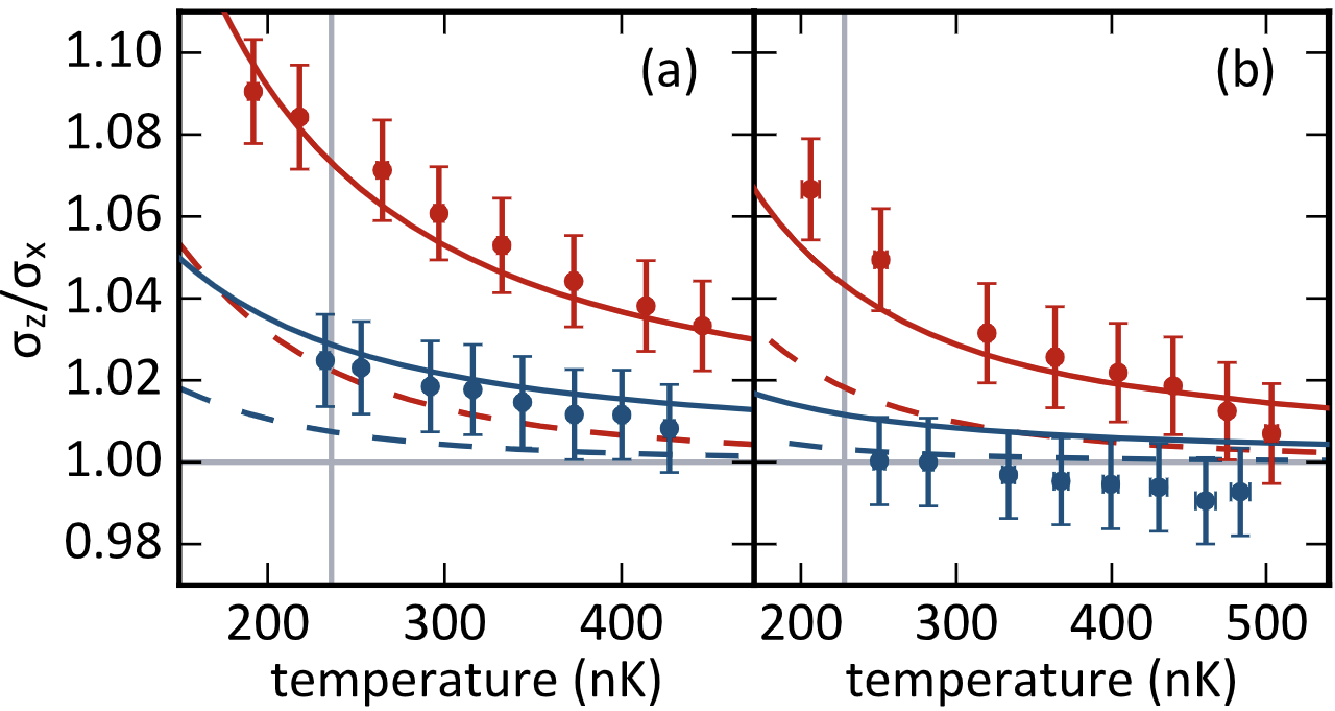}
  \caption{ The anisotropic aspect ratio of dipolar thermal gases  after 16 ms of TOF. Aspect ratio for (a)  $^{162}$Dy and (b) $^{164}$Dy. Red (blue) is for magnetic field along $\hat{z}$ and $\hat{y}$.  The solid (dashed) curves are the full (partial, MF-only) theory developed in Ref.~\cite{Tang2016aeo}. The Bose-condensation temperature is marked by the vertical grey line. From Ref.~\cite{Tang2016aeo}.\label{Fig:anisoexp}}
\end{figure}
Tang \textit{et al.}~\cite{Tang2016aeo} theoretically and experimentally studied the anisotropic expansion of a thermal dipolar Bose gases of \textsuperscript{164}Dy and \textsuperscript{162}Dy just above their degeneracy temperature. Each gas, after TOF expansion, exhibits an aspect ratio that depends on the polarisation angle of the dipoles, as had already been noted in dBECs~\cite{Lahaye2009tpo}; see Fig.~\ref{Fig:anisoexp}. The DDI strength of Dy is sufficient to cause the magnetostriction of even a dilute thermal gas. 
To predict the experimental aspect ratios versus temperature and dipolar angle, the authors developed a theory of the expansion that accounts for the Hartree-Fock MF interactions, Bose-enhanced scattering, and hydrodynamic effects that partially cross-dimensionally rethermalise the gas during the expansion.  By doing so, the authors were able to quantitatively match the theory to the data.  The theory fits provided a method to extract the gas temperature and is a relatively simple method for determining the scattering length of the gas, even near a FR; see Sec.~\ref{scatlengthmeas} for the results of the $\as$ measurements using this technique, which are compatible with previous measurements.
Moreover, the momentum distribution deformation scales with the  ratio $\eta=(n\lambda_{\rm th}^3)\times(k_{\rm th}\add)$, as expected, and arises primarily from the two-body collisions and the direct Hartree contribution to the MF energy. 

A similar effect arises in degenerate Fermi gases, although with a few fundamental differences. The kinetic energy is dominated by the Fermi energy, so the distortion in momentum space is smaller. This Fermi surface deformation stems from the exchange contribution to the mean interaction energy, which does not vanish in  the degenerate Fermi many-body state due to the Pauli exclusion principle; see Sec.~\ref{subsec:fermi_intro} and the discussion in Sec.~\ref{subsec:fermi}.

\subsubsection{Elementary excitations of dBECs}\label{subsec:excitationBEC}
In the previous section, we were concerned with the influence of the DDI at equilibrium: i.e., the magnetostrictive effect on the gas wavefunction in real space, and its consequences on the free-space expansion dynamics, which allows observations of momentum-space magnetostriction. In contrast, elementary excitations around equilibrium provide a window into  dynamical behaviour of the dipolar quantum gases. Due to the sensitivity of spectroscopic measurements, experimental studies of DDI effects on elementary excitations even at low $\edd$ are possible. Besides, stringent modifications of the excitation spectrum at $\edd \sim 1$ were also evidenced.

\myparagraph{Elementary excitations in a (homogeneous) dBEC: from phonons to free-particles.}\label{subsec:exc_hom}

$\bullet$ \textbf{Continuous Bogoliubov spectrum:}

Elementary excitations of BECs were introduced in Sec.~\ref{excitations_intro} in the uniform BEC case. In presence of DDI, the well-known Bogoliubov spectrum, obtained by linearising the GPE around the ground state, with its linear phonon branch followed by the quadratic free-particle dependence, is modified. By combining Eqs.~\eqref{vddtf} and \eqref{Bog}, the spectrum of a uniform 3D dBEC of density $n$ is~\cite{Santos2000BEC} 
\begin{equation}
		\hbar\,\omega(\boldsymbol k)=\sqrt{\frac{\hbar^2k^2}{2m}\left(\frac{\hbar^2k^2}{2m}+2\,gn\,\left(1+\edd(3\cos^2\theta_k-1)\right)\right)},\label{Eq:Dispersion}
\end{equation}
where $\theta_k$ is the angle between the direction of excitation propagation and the dipole orientation. The sound velocity for BECs with only contact interactions ($\edd =0$) is $c_0=\sqrt{gn/m}$ and  acquires an angular dependence in presence of the DDI, $c(\theta_k)=c_0\sqrt{1+\edd(3\cos^2\theta_k-1)}$. Excitations propagating along the dipoles' direction ($\theta_k=0$) have the highest sound velocity (i.e., they are stiff modes), while those propagating in the perpendicular plane ($\theta_k=\pi/2$) have the lowest velocity (i.e., they are softer modes). This anisotropy of the dispersion relation is one of the main feature of dBECs with respect to collective excitations and has important physical consequences; namely, the anisotropic superfluid critical velocity (see below) and a mechanical instability with anisotropic nature manifest for gases with $\edd\geq1$; see Sec.\ref{subsec:Stability}. 

$\bullet$ \textbf{Bragg spectroscopy:}

\begin{figure}[hbt]
  \includegraphics[width=0.98\columnwidth]{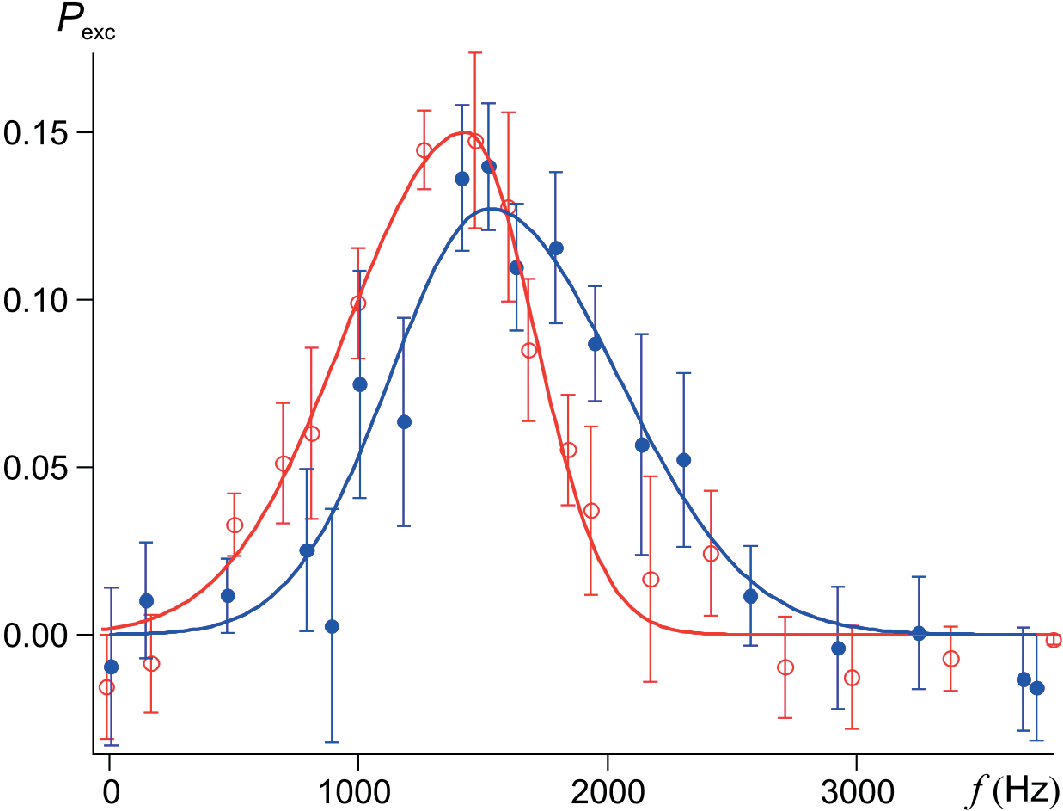}
 \caption{Bragg excitation spectra of a \textsuperscript{52}Cr BEC, with the magnetic field in two perpendicular directions. Lines are fit to the data. From \cite{Bismut2012aes}.}\label{Fig:Bismut2012}
\end{figure}

Using the standard method of two-photon Bragg spectroscopy (see, e.g., Refs~\cite{Stenger1999sei,StamperKurn1999eop,Steinhauer2002eso, Vogels2002eoo,Ozeri2002doo}) on a dBEC of \textsuperscript{52}Cr, Bismut \textit{et al.}~\cite{Bismut2012aes}  probed the dispersion relation of such a trapped dBEC away from FRs. In practice, the finite size of the trapped samples sets a lower limit on the momentum at which one can probe and compare against Eq.~\eqref{Eq:Dispersion} (which holds in the homogeneous case). This limit comes from the fact that  the BEC smallest size ($R_{\rm TF\,min}$) must be larger than the excitation wavelength: $k\,R_{\rm TF\,min}\gg1$. In this limit, the local density approximation can be used, allowing one to take into account the BEC's inhomogeneity. Bismut \textit{et al.} observed a Bragg dispersion peak that depends on the relative orientation between the Bragg momentum and the magnetic field angle, in very good agreement with theory; see Fig.~\ref{Fig:Bismut2012}. At low momentum, a departure from the homogeneous expectation was observed. Numerical simulations taking finite-size effects into account reproduce the evolution of the measured dispersion relation and its anisotropy.\par

$\bullet${\textbf{Critical superfluid velocity measurement:}}
\label{subsec:critvel}

The DDI effect on the dBEC's excitation spectrum implies other changes to the BEC's physical behaviour. A prime example lies in the change to the superfluid critical velocity. The famous Landau's criterion relates the critical velocity for an impurity moving in the direction $\hat{\boldsymbol{v}}$ in an (infinite and uniform) superfluid to its dispersion relation: $v_{\rm c}=\underset{\boldsymbol{k}}{{\rm min}}\,(\frac{\omega_{\boldsymbol{k}}}{\hat{\boldsymbol{v}}\cdot\boldsymbol{k}})$~\cite{Landau1941tto,Yu2017lcf}. In the homogeneous case discussed above, Eq.~\eqref{Eq:Dispersion} implies that the critical velocity becomes anisotropic in a dBEC, depending on whether the excitation is applied along or perpendicular to the dipole orientation~\cite{Ticknor2011asi,Yu2017lcf}. We note that the critical velocity does not generally match the speed of sound, even in the homogeneous case. This is because the dissipating excitations can occur in a different direction than the impurity's motion~\cite{Yu2017lcf}. The critical velocity is thus systematically smaller than the speed of sound in the direction of motion. For  anisotropic confinement, as discussed in Sec.\,\ref{subsec:roton}, the critical velocity should also be affected by the existence of low-energy high-momentum modes, such as the roton mode~\cite{Wilson2010csv,Ticknor2011asi}.

\begin{figure}[hbt]
  \includegraphics[width=0.98\columnwidth]{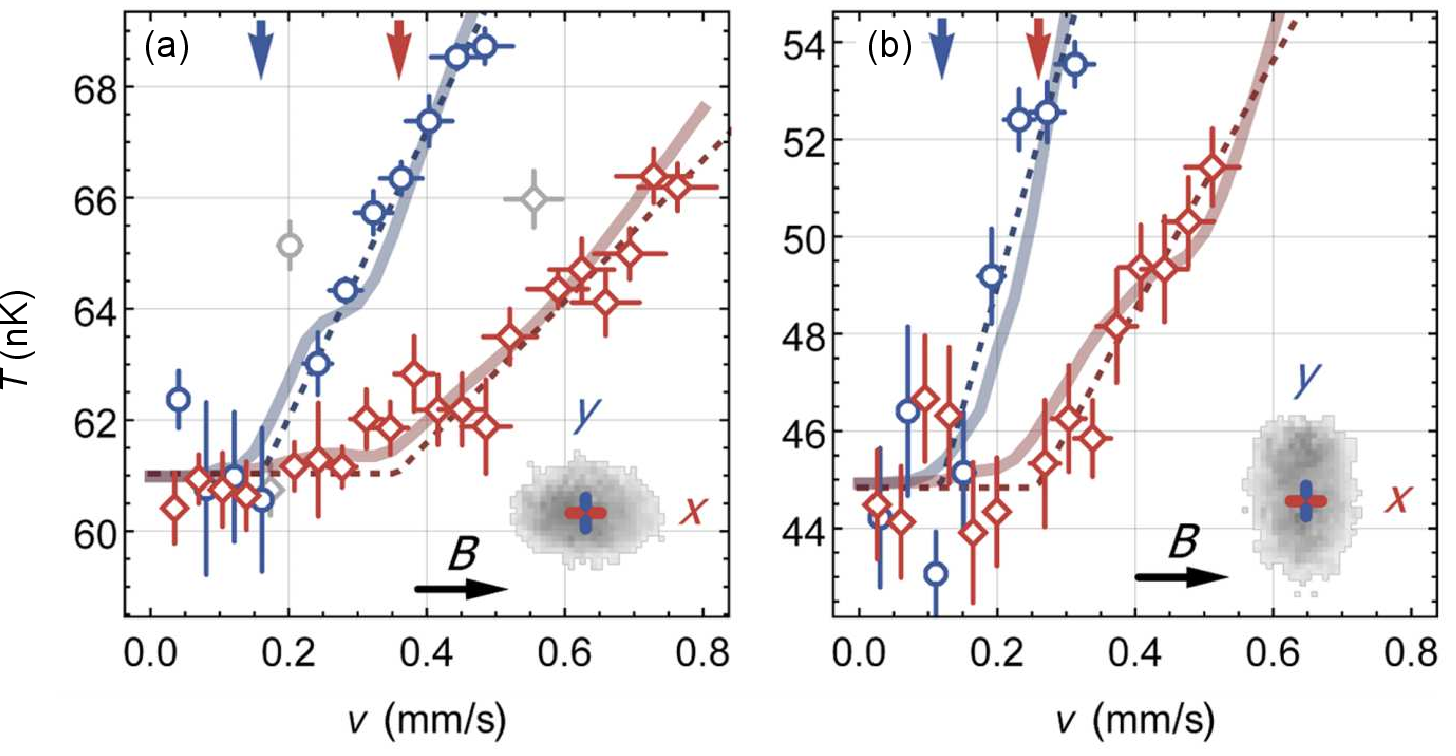}
 \caption{Temperature of the $^{162}$Dy dBEC after applying a stirring protocol.  The stirring beam is moved along $x$ (red squares) or $y$ (blue circles). The dipoles are polarised along $x$. The trap is cylindrically symmetric in $x$ and either (c) elongated  or (b) made narrow in this direction; see insets with example of in situ images. A piecewise linear fit (dashed lines) quantifies the anisotropy of the critical velocities (arrows) with respect to the dipole orientation. For both trap geometries, the velocity, $v_\perp$, for a perpendicular excitation is small than for a parallel one, $v_{||}$,  with $v_\perp=0.16(2)$mm/s ($0.12(3)$mm/s) and $v_{||}=0.36(3)$mm/s ($0.26(4)$mm/s) in b (c). Numerical simulations of the GPE (solid lines) show excellent agreement with the experiment. Adapted from \cite{Wenzel2018acv}.}\label{Fig:AnisotropVel}
\end{figure}

Superfluid velocity measurements were performed on a $^{162}$Dy dBEC using a local light defect driven linearly along one axis~\cite{Wenzel2018asb}. These showed that the anisotropy of the critical velocity as well as of the heating rate above the critical velocity  are in excellent agreement with dynamical simulations based on the GPE; see Fig.\,\ref{Fig:AnisotropVel}. Large corrections compared to the homogeneous predictions are observed and mainly attributed  to inhomogeneous density effects, as corroborated by numerical simulations of the GPE.

\myparagraph{Low-lying collective modes of trapped dBECs and oscillation measurements}
For a trapped gas, if one lowers the excitation momentum down to the regime where the corresponding wavelength is on the order of the cloud size, then one reaches the regime of low-lying excitations where the spectrum is discrete and momentum is not a good quantum number. The low-lying modes are typically \textit{surface} modes, implying an (out-of-phase) oscillation of radii in different directions. These modes typically have a compressional character, yielding a density (and thus interaction) dependence.  Signatures of the DDI in these systems have been theoretically studied  \cite{Yi2001tco,Yi2002pde,Goral2002gse,ODell2004eho,Giovanazzi2007coo}. MF methods to extract the collective modes nature and frequency rely on expanding the energy functional Eq.~\eqref{Eq:Energy}, or corresponding hydrodynamic equations, around the stationary solution. This can be done semi-analytically by either applying the TF approximation or using the Gaussian ansatz; see Secs.\,\ref{subsec:magnetostriction} and \ref{subsec:global_stability}.

\begin{figure}[hbt]
  \includegraphics[width=\columnwidth]{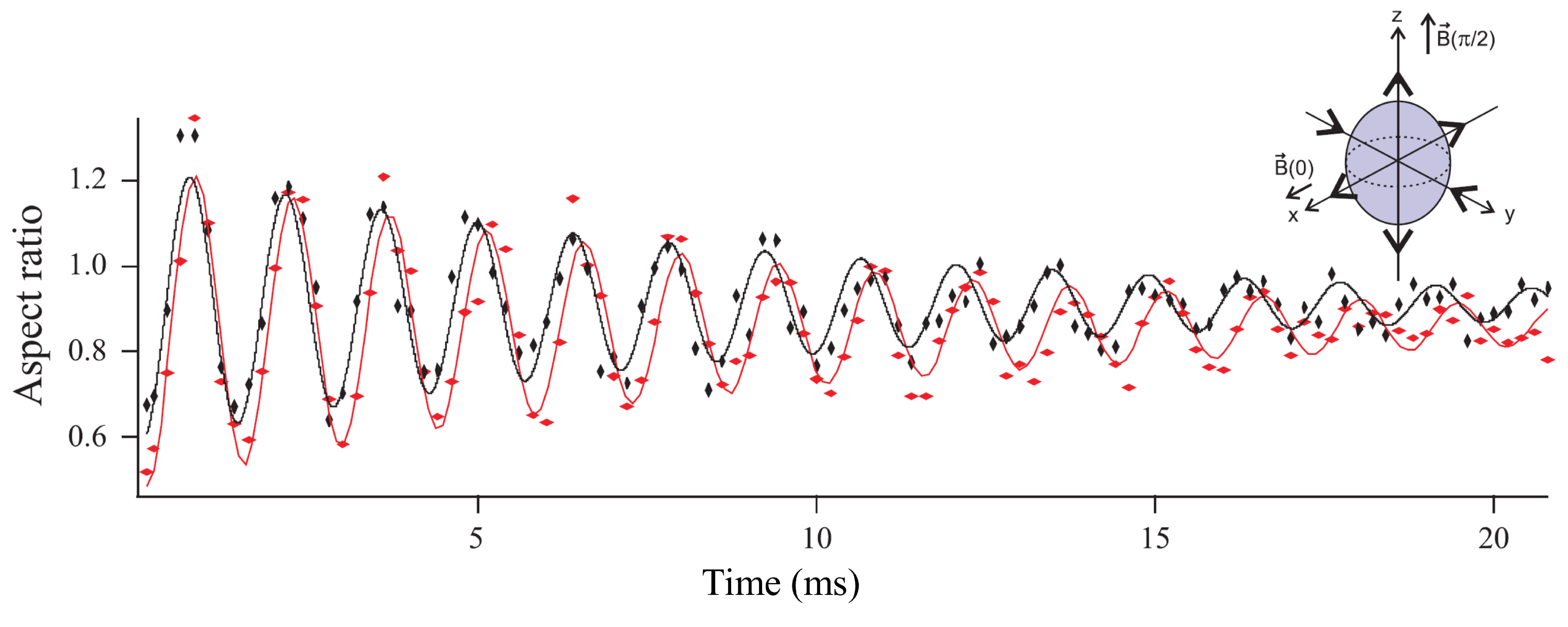}
  \caption{Time evolution of the aspect ratio of a dBEC oscillating due to the excitation of a low-lying surface collective mode (taken after TOF expansion). The two different colours show the time evolution for two different angles of the $B$ field, showing the dependence of the frequency of the collective modes on the dipoles direction due to the DDI. From~\cite{Bismut2010ceo}.}\label{Fig:Bismut2010}
\end{figure}

Experimentally, Bismut \textit{et al.}~\cite{Bismut2010ceo} studied the influence of the DDI on such modes with a \textsuperscript{52}Cr BEC. They investigated the second-lowest-lying mode, a quadrupole mode in a non-axisymmetric trap. The experimental results demonstrate that the collective mode frequency is dependent on the relative orientation of the dipoles with the trap axes, and that the frequency shift is dependent on the trap geometry, see Fig.~\ref{Fig:Bismut2010}. The experiment is in good agreement with a TF approximation theory, which neglects the kinetic energy of the atoms. When lowering the atom number below a few thousand,  the frequency shift is clearly reduced, demonstrating the importance of quantum pressure for very small samples. Quantum pressure can be taken into account by either performing full numerical simulations of the GPE, or by using a Gaussian variational ansatz for the BEC wavefunction.

\myparagraph{Spectrum of elementary excitations in trapped dBEC, roton mode.}
\label{subsec:roton}

As highlighted in Eq.\,\eqref{Bog}, the DDI contributes to  the continuous excitation spectrum via an effective coupling strength $\widetilde{\Vdd}(\vk)$, which adds to the constant contact contribution $g$. As observed in Eq.~\eqref{Eq:Dispersion}, in the 3D homogeneous case, the DDI contribution, simply given by its Fourier transform (Eq.~\eqref{vddtf}), yields an orientation-dependence, but no $k$-norm dependence. When combining the DDI with an anisotropic confinement, even more striking behaviours of the dBEC's excitation spectrum arise. Here, a $k$-norm dependence may arise from the interplay of the DDI with the trap's natural length scales. 
This $k$ dependence yields qualitative differences in the dispersion relation compared to that of a non-dipolar BEC and of a uniform dBEC. In this section, we keep the notation of $\widetilde{\Vdd}(\vk)$ for the DDI effective coupling strength, which determines the excitation dispersion relation, even if the gas in non-uniform and this differs from the Fourier transform of Eq.~\eqref{vddtf}.

A particularly interesting case is when the dispersion relation $\epsilon(k)$ of a dBEC becomes non-monotonic, presenting a local maximum (maxon) followed by a local minimum (roton). Such a qualitative change in the dispersion relation underlies important new physical behaviour. To get a first insight into such changes, it is interesting to highlight that such dispersion relation  resembles the celebrated dispersion relation of superfluid helium~\cite{Landau1941tto,Landau1947ott,Henshaw1961mam}. Here, roton excitation were first speculated to explain the exotic  macroscopic properties of this superfluid~\cite{Landau1941tto,Landau1947ott,Feynman1955}, long before their observation~\cite{Henshaw1961mam}. Thanks to its low energy, the roton excitation strongly influences the response of the superfluid to small excitations. Furthermore, because of its large momentum, the roton underlies the tendency of the fluid to crystallise (at the wavelength corresponding to its inverse momentum)~\cite{Nozieres2004itr,Kirzhnits1971cco, Schneider1971tot} (although one should note that the phase transition to solid helium is not due to Roton softening). 

$\bullet$ \textbf{Roton excitation spectrum in anisotropic semi-infinite dBECs:}

In 2003, a dispersion relation presented a roton minimum was predicted to occur in anisotropically trapped weakly interacting dipolar gases, first in the context of light-induced DDI by O'Dell \textit{et al.}~\cite{ODell2003rig} and, shortly after, in that of magnetic or electric dipolar system by Santos \textit{et al.}~\cite{Santos2003rms}. These seminal works consider semi-infinite trapping geometries, i.e., infinite along one  ($\omega_z=0$)~\cite{ODell2003rig} or two ($\omega_\rho=0$)~\cite{Santos2003rms} directions of space, and harmonically confined along the others. This treatment allows to account for anisotropy effects while facilitating theoretical treatment, yielding semi-analytical expressions of $\widetilde{\Vdd}(\vk)$ within the TF approximation, and providing an intuitive picture of the effect.

The occurrence of a roton minimum arises from $\widetilde{\Vdd}(\vk)$  becoming attractive at large $k$. In the quasi-infinite geometries of Refs.~\cite{Santos2003rms,ODell2003rig}, the confinement acts 
to limit the attractive contribution of the DDI: The attraction dominates over the repulsive contribution only if the momenta (along an unconfined direction) have a norm larger than the inverse characteristic confinement length $\ell_z$. In this way, for excitations along the unconfined directions, $k\ll 1/\ell_z$ yields $\widetilde{\Vdd}(\boldsymbol k)>0$, while, for  $k\gtrsim 1/\ell_z$, $\widetilde{\Vdd}(\boldsymbol k)<0$. 
Therefore, the DDI stiffens the dispersion relation in the phononic regime while it bends it down for large $k$. Because of the additional contribution of kinetic term $\frac{\hbar^2k^2}{2m}$ which ultimately dominates at very large $k$, the effect of $\widetilde{\Vdd}(\boldsymbol k)<0$ is the strongest at $k\sim 1/\ell_z$ and, for weak enough $s$-wave coupling strength $g$, a minimum arises at $\kr \sim 1/\ell_z $, matching a roton mode. Here the roton wavelength is typically set by the confinement along the direction of attractive DDI. 
Furthermore, the roton energy $\epsmin$ can be lowered by increasing the density or increasing $\edd$. When $\epsmin=0$, Ref.~\cite{Santos2003rms} finds $\kr=\sqrt{2}/\ell_z$, independent of the density and the interaction parameters $g,\,\edd$.

$\bullet$ \textbf{Roton excitation spectrum in anisotropic fully trapped dBECs:}

Numerous subsequent theoretical works describe the roton in dipolar gases confined in finite geometries 
(Eq.~\eqref{Eq:trap} with $\lambda\gg 1$). They study 
how the roton spectrum of the infinitely elongated geometries described above survive in the fully trapped case and how the steady state and dynamical behaviour of dBECs are affected~\cite{Giovanazzi2004iat,Ronen2007raa,Martin2012sas,Blakie2012rsi,JonaLasinio2013rci,Bisset2013fri,Bohn2009hda,Parker2009sfd,Wilson2008mot,Nath2010fpi,Wilson2010csv,Klawunn2011lan,Bisset2013rei,Blakie2013daf,JonaLasinio2013tof,Natu2014dof}. In the case where the trap is tightly confining along the dipoles and sufficiently anisotropic (e.g., $\lambda\gg1$), a roton-like spectrum arises for large enough $\edd$. The relevant range of $\edd$ depends on the exact trap geometry. While, in the inhomogeneous case, the momentum is not longer a good quantum number, the rotonic properties of the spectrum are most directly revealed by the behaviour of the dynamic structure factor. This factor $S(k,\omega)$ conveniently describes the system's response to a Bragg excitation at momentum $k$ and frequency $\omega$~\cite{Blakie2012rsi}.  Furthermore, in harmonically trapped geometries, the elementary mode contributing to the roton minimum in the dynamic structure factor has an amplitude that vanishes away from the dBEC centre~\cite{JonaLasinio2013rci}. This confinement effect can be understood, within the local density approximation, as the roton energy decreasing with increasing density. This translates into a finger-like feature in the (discrete) dispersion relation~\cite{Bisset2013rei}. Other signatures of the existence of the roton mode in finite dBECs include anomalously large density fluctuations at the roton wavelength in an equilibrium gas~\cite{Klawunn2011lan,Bisset2013fri,Blakie2013daf}, strong correlation emerging from an interaction quench~\cite{Natu2014dof}, and peculiar structures in the collapse dynamics~\cite{Parker2009sfd,Bohn2009hda}; see also Sec.~\ref{subsec:local_instab}. 

When the trap anisotropy is reduced (e.g., $\lambda \gtrsim 1$) and $R_{\perp} \gtrsim R_{z}$, 
the existence of a roton feature, occurring at large $k$ compared to $1/R_{\perp}$, becomes progressively lost. Indeed, the argument of a finite $k$ scale for the activation of the attractions starts to break down. Yet the interplay between DDI and geometry persists, now including also finite-size effects. This yields other interesting features in the excitation spectrum, in particular the so-called angular roton mode that has a momentum $k\sim 1/R_{\perp}$, but a non-zero angular momentum~\cite{Ronen2007raa,Wilson2009aco}; see also discussion in Sec.~\ref{subsec:local_instab}.  Finally, we note that while most of the theoretical work has focused on cylindrically symmetric geometries about the dipole axis, extensions to transversely anisotropic geometries have also been discussed~\cite{Martin2012sas}.

\begin{figure}[htbp]
  \includegraphics[width=\columnwidth]{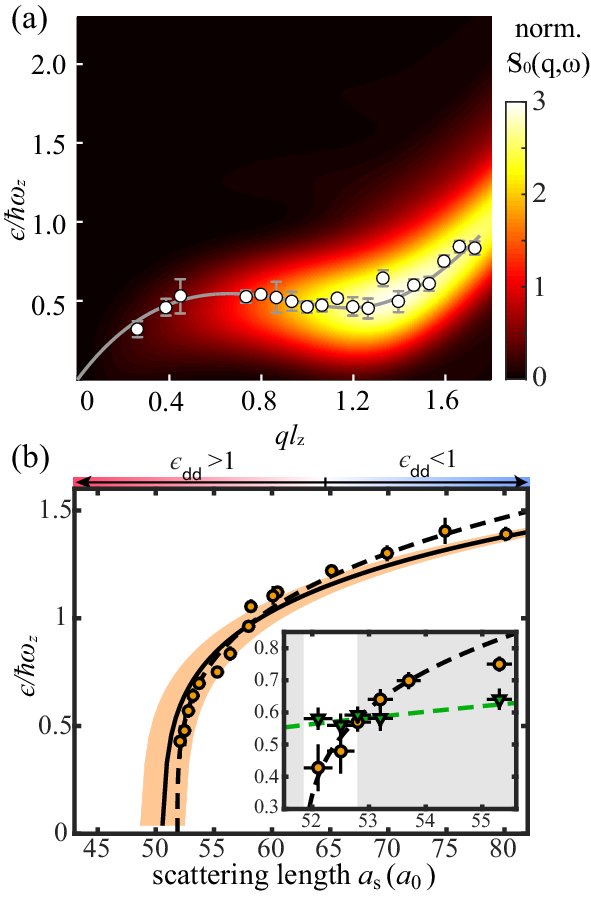}
  \caption{Roton excitation spectrum measured with a $^{166}$Er dBEC of cigar shape (cigar along $y$, dipoles along $z$). (a) Measured dispersion relation (dots) compared to predicted response for the Bragg measurement (colorcode) at $a=52.5a_0$ 
  [theory at $51.6a_0$]. (b) Measured $\epsilon(\kr)$ vs.\, scattering length (circles) and corresponding theory prediction. The shadings show the calculations over the prediction interval of $\as$. The dashed line shows the power-law fit to the experiment. Inset: zoom-in around the instability threshold comparing $\epsilon(\kr)$ (circles) to $\epsilon(k_{\rm max})$ of the maxon [local maximum] (triangles), respectively with the power-law fit and a guide to the eye (dashed lines). The region where $0\leq\epsilon(\kr)\leq\epsilon(k_{\rm max})$ is highlighted with a white background. Adapted from \cite{Petter2019ptr}.\label{Fig:Petter2019}}
\end{figure}

$\bullet$ \textbf{Experimental measurements of the roton spectrum:}

The roton mode has remained elusive in Cr experiments due to the weak dipolar character of this species. 
That weakness makes the $\as$-range of existence of a potential roton minimum very narrow, as well introduces strong three-body losses when tuning within this small $\as$ -- see also Sec. \ref{Sec:DCQD}.  This leads to a fast change of the BEC's properties, including its excitation spectrum. The roton mode was experimentally observed in a $^{166}$Er BEC~\cite{Chomaz2017oot, Petter2019ptr} using a cigar-shaped trap geometry with the dipoles aligned along a tightly confining direction. This geometry, simplifying experimental observations, differs from most of the early theoretical works on roton excitations, which rely on cylindrical symmetry around the dipole axis. The observation of Ref.~\cite{Chomaz2017oot}, relying on instability dynamics, will be described in Sec.~\ref{subsec:local_instab}.  In Ref.~\cite{Petter2019ptr}, Petter \textit{et al.}~\cite{Petter2019ptr} reported on roton spectrum measurements based on Bragg spectroscopy, similar to Sec.\,\ref{subsec:exc_hom}. The scattering length was tuned close to instability and the excited momentum, along the trap axis, varied from $k\ll 1/\ell_z$ to $k\sim 2/\ell_z$, thus probing the full phonon-maxon-roton dispersion relation; see Fig.\,\ref{Fig:Petter2019}(a).   
When increasing $\edd$, a preferential softening of $\epsilon(k)$ at large $k$ is observed, finally forming a minimum at $k=\kr\sim 1.3/\ell_z$; see Fig.\,\ref{Fig:Petter2019}(a). The minimum is observed only in a narrow range of scattering lengths  and the roton gap shows a fast decrease with $\edd$ toward instability ($\epsilon(\kr) \approx 0$), see Fig.\,\ref{Fig:Petter2019}(b). The Bragg measurements also proved, through the increase of the Bragg response, the enhancement of density-density correlations at $\kr$ as the roton instability is approached, indicative of the system tendency to crystallise. 
More recently, experiments performed on Dy BECs revealed roton excitations by a direct analysis of the in-situ density fluctuations~\cite{Hertkorn2021dfa}, following early theoretical studies~\cite{Klawunn2011lan,Bisset2013fri,Blakie2013daf}. Fluctuation analysis provides access to the static structure factor. It  also allowed to identify the two degenerate roton modes, which correspond to symmetric and antisymmetric density patterns to the trap centre.
The study of roton excitations via the induced density fluctuations was extended beyond the cigar-shaped case, to oblate systems, in an independent set of experiments~\cite{Schmidt2021rei}. Here the  the distinct softenings of several radial and angular rotons were revealed. As introduced in the previous paragraph, radial rotons correspond to the standard situation of a mode of large radial momentum $\langle k\rangle$ and radial symmetry, angular rotons distinctly have low $\langle k\rangle$ but large angular momentum and show azimuthal patterns~\cite{Wilson2009aco,Bisset2013rei} and reveal as such in the density fluctuations.

\subsubsection{Mean-field stability of dBECs}\label{subsec:Stability}

Beyond the properties of the ground state and its collective excitations in a stable regime, the DDI also impacts the very stability of the state, in particular making it  geometry dependent. In this section, we  review in detail the effect of the DDI on the stability of the quantum gases: first, how the DDI makes  stability depends on the trap geometry; second, how  it affects the stability of a cloud  via long-range interactions between neighbouring clouds; and finally how  the DDI introduces, beyond the global stability condition, a distinct (local) collapse mechanism in some special geometries. We will see how the later case relates to the softening of excited modes distinct from the lowest lying ones, arising from the interplay of DDI and either anisotropic geometries or, additionally, finite-size effects.

\myparagraph{Global mechanical stability for a single BEC}\label{subsec:global_stability}

Three-dimensional, homogeneous BECs under contact interactions are mechanically stable for positive compressibility~\footnote{$\chi=-\frac1V\frac{\partial V}{\partial P}=\frac1n\frac{\partial n}{\partial P}=\frac{1}{n^2}\frac{\partial n}{\partial \mu}$} $\chi=\frac{1}{nmc_0^2}>0$ (where $c_0$ is the speed of sound; see Sec.\,\ref{subsec:exc_hom}), they are thus unstable for $\as<0$~\cite{Stringari2016bec,Kagan1998ceb}. Finite-sized~\cite{Eigen2016oow} and harmonically trapped~\cite{Bradley1997BEC,Roberts2001cco} BECs can be stabilised by quantum pressure for small negative scattering length. As dBECs experience competing interactions, their mechanical stability criterion is more complex. Even at positive scattering length, this mechanical instability occurs for $\edd=1$ as the sound velocity perpendicular to the dipole orientation then cancels, $c^2|_{\theta_k=\pi/2}=0$; see Sec.~\ref{subsec:excitationBEC}. 
However, this holds  for only infinite, homogeneous, and isotropic BECs. Because of the anisotropic character of the DDI, the anisotropy of real samples must be considered to understand their stability.

$\bullet$ \textbf{Stability within the TF approximation:}

\begin{figure}[hbt]
  \includegraphics[width=\columnwidth,trim={7cm 0 0 0}]{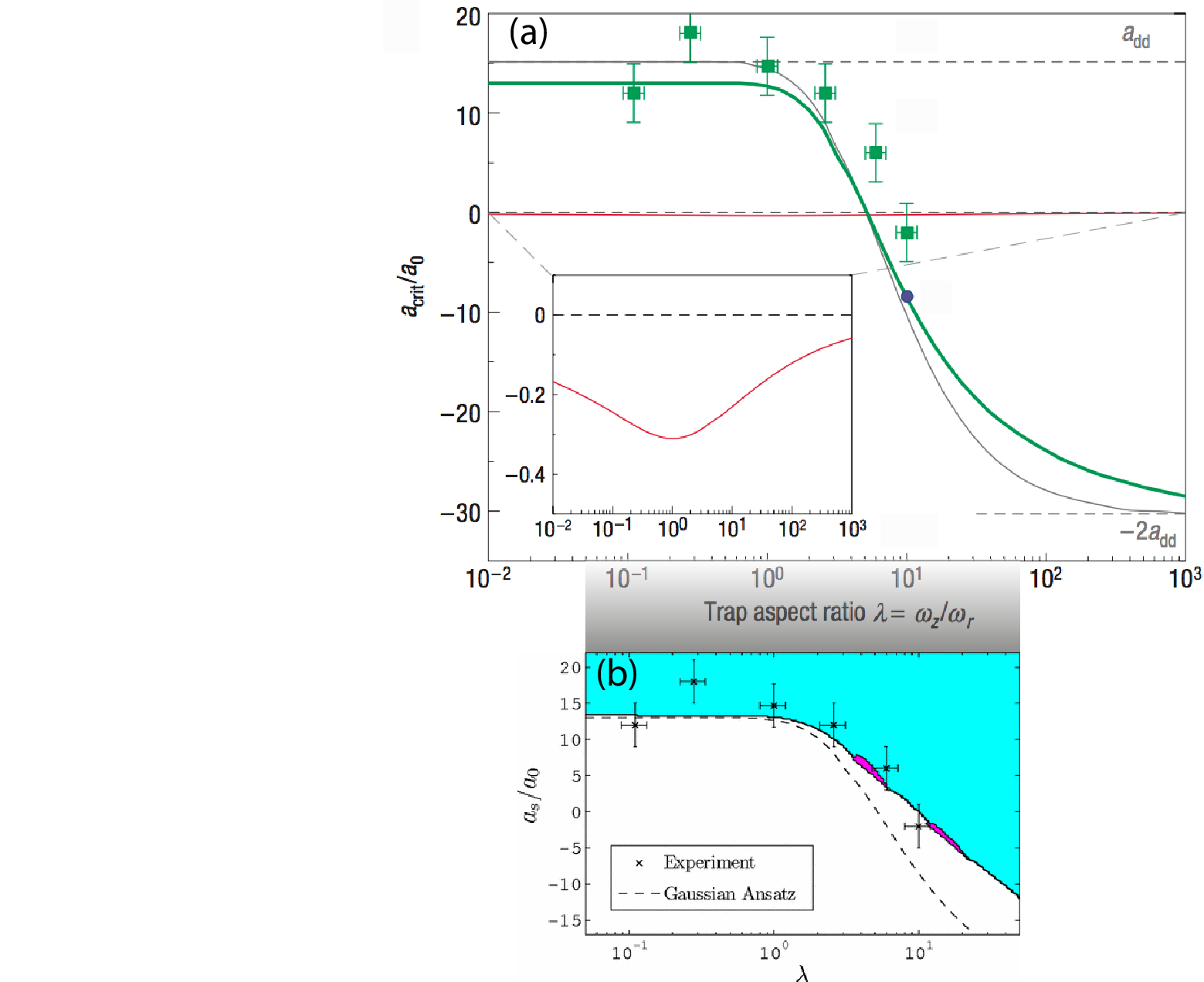}
  \caption{Stability diagram of a $^{52}$Cr dBEC in a cylindrical trap of aspect ratio $\lambda$ versus $\as$. (a) Taken from \cite{Koch2008soa}. The thin grey line is the result of the condition $\as<\add \fdip(\kappa)$, while the green line is the result of the Gaussian ansatz analysis. The red line is for $\edd=0$.  In the large N limit this diagram is universal, i.e., it also holds for electric dipoles. (b) Same data as (a), now compared to either a Gaussian ansatz (dashed line) or a numerical solution, minimising Eq.~\eqref{Eq:Edd} (solid line). From \cite{Wilson2009aco}}\label{Fig:Koch2008}
\end{figure}

Global mechanical collapse can be alternatively understood from an energy argument. The energy density of MF interactions, when attractive, scales like $E_{\rm MF}\sim-n^2$ and is thus minimised for infinite density. Using the TF approximation and neglecting the kinetic energy contribution, $E_{\rm MF}<0$  leads to a singularity of the ground state density, thus giving an instabibility. From Eq.~\eqref{Eq:Edd}, the sign of the MF dipolar energy of a cylindrical dBEC depends on the cloud aspect ratio $\kappa$. Taking into account both interactions, one can show that the total MF interaction energy scales like 
\begin{equation}\label{eq:E_MF_TF}
	E_{\rm MF}\sim gn_0N(1-\edd \fdip(\kappa)).
\end{equation}
Imposing an attractive MF interaction, $E_{\rm MF}<0$, leads to the instability condition $\as<\add \fdip(\kappa)$. Solving for $\kappa(\lambda)$ through (\ref{Eq:Kappa}) leads to a stability diagram as a function of $\as$ and $\lambda$. Intuitively, for very prolate traps, $\lambda\ll1$ and the dipoles are aligned head-to-tails $\kappa\ll1$ while the DDI is mostly attractive ($\fdip(\kappa)\approx1$). Thus, $\as$ must surpass $\add$ for stability. For the opposite case of very oblate traps ($\lambda\gg1$), dipoles repel each other twice more strongly $\fdip(\kappa)\approx-2$, and $\as$ must be large and negative to reach an effective attractive MF energy, $\as<-2\,\add$. This simple reasoning that considers only the MF interaction leads to a prediction that well describes  experimental observations \cite{Koch2008soa}; see Fig.~\ref{Fig:Koch2008}(a).

$\bullet$ \textbf{The Gaussian ansatz for a BEC near collapse:}

However, near collapse, kinetic energy effects also play a role, especially for low atom numbers. One must depart from the TF approximation, keeping all terms in Eq.~\eqref{Eq:Energy}. The variational Gaussian ansatz provides a semi-analytical approach 
\begin{equation}\label{Eq:psi_gauss}
	\psi(\boldsymbol r)=\frac{N}{\pi^{3/4}\bar\sigma^{3}}e^{-\sum_{j=(x,y,z)} \frac{i^2}{2\sigma_j^2}},
\end{equation}
inserted into Eq.~\eqref{Eq:Energy}, the energy is minimised with respect to the variational parameters $\sigma_{x,y,z}$, describing the BEC sizes along the trap axes $x,y,z$. Here, $\bar\sigma=(\sigma_{x}\sigma_{y}\sigma_{z})^{1/3}$. Once the integral is carried out, the different contributions read: 
\begin{equation}
E_{\rm k} = N\,\frac{\hbar^2}{4\,m}\sum_i\frac{1}{\sigma_i^2}
\end{equation} for the kinetic energy, 
\begin{equation}
E_{\rm trap} = N\,\frac m4\sum_i\omega_i^2\sigma_i^2
\end{equation}
for the external trapping energy, and the MF interaction energy reads
\begin{equation}\label{eq:E_MF_gauss}
E_{\rm MF}=N^2\,\frac{g}{2(2\pi)^{3/2}\bar\sigma^3}(1-\edd\fdip(\kappa_x,\kappa_y))
\end{equation}
where we allowed non-axisymmetric traps. The expression for the generalised $\fdip$ function in terms of the two aspect ratios $\kappa_{x,y}=\sigma_{x,y}/\sigma_z$ was first worked out in \cite{Giovanazzi2006edo}. One can then minimise the total energy $E=E_{\rm k}+E_{\rm trap}+E_{\rm MF}$ to find the ground-state size $\sigma_{x,y,z}$. In general, for $\edd>1$, there is always a global singularity with $\kappa_{x,y}\ll1$ and $E\rightarrow -\infty$, but a local metastable minimum exists as well and sets the stability condition. While they coincide for very large $N$, at finite $N$ the stability diagram from the Gaussian ansatz differs from the one derived using a TF approximation, see, e.g., Fig.~\ref{Fig:Koch2008}(a). At low trap aspect ratios $\lambda\ll1$, the kinetic energy acts to stabilise the gas against collapse, as in the contact case. However, for $\lambda\gg1$ the effect is opposite, kinetic energy increases the smallest size, which adds a little attraction via dipolar effects and destabilises the gas. 

$\bullet$ \textbf{First measurements of the instability threshold:}

The instability threshold was studied in $^{52}$Cr BECs of about 25,000 atoms in tunable traps~\cite{Koch2008soa}: When quenching $\as$ downward using a magnetic FR, the authors observed an abrupt and quick disappearance below a critical value, $a_{\rm crit}$, of the BEC peak in TOF images. The experimental values of $a_{\rm crit}$ as a function of $\lambda$ were found to agree well with the Gaussian ansatz predictions; see Fig.~\ref{Fig:Koch2008}. For the first time, the stability and thus the very existence of a BEC was ensured solely by dipolar interactions, and a purely dipolar BEC with $\as=0$ was obtained~\cite{Koch2008soa}.

\myparagraph{Stability of dBEC assemblies}\label{subsec:1dstab}

\begin{figure}[hbt]
  \includegraphics[width=\columnwidth]{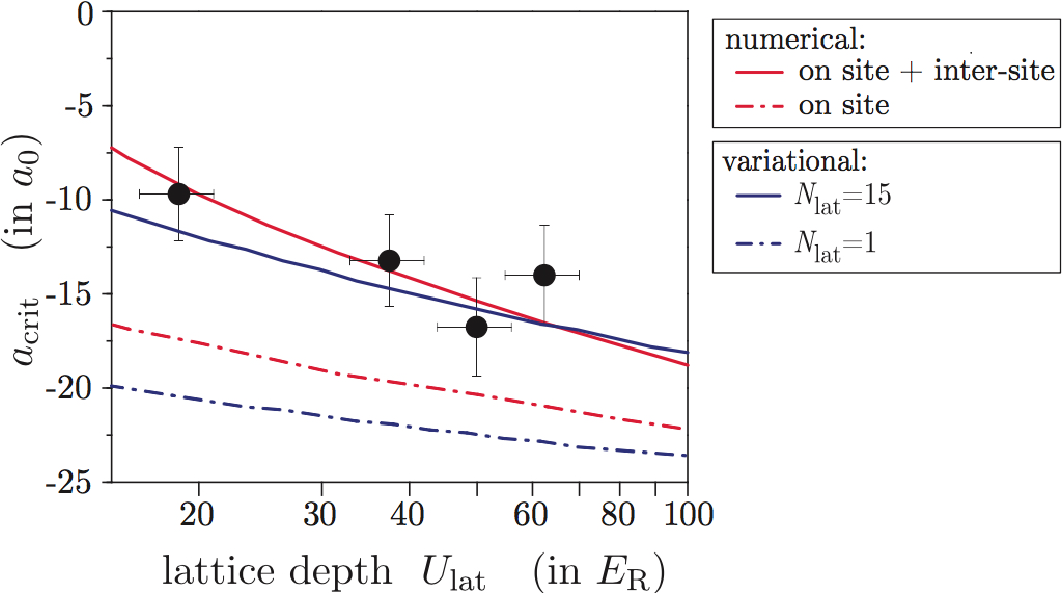}  
  \caption{Stability diagram of a dBEC in an optical lattice as a function of lattice depth and scattering length. The experimental data is compared with theory including (excluding) the influence of neighbouring traps as full (dot-dashed) lines. Red lines correspond to a full numerical solution of the GPE (Eq.\,\eqref{GP}), while blue lines correspond to a Gaussian variational ansatz for the on-site density distribution. From S.~M\"uller, PhD thesis, Stuttgart University.}\label{Fig:Muller2011}
\end{figure}

Consider a stack of pancake-shaped BECs, realised with a one-dimensional optical lattice. The DDI being long-range, the total energy of a given layer contains contributions from interactions with  neighbouring layers. Since the stability of the system is ensured by the existence of an energy minimum, nearest-neighbour interaction can modify the stability diagram. Indeed, the first observation of nearest-neighbour dipolar effects was reported in \cite{Fattori2008mdi} via dephasing of Bloch oscillations in a potassium optical lattice interferometer. Fattori~\textit{et al.} then calculated the inter-site interaction using a Gaussian ansatz for an individual layer. The contribution is attractive if the dipoles are aligned with the lattice direction. This negative contribution modifies the energy landscape and can suppress the local minimum, destabilising the gas. This can be understood in the following way: The neighbouring layers attract the atoms resulting in an effective repulsive potential along this axis. The pull is stronger at the radial centre where the neighbouring density is highest leading to a stronger effective radial trapping. The total effect is destabilising the BEC. The instability threshold in scattering length is thus higher. M\"uller \textit{et al.} \cite{Muller2011soa} have observed this effect with a \textsuperscript{52}Cr gas in a one-dimensional optical lattice, obtained with a retro-reflected 1064~nm beam. The difference in critical scattering length with respect to a theory neglecting the neighbouring layers was as high as $8\,a_0$, in agreement with expectations; see Fig.~\ref{Fig:Muller2011}.\par
All arguments given above consider the stability of a dBEC at equilibrium inside a trapping potential (harmonic or optical lattice). The in-situ density distribution determines the MF stability. However, even for a stable in-trap density, it is possible that the release from an optical lattice into free-space modifies the distribution in a way that makes it mostly attractive and then induces collapse. This deconfinement-induced collapse has been observed experimentally in Ref.~\cite{Billy2012dic}. Experiments were performed in an optical-lattice along the dipoles axis. They showed that, for large lattice depths, no losses were observable in-trap. However the release from the trap leads to a collapse visible in the d-wave shape of the density distribution, see also Sec.~\ref{Sec:DCQD}. Simulations of the GPE confirmed that no atomic losses were expected in trap, but were induced by the deconfinement and following collapse. 

\myparagraph{Modulational instabilities}\label{subsec:local_instab}

$\bullet$ \textbf{Modulational instabilities predicted from DDI:}

The stability criterion developed above, based on the existence of a local minimum in the energy landscape as a function of the BEC widths ($\sigma_{x,y,z}$), is only partial. It  allows only  shape-conserving perturbations to the BEC profile. However, other perturbations containing local density modulations are not taken into account by this criterion. The existence of instabilities driven by higher lying modes in dBECs was first noted in the seminal paper~\cite{Santos2003rms} predicting a roton-type dispersion relation in anisotropic semi-infinite dBECs, see Sec.~\ref{subsec:roton} and ref.~\cite{Giovanazzi2004iat}. 
Tuning the parameters of the dBEC  (i.e., its scattering length and trapping geometry) can lead to a softening of this roton minimum ($\omega$ reaching zero and becoming imaginary). Then, the dBEC becomes unstable and, in the early dynamics, its population gets transferred to the roton mode leading to the onset of density waves and, hence, a type of modulational instability. 

The combined effect of axial trapping and DDI is also present in finite-size dBECs, as long as a tighter confinement is applied along the dipoles; see also Sec.~\ref{subsec:roton}. This mechanism was also shown to favour local density modulations in the equilibrium profile of the dBEC itself, i.e., as a result of finite-size effects~\cite{Ronen2007raa,Ronen2007dbe,Dutta2007gss}. In this case, an instability can be triggered when $\edd>1$. It differs from a global instability in that several local attractors gather the atomic density leading to an ensemble of local collapses~\cite{Parker2009sfd,Wilson2009aco,Bohn2009hda}. This kind of instability can be seen as the softening ($\omega^2<0$) of a mode that is not one of the lowest-lying surface and monopole modes: one quantum number is not the lowest possible one, for instance the momentum $\langle k\rangle$ or the angular momentum $m$ \cite{Wilson2009aco,Bisset2013rei}. Yet, due to confinement, this mode still belongs to the discrete part of the spectrum where the wavenumber is not a good quantum number : $\langle k\rangle\times R_{\rm \perp}\lesssim1$ where $R_{\rm \perp}$ is the typical size of the BEC in the plane perpendicular to the magnetic field and $\langle k\rangle$ is calculated for the given mode. In particular, in narrow regions of the $(\lambda,N)$-space, BECs with biconcave shapes have been predicted by minimising the energy functional of the GPE (i.e., in the MF regime) and their collapse are driven by angular roton 
~\cite{Ronen2007dbe,Dutta2007gss,Parker2009sfd,Wilson2009aco,Schmidt2021rei}.
The softening of this intermediate-low-lying modes also yields a type of modulational instability. 

We note that, in ultracold atom experiments, modulational types of instability were first predicted for contact interacting gases~\cite{Kevrekidis2004pfd} and investigated in this case~\cite{Nguyen2017fom, Everitt2017ooa}. These modulational instabilities are in fact of a different nature to the one described above for dBECs. Indeed, in the contact case, the lowest lying mode is always soft at the instability (imaginary $\omega$). The modulational instability here arises when other modes are also unstable and have a larger growth rate. 
The mode with the highest growth rate has a finite $k$, which is then favoured rather than global collapse~\cite{Salasnich2003mia,Carr2004ssf}.

\begin{figure}[htbp]
  \includegraphics[width=\columnwidth]{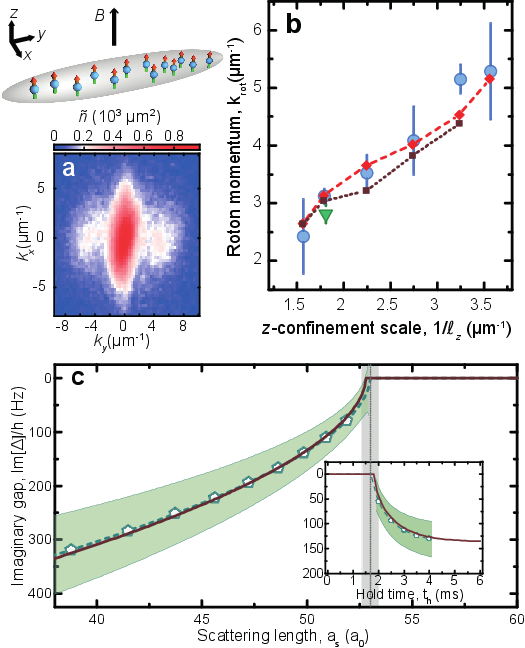}
  \caption{Roton mode in a $^{166}$Er dBEC of cigar shape (cigar along $y$, dipoles along $z$). (a) Average momentum distribution showing the appearance of two side peaks at finite $k_y$ (along the cigar elongation) after a quench to lower $\as$. 
  (b) Extracted $\kr$ for different traps, as a function of $1/\ell_z$, $\ell_z$ the confinement length along the dipoles. Data (dots) are compared to a semi-analytical theory (dotted line) and numerical simulations of the GPE (dashed line). (c)  Imaginary roton gap, $\Delta$, versus $\as$, extracted from the growth rate of the momentum-peak population and comparison to the results of the semi-analytical theory (solid line). The inset show the same data but as a function of the time after the $\as$-quench. \cite{Chomaz2017oot}.\label{Fig:Chomaz2018}}
\end{figure}

$\bullet$ \textbf{Experimental evidence for local collapse:}

In dBECs, the role played by local collapse was first noted by an extensive theoretical analysis of the data from Ref.~\cite{Koch2008soa} by Wilson \textit{et al.}~\cite{Wilson2009aco}, going beyond the Gaussian ansatz (Eqs.~\eqref{Eq:psi_gauss}-\eqref{eq:E_MF_gauss}) and solving exactly the GPE (Eq.~\eqref{GP}). Their predictions, showing a better agreement with the measured stability threshold at large  trap anisotropy with $\lambda\gtrsim 3$ (Fig.~\ref{Fig:Koch2008}(b)), implies the occurrence of local collapse. Yet the instability mechanism was not experimentally resolvable and later studies of the collapse dynamics also let modulational instabilities remain elusive, see Sec.~\ref{subsec:collapse}~\cite{Metz2009cco}.

More recently, the impact of such instabilities was experimentally studied on finite-size anisotropically-trapped dBECs of \textsuperscript{164}Dy, either in pancake traps \cite{Kadau2016otr} or in cigar shaped ones \cite{FerrierBarbut2016ooq,FerrierBarbut2016lqd} with a tight confinement along the dipoles. Here, after quenching $\as$ down, remarkable long-lived in-situ density structures were observed. 
The observed density structures have been attributed to the occurrence of a modulational instability following the quench~\cite{Kadau2016otr}.  
 Because of the limited system size of the original pancake-shaped geometry in Ref.~\cite{Kadau2016otr}, the instability is expected to be driven by an angular mode with $\langle k\rangle\times R_{\rm BEC}\sim 1$ and $m>1$~\cite{Wachtler2016qfi}, as also experimentally evidenced in a recent set of experiment~\cite{Schmidt2021rei}.
Following their first observation, the authors of the above-cited experimental works characterised the absence or existence of a modulational instability as a function of the trap geometry via its signature on the final density distribution; i.e., the observation of single or multiple droplets long after the quench~\cite{FerrierBarbut2018ooa}. They identify a critical trap aspect ratio  $\lambda=1.87(14)$ (compared to the dipole orientation, $z$) above which the modulational instability exists; see also Sec.~\ref{subsec:phasediagram_droplet}. The instability itself was not experimentally investigated in these works as they focused on the long-time behaviour, which results from a subsequent intricate non-linear dynamics loosing track of the unstable mode driving the collapse. The authors found that these final density distributions were surprisingly stable, which revealed an unpredicted stabilisation mechanism. This discovery set the ground for a new paradigm of quantum fluids, which will be discussed in the next Section~\ref{Sec:DCQD}.

$\bullet$ \textbf{Experimental investigation of the roton instability:}

As introduced in Sec.\,\ref{subsec:roton}, Ref.~\cite{Chomaz2017oot} first observed the roton excitation by probing the instability dynamics of a large, cigar-shaped dBEC of \textsuperscript{166}Er with transverse magnetisation. Chomaz \textit{et al.} performed a fast interaction quench and studied the short-time evolution of dBEC.  The authors reported  the transient appearance of remarkable structure in the momentum distribution of the dBEC with a high-amplitude central peak framed along the cigar-long axis by two lower-amplitude symmetric side peaks, see Fig.~\ref{Fig:Chomaz2018}(a). Based on a Bogoliubov picture relevant for the short-time dynamics, this was interpreted as the coherent population of finite-momentum excitation modes thanks to its privileged dynamical softening to imaginary energies, indicative of an unstable roton mode. By further studying the peak position and the time evolution of its population, the authors demonstrated the characteristic scalings $\kr\sim 1/\ell_z$ and  $\omega_{\rm rot}\propto (\as-\as^*)^{1/2}$ for the unstable regime ($\as<\as^*$); see Fig.~\ref{Fig:Chomaz2018}(b-c). These observations are in agreement with theory predictions based on an analytical model as well as on GPE simulations.

\subsection{Dipolar quantum-degenerate Fermi gases}
\label{subsec:fermi}

\begin{figure}[htbp]
  \includegraphics[width=0.95\columnwidth]{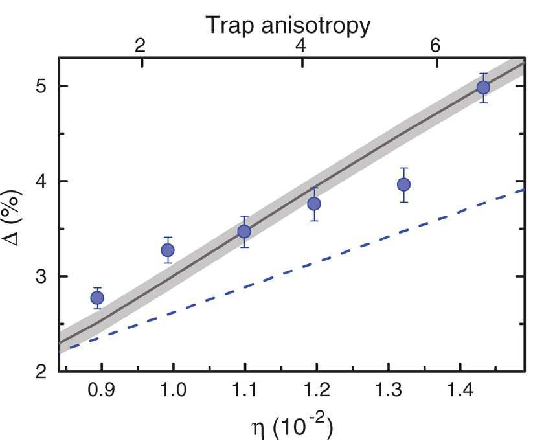}
  \caption{Anisotropy $\Delta=1-{\rm AR}$ (${\rm AR}$ is the cloud aspect ratio) of the time-of flight distribution of a degenerate Fermi gas of $^{167}$Er versus the parameter $\eta=E_{\rm dip}/E_{\rm F}$. Data (dots) are compared to the predicted Fermi surface deformation (dashed line) as well as the prediction accounting for the interaction effects during the TOF. From \cite{Aikawa2014oof}.\label{Fig:Aikawa2014}}
\end{figure}

Degenerate Fermi gases of polarised dipolar fermions constitute an interesting system in which identical fermions directly interact. Alkali-metal polarised fermions are usually non-interacting, as, away from FRs, the short-range interactions can be neglected due to the Pauli exclusion principle; see~\ref{scatttheory}. Because of  universal dipolar scattering (see Sec.~\ref{elasticdipolarscattering}), dipolar fermions do interact and thus offer a unique possibility to explore physics combining the effects of Fermi statistics and interactions. In particular, this system offers new prospects for the creation of novel Fermi liquids~\cite{Sieberer2011cms,Lu2012flo}, anisotropic superfluids \cite{Baranov2002spi}, interlayer superfluids in optical lattices~\cite{Pikovski2010isi,Potter2010sad,Baranov2011bso}, or topological $p_x+ip_y$ phases~\cite{Cooper2009sts}.  Interesting prospects relate to higher-orbital  BCS pairing. This is permitted in dipolar DFGs (dDFGs) thanks to the partially attractive character of the DDI. Such a high-orbital BCS transition would induce exotic superfluid behaviour.  However, it requires a very low temperature to be observed~\cite{Baranov2008tpi,Baranov2012cmt}, beyond current possibility; see Sec.~\ref{sec:amgnetic_atoms}.
 
In dDFGs, the DDI competes with the Fermi energy, $E_{\rm F}=\frac{\hbar^2k_{\rm F}^2}{2m}$. 
The figure of merit is given by the ratio $nS^2d^2/E_{\rm F}$, corresponding to the ratio $\eta= E_{\rm dip}/E_{\rm k}$ also used when considering thermal gases in Sec.~\ref{subsec:thermal_anis}. For small values, $\eta$ can be  rewritten, using the density scaling of the ideal Fermi gas as a function of the characteristic dipolar length and the Fermi momentum, $\eta = A k_{\rm F}\add$, where $A$ is a numerical coefficient. In the homogeneous case, $A=1/\pi^2$. In experimental systems of magnetic atoms, $k_{\rm F}$ is a few tens of $\um^{-1}$ and $\add$ up to a few tens of nm, so $\eta$ is typically of the order of a few percent. This small value makes the observation of many-body dipolar phenomena more involved in a dDFG as compared to a dBEC, due to the large kinetic energy stored in the system (similar to the thermal case).  

On the other hand, the small $\eta$-value enables a semiclassical (Hartree-Fock) treatment of the physics of the dDFG. Within the Hartree-Fock theory, introduced in Sec.~\ref{subsec:fermi_intro}, the DDI contributes to the total mean energy of the system via two terms: the Hartree direct term, which is the usual MF term, and the Fock exchange interaction, which comes from the required antisymmetrisation of the wave-function. 

$\bullet$ \textbf{Fermi surface deformation:}

Similar to the Bose case presented in Sec.~\ref{subsec:bose}, the first effect of the DDI lies in a deformation of the ground state compared to the non-interacting case (magnetostriction). For a DFG, this deformation has a different flavour than for Bose gases, because of the predominant role played by the Fock term, which, it should be highlighted, has a pure quantum origin: While the Hartree term dominates the ground-state distortion in position space as in the bosonic case, it is now the exchange term which gives a dominant contribution to its distortion in momentum space; that is to say, a distortion of the Fermi sea itself. This effect has been extensively studied theoretically \cite{Fregoso2009fgs,Fregoso2009bnp,Sogo2009dpo, Miyakawa2008psd,Baillie2012mae,Wachtler2013lle,Veljic2017tof}. 

The deformation of the Fermi sea under the effect of the DDI has been experimentally observed in an Er DFG via TOF expansion~\cite{Aikawa2014oof}. The experiment was performed in a harmonic trap in which no well-defined Fermi surface exists, but interactions do deform the Fermi sea momentum distribution. In addition, interaction effects during the expansion weakly break the one-to-one correspondence between momentum and TOF distributions. Collisions during TOF increase the deformation, but still the bulk of the TOF distribution deformation stems from the momentum distribution; see  Fig.~\ref{Fig:Aikawa2014}. 

The TOF aspect ratio of the cloud is observed to rotate with the quantisation axis of the dipoles and its value is in agreement with theory. A linear dependence of the deduced Fermi-surface-deformation amplitude on $\eta$ has been verified experimentally as shown in Fig.~\ref{Fig:Aikawa2014}. Additionally, the effect has been seen to disappear  at higher temperature $k_{\rm B}T/E_{\rm F}\gtrsim 1$. Complementary measurements and theory have been later reported in Ref.\,\cite{Veljic2017tof}. Here, the Fermi-surface-deformation amplitude could be reconstructed from the observed TOF aspect ratio for an arbitrary gas geometry. It was also shown that the Fermi surface deformation does not always rigidly rotate with the dipoles' orientation but can also change in amplitude under the effect of the trap anisotropy. Yet such an effect remained elusive in the Er system due to the too weak DDI and moderate achievable trap anisotropies in the DFG regime.

Observation of many-body physics in spin-polarised DFGs has been for now limited to the  above described Fermi surface deformation. However, many other interesting aspects may arise, even without reaching more degenerate samples. Following similar trends as for the bosonic case, one can mention the effect of the DDI on the excitation spectrum and in combination with (highly) anisotropic confinement~\cite{Baranov2012cmt}. Other directions includes the special character of impurity physics in dDFGs, the physics of multilayered or multitube systems, etc.

%% file: Section5/5_instabilityanddroplets.tex
\section{Dipolar collapse and quantum stabilised states of Bose gases}\label{Sec:DCQD}
In Sec.~\ref{subsec:Stability}, we discussed the instabilities of dBECs. In the framework of MF theory, they lead to singularities in the density ($n\to\infty$) and thus in the energy $|E|\to\infty$ for the many-body ground-state. This unphysical conclusion comes from the fact that the simple theory framework developed in Sec.\ref{Sec:RepulsiveGases} (see also Sec.~\ref{subsec:mf}) breaks down at high density. When the density is increased, two effects should be additionally taken into account.\par
\begin{itemize}
\item First, at high density the approximation neglecting few-body collisions beyond binary ones breaks down. At the next order, three-body collisions should be considered. Their inelastic contribution leads in particular to atomic losses at high density \cite{Stenger1999sei,Roberts2000mfd,Weber2003tbr}. Thus density is not locally conserved and obeys the equation $\partial_tn+\nabla\cdot(n\vec v)=-L_3n^3$, where $L_3$ is the recombination loss constant. One can effectively add a non-conservative term in the GPE which reproduces the above equation, this term reads 
\begin{equation}
	i\hbar\partial_t\psi|_{\rm 3b}=-i\frac\hbar2L_3|\psi|^4\psi.\label{Eq:LossesGPE}
\end{equation}
We also note that three-body interactions can additionally lead to a conservative term and, very early on, its effect has been theoretically considered, yielding the prediction of a liquid phase~\cite{Gammal2000abe,Bulgac2002dqd}. Despite the fact that this mechanism has regained recent theoretical interest in the case of dBEC~\cite{Xi2016dfi,Bisset2015coa,Blakie2016poa}, it has remained elusive in current experimental setups~\cite{FerrierBarbut2016lqd}. 
Conservative three-body interactions will then be neglected in the following.

    \item Second, strong enough interactions yield corrections to the population (so-called quantum depletion) and to the energy of the BEC mode. This can be described within Bogoliubov theory, where one still considers only two-body interactions and assumes a macroscopic, yet not complete, occupation of the condensate mode (i.e. single-particle ground state). Expanding to second order in powers of the non-condensed population (i.e. population in the single-particle excited states) yields a quadratic Hamiltonian approximating the many-body one, which can be diagonalised~\cite{Bogoliubov1947ott,Stringari2016bec}. At this order, the ground state energy is given by the zero-point energy of the elementary excitations that diagonalize the  quadratic Hamiltonian, which correspond to elementary excitations. 
    This ground-state energy is shifted compared to the MF energy that matches that of a pure condensate (zero  population in excited single-particle states). 
    The correction to the MF results are thus referred to as quantum fluctuations effects, i.e. coming from the fluctuations of the vacuum of the excitations.

The resulting modification to the equation of state for a non-dipolar BEC was first calculated by Lee, Huang and Yang in 1957~\cite{Lee1957mbp,Lee1957eae}. The energy of a homogeneous repulsive Bose gas then reads $E=\frac g2\frac{N_0^2}{V}(1+\frac{128}{15\sqrt\pi}\sqrt{n_0\,\as^3})$ where $N_0\;(n_0=N_0/V)$ is the number (density) of atoms in the BEC occupying a volume $V$. The first term of the sum is the usual MF energy, the second one corresponds to the first-order beyond-MF (BMF) correction, so-called Lee-Huang-Yang (LHY) correction. The strength of the LHY correction is set by the \emph{gas parameter} $n_0\as^3$. 

Adding the DDI modifies the spectrum of the elementary excitations, see Sec.~\ref{subsec:excitationBEC}, Eq.~\eqref{Eq:Dispersion}, and thus should modify their zero-point energy. 
This was calculated in \cite{Schutzhold2006mfe,Lima2011qfi,Lima2012bmf}, giving
\begin{equation}
	E=\frac g2\frac{N_0^2}{V}\left(1+\frac{128}{15\sqrt\pi}\sqrt{n_0\,\as^3}\,Q_5(\edd)\right),\label{Eq:EnergyBMF}
\end{equation}
where the function $Q_l(x)=\int d\theta\sin\theta(1+x(3\cos^2\theta-1))^{l/2}$ results from angular averaging of Eq.~\eqref{Eq:Dispersion}. Since the dispersion relation becomes imaginary at some angles for $\edd>1$, so does $Q_5$. Therefore the energy \eqref{Eq:EnergyBMF} is formally defined only for $\edd\leqslant1$. However, the imaginary part of $Q_5$ remains very low for $\edd\lesssim 3$. Then, one might ignore it and use Eq.~\eqref{Eq:EnergyBMF} for  $\edd\gtrsim 1$ without  a complete breakdown of the theory.  

In addition to this energy correction, the computation of the many-body ground state (total density $n$) within Bogoliubov theory, also gives access to the corresponding quantum depletion density $\delta n=n-n_0$. For a homogeneous dBEC, one finds $\delta n/n_0=\frac{8}{3\sqrt\pi}\sqrt{n_0\as^3} Q_3(\edd)$~\cite{Schutzhold2006mfe,Lima2011qfi,Lima2012bmf}, recovering $\delta n/n_0=\frac{8}{3\sqrt\pi}\sqrt{n_0\as^3}$ in the contact interacting case~\cite{Lee1957mbp,Lee1957eae}. One sees that the main assumption behind the Bogoliubov theory, namely a dominant population of the zero-momentum state, holds for small gas parameters $\sqrt{n_0\as^3}\ll 1$. In the dipolar case, a similar discussion on the $\edd$-range takes place as for the energy correction. This justifies a-posteriori the validity regime given in Sec.~\ref{Sec:RepulsiveGases}  for the MF theory, namely $n\as^3 \ll 1$ and $\edd\leqslant1$.   

The above results hold true only for infinite, isotropic, homogeneous BECs. The connection to experimental systems is done through two further approximations to estimate Eq.~\eqref{Eq:EnergyBMF}. First, one might neglect the quantum depletion assuming  that all the atoms are in the BEC $n_0=n$. One can therefore write an equation only for the BEC density and ignore its coupling to other modes. This approximation is mostly valid for experimental conditions so far where $\delta n/n_0$ is below a few percents. Second, one might use the local-density approximation (LDA) for the calculation of the equation of state. This assumes that the density varies sufficiently slowly to calculate locally, for a given density $n$, the energy shift. The derivation of Eq.~\eqref{Eq:EnergyBMF} involves an integral over all momenta $k$, which 
is dominated by the contribution at $k\simeq\xi^{-1}$ where $\xi$ is the healing length \cite{Stringari2016bec}. Typically, $\xi$ relates to the sound velocity $c_0$ via $\xi=hm/c_0$. For dBECs, following Eq.~\eqref{Eq:Dispersion} (see Sec.~\ref{subsec:excitationBEC}), one can only define an 'angle dependent' healing length and this diverges for $\edd=1$ along the dipole orientation. At first sight, thus, the LDA is never applicable for $\edd\geqslant 1$. We will see below that the use of the LDA can still be justified for typical experimental samples at $\edd\gtrsim1$, see Sec.~\ref{subsec:theory_droplet}. Thus, within both approximations, the local energy shift can be calculated, and identically the chemical potential shift  $\mu_{\rm BMF}=\frac{\partial E_{\rm BMF}}{\partial N}=\frac{32}{3\sqrt\pi}\,gn\,\sqrt{n\as^3}Q_5(\edd)$. To solve the equation of motion for a non-homogeneous system, one can then extend the GPE by this extra chemical potential \cite{Wachtler2016qfi,Bisset2016gsp,Saito2016pim}:
\begin{equation}
	i\hbar\partial_t\psi|_{\rm BMF}=\frac{32}{3\sqrt\pi}\,g\,\as^{3/2}Q_5(\edd)|\psi|^3\psi.\label{Eq:BMFGPE}
\end{equation}
\end{itemize}

\begin{figure}[hbt]
  \includegraphics[width=\columnwidth]{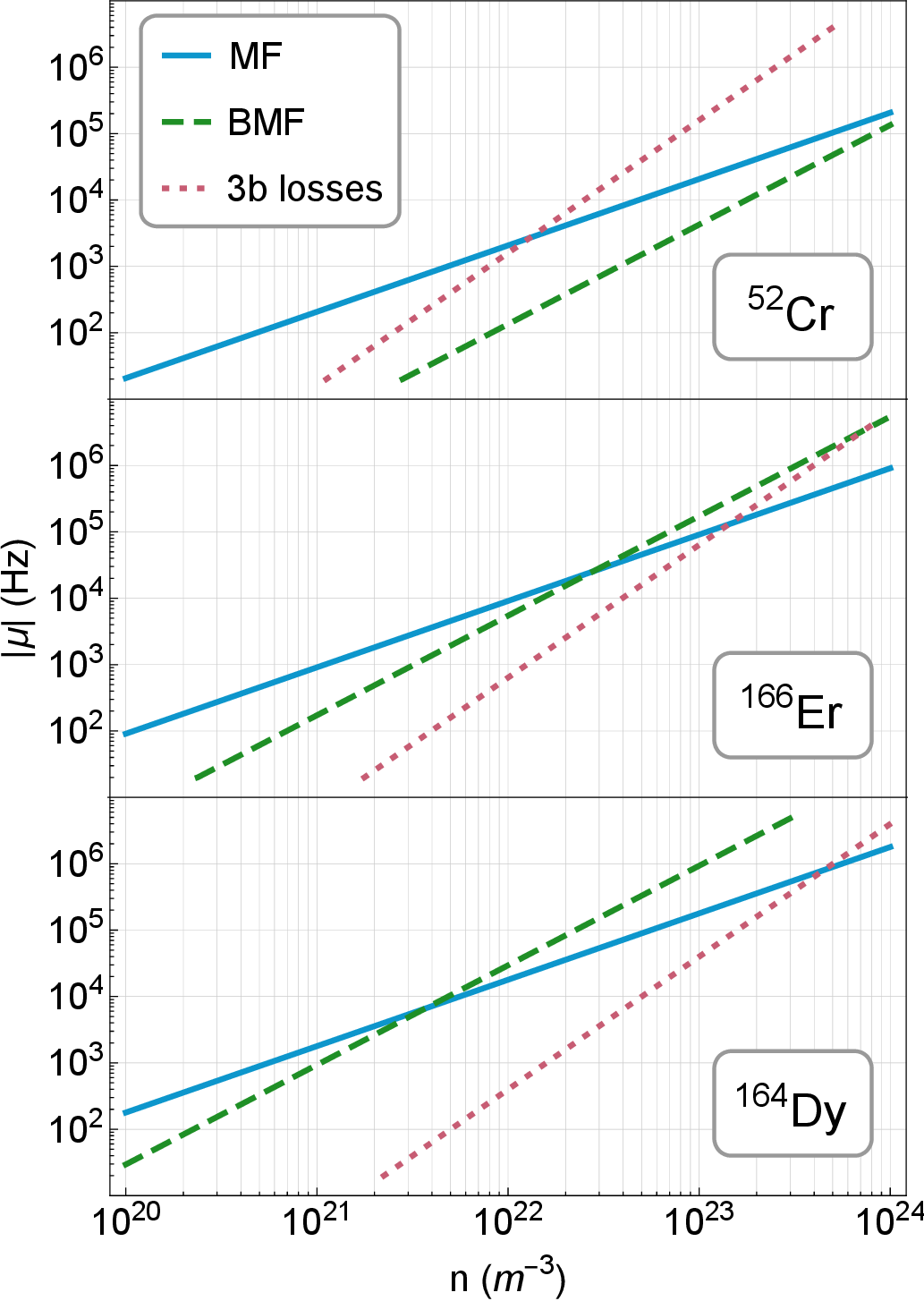}
  \caption{Absolute value $|\mu|$ of the different non-linear terms of the extended Gross-Pitaevskii equation, assuming homogeneous density, and $\edd=1.5$. The (attractive) mean field dominates at low density. For \textsuperscript{52}Cr three-body losses (Eq.~\eqref{Eq:LossesGPE}) first become stronger, while for \textsuperscript{166}Er and \textsuperscript{164}Dy, BMF effects (Eq.~\eqref{Eq:BMFGPE}) first take over.}\label{Fig:Comparisons}
\end{figure}

Now that we have effective terms to be added to the GPE for three-body recombination (Eq.~\eqref{Eq:LossesGPE}) and BMF (Eq.~\eqref{Eq:BMFGPE}) effects, we can compare their magnitudes to the MF chemical potential (see Eqs.~\eqref{GP}-\eqref{meanfield}) to know which one comes first when the density increases significantly following an instability. To that end, one must know $a_{\rm dd}$, $\as$ and $L_3$. $a_{\rm dd}$ is fixed for a chemical species, see Sec.~\ref{subsec:compareSystems}. Since the collapse occurs for $\edd\geq1$ with the exact value depending on the particular trap geometry, we will fix a typical value of $\edd=1.5$, which further fixes $\as$ to the values $a=10\,a_0$ (\textsuperscript{52}Cr), $44\,a_0$ (\textsuperscript{166}Er), $87\,a_0$ (\textsuperscript{164}Dy). The parameter $L_3$ must be measured experimentally. In Figure \ref{Fig:Comparisons} one can see the absolute values of three different contributions as a function of density\footnote{assuming for simplicity an elongated density distribution, with a central MF chemical potential $\mu\simeq g\,n\,(1-\edd)$} for the three dipolar atoms \textsuperscript{52}Cr, \textsuperscript{166}Er and \textsuperscript{164}Dy. The values of $L_3$ are extracted from Refs.~\cite{Lahaye2007sde,Chomaz2016qfd,FerrierBarbut2016ooq}. For the above given values of $\as$, set using Feshbach tuning via the  FRs specified in the refs mentioned above, we have $L_3\simeq2\times10^{-40}$ for \textsuperscript{52}Cr, $\simeq8\times10^{-41}$ for \textsuperscript{166}Er, and $\simeq5\times10^{-41}$ for \textsuperscript{164}Dy. Note that $L_3$ depends on the few-body-physics details and thus can vary with the specific FR used to tune $\as$. In Figure \ref{Fig:Comparisons}, one observes different hierarchies of mechanisms depending on the atomic species.  

In the case of Cr, the weakness of $a_{\rm dd}$ leads to a weak $\as$ beyond instability and thus weak MF and BMF effects. As a consequence the three-body losses will have a much stronger effect and the BMF effect can be safely ignored from the dynamics. In this case, the instability leads to a so-called d-wave collapse, that we review in Sec.~\ref{subsec:collapse}.\par
For Dy and Er, the BMF can be at par with the MF effects before the three-body losses destroy the sample. This leads to an entirely different dynamics, as well as new ground states, stabilised beyond the MF instability. We focus on this physics in Secs.~\ref{subsec:droplets}-\ref{subsec:supersolidity}, see also Ref.~\cite{Boettcher2020nso}.

\subsection{Dipolar collapse}\label{subsec:collapse}
We focus here on the case of \textsuperscript{52}Cr where the BMF effects are negligible (see above). A collapse occurs once the instability threshold is crossed. Experiments have focused on crossing the instability line (Fig.~\ref{Fig:Koch2008}) by lowering $\as$. Once a collapse occurs, the density increases until the three-body losses become dominant over MF attraction (Fig.~ \ref{Fig:Comparisons} top). This leads to a strong density depletion at the density maximum. Then, the MF attraction is reduced, down to a point where the kinetic energy, which is itself increased by the strong localisation of the wavefunction close to the collapse centre, becomes stronger. At that point, the remnant BEC fraction is strongly expelled, the BEC 'explodes'. This can be seen as a reflection of the wavefunction at the collapse centre. The explosion dynamics is thus expected to reflect the symmetry of the collapse. Being induced by the DDI, the collapse is not isotropic. The dipolar energy is minimised by decreasing the gas aspect ratio $\kappa_{x,y}$ (see Eqs.~\eqref{eq:E_MF_TF} and \eqref{eq:E_MF_gauss}), and so one might expect that 
the BEC collapses radially, i.e. shrinks and later explodes in the directions transverse to the dipole orientation. An interplay with the initial anisotropy of the cloud is  also expected.\par
\begin{figure}[hbt]
  \includegraphics[width=\columnwidth]{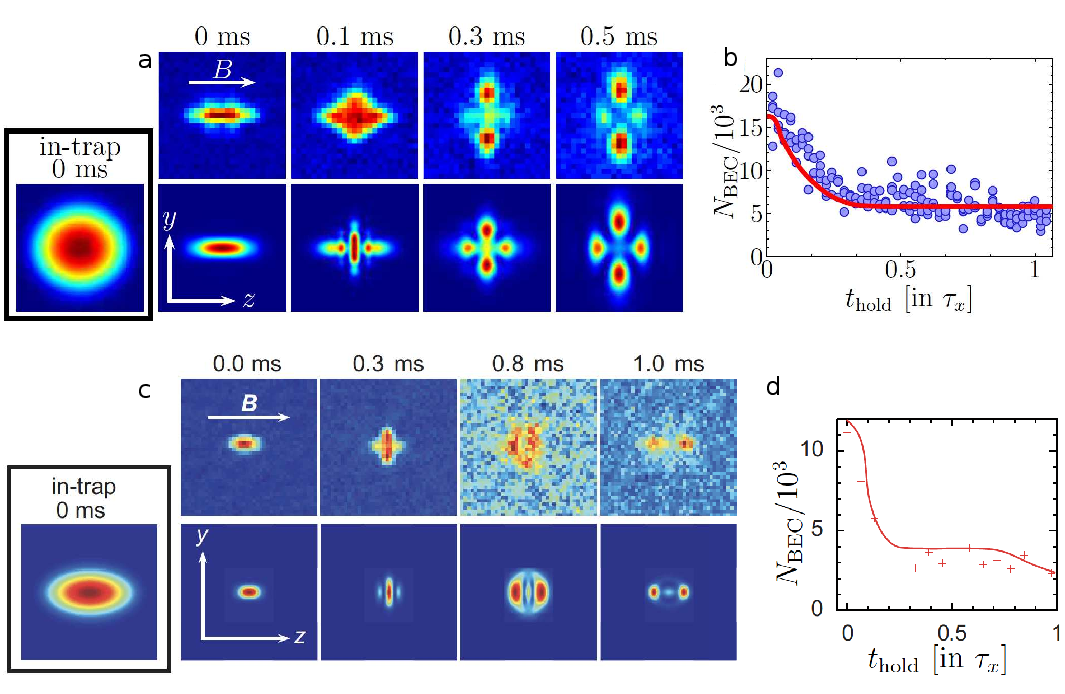}
  \caption{Collapse dynamics of a Cr BEC, (a-b) in a trap $(\nu_x,\nu_y,\nu_z)=(660,400,530)\,$Hz for which the initial BEC is nearly isotropic, adapted from~\cite{Lahaye2008dwc} and (c-d) in a trap $(\nu_x,\nu_y,\nu_z)=(650,520,400)\,$Hz for which the initial BEC is prolate, adapted from~\cite{Metz2009cco}. Dipoles are oriented along $z$. (a) and (c), time-of-flight absorption images in the $yz$ plane (upper rows) and corresponding simulated column density distributions (lower rows) for different hold time after the quench of $\as$ (values given above). Time of flight is 8\,ms. The initial scattering length is around 30-35\,$a_0$ while the final around 5-8\,$a_0$, well below $a_{\rm crit}$. The in-trap inset shows the initial BEC aspect ratio $\kappa_y$. (b-d) measured (dot) and simulated atom number as a function of the hold time after the quench, $t_{\rm hold}$, in unit of $\tau=\tau_x =1/\nu_x=1.5\,$ms. All simulations rely on a GPE including three-body loss effects, i.e. Eq.~\eqref{GP} plus Eq.~\eqref{Eq:LossesGPE}.}\label{fig:Cr_collapse}
\end{figure}
The first experiments probing the collapse dynamics of a \textsuperscript{52}Cr dBEC were performed starting from a nearly isotropic BEC \cite{Lahaye2008dwc}. The collapse was induced by a fast quench of $\as$ down to an unstable value. Imaged after time-of-flight, the expanding cloud displays a strong anisotropy, see Fig.~\ref{fig:Cr_collapse}(a).  As the collapse is induced in the radial direction, the relevant timescale for the collapse, $\tau$, is set by the largest radial trapping frequency, $\tau=\min[1/\nu_x,1/\nu_y]$. Strong atom losses are observed to occur at the initial time of the collapse (fraction of $\tau$), before the wavefunction is reflected away, see Fig.~\ref{fig:Cr_collapse}(b). A cloverleaf-like pattern is observed in the time-of-flight images after holding $0.2$ to $0.5\tau$. This pattern, reminiscent of the d-wave symmetry of the DDI,  gave the name \textit{d-wave collapse}. The density distribution is then observed to refocus after holding $\tau$. This is because of the presence of the harmonic trap. The observed dynamics is very well reproduced by simulations of the GPE (Eq.~\eqref{GP}) additionally including the three-body loss term (Eq.~\eqref{Eq:LossesGPE}). The lost fraction of condensed atoms depends on the in-trap wait time after the collapse start, also very well described by the GPE simulation results, see Fig.~\ref{fig:Cr_collapse} (a-b). 

The fact that the expanding cloud remains condensed was investigated by collapsing several independent BECs in different sites of an optical lattice \cite{Metz2009cco}. Following the collapse, the expanding clouds overlapped, exhibiting very clear interference fringes at long time. This proved the local coherence of the individuals expanding clouds, and thus that condensation was not fully destroyed by the collapse.\par
In anisotropic traps, not only the DDI but also the trap geometry impacts the collapse symmetry. In this way, very different patterns of the expanding BEC can be generated, evidencing the rich interplay between trap geometry and DDI in dBECs. This effect has been extensively studied in Ref.~\cite{Metz2009cco}, where the authors studied different cases from very prolate to very oblate geometry. They also investigated the cross-over from a prolate geometry to the symmetric one of Ref.~\cite{Lahaye2008dwc}, as shown in Fig.\ref{fig:Cr_collapse}(a,c). While the two traps considered here have very similar frequencies within a simple permutation $\nu_y \leftrightarrow\nu_z$, they display significantly different dynamical behavior. This ultimately demonstrates the role of the anisotropic DDI in the collapse dynamics.

\subsection{Theoretical description of dipolar quantum stabilised states}\label{subsec:theory_droplet}

\subsubsection{Simple description of dipolar quantum stabilisation}
The phenomenology of MF unstable, strongly dipolar BECs of Er and Dy is fundamentally modified by the BMF effects. {These effects can affect not only the dynamics of the BEC when driven to an unstable regime but also modify its stability itself, as it introduces a stabilisation mechanism. Indeed, the LHY correction of Eq.~\eqref{Eq:BMFGPE} provides an additional conservative potential, which acts as an effective higher-order (in three-dimension, effective 5/2-body interaction) interaction term in the GPE and which remains repulsive within its whole validity range~\footnote{Provided the imaginary part of $Q_5(\edd)$ can be neglected.}. This higher-order repulsive term has a density dependence of a higher power than MF (2-body interaction). In the attractive mean field regime, it thus could be thought to stabilise a high-density state for which BMF repulsion compensates MF attraction.} 

Let's first consider the simplest case of single-peaked ground state --- standard BEC in the MF stable regime and later called "droplet" in the MF unstable regime. Using a gaussian ansatz for its wave function $\psi$ (see also Eqs.~\eqref{Eq:psi_gauss}-\eqref{eq:E_MF_gauss}), one can calculate the MF and BMF contributions to the ground-state energy per particle and single-out the key ingredients that lead to stability:
\begin{eqnarray}
	\frac{E_{\rm MF}}{N}&=\frac{1}{2^{3/2}}\,\frac g2\,n_{\rm c}\,(1-f_{\rm dip}(\kappa)\,\edd)\\
	\frac{E_{\rm BMF}}{N}&=\left(\frac25\right)^{3/2}\,\frac g2\,n_{\rm c}\,\frac{128}{15\sqrt \pi}\,\sqrt{n_{\rm c} \as^3}\,Q_5(\edd)
\end{eqnarray}
where $n_{\rm c}$ is the central density. For $f_{\rm dip}(\kappa)\edd>1$, $E_{\rm MF}<0$ while $E_{\rm BMF}>0$. 
$E_{\rm BMF}$ has a stronger dependency in the density by a power $n_{\rm c}^{1/2}$ compared to $E_{\rm MF}$, therefore, by increasing $n_{\rm c}$, the ground-state energy can be minimized at a finite value, corresponding to a finite density. This stabilisation mechanism, relying on the mere effect of quantum fluctuation, is what we coin dipolar quantum stabilisation. 

\subsubsection{Simple description of ultra-dilute liquid state.}\label{subsec:theory_droplet_simple} 

Remarkably, in presence of MF attraction, the single-peaked state considered above can be stabilised even in the absence of trapping, by competing  MF attraction and the BMF repulsion, the equilibrium between these two forces fixing the peak density $n_{\rm c}$ (MF attraction dominates at low density, and BMF repulsion at high density).

Such a stabilisation mechanism is reminiscent of a liquid phase of matter. In ordinary liquids, the weak attraction on large distance (e.g. of van der Waals or covalent types) is counterbalanced by a strongly repulsive core arising from the electromagnetic forces and the Pauli exclusion between the atoms' (or molecules') electron cloud. This stabilises the liquid at a large density at which the repulsion becomes effective $n_{\rm c}{a_0^3}\sim{1}$.    
The stabilisation resulting from competing MF and BMF effects in strongly dipolar gases can similarly be seen as resulting in a liquid state. 

Considering for now such an untrapped system, and neglecting kinetic energy, the equilibrium peak density can be estimated by imposing $\frac{\partial E}{\partial n_{\rm c}}=0$, with $E=E_{\rm MF}+E_{\rm BMF}$. One gets
\begin{equation}
	n_{\rm c}\sim\frac{1}{\as^3}\left(\frac{f_{\rm dip}(\kappa)\,\edd-1}{Q_5(\edd)}\right)^2.\label{Eq:SimpleDensityScaling}
\end{equation}
We note that Eq.~\eqref{Eq:SimpleDensityScaling} also provides a good estimate of the peak density in a trapped stabilised state, as long as one can neglect contributions of the kinetic and external trapping energies. 

Interestingly, from Eq.~\eqref{Eq:SimpleDensityScaling} one can read-off that a stabilisation of the density occurs before reaching the dense regime $n_{\rm c}{\as^3}\sim{1}$, thus forming "ultradilute" liquid states. 
Two main effects enable to maintain the diluteness: 
The equilibrium density is \textit{reduced} by the fact that there are two \textit{competing MF interactions}, resulting in an effective MF attraction of much smaller amplitude than each of the two contributions (numerator in Eq.~\eqref{Eq:SimpleDensityScaling}). 
Second there is an \textit{amplified BMF} contribution that leads to a \textit{further reduction} of the density (denominator in Eq.~\eqref{Eq:SimpleDensityScaling})~\footnote{At $\edd=1.5, \Re[Q_5(\edd)]\simeq3$, leading to a reduction of the density by roughly an order of magnitude.}. 

The relatively low stabilising density resulting from these two ingredients protects the sample against an immediate destruction by three-body recombination. The dipolar collapse observed with Cr is thus prevented. Instead the BEC is stabilised via the effect of its quantum fluctuations, resulting in a distinct phase of liquid-like properties, the clouds formed are thus named quantum (dipolar) droplets. These droplets are denser than the MF-stable gaseous BEC, yet much more dilute than ordinary liquid (by $\approx8$ orders of magnitude). They offer a distinct paradigm of quantum fluid where the BMF effects are predominant yet tractable, see e.g.~Secs.~\ref{subsec:theory_droplet},\ref{subsec:droplet_props}.

We note that these exact same ingredients are present in another experimental system, namely mixtures of contact-interacting BECs with repulsive intra- and attractive inter-species interactions~\cite{Cabrera2018qld,Cheiney2018bst,Semeghini2018sbq}. The stabilisation mechanism of bosonic mixtures was in fact proposed prior to the observations on dipolar BECs by D.~Petrov \cite{Petrov2015qms}. For more information on these systems, see Ref.~\cite{Boettcher2020nso}.

Finally, we note that, despite the fact that we have for now neglected trapping effects in this first description, they may also play a crucial role in the newly stabilised quantum states. In the case of dipolar atoms, this effect is not only quantitative (energy shifts) but also qualitative. Indeed, the interplay between trap anisotropy and DDI yields new features already in the MF-stable BEC, as reviewed in Secs.~\ref{subsec:roton},\ref{subsec:local_instab}. These new features in the excitation spectrum, that affect the MF instability, when combined with quantum stabilisation, may yield new ground states. This effects will be the focus of Sec.~\ref{subsec:supersolidity}. 

\subsubsection{Toward a quantitative theory description:  extended Gross-Pitaevskii equation.}
\label{subsec:eGPE}
Following the experimental observation of quantum droplets that we will more thoroughly describe in Sec.~\ref{subsec:droplet_assemblies}, several theory works~\cite{Wachtler2016qfi,Bisset2016gsp,Wachtler2016gsp} developed the framework based on an extension of the GPE (eGPE) to include the first order BMF effects. As introduced at the beginning of this chapter~\ref{Sec:DCQD}, this consists in extending Eq.~\eqref{GP} with the term Eq.~\eqref{Eq:BMFGPE} (and potentially Eq.~\eqref{Eq:LossesGPE}). This perturbative approach is justified because the sample remains dilute with $na_s\ll 1$. Here the BMF correction comes into play not because the gas parameter becomes large but rather because the overall MF terms are tuned small while both contact and dipolar interaction remains individually large. Then the BEC quantum depletion remains limited. 

We note that the theory model relies on the use of the LDA for the LHY term. This approximation is justified if the dominant contribution to the LHY term have momentum larger than the inverse size of the ground state. Physically, the LHY correction, which corresponds to the zero point motion of the elementary excitations (see above), are dominated by contributions of the \emph{hard} modes, i.e. the most energetic ones. 
Crucially, when decreasing $\as$ (as to drive the MF-instability), the modes whose excitation occurs perpendicular to the dipoles get softer (and lead to the instability) while the one along the dipole get harder. The latter will then dominate the LHY correction. Remarkably, the BEC close to instability and the quantum stabilised states get very extended in the direction of the dipole, under the effect of magnetostriction (see Sec.~\ref{subsec:magnetostriction}). Therefore, the LDA may remain a surprisingly good approximation in this regime, and even beyond the instability threshold. 

Quantitatively, W{\"a}chtler et Santos~\cite{Wachtler2016qfi} show that, using an anisotropic momentum cutoff, with an anisotropy matching the anisotropy of the droplet itself, at 1\% of the inverse of the droplet extent, one still recovers most of the BMF effects (80\%). This proves that the LDA is a good qualitative, and even quantitative, approximation, as long as the droplet remain elongated enough. The contribution of the long-wavelength modes may lead to small corrections, yet the authors note that this would mainly modify the prefactor of the LHY correction while the scaling should remain that of Eq.~\eqref{Eq:BMFGPE}. This domination of the LHY correction by hard modes which makes relevant the use of the LDA even in small sized quantum-stabilised states is also found in quantum droplets of bosonic mixtures as originally proposed by Petrov \cite{Petrov2015qms}, see also Refs.~\cite{Cabrera2018qld,Cheiney2018bst,Semeghini2018sbq,Boettcher2020nso}.

{The relevance of eGPE framework was later quantitatively studied both in theory and experiments. Results from the eGPE and quantum Monte-Carlo (QMC) simulations of dipoles with hard-sphere repulsion were compared in Ref.~\cite{Saito2016pim} and found to be in good quantitative agreement. In experiment, the degree of agreement was found to vary depending on the exact settings, in particular on the gas geometry and on the density, see e.g. Refs.~\cite{Chomaz2016qfd,schmitt2016sbd,FerrierBarbut2018smo,Chomaz2017oot,Petter2019ptr,Chomaz2019lla,Tanzi2019ooa,Boettcher2019tsp} and later discussions. Typical discrepancies could be simply accounted by shifting the scattering length value by a few, and up to a few tens, of percents. More recently, a quantitative study compared eGPE theory, diffusive QMC simulations using finite-range interactions, and experiments~\cite{Boettcher2019qci}. A good agreement of the QMC results and the experiments was found, whereas the eGPE results are systematically shifted. Boettcher \textit{et al} elaborate on the eGPE mismatch and show that it may be accounted by a more sophisticated description of the scattering, in particular by accounting for the effects of finite collision energy on the scattering properties, due to the finite temperature of the samples. Such effects are known to result in an effective renormalisation of the dipolar length~\cite{Oldziejewski2016pos}, see also Sec.~\ref{scatttheory}. By including such corrections in the eGPE yields a better agreement with the experimental data.  Note that corrected eGPE and QMC constitutes two complementary theories that are able to describe experimental observations by accounting for different effects. The corrected eGPE accounts for temperature effects on the scattering properties but only includes quantum fluctuations at a perturbative and approximate level. QMC fully accounts for the quantum fluctuations, yet neglects thermal effects. The two theories account for different effects, and suggest that different sources of corrections with respect to the standard eGPE theory may be relevant. A full theoretical modelling, accounting for the different effects at once, and revealing their repsective role, is yet missing.   
}  

\subsection{Dipolar Quantum droplets}\label{subsec:droplets}

We now divide the discussion of quantum-stabilised states and their observations in two sections. In the present section, we will discuss the regime where the underlying ground state present no self-modulation, i.e it has only one density peak and forms either a standard BEC (repulsive MF) or a droplet state (attractive MF). In experiments, the droplets could there be produced either in assemblies of independent droplets, forming then a metastable state, or individually. We will more specifically focus on the latter case, which allows for a detailed characterisation of the state's properties, in presence and in absence of external trapping. In a second part, we will focus on the generation of self-modulated states in dipolar gases with anisotropic confinement. We will particularly focus on the global coherence and superfluid properties of the modulated states, and discuss the existence of a so-called supersolid phase, where solid and superfluid orders coexists, see Sec.~\ref{subsec:supersolidity}. 

\subsubsection{{single droplet ground state in an external trap, eGPE phase diagram}}\label{subsec:phasediagram_droplet}

\begin{figure}[hbt]
  \includegraphics[width=\columnwidth]{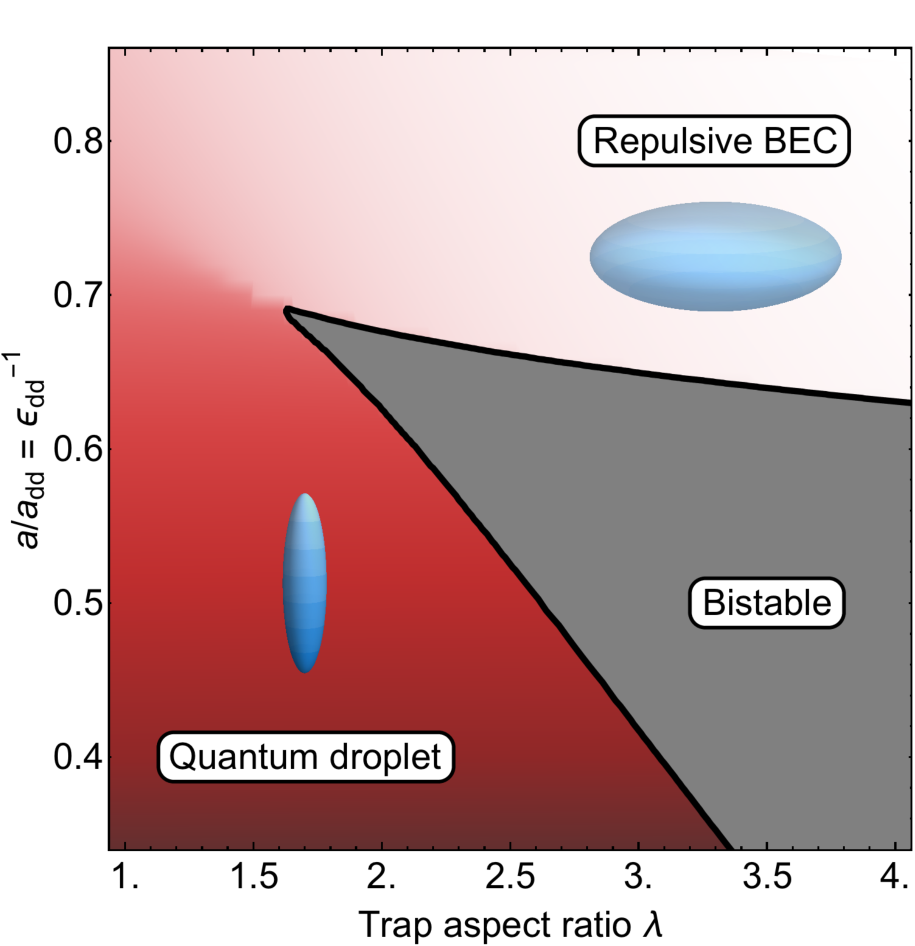}
  \caption{Eigen-state phase diagram of a dipolar BEC of $N=10000$ atoms in a cylindrical harmonic trap of aspect ratio $\lambda=\omega_z/\omega_r$. This diagram is obtained applying a gaussian ansatz to the energy functional \ref{Eq:Energy}, for a fixed value of $N\add/a_{\rm ho}\simeq79$, this corresponds for Dy (Er) to a mean trap frequency of 80 Hz (312 Hz). The colouring shows the ground-state density in logarithmic scale and arbitrary units.}\label{Fig:phaseDiag}
\end{figure}

Following the seminal work establishing a theory description of the quantum-stabilisation mechanism, theoretical works tackle the question of the ground-state phase diagram of a dipolar gas and in the presence of the newly discovered stabilising term~\cite{Bisset2016gsp,Wachtler2016gsp}. 

The presence of an anisotropic harmonic trap with aspect ratio $\lambda=\omega_z/\omega_r$ (see also Eq.~\eqref{Eq:trap}) was considered, yet cylindrical symmetry around the dipoles direction $\boldsymbol{z}$ was assumed. Mean trap frequencies and atom numbers following the experimentally relevant values were considered. In these works, only single-peaked ground states were predicted (i.e. no droplet assemblies, see Sec.~\ref{subsec:supersolidity_preliminary}). When decreasing $\as$ at fixed atom number and trap geometry, the ground state was found to change from a low-density phase, matching a standard MF-stabilised BEC, 
to a high-density phase, stabilised by the LHY term, forming a ``quantum droplet". The low-density BEC has a geometry roughly following that of the trap, while the droplet state is not. In particular, the droplet is always elongated in the direction of the dipoles, which is why it is sometimes called "filament". 
For traps elongated along the dipoles ($\lambda\gtrsim1$), the ground state smoothly evolves from the low-density MF-repulsive BEC to the high-density MF-attractive quantum-droplet phase when decreasing $\as$. This smooth crossover can be apprehended as the trap elongation along the dipole direction enables the two states to have similar geometries (elongated along the dipoles), so that one can continuously evolve into the other, while the BMF term suppresses the dipolar collapse described in Sec.~\ref{subsec:collapse}. In the opposite case of large $\lambda$, a discontiuous transition between the two states is found, with an intermediate region of bistability, where the two states form local energy minima. This bistability can be apprehended from the following argument: In such traps, a low-density state can be stabilised even for $\edd>1$ as the trap asymmetry forces the dipoles to lie side-by-side enhancing the MF DDI repulsion. 
This is exactly how Cr BECs have been stabilised at low $\as$, as described in Sec.~\ref{subsec:Stability}, and the addition of the BMF term only weakly modifies this behaviour. On the other hand, the BMF term stabilises an other solution which is elongated along the dipoles and where MF DDI is attractive. The bistability occurs as there is an intermediate regime of $\as$ where both solutions are local energy minima, while in the absence of the BMF term (as relevant for Cr) in this region the MF repulsive BEC is metastable. 

The theoretical phase diagram is represented in figure~\ref{Fig:phaseDiag}. The exact position of the boundaries and critical point depend on the exact experimental parameters (number of atoms, mean trap frequency, ...). The crossover and the bi-stable region are separated at a critical aspect ratio $\lambda_{\rm c}$, typically $\lambda_{\rm c}\approx 1.5-2$. In the remainder of this section, we review the extensive experimental exploration of this phase diagram.

We note that the description given above does not give a full picture of the dipolar gas phase diagram in presence of quantum stabilisation. This description encompasses the behaviour in a low-atom-number regime. For larger atom numbers (and/or tighter traps), distinct ground states may arise, and in particular spontaneous density modulation may occur, or, in other words, ground states bearing several droplets may be found. This regime, achieved more recently in experiments, will be the topic of Sec.\,\ref{subsec:supersolidity}.

\subsubsection{First observations of droplet states: {metastable droplet assemblies and the quest for the stabilisation mechanism}}
\label{subsec:droplet_assemblies}
\begin{figure}[hbt]
  \includegraphics[width=\columnwidth]{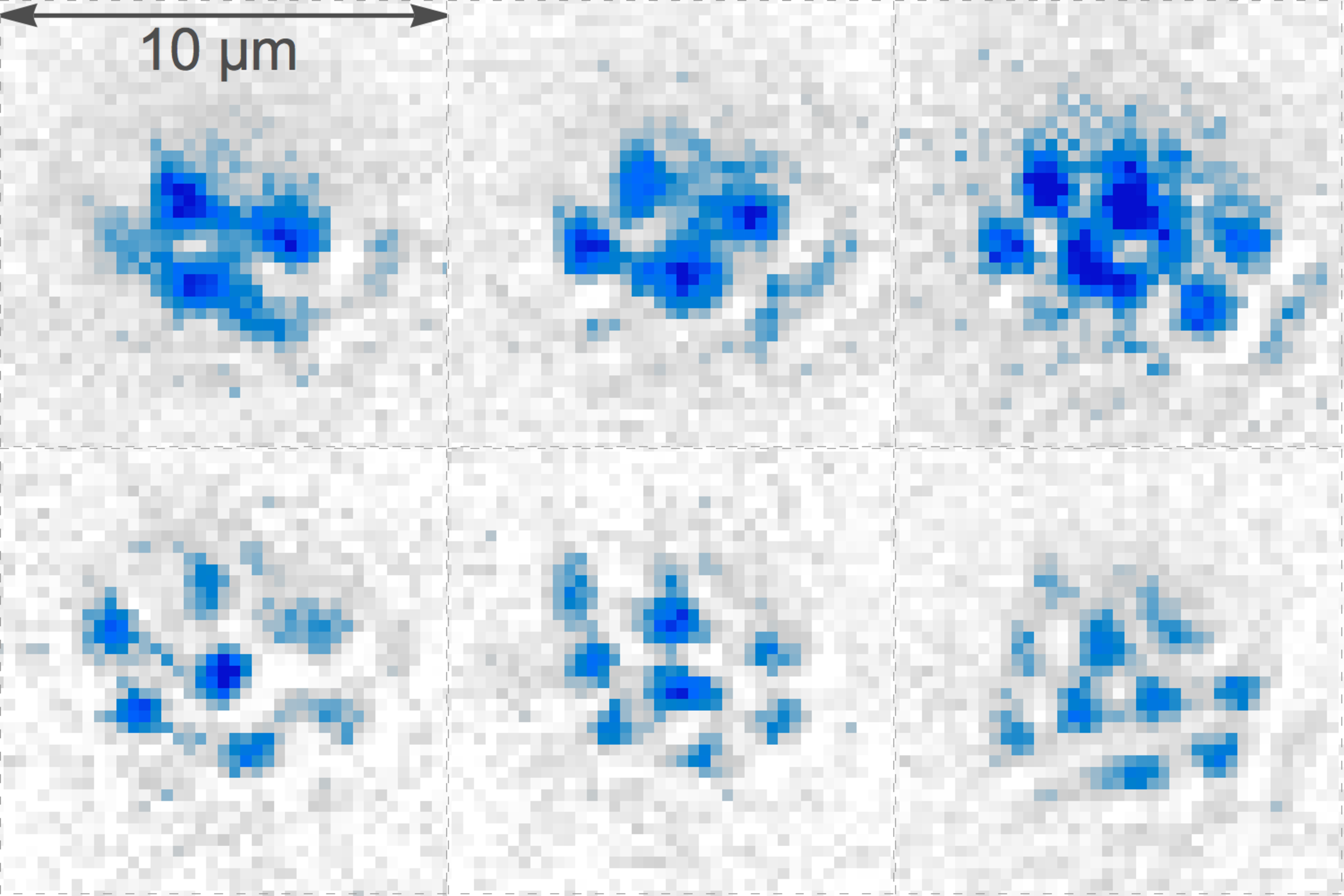}
  \caption{Ordered droplet ensembles observed in the MF-unstable region of the phase diagram. The magnetic field points put of the plane. Adapted from \cite{Kadau2016otr}.}\label{Fig:ensembles}
\end{figure}

The first evidence for the absence of dipolar collapse and the existence of a stabilisation mechanism was reported by Kadau \textit{et al.}~\cite{Kadau2016otr} on \textsuperscript{164}Dy, before any of the theoretical development described in Secs.~\ref{subsec:theory_droplet}-\ref{subsec:phasediagram_droplet}. 
Starting from a pancake-shaped BEC of $^{164}$Dy atoms in a trap with aspect ratio $\lambda\simeq3$, $\as$ was lowered down to the mean field unstable regime. In this geometry, we note that a modulational instability is expected, see Sec.~\ref{subsec:local_instab}. Below the instability threshold, ordered droplet ensembles were observed in the in-situ radial density distribution of the cloud, as can be seen in Fig.~\ref{Fig:ensembles}. Contrary to the expectations at the time, these droplets were observed to have very long lifetimes of several hundreds of ms, simply limited by atom loss due to three-body recombination. 

As shown in the early theory works \cite{Wachtler2016qfi,Bisset2016gsp}, the observation of multiple droplets arises from a crossing of the boundary between the BEC region and the bi-stable region of the phase diagram. In this regime, a continuum of metastable states, with different numbers of droplets, is expected to exist. These works suggest that, following the fast change of the interaction strength, the system ends up in one of these metastable states, explaining the observed long lifetime of the crystal as well as the shot-to-shot variability of the structures. 
Experimentally, the system's bistability was evidenced via the hysteresis in the appearance and disappearance of the droplet ensembles when varying $\as$ down and back up, thus supporting the theoretical picture. As observed after a long hold time, the arrangement of the droplets was found to result from the repulsion between the corresponding macroscopic dipoles (see also Refs.\cite{FerrierBarbut2016ooq,FerrierBarbut2016lqd}). 

A following set of experiments~\cite{FerrierBarbut2016ooq} showed that the droplets have an elongated shape along the dipoles with an axial length of a few micrometers and a radial size estimated to half a $\um$ or less (this lies below the imaging resolution). Furthermore, matter-wave interference fringes observed from several expanding droplets proved them to be individually superfluid. They were observed to have extremely low expansion velocities when released into a waveguide.  In Ref.~\cite{FerrierBarbut2016ooq}, Ferrier-Barbut \textit{et al.}, inspired from the work of Petrov~\cite{Petrov2015qms} on BEC mixtures, proposed that BMF effects act as a stabilisation mechanism. By comparing additional expansion and lifetime measurements to a model following the lines given in Sec.~\ref{subsec:theory_droplet_simple} (neglecting kinetic and trapping effects), they validate this idea and invalidate a possible mechanism based on three-body forces proposed in Refs.~\cite{Xi2016dfi,Bisset2015coa,Blakie2016poa}, see also the earlier works of Refs.~\cite{Bulgac2002dqd,Gammal2000abe}. In particular, the lifetime $\tau$ of the sample, being set by the three-body loss processes, gives information on the sample's density  via $\tau=L_3\langle n^2\rangle$ (see Eq.~\eqref{Eq:LossesGPE}). The estimated density scaling of density as a function of $\as$ deduced from measuring $\tau$ as a function of $\as$ was found to reasonably agree with the simple density scaling of Eq.~\eqref{Eq:SimpleDensityScaling}, and disagree with the scaling expected from  three-body forces stabilisation. 

In this setup, the bistability, yielding excited metastable states of multiple droplets after an interaction quench, prevents the formation of a large single droplet in the ground state. This feature has limited the study of the properties of the liquid-like state. Shortly after, it was found out, in particular thanks to the theory development described in Sec.~\ref{subsec:phasediagram_droplet}, that this issue can be resolved by taking advantage of the "crossover region" of the phase diagram, i.e. changing the trap geometry,  as we will now discussed.

\subsubsection{Crossover from a Bose-Einstein condensate to a large quantum droplet.}
\label{subsec:single_droplet}

\begin{figure}[hbt]
  \includegraphics[width=\columnwidth]{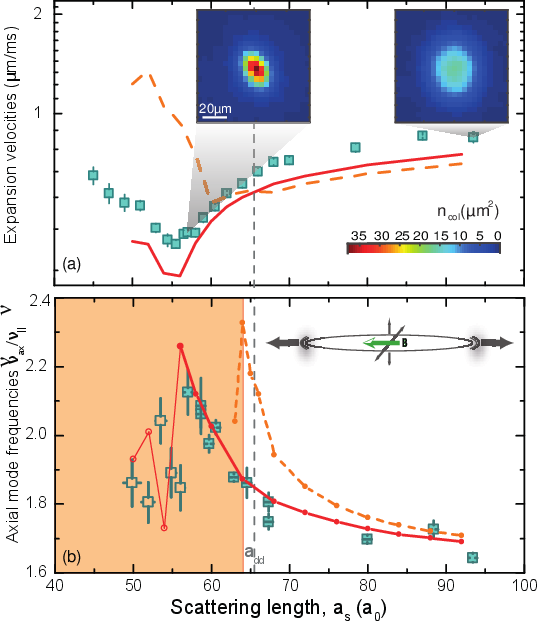}
  \caption{Examples of properties of a \textsuperscript{166}Er dBEC in the crossover from a BEC to a single macrodroplet reported in \cite{Chomaz2016qfd}. Here $\as$ is varied in 10\,ms to its final value, in abscissa. (a) mean expansion velocity measured in 16 to 28\,ms free-expansion after 5\,ms holding in trap. The insets show measured distributions in the plane transverse to the dipole orientation for 28\,ms of free-expansion, and also evidence the absence of collapse dynamics in the low $\as$-regime. (b) The axial mode of the dBEC, whose character is illustrated in inset, is excited by transiently decreasing the confinement frequency along the dipole, $\nu_{||}$, and the mode frequency, $\nu_{\rm ax}$, measured by recording the time-evolution of the cloud size along the dipole. In (a) and (b), the data (squares) are compared to simulations from real time evolution of the eGPE including (red solid line) or not (orange dashed line) the LHY correction (Eq.~\eqref{Eq:BMFGPE}). In (b),  $\nu_{\rm ax}$ cannot be reliably extracted for quenches to $\as\leq 56 a_0$, nor from the experiment (open squares) neither from the eGPE theory (open circles, thin line).
  }\label{Fig:singleDrop}
\end{figure}

Following the first droplet observations~\cite{Kadau2016otr,FerrierBarbut2016ooq,FerrierBarbut2016lqd} and the resulting theoretical development~\cite{Bisset2016gsp,Wachtler2016gsp}, Chomaz \textit{et al.} tested the universality of the stabilisation effect (which relies on the sole quantum-mechanical nature of the fluid, provided that the interactions are strong enough), by realising the first quantum droplet of a distinct chemical species, using $^{166}$Er~\cite{Chomaz2016qfd}. They also used a complementary geometry to the previous Dy observations, using a cigar-shaped trap with $\lambda\ll1$, and thus they observed a smooth crossover from a BEC to a single large droplet of Er atoms containing all the condensed atoms. 
Because the created droplet is isolated, the authors could study its properties such as its elementary excitation, and its expansion dynamics, see Fig.\,\ref{Fig:singleDrop}.  The authors performed a systematic comparison of the measurements with the eGPE predictions (Eq.~\eqref{GP} including also Eqs.\eqref{Eq:BMFGPE}-\eqref{Eq:LossesGPE}, see Sec.~\ref{subsec:theory_droplet}) using independently measured values of  $\as$ and $L_3$. The quantitative agreement reached 
 confirmed the stabilisation scenario and validated the eGPE framework in this setting. 
We note again, that quantum droplets have later been observed in mixtures of contact-interacting BECs \cite{Cabrera2018qld,Semeghini2018sbq} following Petrov's initial proposal, establishing the quantum-stabilisation mechanism as universal even across different interaction types, provided that competing interactions lead to a balance of MF and BMF contributions.

Finally, the two regimes of bi-stability and crossover where experimentally connected in \cite{FerrierBarbut2018ooa}, where  similar experiments were performed using \textsuperscript{164}Dy in traps of variable aspect ratio $\lambda$. The formation of a single droplet resulting from lowering $\as$ in the crossover was observed for small $\lambda$ up to a critical aspect ratio $\lambda_{\rm c}$ which marked the onset of a modulational instability as in \cite{Kadau2016otr}. The experimental value of $\lambda_{\rm c}$ is in agreement with the expected critical aspect ratio marking the separation between the crossover and bi-stable regions.

\subsubsection{Droplet properties and their signature in the collective modes}
\label{subsec:droplet_props}

Following the first experimental observations, several works investigated the specific properties of the droplet states properties. 

Theoretically, it is interesting to highlight that, while the trap plays a role in the phase diagram and the transition between the different phases (see Sec.\,\ref{subsec:phasediagram_droplet}), it plays a much lesser role on the properties of the single droplet state itself, at least for large enough atom number and $\edd$.
By studying full ground-state solution from the eGPE, Refs.~\cite{Wachtler2016gsp,Bisset2016gsp,FerrierBarbut2016lqd} show for instance a very weak dependence of the droplet density on the trapping potential. 
At very low atom number (i.e. on the order of a few thousand), the one-body kinetic energy is non-negligible, and its interplay with the interaction energies leads to a dependence of the central density on $N$. However, at high atom number, the interaction energies fully dominate. Following the simple description of Sec.~\ref{subsec:theory_droplet_simple}, the density then reaches a saturation value independent of $N$, given only by the balance of MF and BMF energies. This marks the low compressibility typical of a liquid phase, and which stems here from the high energy cost of increasing density due to the BMF term. While early experiments have mostly explored the low atom regime, first measurements showing the onset of this density saturation have been recently reported in Ref.~\cite{Boettcher2020nso}.

Besides its evidence via density saturation, the low compressibility of the dipolar quantum droplets has been revealed in different sets of measurements, in particular focusing on the collective modes. First investigations were preformed by Chomaz \textit{et al.}, studying a particular collective mode, with compressional character, of a quantum gas of \textsuperscript{166}Er in the BEC-droplet crossover~\cite{Chomaz2016qfd}. 
As the scattering length is lowered through the crossover, a steep increase in the mode mode frequency was observed, changing by $50\%$ when varying $\as$ by less than $20\%$ while the trap frequency remained constant, see Fig.\,\ref{Fig:singleDrop}. This shows that the compressibility quickly drops when going from the BEC to the droplet phase. 

In addition to being weakly compressible, a dipolar quantum droplet is anisotropic. This stems of course from the anisotropy of the DDI with respect to the external magnetic field axis. The consequence of this external breaking of rotational invariance is the existence of a collective mode, deemed the scissors mode. It corresponds to a rigid-body angular oscillation around the magnetic field direction. This scissors mode has been observed first in atomic nuclei \cite{LoIudice1978nic,Lipparini1983imr,Enders1999cao} and in contact-interacting BECs in anisotropic traps \cite{GueryOdelin1999sma,Marago2000oot}. In the presence of the DDI, the rotational symmetry is broken even in isotropic traps. This was showed to result in a scissors mode for MF dipolar BECs in Ref.~\cite{VanBijnen2010cef}. This mode is well-defined only for angular oscillations in the $xz$-plane with an amplitude lower than  $\theta_{\rm max}=(\langle z^2\rangle-\langle x^2\rangle)/(\langle z^2\rangle+\langle x^2\rangle)$ where $z$ is the field direction \cite{GueryOdelin1999sma}. It is thus very challenging to observe with dipolar MF BECs, but was observed in dipolar quantum droplets of \textsuperscript{164}Dy thanks to their considerable anisotropy \cite{FerrierBarbut2018smo}. In conclusion, studies of the collective modes of the liquid state are a sensitive probe for its properties \cite{Baillie2017ceo} and will likely be pushed further. 


\subsubsection{Self-bound droplets and liquid-gas phase diagram}\label{subsec:droplet_selfbound}
\begin{figure}[hbt]
  \includegraphics[width=\columnwidth]{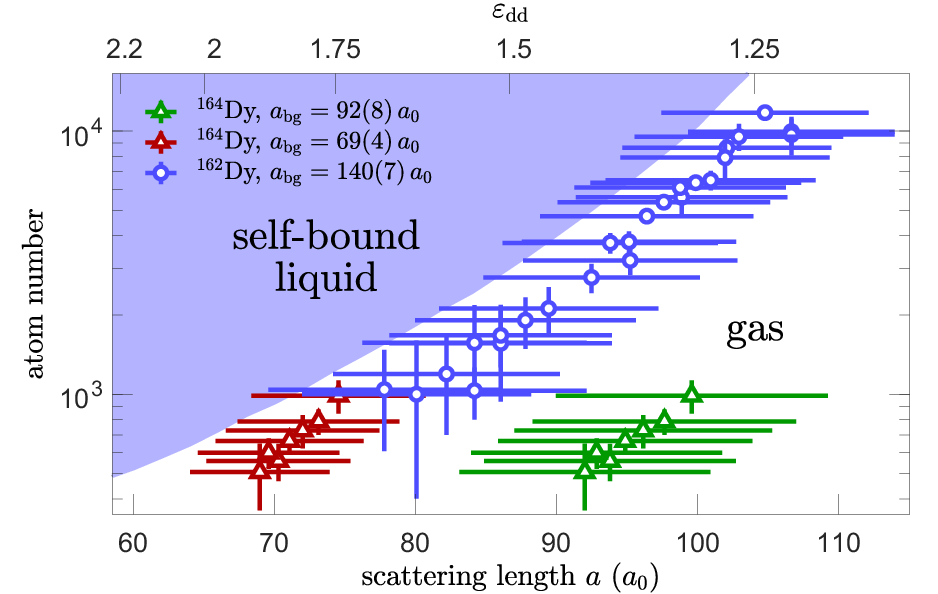}
  \caption{Measured self-bound liquid to gas phase diagram of Dy atoms. 
  The theory curve results from simulations of the eGPE. Data are shown for both \textsuperscript{164}Dy and \textsuperscript{162}Dy. For \textsuperscript{164}Dy, the background scattering length had to be adapted from the literature value ($92(8)\,a_0$, green circles) to a lower value ($69(4)\,a_0$, red diamonds), see also discussion in Secs.~\ref{subsec:theory_droplet} and \ref{subsec:droplet_props}. The horizontal error bars are systematic errors coming from uncertainties on FRs positions and widths. Adapted from Refs.~\cite{Boettcher2019qci,schmitt2016sbd}.}\label{Fig:liquidGasPhaseDiag}
\end{figure}

\begin{figure}[hbt]
  \includegraphics[width=\columnwidth]{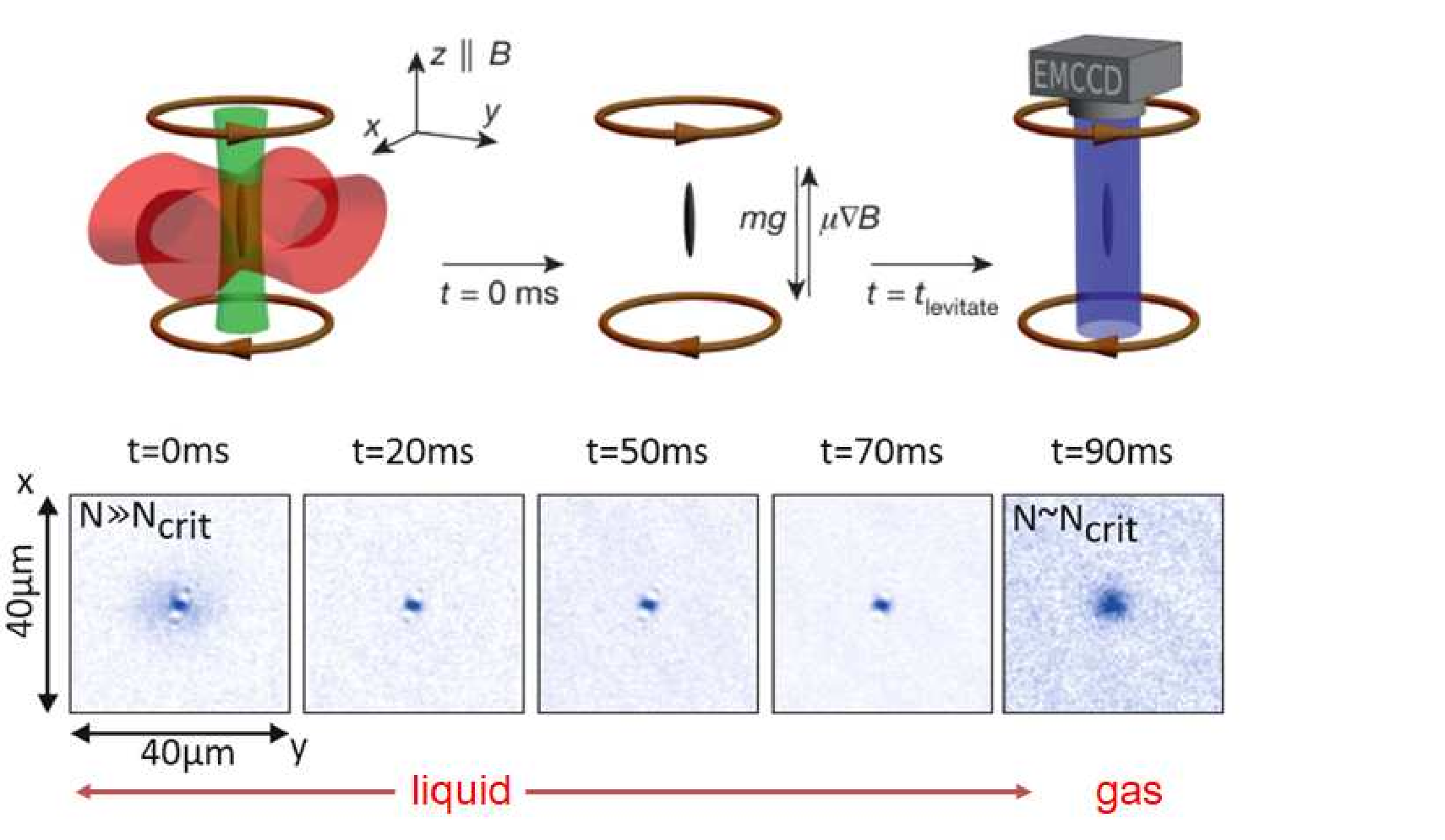}
  \caption{Single self-bound droplets without a trapping potential can be produced by preshaping the droplet in an appropriate trap which is then switched off. The droplet stays bound and floating in the chamber until eventually the losses reduce the atom number below a critical value where no bound states  exist any more. That manifests the transition to a gaseous phase where the atoms expand like a gas. Adapted from Ref.~\cite{schmitt2016sbd}.}
  \label{Fig:dropletForm}
  \end{figure}

The possibility of creating a single large quantum droplet by tuning the trap aspect ratio, as described in Sec.~\ref{subsec:single_droplet},  was key in the demonstration of the self-bound liquid nature of this phase. A quantum droplet living in equilibrium between the repulsive BMF and the attractive MF interactions, the question of the necessity of the trap for the existence of this phase then arises. Two theory papers demonstrated indeed the existence of a self-bound state in the absence of a trap, characterised by a non-vanishing peak density in infinite volume \cite{Wachtler2016gsp,Baillie2016sbd}. 
This defines a liquid state, existing from the mere balance between high-density repulsion and low-density attraction. 
However, this liquid is in essence quantum, which provides it with macroscopic new properties. First, it remains phase-coherent within a droplet. Second, the self-binding potential resulting from MF attraction from other atoms within the droplet must counter-act kinetic energy. As a consequence if the volume is too small, kinetic energy prevents the existence of a bound state. This yields a phase diagram as a function of atom number and scattering length, with a single line separating the self-bound state from a gas at low atom number and high scattering length, see Fig.~\ref{Fig:liquidGasPhaseDiag}. 
The scattering length dependence of the atom number value on the separation line is sharp. Indeed, using scaling arguments, one can show that the minimal atom number for the existence of the self-bound state scales as $N_{\rm c}\sim(1-\edd\,f_{\rm dip}(\kappa))^{-2/5}$.
First indications of a self-bound behaviour were already evidenced in the early work~\cite{FerrierBarbut2016ooq} through the absence of expansion in a waveguide and record-low expansion velocities in free space. Yet, the self-bound character could not be observed because of the limited atom number. Because of the large loss rate occurring in Er droplet (see Fig.~\ref{Fig:Comparisons}), the evidences of the self-bound behaviour of the macro-droplet of Ref.~\cite{Chomaz2016qfd} was partial, limited by the state lifetime. 
Here, only a slowing down of the expansion dynamics was observed. An unambiguous self-bound behaviour was then observed on \textsuperscript{164}Dy in Ref.~\cite{schmitt2016sbd}. Here a single droplet was created using an axially elongated dBEC~\footnote{the trap has yet  a prolate shape with $\lambda=1.3$. The axial elongation of the dBEC comes from magnetostriction effects, see Sec.~\ref{subsec:magnetostriction}} of 3000 Dy atoms and cruising the crossover to the single droplet state, similarly to Ref.~\cite{Chomaz2016qfd}. The trap was then smoothly released while the atoms were levitated against gravity using a magnetic-field gradient (see Fig.\,\ref{Fig:dropletForm}). Unexpanded droplets were observed up to 90\,ms of levitation. Using the atom loss as a probe, the liquid-to-gas transition was mapped out in the $(\as,N)$-space. This phase-diagram was later expanded to larger atom number by using samples of \textsuperscript{162}Dy atoms~\cite{Boettcher2019qci}, see Fig.~\ref{Fig:liquidGasPhaseDiag}. The experimental points of the two isotopes match well together and agree with eGPE predictions, at the expenses of a substantial renormalisation of the background scattering length of the \textsuperscript{164}Dy isotope, by a factor of 3/4. The need of such an important change of $\abg$ rose questions about the validity of the eGPE treatment in particular in the low-atom-number and large-$\edd$ regimes(
see also the discussion in Sec.\,\ref{subsec:theory_droplet}), or again of the methods employed to extract $\abg$ (see Sec.~\ref{subsec:fesbachsec}), in  particular in the Dy case. 

To conclude, the self-bound liquid-gas phase diagram, being simply due to the quantum mechanical nature of the system, is not restricted to the dipolar case. It may in fact be generally expected in quantum gases where two interactions of different origin compete, and as such it has been observed in Bose-Bose mixtures~\cite{Cabrera2018qld,Semeghini2018sbq}. It is interesting to note that in lower dimensions 
this phase diagram is completely modified. 
This has been the topics of several theoretical works, both for dipolar and mixtures systems, see e.g.~\cite{Astrakharchik2016uld,Edler2017qfi}. 


\subsection{Dipolar Supersolids}
\label{subsec:supersolidity}


In Section~\ref{subsec:droplets}, we have reviewed how the quantum-stabilisation mechanism discovered in 2016 (see Sec.\,\ref{subsec:theory_droplet}), may yield new ground states beyond the MF instability. There the ground state were limited to single droplet, as relevant for the early experiment, due to the small atom numbers and shallow traps (independent on the trap geometry, see Sec.~\ref{subsec:phasediagram_droplet}). Here we review how the same stabilisation mechanism yield to formation of self-modulated and in particular supersolid ground state, as well as their experimental investigations.

\subsubsection{Preliminary works}\label{subsec:supersolidity_preliminary}


Supersolidity is a paradoxical phase of matter in which the antithetical properties of  crystal arrangement and of superfluid flow coexist. It has been suggested more than half a century ago as a paradigmatic manifestation of a state in which two continuous symmetries of distinct nature are simultaneously broken~\cite{Penrose1956bec,gross1957uto,Boninsegni2012csw}. Originally predicted in quantum solids with mobile bosonic vacancies \cite{Andreev1969mbp,Chester1970sob,Leggett1970cas}, the search for supersolidity has 
spread in many different fields of physics.
Observation of supersolidity in helium was claimed~\cite{Kim2004poo} but the claim was withdrawn by the same authors a few years later~\cite{Kim2012aos} and the quest for supersolidity in helium is still open \cite{Boninsegni2012csw,Balibar2010teo}. The possibility of supersolid states in quantum gases were theoretically proposed long ago, see Refs.~\cite{Giovanazzi2002dmo,Mora2007qmo,Buechler2007sc2,Astrakharchik2007qpt,Pupillo2010scg,Golomedov2011mso,Boninsegni2012spo,Boninsegni2013mdq,Moroni2014coi,Lu2015sds,Ancilotto2019sbo}; this possibility linking back to the seminal work from E. Gross~\cite{gross1957uto} on assemblies of bosons with momentum-dependent interactions, see also Ref.~\cite{Kirzhnits1971cco,Pomeau1994doa}.   

{Besides the supersolids made from dipolar quantum fluids alternative approaches to supersolidity with ultracold atoms include miscible two-component BECs with SOC as well as optically pumped superfluids in a cavity that mediated interactions via the scattered light~\cite{Leonard2017sfi,Li2017asp}. All concepts have in common a momentum-dependent interaction that leads to a minimum at finite momentum in the dispersion relation on the superfluid side of the phase transition.  This is called the roton minimum. In both the dipolar supersolid and the SOC systems, two branches of the excitation spectrum exist on the supersolid side of the phase transition.  They correspond to the Goldstone modes of the two double symmetry-breaking processes, the breaking of the translational symmetry and the breaking of the U(1) phase invariance of the condensate. More recently, a continuous U(1) translational symmetry was created in system of multimode cavity light coupled to a BEC. The breaking of this symmetry resulted in a supersolid with transverse vibrations exhibiting a Goldstone dispersion of phonon excitations~\cite{Guo2021aol}. We will now focus the discussion on the simultaneous quest for a supersolid supporting crystal phonon excitations like a real solid in a dipolar gas only.}

This authentic supersolid behaviour can be expected if the simultaneous breaking of the two symmetries arise from the intrinsic interactions between the particles. As already highlighted by E. Gross, the relevant platforms for observing supersolidity are quantum gases with inter-particle interactions yielding large-momentum attraction~\cite{gross1957uto,Kirzhnits1971cco}. Practical examples are Rydberg-dressed potentials, spontaneous or light-induced dipole-dipole interactions in confined geometries~\cite{Mora2007qmo,Buechler2007sc2,Astrakharchik2007qpt,Pupillo2010scg,Golomedov2011mso,Boninsegni2012spo,Boninsegni2013mdq,Moroni2014coi,Lu2015sds,Ancilotto2019sbo}. In these settings, a roton-type excitation is induced in the superfluid' excitation spectrum by the large-$k$ attraction (see also Sec.~\ref{subsec:roton}). Its full softening may indicate the transition to a density-modulated state as the roton signals an intrinsically favoured length scale and has been seen as a precursor of crystallisation~\cite{Kirzhnits1971cco,Pomeau1994doa}. The observation of a roton mode in dBECs confined in cigar-shaped geometries~\cite{Chomaz2017oot,Petter2019ptr} was key to point promising grounds to observe supersolidity~\cite{Ancilotto2019sbo}, see also Sec.~\ref{subsec:roton}.

A recurrent hindrance to the interaction-driven formation of supersolid in quantum gases lies in the predicted MF collapse of the gas at the roton instability, see Secs~\ref{subsec:local_instab},\ref{subsec:collapse} and Refs.~\cite{Shlyapnikov2006coc,Komineas2007vli}. Protocols to stabilise the gases beyond its instability thanks to the engineering of higher-order (three-body) interaction potentials have been proposed~\cite{Lu2015sds}. However, these protocols have not been implemented in experiments so far. The discovered BMF stabilisation mechanism provides the key ingredient for an intrinsic stabilisation.

In 2017, theoretical works based on the eGPE~\cite{Wenzel2017ssi,Baillie2017dcg} demonstrated that not only a single droplet (as described in Sec.~\ref{subsec:phasediagram_droplet}) but also assemblies of multiple droplets, or in other words density-modulated states, could constitute the ground state of dipolar quantum gases. This was found for both pancake and cigar geometries, by simply using tighter trap and/or larger atoms numbers than in the previous works of Ref.~\cite{Bisset2016gsp,Wachtler2016gsp,Baillie2017ceo}. Similar observations were also achieved via quantum Monte Carlo simulations with large densities~\cite{macia2016dot,cinti2017caq}. Starting from the picture of a single droplet state, the physical argument to the formation of multiple-droplet ground-states can be formulated as follow: The liquid phase has a weak compressibility that yields a sharp increase in energy at high density, see also Sec.~\ref{subsec:droplet_props}. If one compresses an (isotropic) liquid in one or two directions, it then deforms to keep a constant density at the small cost of increasing its surface energy. Yet for the dipolar quantum liquid, the DDI dictates an anisotropy, and deformation costs a high amount of (dipolar) energy. Therefore, when a dipolar quantum droplet is compressed along its long axis, it might be energetically favourable to split it in several droplets (of smaller radial sizes), recovering thus an anisotropy closer to the one naturally imposed by the DDI. Ground states can thus form spontaneous density modulation. Calculations with periodic boundary conditions along one direction were also performed, see Ref.~\cite{Wenzel2017ssi}, proving that a continuous translation symmetry can indeed be broken (at the thermodynamic limit).  These predictions indicate possibilities for quantum-stabilised dipolar ``crystallised" ground states and in particular open the doors for quantum-stabilised dipolar  supersolids.

The work of Ref.~\cite{Wenzel2017ssi} also experimentally produced spontaneously density-modulated states in cigar-shaped Dy gases that show close features to the expected ground states yet global phase coherence was found to be absent. 
The lack of phase coherence was interpreted based on a Josephson-Junction formalism, which describes the maintenance of a phase relation via tunnelling processes between the individual droplets, and its preclusion via quantum or thermal fluctuations. In this experimental realisation, the tunnelling rate between the individual droplets, or in other words, the wave-function overlap between them, was not enough to lock the droplets in phase. The authors of Refs.~\cite{Wenzel2017ssi,Baillie2017dcg} also showed that assemblies of droplets with sizeable wave-function overlap were theoretically possible by either increasing the trap frequencies or the atom numbers compared to the current experimental configurations.

\subsubsection{First experimental evidences of supersolid behaviors}\label{subsec:supersolidity_first}
\begin{figure}[hbt]
  \includegraphics[width=\columnwidth]{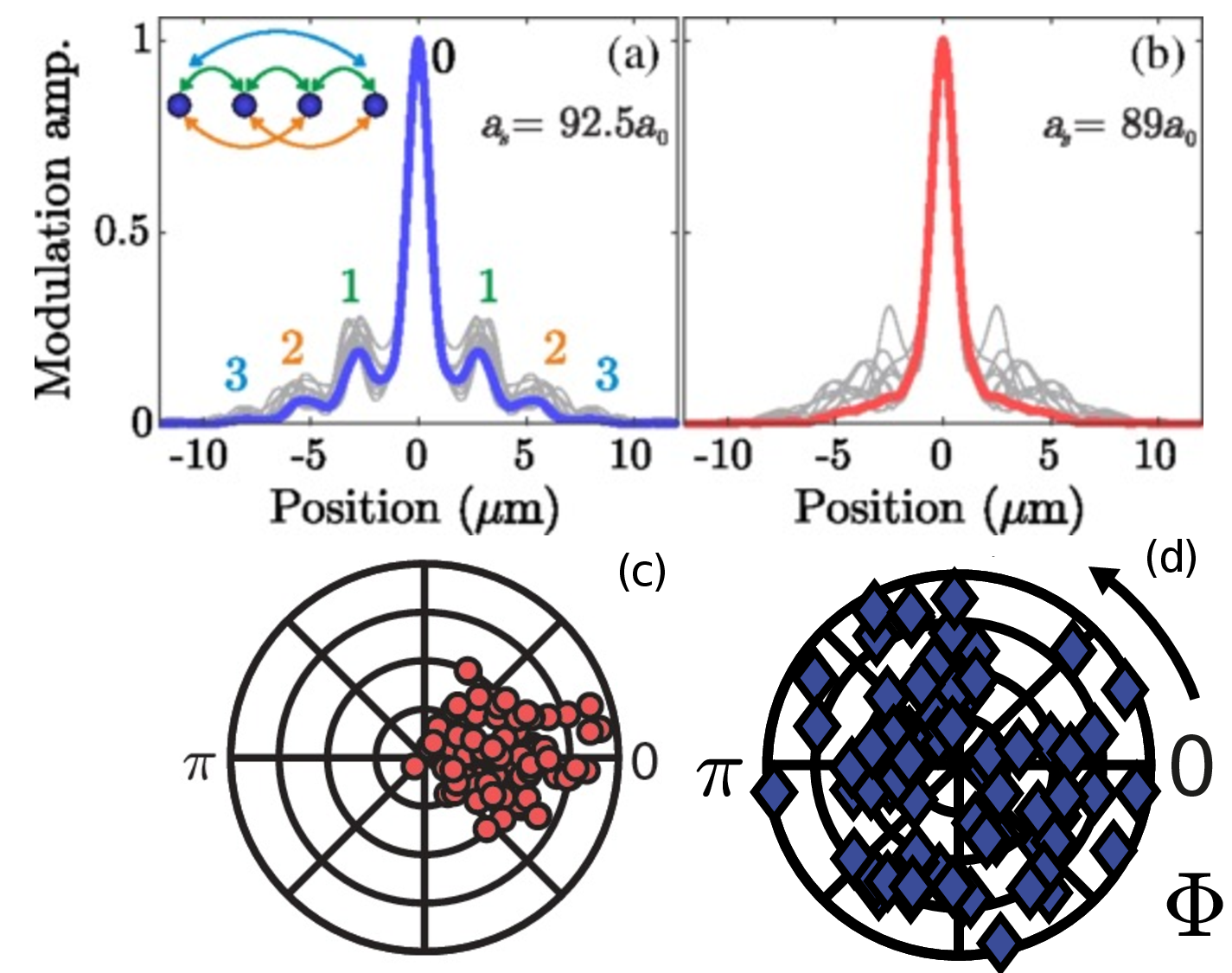}
  \caption{Examples of statistical analysis of the self-interference patterns of dipolar quantum gases showing hallmarks of supersolidity (a,c) (or normal solidity (b,d)), i.e., density modulations with (or without) global coherence. (a), (b) Norm of the Fourier transform of the integrated density profiles from the individual TOF images (grey lines) and the norm of their complex average (thick, blue or red, line) in the supersolid case and the insulating array of droplets case.   The existence of side peaks in the norm of the complex average evidence coherence in between density peaks (see inset). Adapted from ref.\,\cite{Boettcher2019tsp}.  (c),(d)  Representation in the complex plane of the values of the Fourier transform of the TOF profiles at the position of the first side peak (see a). The limited spread of the phases $\Phi$ evidences global phase coherence. Adapted from Ref.\,\cite{Ilzhofer2021pci}.
  }\label{fig:sss_interferences}
\end{figure}

At the end of 2018, the progressive understanding of the many key features of Bose gases of highly magnetic atoms combined in an acute picture. These features include the discovery of the BMF stabilisation in such gases (see Sec.~\ref{subsec:droplet_assemblies}), the observation of the roton mode and its softening in cigar-shaped clouds (see Sec.~\ref{subsec:roton}-\ref{subsec:local_instab}), the possibility of density-modulated ground states, matching an assembly of quantum-stabilised droplets (see Sec.~\ref{subsec:supersolidity_preliminary}), and the need of making the droplets more extensively overlap in order to maintain the global phase coherence in such assemblies (see Sec.~\ref{subsec:supersolidity_preliminary}). 
Building on this knowledge, a set of experiments~\cite{Tanzi2019ooa,Boettcher2019tsp,Chomaz2019lla} observed hallmarks of supersolid behaviours in cigar-shaped gases of \textsuperscript{162}Dy, \textsuperscript{166}Er, as well as \textsuperscript{164}Dy atoms.
Accompanying theoretical works~\cite{Boettcher2019tsp,Chomaz2019lla,Ancilotto2019sbo,Youssef2019psi,Zhang2019saa} related these observations to an underlying supersolid ground state of the gas, see Sec.~\ref{subsec:theory_sss}.

The three experimental works proved two hallmarks of supersolidity in experiments, namely the simultaneous occurrence of a spontaneously formed density modulation and of a global phase coherence. 
Density modulation was revealed by the occurrence of modulated patterns in the gas absorption images, either in situ~\cite{Boettcher2019tsp} or in time of flight. Here the density patterns form from the self interference of the gas via free expansion of the in situ modulation. The global coherence was demonstrated by a statistical analysis of multiple repetitions of this self-interference TOF patterns. Global coherence is here marked by the stability of the interference patterns, differentiating supersolids from an incoherent array of droplets. Global coherence was mostly quantified through an analysis of the complex values of the Fourier transform of the TOF density profiles, see Fig.~\ref{fig:sss_interferences}.  
We note that the coexistence of density modulation and phase coherence does not prove the superfluidity of the state, which can be demonstrated only by probing dynamic-related properties of the state, see Sec.~\ref{subsec:supersolid_dynamics}. 

All three experiments used relatively shallow cigar-shaped traps with transverse magnetisation, similar to Ref.~\cite{Chomaz2017oot}. The spontaneous density modulation occurred only along one axis, the long axis of trap.
The coexistence of density modulation and phase coherence was observed in narrow ranges of scattering length values, of a few $a_0$ wide, and survives a few tens~\cite{Tanzi2019ooa,Boettcher2019tsp,Chomaz2019lla} up to a few hundreds of ms~\cite{Chomaz2019lla} in these first experiments, later extended up to few seconds~\cite{Sohmen2021bla}. 
Building on this long lifetime, Ref.~\cite{Chomaz2019lla} additionally established a different route than standard interaction tuning towards supersolid states, which is based on direct evaporative cooling starting from a thermal state, see also~\cite{Sohmen2021bla}.

In all three works, the states with supersolid properties could be achieved by ramping down the scattering length starting from a stable BEC, using a slower ramp and a finer tuning of $\as$ than Ref.~\cite{Chomaz2017oot}. In the earlier works of Refs.~\cite{FerrierBarbut2016lqd,Wenzel2017ssi} observing droplet assemblies in cigar shaped traps, the rougher tuning of $\as$ as well as the lower initial BEC atom numbers are also thought to have prevented the observation of supersolid properties (see also Secs.~\ref{subsec:droplet_assemblies}, \ref{subsec:supersolidity_preliminary}). Finally, in the seminal work of Ref.~\cite{Kadau2016otr}, a distinct pancake-shaped geometry as well as smaller initial BEC atom numbers were used. Several following theoretical as well as experimental works indicates that the use of a cigar-shaped geometry was crucial for more easily achieving supersolid states in experiment. In this geometry, the MF instability is driven by the softening of a single (doubly-degenerate) roton mode dictating the dominant wavelength of the density fluctuations~\cite{Chomaz2017oot,Petter2019ptr,Hertkorn2021dfa}. In contrast, the pancake case is more complex, with several radial and angular rotons, corresponding to different structures of density fluctuations, simultaneously softening~\cite{Hertkorn2021pfi, Schmidt2021rei}. Furthermore, in the cigar-shaped case, it is expected that the transition to a supersolid state can occur continuously, with the supersolid modulation directly connecting to the softened roton mode, see e.g.~\cite{Blakie2020sia}. 

\subsubsection{eGPE phase diagram beyond the single-droplet regime: droplet assemblies and supersolid states.}\label{subsec:theory_sss}
\begin{figure}[hbt]
  \includegraphics[width=\columnwidth]{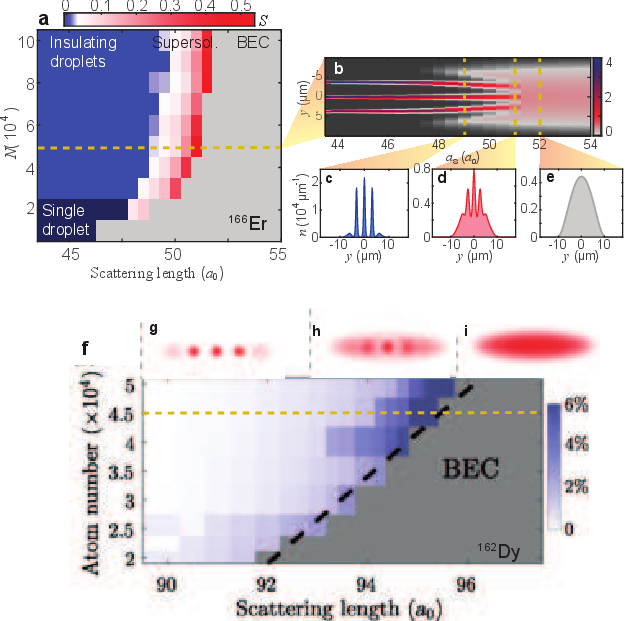}
  \caption{Phase diagrams, calculated in \cite{Chomaz2019lla} (a) and \cite{Boettcher2019tsp} (e) for the relevant experiments using Er, Dy respectively. 
  (b-e) ((g-i)) corresponding insitu axial density profile for $N=5\times10^4$ ($N=3.5\times10^4$). 
  Besides the BEC and single-droplet phases, identified in Fig.\,\ref{Fig:liquidGasPhaseDiag}, density-modulated ID and SSP ground states are found. The SSP is identified by a non-fully contrasted density modulation. It is sandwiched in a narrow $\as$-range in between a regular BEC and a crystal of independent droplets (ID).}\label{fig:sss_pds}
\end{figure}

{Following the observations of the roton mode population in Ref.~\cite{Chomaz2017oot} and shortly preceding the 
works of Refs.~\cite{Tanzi2019ooa,Boettcher2019tsp,Chomaz2019lla}, Rocuzzo and Ancilotto~\cite{Ancilotto2019sbo} theoretically explored the phase diagram of an Er quantum gas in an infinite cigar-shaped geometry with periodic boundary conditions. They relied on the eGPE framework developed in the context of the quantum droplet studies (see Sec.~\ref{subsec:theory_droplet}) and calculated the ground state as a function of $\as$. In this setting, they demonstrated the existence of a supersolid phase (SSP) in an intermediate range of $\as$, separating an array of insulating droplet (ID) at low $\as$ and a regular BEC at large $\as$. Both the SSP and the ID are density modulated ground states stabilised by quantum fluctuations, similar to the single-droplet phase described in Sec.~\ref{subsec:phasediagram_droplet}. The SSP distinguishes itself by bearing a density modulation of finite contrast.  In addition, Rocuzzo and Ancilotto directly computed, via dynamical simulation, the superfluid density of the density-modulated states, thus rigorously establishing the connection between the occurrence of non-fully-contrasted density modulation and non-zero superfluid fraction. This work proved  the relevance of the SSP in cigar geometries, beyond finite-size effects, see also Ref.~\cite{Blakie2020sia}.  }

Together with their experimental results, both Refs.~\cite{Boettcher2019tsp,Chomaz2019lla} reported on eGPE calculations of the ground-state phase diagrams in the finite experimental geometries, see Fig.\,\ref{fig:sss_pds}(a,f). 
 The ground states were calculated as a function of the atom number $N$ and the scattering length $\as$ for a given trap and atomic species. At low $N$, a transition occurs from the regular BEC to the single-droplet phase when decreasing $\as$, as described in Sec.~\ref{subsec:phasediagram_droplet}, see also  Refs.~\cite{Bisset2016gsp,Wachtler2016gsp,Chomaz2019lla}. 
In contrast, for large-enough atom numbers, three different regimes could be identified similar to the infinite case results of Ref.~\cite{Ancilotto2019sbo}: When decreasing $\as$, the regular BEC state transitions first to a density-modulated state of finite contrast, identifying a SSP, and then to a density-modulated state of contrast almost unity, forming an ID array, see Fig.\,\ref{fig:sss_pds}(b-e,g-i). 
We note that the atom number for the occurrence of  density-modulated states versus single-droplet depends on the atomic species (being smaller for Dy than for Er), and on the trap, both on its overall tightness and on its shape. 
The exploration of the most favourable parameters for supersolidity as well as achieving such a phase in different settings 
are interesting directions that are currently under investigation, see e.g. Sec.~\ref{subsec:supersolid_rich}.

Based on the eGPE framework, Refs.~\cite{Boettcher2019tsp,Tanzi2019ooa,Petter2021bso} also reported on simulations of the real-time evolution induced by the finite scattering-length ramp used to reach the various  BEC/SSP/ID states in the experiments. These show that in the SSP regime, the dynamical state has strong similarities with the expected ground state and shows limited phase fluctuations. In contrast, when decreasing the scattering length lower, to an ID regime, the dynamical state deviates from the ground-state expectation and show large phase fluctuations.

\subsubsection{Toward probing the dynamical response and superfluidity of dipolar supersolids}\label{subsec:supersolid_dynamics}

Besides the study of the static properties of the state, as the hallmarks reported in the first studies (see Sec.~\ref{subsec:supersolidity_first}), a huge interest is drawn by the special dynamical properties associated to supersolidity. Indeed, such a dynamics intrinsically connects to the superfluid character of the supersolid and would characterise the related rigidity of its phase. The study of the dynamical properties covers a wide range of phenomena and concepts, spanning from the study of the spectrum of elementary excitations, to that of transport properties, or of the response to an external rotation for instance. Supersolidity brings exotic characteristics to these different features. 

\begin{figure}[hbt]
  \includegraphics[width=\columnwidth]{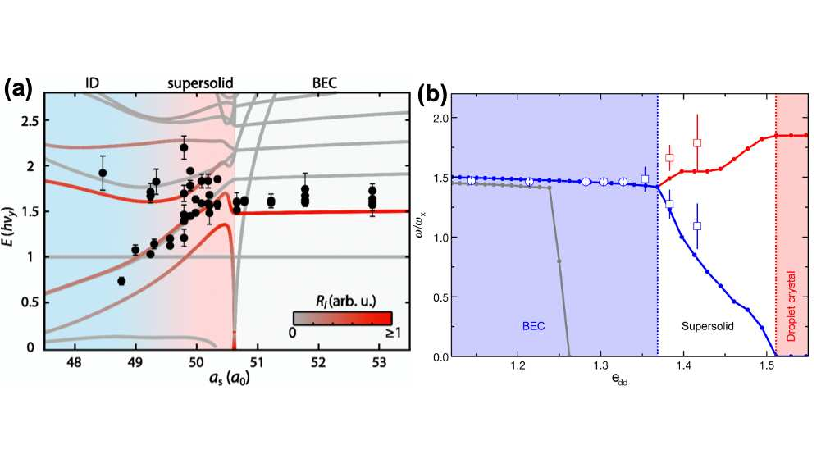}
  \caption{Observed response frequencies for quadrupole-type excitations in the BEC-SSP-ID phase diagram, from (a) an Er sample, taken from~\cite{Natale2019eso}, and (b) a Dy sample, from~\cite{Tanzi2019ssb}. Theory predictions from (a) Bogoliubov theory, (b) real-time simulations, based on eGPE theory are shown. In (a) all the elementary excitations, $l$, are shown and the colour highlight their expected response $R_l$ to the performed trap excitations. }\label{Fig:excitations_ssp}
\end{figure}
The first of these aspects to be explored in dipolar gas experiments concerns the properties of the spectrum of elementary excitations of a supersolid  at low momenta. Here, not one but two branches of low-energy (gapless) ``phonons" modes~\cite{Andreev1969mbp,saccani2012eso, macri2013eeo, Rossotti2017qcb,Macia2012eas, Ancilotto2019sbo} are expected, the different branches bearing excitations of different characters. Following Goldstone's idea~\cite{goldstone1961ftw}, the number of gapless branch connects to the number of spontaneously broken symmetries -- two in the case of the supersolid:  one branch holds phonons of the crystal and relate to a broken translational symmetry, while the other branch holds phonon of the superfluid (phase phonons) and relate to a broken gauge symmetry.  The phase branch corresponds to the branch of lower velocity, and, when evolving through the BEC-SSP-ID phase diagram, it is observed to soften and its weight in the response to decrease, up to vanishing in the ID case. On the contrary the crystal branch slightly harden. These properties of the supersolid excitation spectrum were theoretically investigated in various systems~\cite{saccani2012eso, macri2013eeo, Rossotti2017qcb,Macia2012eas}. The excitation spectrum behaviour in dipolar supersolids was further theoretically investigated in various works~\cite{Ancilotto2019sbo,Natale2019eso,Petter2021bso,Hertkorn2019fot}.

\begin{figure}[hbt]
  \includegraphics[trim=70 0 50 0,clip, width=\columnwidth]{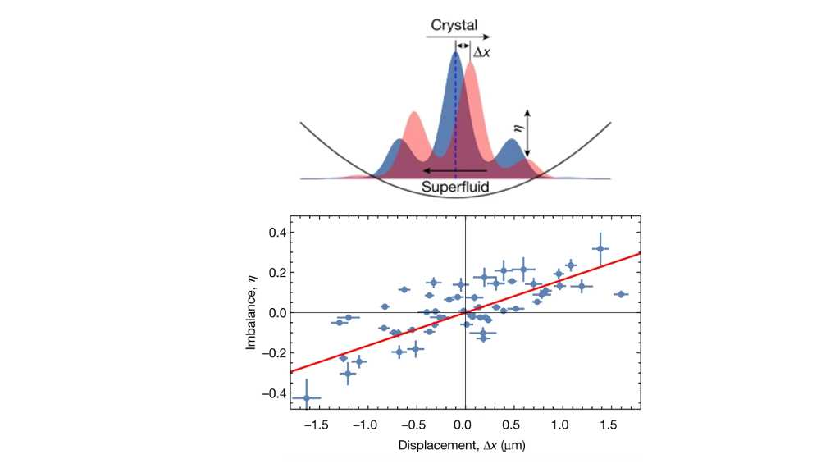}
  \caption{Dynamics induced by the lowest lying Goldstone mode of a supersolid (top) and observed correlation between displacement and imbalance in the in situ density patterns of experimental realisations of Dy supersolids at $97.6 a_0$ (bottom). Adapted from Ref.~\cite{guo2019tle}}\label{Fig:goldstone_ssp}
\end{figure}

Several experiments show signatures of these two branches~\cite{Natale2019eso,Tanzi2019ssb,guo2019tle,Hertkorn2021dfa}. Due to the finite system size, the phonon branches are here discretised (see also Sec.~\ref{subsec:excitationBEC}), and the experiments probe the response of some specific low-lying modes of the trapped dipolar supersolid states. In Refs.~\cite{Natale2019eso,Tanzi2019ssb}, the evolution of the BEC's lowest-lying quadrupole mode within the BEC-ID-SSP phase diagram is probed, see Fig.\,\ref{Fig:excitations_ssp}. In the BEC regime, the system responds at a single and roughly constant frequency, and when reaching the SSP regime, several frequencies are observed in the system's response. Furthermore, these frequencies are found to organise in two branches as a function of $\as$. One branch is softening and one is hardening when decreasing $\as$. These properties are indicative of the emergence of modes of dominant phase and crystal characters, respectively. Ref.~\cite{Tanzi2019ssb} identifies the different modes via their distinct signatures in the time-evolution of the gas's self-interference patterns. Ref.~\cite{Natale2019eso} identifies resonant frequencies via an unbiased principal component analysis of the self-interference patterns and observes a mixed character in the associated principal component structures. The different modes' characters between Refs.~\cite{Natale2019eso,Tanzi2019ssb} may be attributed to the different numbers of density peaks in the underlying supersolid states. Distinctly, Guo \textit{et al.}~\cite{guo2019tle} probe the lowest lying mode of the trapped supersolid, which has a frequency lying below the dipole mode and is a Goldstone mode of phase character, see Fig.~\ref{Fig:goldstone_ssp}. The frequency of the mode itself is not probed due to its low value, instead signatures of a spontaneous population of the low-energy excited mode are observed via a statistical analysis of the in situ density patterns. A peculiar feature of this mode is that the motion it induces preserves the centre of mass position thanks to superfluid flow. This implies a correlation between the array's displacement and the population imbalance between the density peaks, see Fig.~\ref{Fig:goldstone_ssp}. By repeatedly producing and imaging steady-state samples, Ref.~\cite{guo2019tle} evidenced such correlations in the limited $\as$ range where the state is supersolid. The presence of such correlations points to the existence of the phase Goldstone mode and implies an underlying superfluid flow. An interesting point to note is that the emergence of this Goldstone mode in the excitation spectrum connects to the softening of the antisymmetric roton mode from the BEC~\cite{guo2019tle}, see also Sec.~\ref{subsec:roton}. In contrast, the symmetric roton mode connects to an amplitude mode, also called Higgs mode, which sharply hardens when moving away from the SSP-BEC transition in the SSP~\cite{Hertkorn2019fot}. More recently, a detailed study of the in situ density fluctuations in the SSP also revealed the existence of density and crystal phonons~\cite{Hertkorn2021dfa}, see also Sec.~\ref{subsec:roton}. 

Besides the study of low-lying excitations, investigation of higher energy modes of supersolids have been recently undertaken. This includes studies of the response to rotational excitations~\cite{Tanzi2021eos}, or to large-momentum excitations~\cite{Petter2021bso}. Rotation excitations in Ref.~\cite{Tanzi2021eos} probe a scissor-type excitation, similar to that described in Sec.~\ref{subsec:droplet_props} for droplet states. Here the observed scissor-mode frequency decreases at the transition from BEC to modulated states. This indicates an increase of the state's momenta of inertia. It is however not straightforward to relate this behaviour to the superfluid fraction of the state. 
Ref.~\cite{Petter2021bso} probes the scattering of quasi-free particles via Bragg excitations (see also Sec.~\ref{subsec:excitationBEC}) along the BEC-SSP-ID phase diagram. Here a decrease in the response amplitude is observed when crossing from BEC to modulated states. From impulse-approximation theory, this decrease probes a reduction of the population of the zero-momentum state. The  observed decrease surpasses expectations obtained from the eGPE steady states. This was related to coherent phase evolution, induced by the dynamical crossing of the quantum phase transition~\cite{Chomaz2020pts}. 

Besides studies of specific excited modes, the supersolid's dynamical properties have also been unveiled in the time evolution induced by parameter quenches or ramps. This includes interaction quenches~\cite{Ilzhofer2021pci} and evaporative cooling ramps~\cite{Sohmen2021bla}. Ref.~\cite{Ilzhofer2021pci} probes the dephasing and rephasing dynamics of the gas' global coherence when dynamically crossing the ID-SSP transition. The observed dynamic was understood by comparing to a simple Josephson-Junction-Array model where the Josephson tunnelling amplitude was quenched, while dissipation is empirically introduced, using a Langevin formalism. This model shows a good qualitative agreement with experiments, while quantitative discrepancies are attributed to the inherently soft nature of the supersolid's crystal and the effects brought in by the possible excitation of the crystal's phonons. Studying the formation dynamics of supersolids via direct evaporative cooling (ramp of optical dipole traps), Ref.~\cite{Sohmen2021bla} shows that density modulation appears before global phase coherence, but after local phase coherence, is established. Reference~\cite{Sohmen2021bla}  also studies the subsequent decay process of the supersolids, occurring spontaneously under the effect of three-body recombination. In this decay, in contrast to the formation dynamics, global coherence is observed to survive longer than density modulation, while the temperature remains roughly constant. Furthermore, in this process, the strength of the density modulation, for a fixed value of the number of locally coherent atoms, is observed to depend on the temperature, being weaker when colder, see also ref.~\cite{Mazets2004doa}.  

The various dynamical studies performed up to now provide valuable insights into the supersolid state special behaviours, including direct and indirect signatures of the existence of phase and crystal phonons (near $k=0$) and their mixing, of superfluid flow and transport, or of coherent and incoherent phase dynamics in the experimentally produced states. Interesting prospects include direct measurements of the superfluid properties, and in particular of the superfluid density of supersolids, and the creation of vortices in the supersolid rotation, see e.g.~\cite{Roccuzzo2020roa,Ancilotto2021vpi}.

\subsubsection{Supersolids with richer crystalline patterns and two-dimensional character}
\label{subsec:supersolid_rich}
As discussed in Sec.~\ref{subsec:supersolidity_first}, the first observations of supersolidity in dipolar gases have been conducted in elongated and relatively shallow, cigar-shaped traps, resulting in supersolids in which (i) the breaking of the translational invariance occurs only along one dimension, (ii) the number of density modulations found in the finite experimental system is small (typically a handful) and thus finite size effects might be important. Yet, dipolar supersolid states with larger and/or more complex crystalline structures, and in particular where crystallisation occurs in two directions of space have attracted intense interest. Theoretically, a series of works have focused on the phase-diagram and excitation spectra of two-dimensional supersolids in isotropic and anisotropic traps ~\cite{Zhang2021pos,Hertkorn2021sit, Norcia2021tds,Bland2021tds,poli2021msi}, on the possibility of creating vortex excitations~\cite{Roccuzzo2020roa,Ancilotto2021vpi,Gallemi2020qvi}, and on the emergence of exotic crystalline structures ~\cite{Zhang2021pos,Hertkorn2021sit,Hertkorn2021pfi}.
Very recently, two-dimensional supersolidity has been observed in experiments with Dy atoms, using anisotropic traps~\cite{Norcia2021tds}. As a function of the anisotropy of the trap in the directions perpendicular to the atomic dipoles, experiments have demonstrated evaporative phase transitions to ground states of various supersolid patterns from a linear chain, to a zig-zag crystalline structures~\cite{Norcia2021tds}, and finally to a hexagonal structures in circular traps~\cite{Bland2021tds}. In a related effort, two-dimensional angular roton modes, analogous to the linear roton mode in elongated traps~\cite{Santos2003rms, Chomaz2017oot, Petter2019ptr}, has been observed in radially extended traps~\cite{Schmidt2021rei}, see also Sec.~\ref{subsec:roton} for more details. 
By further tuning the interplay between trapping configurations, interaction parameters, and atom numbers, even more exotic ground states are expected to appear, such as honeycombs, ring and  labyrinth-like supersolid phases ~\cite{Zhang2021pos,Hertkorn2021sit,Hertkorn2021pfi}.
With the considerable interest that the discovery of dipolar supersolids have generated, we expect a quick development of the field and a blossom of works probing the various intriguing properties of this phase as well as exploring its possibility in different settings. 

%% file: Section6/6_Spinors.tex
\section{Spin physics with highly magnetic atoms}\label{spinorsection}

In the previous Secs.~\ref{Sec:RepulsiveGases} and~\ref{Sec:DCQD}, we presented the many-body physics arising in quantum gases of magnetic atoms fully polarised in their lowest Zeeman state, under the influence of the elastic DDI between the aligned atomic dipoles. In the following, we discuss the physics of ultracold gases of magnetic atoms when the spin degrees of freedom are free and taken into account. 
Of course, short-range interactions among atoms can also be spin-sensitive, see Secs.\ref{subsec:DDI_spin_intro},\ref{subsec:fesbachsec}: magnetism is typically driven by the interplay of spin-dependent long-range dipolar forces and  spin-dependent short-range forces. This interplay intrinsically takes on a very different flavor if the  atoms are free to move and collide, or if the experiment is performed with atoms confined in an optical lattice wherein the spins are localised at each node of a 3D array. In the latter case, the system realises spin-lattice models, where spins interact both at a distance, directly through DDI, and by short-range interactions via second-order processes in tunnelling.  These so-called super-exchange interactions will be reviewed in Sec.\,\ref{sec:lattice}, while the current section focuses on the situation without a lattice potential.

After introducing a few important concepts of the spin physics with ultracold gases, including that which is unique to highly magnetic atoms in Sec.\,\ref{sec:spin_intro}, we will review the experimental achievements revealing the impact of the DDI on spinor physics and on two-component Fermi mixtures; see Secs.\,\ref{sec:constantmagn} and~\ref{freemag}. The latter two situations can be distinguished by whether dipolar relaxation can be neglected, resulting in a system in which the total magnetisation is conserved in the sample; see Sec.\,\ref{sec:constantmagn}, or whether the demagnetisation is a key mechanism of the physics at play, yielding gases with free magnetisation  and an intrinsic spin-orbit coupling; see Sec.~\ref{freemag}. Finally, we will discuss the engineering of spin-dependent Hamiltonians using light fields and the unique contributions provided by  magnetic atoms in this respect; in Sec.\,\ref{sec:lightspincoupling}. 
We will examine the engineering an artificial spin-orbit coupling, realising a special class of artificial gauge fields, and the engineering of spin-spin interactions that yield entangled quantum spin states. 

\subsection{Introduction to spinor physics}\label{sec:spin_intro}

\subsubsection{Magnetic atoms: a large composite spin}\label{subsec:large_spin} Unlike electrons, atoms possess a composite spin that may be large compared to electrons' $S=1/2$. This enriches spinor physics. For alkali atoms, the total spin arises from the coupling of the spin-1/2 electron to its orbital angular momentum $L$ and the nuclear spin $I$. This large spin allows for the study of a new type of quantum fluid, involving the interplay between magnetism and superfluidity~\cite{StamperKurn2013sbg}. The study of degenerate quantum gases with a large spin degree of freedom $s>1/2$ initially focused on bosonic atoms, i.e., Bose spinor gases~\cite{Kawaguchi2012sbe}.  Experiments  explored the spinor physics of $F=1$ Na atoms and $F=1$ and $2$ Rb atoms~\cite{StamperKurn2013sbg}. More recently, these studies were extended to spinor fermions using the $F=9/2$ state of fermionic K~\cite{Krauser2014gso}. 

Strongly magnetic atoms such as Cr and Lns are multi-electron systems, which can possess an even larger total spin in the ground state than alkalis. For example, bosonic Cr atoms have an electronic spin $S=3$ in the ground state ($L=0$), while bosonic Dy and Er atoms have $L=6,\,S=2$ and $L=5,\,S=2$ respectively. Including the hyperfine structure for the Fermi isotopes, one can reach, for example, $F=21/2$ using $^{161}$Dy atoms. Besides the richer physics at the many-body level, such large-spin atoms have also attracted  attention for studying highly non-classical behaviours involving quantum coherence and entanglement, either at the many, few, or even single-particle levels \cite{Fernholz2008sso,Chalopin2018qes,Evrard2019ems}. These prospects may have important implications and applications for quantum information processing, sensing, or metrology purposes. In particular, we note that a single magnetic atom provides the realisation of large spin states (length $F$) in an Hilbert space of relatively moderate size ($2F+1$) and increasing linearly with the spin size. In contrast, a spin of similar length realised by a set of spin-$1/2$ particles lives in an Hilbert space whose size increases exponentially with the spin length. The reduction of the Hilbert space offered by magnetic atoms is beneficial to temper the effect of decoherence in the system.

We note that, in atomic gases, research has also explored small-spin systems, e.g., those involving only two spin states to form an (effective) spin-1/2 system. Such a system can be accessed either by considering a hyperfine level of $F=1/2$ (as for fermionic Li and K~\cite{OHara2002ooa,Bourdel2003mot,Regal2004oor,Zwierlein2004cop,Bartenstein2004ceo}) or by isolating two states of a larger-$F$ level; see e.g.~\cite{Trotzky2008tro,Anderlini2007cei}. For magnetic atoms, however, isolation is complicated by the spin-changing DDI; see also Sec.\,\ref{Drelaxationsec}. Nevertheless, as we have seen in Sec.\,\ref{fsuppress}, thanks to the quantum statistics, some subspaces can be effectively preserved in spin mixtures of ultracold fermions on long time scales because decay rates are strongly suppressed~\cite{Burdick2015fso}. This is in particular the case for the subspace formed by the two lowest spin states of a fermionic magnetic atom. This feature has been used in experiments to study effective spin-1/2 dipolar systems under the effect of an artificial spin orbit coupling~\cite{Burdick2016lls} and in bulk~\cite{Baier2018roa}.

\subsubsection{Spin-dependent one-body Hamiltonian for spinor gases}\label{subsec:spin_1bdpart}

In order to study pristine spinor physics, i.e., to isolate the effects arising  solely from spin-dependent interactions, it is convenient to first consider  a spin-independent one-body Hamiltonian (typically the sum of a kinetic term and external potential; see Eq.~\eqref{GP}). Using a spin-space generated from the hyperfine level of an atomic species, the kinetic part of the Hamiltonian is spin independent. Then one need only consider the conservative trap confining the atoms. This potential should  be independent of the magnetic sublevels. For alkali atoms, this is typically obtained by using optical dipole traps created by far-off-resonant focused laser beams, in which case the AC-Stark shift is approximately independent of the magnetic sublevel~\cite{Grimm2000odt}. 
As discussed in Sec.\,\ref{subsec:atomlightLn}, the situation of magnetic atoms differs. Both for Cr and for Lns, the large electronic spin/angular momentum results in large vector and tensor parts of the atomic polarisability. This yields a significant dependence of the light shifts on the magnetic sublevel, even when using light far away of optical transitions. Typically, this is  not negligible in experiments exploring dipolar magnetism.  It is then expected that spinor phases will be driven by an interplay between spin-dependent  interactions, the Zeeman effect, and the tensor light-shift.

\subsubsection{Spin-dependent interactions for large-spin atoms}
\label{subsec:spin_int}

To study of spin physics with large-spin atoms it is important to recall that the interaction between the atoms can depend on the spin channel $\mathcal{S}$ of the collision. Here, $\mathcal{S}$ describes the total spin of  two atoms and  is preserved during the collision in the absence of a DDI, see Secs.\,\ref{subsec:fesbachsec} and \ref{sec:scatt}. 

Quantum statistics plays a major role in selecting which molecular potentials should be taken into account. Generally speaking, regardless of the statistics of the particles involved, $\mathcal{S}+l$ must be even, where $l$ is the relative angular momentum; see also Sec.~\ref{scatttheory}. For short-range interactions, $s-$wave scattering dominates at low collision energy. Under these circumstances, only even $\mathcal{S}$ need to be considered, both for bosons and fermions.

The different molecular potentials have identical multipole expansion parameters at long distance where only electrostatic interactions contribute. However, at short distance, where electronic orbitals may overlap, quantum statistics also plays a pivotal role.  This is because the molecular potentials associated with different spin channels are often quite different. Therefore, different spin channels $\mathcal{S}$ typically correspond to different scattering lengths $a_{\mathcal{S}}$. This is clearly true in the case of Cr because of the large electronic spin of the atom.  This leads to seven very different molecular potentials, as calculated in Ref.~\cite{Pavlovic2004cpo} and confirmed by experiments \cite{Werner2005oof,Pasquiou2010cod}. Scattering lengths are also (most likely) spin dependent in Lns. Because the outermost electronic shell has zero spin (identical to the case in Yb), all molecular potentials have a similar shape even at short distances. However, slight differences between the molecular potentials (due to, e.g., orbital anisotropy of unfilled submerged $f$-shell) are most likely sufficient to lead to spin-dependent contact interactions~\cite{Kotochigova2014cib}. As highlighted in Sec.~\ref{subsec:fesbachsec}, scattering properties of Ln are extremely difficult to predict, while, on the experimental side, spin-dependent scattering lengths have remained unexplored in these atoms; see also Sec.~\,\ref{subsec:fesbachsec}.

In gases of large-spin magnetic atoms, there is, in addition to spin-dependent contact interactions, the DDI that must be taken into account. Dipolar interactions are not only spin-dependent, but they also induce direct spin-orbit coupling (SOC); see Eqs.~\ref{eq:vddz}-\ref{relaxationterms}. 
Indeed, when including the spin degree-of-freedom, the DDI leads to two extra terms in the Hamiltonian.  These are in addition to the bare (yet also spin-dependent) elastic term Eq.~\eqref{eq:vddz} that we have considered up till now. The first term is the so-called spin-exchange term of Eq.~\eqref{eq:vddex}: it changes the spins of atoms at long distances while conserving the total longitudinal magnetisation of the pair of particles. The second term is the relaxation term of Eq.~\eqref{relaxationterms}. This leads to a change in magnetisation and converts spin angular momentum into orbital angular momentum to conserve the total angular momentum in the system, thus providing a form of SOC. Due to this peculiar SOC, and to the long-range coupling between spins,  large-spin systems thus allow the exploration of magnetism beyond  paradigms inherited from solid-state physics in addition to beyond what could be achieved with ultracold atomic gases up till now. Experiments can be performed both with bosonic and fermionic isotopes (see Sec.~\ref{sec:amgnetic_atoms}), greatly enhancing the scope for new magnetic behaviour. 

\subsubsection{Mean-field spinor physics with spin-dependent contact interactions}
\label{subsec:spin_contact}
Because of the spin-dependence of the contact interactions described above, the Hamiltonian of a spinor gas depends on the spin $s$ of the interacting particles  even in the absence of a DDI.  The Hamiltonian becomes increasingly complicated for increasing $s$~\cite{Kawaguchi2012sbe}. It is useful to examine the simplest case for what can be considered as a large-spin system, i.e., one beyond the spin-1/2 case. This is the case of $s=1$ atoms. In this case, only two spin channels exist for the short-range interactions, $\mathcal{S}=0$ and $\mathcal{S}=2$, with the corresponding scattering lengths $a_{\mathcal{S}=0}$ and $a_{\mathcal{S}=2}$. In the Hamiltonian, given by Eq.~\eqref{mfspinor} within the mean-field picture, the difference between the two scattering lengths is the key ingredient for spinor physics, giving rise to a so-called $c_1$-interaction term with $c_1 \propto \left(a_2 - a_0 \right)$. 

This spin-dependent interaction term for example results in spin-exchange dynamics that is experimentally apparent by monitoring the evolution of the population in the different Zeeman states as a function of time.  The spin-exchange processes correspond to the Forster-like exchange of a spin excitation $(m_s=0,m_s=0) \rightarrow (m_s=-1,m_s=1)$  and are driven in the mean-field picture at a rate:
\begin{equation}
\Gamma_{\rm exc} \propto \frac{4 \pi \hbar^2}{m} n (a_{\mathcal{S}=2}-a_{\mathcal{S}=0}). \label{Gamma_exchange}
\end{equation}
Spin dynamics has been widely studied in the community for both Bose and Fermi quantum degenerate and thermal gases~\cite{StamperKurn2013sbg,Schmaljohann2004dof,kronjager2006mts,Sadler2006ssb,Liu2009qpt,Jacob2012pdo,pechkis2013sdi,Krauser2012cmf, Krauser2014gso, Evrard2021fmb}. More generally, the study of excitations within a spinor condensate is expected to be very rich, with a number of possible topological excitations observable, such as line defects, point defects, skyrmions~\cite{choi2012oot}, and knots~\cite{Kawaguchi2012sbe}.

Most importantly, spin-exchange processes associated with contact interactions conserve the total longitudinal magnetisation of the pair of particles. This conservation stems from the isotropy of contact interactions. A very important practical consequence is that the linear Zeeman effect is gauged-out and does not contribute to the dynamics or to the phase diagram. Spinor phases are rather determined by the interplay between spin-dependent contact interactions and the quadratic Zeeman effect~\cite{StamperKurn2013sbg}, leading to quantum phase transitions~\cite{Sadler2006ssb,Liu2009qpt,Jacob2012pdo}. At low magnetic fields, where the quadratic Zeeman effect may be neglected, the spinor ground state strongly depends on interactions. For $s=1$, when $a_2<a_0$, the BEC is ferromagnetic (favouring collisions in the $S=2$ channel), while the predicted ground state when $a_0<a_2$ is polar~\cite{Ho1998sbc,Ohmi1998bec}. On the other hand, a large positive quadratic Zeeman effect favours the polar state. As the spin increases, the wealth of possible phases makes spinor quantum gases a promising area of research~\cite{Kawaguchi2012sbe}, including for the study of non-Abelian spinors~\cite{Lian2012sfn}.

\subsubsection{Beyond-mean-field spinor physics with spin-dependent contact interactions} One of the important features of spinor gases is that quantum correlations can naturally occur, allowing quantum fluctuations to play an important role. For example, spin-exchange dynamics vanish at the mean-field level when a spinor condensate is prepared in the initial state $m_s=0$. Quantum fluctuations then drive the onset of spin dynamics, with analogies to parametric amplification~\cite{Bookjans2011sqs,Klempt2012pao,Evrard2021fmb}. Atomic quantum optical effects lead to the generation of entanglement~\cite{Lucke2014dme}, which may have important applications for atom-based squeezing and atom interferometry~\cite{Kruse2016ioa}. 
 
Quantum fluctuations can also play a key role in determining  the nature of the many-body ground state. A formal mapping to quantum optics was proposed by Law \textit{et al.}~\cite{Law1998qsm}, who showed that the ground state differs from  mean-field predictions~\cite{Ho1998sbc,Ohmi1998bec}, at least for mesoscopic systems. Depending on the sign of the spin-dependent interactions, the ground state is strongly degenerate (with spontaneous symmetry breaking likely), or a condensate of spin-singlet pairs.  So-called `fragmented BECs' have also been studied by Ho and Yip~\cite{Ho2000fas}, and were very recently experimentally produced in Ref.\,\cite{Evrard2020pac}.

While most of the experiments have been performed with bosonic atoms, large-spin fermions are also a  promising direction for research on large-spin systems. First studies have been performed in Hamburg~\cite{Krauser2012cmf, Krauser2014gso}. These large-spin Fermi systems possess increased spin fluctuations due to the large spin~\cite{Wu2006hsa}, new SU(N) symmetries for purely nuclear spins, as it is the case for Yb and Sr~\cite{Hermele2009mio,Tey2010ddb,Taie2010roa,Taie2012asm,Pagano2014aod}, BCS pairing with a non-singlet character~\cite{Ho1999pof}, and greater than two-particle clustering~\cite{Rapp2007csa}.

\subsubsection{How are dipolar interactions expected to impact spinor physics?}\label{subsec:DDI_spin_intro}

The impact of dipolar interactions on spinor gases crucially depends on whether the dipolar field is larger or smaller than the external magnetic field. For a large magnetic field, magnetisation-changing collisions merely lead to atomic losses, and therefore, within the dipole-relaxation-limited lifetime, the experiment is essentially performed at constant magnetisation.  That is, it evolves under terms that conserve the longitudinal magnetisation in the dipolar Hamiltonian. 

Because they are spin-dependent, the DDIs  modify the properties of spinor gases when they are not negligible compared to spin-dependent contact interactions. To judge \textit{a priori} the impact of the DDI, one needs to compare the dipolar length $\add$ to \textit{differences} between scattering lengths~\cite{Lahaye2009tpo}. As a consequence, dipolar effects can be prominent even for the case of alkali metal atoms, such as Na or Rb for which spin-dependent contact interactions are much weaker than spin-independent interactions.  For example, the long-range and anisotropic nature of the DDI introduces a nonlocal nonlinearity that can modify spin textures, as experimentally investigated using alkali atoms~\cite{Vengalattore2008sms,Eto2014ood}. As we will see, dipolar interactions also impact spin dynamics.

The modifications introduced by the DDI on spinor physics are even more pronounced when the magnetic field is not large compared to the dipolar field. Then, the magnetisation-changing terms in the dipolar Hamiltonian cannot be neglected. Unlike contact interactions, the DDI, by allowing changes in magnetisation, yields an intrinsic SOC which breaks rotational symmetry; see Sec.~\ref{subsec:spin_int}. It thus modifies  the classification of the possible spinor phases.  The exact nature of the dipolar spinor ground state phases at low magnetic field, which may depend on the trapping geometry and other spin-dependent forces, is still a matter of theoretical investigation; see for example \cite{yi2004qpd} for the $s=1$ dipolar system in the single mode approximation and Refs.~\cite{Lian2012sfn,Cui2013sgf}. 

Spin-orbit coupling associated with the DDI corresponds to the exchange between orbital angular momentum and spin angular momentum. The DDI can, for example, trigger the demagnetisation of a sample, leading to spontaneous rotation, analogous to the Einstein-de-Haas effect~\cite{kawaguchi2006edh, Santos2006scb, gawryluk2007red, sun2007ote,Gawryluk2011hto,Swisocki2011tdr}. 
Furthermore, because of magnetisation-changing processes,  the dipolar spinor gas is sensitive to magnetic field, which provides new ways to study phase transitions driven by the interplay between spin-dependent interactions and the linear Zeeman effect~\cite{Pasquiou2011sdo}. For example, it is expected that a ferromagnetic spinor dipolar condensate may display spontaneous circulation in the ground state at low magnetic fields~\cite{kawaguchi2006scg}.

In the following, we will describe the experiments that address the role the DDI has played in  spinor physics, distinguishing between the cases where  magnetisation is conserved from those where it is free to evolve. 

\subsection{Effects of dipolar interactions on spinor physics at constant magnetisation}\label{sec:constantmagn}

If the time-scales of dipolar relaxation  are large compared to the other relevant timescales in the system (see \ref{Drelaxationsec}),  one can study spinor physics under the effect of the DDI but at constant magnetisation. Here, the effect of the DDI arises from the elastic term of DDI, Eq.~\eqref{eq:vddz}, and the spin-exchange term,  Eq.~\eqref{eq:vddex}.   The intrinsic spin-orbit coupling introduced above is neglected for now. The spin-dependent elastic DDI term impacts the  spinor physics because of its long-range and anisotropic nature. 

\subsubsection{Dipolar-induced spin textures in BECs}

One of the prominent features of the DDI, associated with its non-trivial momentum dependence shown in Eqs. \eqref{vddtf}, is a tendency to develop spatial structures.  This has already been discussed in the spin-polarised case in Secs.~\ref{subsec:roton},~\ref{subsec:local_instab}, and~\ref{Sec:DCQD}. In the context of spinor dipolar gases, this tendency translates into the possibility of developing various spin-textures~\cite{Kawaguchi2012sbe}, i.e., inhomogeneous distributions of the spin vector. 

The influence of the DDI on spin domains of Bose quantum gases was experimentally studied both in Berkeley~\cite{Vengalattore2008sms} (for a Rb $F=1$ BEC with ferromagnetic contact interactions) and in Tokyo~\cite{Eto2014ood} (for a Rb $F=2$ BEC whose contact interactions disfavour ferromagnetism). In both experiments, spin domains were created using a well-defined magnetic field gradient. In Ref.~\cite{Vengalattore2008sms}, the spin textures decayed toward a spatially modulated structure of spin domains. The crucial role of the DDI was demonstrated by eliminating the DDI by a sequence of rf pulses, in which case the authors observed a suppression of the formation of the short-range domains. In Ref.~\cite{Eto2014ood}, the spin textures observed after propagation within the magnetic field gradient are compared to numerical simulations of the Gross-Pitaevskii equation. This reveals that the observed spatial modulation of the longitudinal magnetisation is due to the spin precession in the effective magnetic field produced by the DDI. Both these results demonstrate that the DDI has considerable effect, even on spinor condensates of weakly dipolar alkali metal atoms.

\subsubsection{Gapped magnon excitations of BECs}
\label{subsec:magnons}
Phonon excitations have been studied in the context of scalar BECs through, e.g., Bragg excitation spectra~\cite{Stenger1999bso,StamperKurn1999eop,Steinhauer2002eso, Vogels2002eoo,Ozeri2002doo,Bismut2012aes}; see also Sec.~\ref{subsec:excitationBEC} for the dipolar case.  Moreover, there exists, in spinor BECs, spinfull excitations that also behave like quasiparticles. Surprisingly, these excitations remain relatively unexplored experimentally. One of the first investigations was performed at Berkeley on a ferromagnetic Rb BEC~\cite{Marti2014cmo}. Magnon excitations  were precisely studied and shown to consist of a standing-wave of spin-excited atoms above a ferromagnetic BEC. While the  Goldstone theorem  predicts that a ferromagnetic state has gapless excitations (for symmetric interactions), a gap was measured and ascribed to the presence of a weak DDI that  breaks rotational symmetry. Perhaps related to this feature, the authors also observed a larger effective mass than expected by mean-field and beyond-mean-field theories involving only $s$-wave interactions. Although these effects remain small due to the weakness of the DDI in alkali atoms, they provide qualitative departures from the paradigms of magnetism in the presence of symmetric short-range interactions.  This constitutes but one illustration of the possibilities dipolar spinor gases offer.

In the case of Cr, trapped magnon excitations were created by applying, on a polarised BEC,  magnetic field gradients perpendicular to the magnetic field axis~\cite{Lepoutre2018csm}. In that experiment, the wavelength of the excitation is not small compared to the cloud's size. The magnons that are created are quantised in energy, and manifest themselves as collective modes that couple the spin and the orbital degrees of freedom. The lowest energy mode, whose frequency is set by the zero-point energy of the BEC, consists of a sinusoidal oscillation of the local spin around its original axis, with an oscillation amplitude that linearly depends on the spatial coordinates. The observations are in excellent agreement with hydrodynamic equations. In the regime the experiment was performed, the observed spin mode has a universal character, independent of the atomic spin and spin-dependent contact interactions, and is therefore rather insensitive to the DDI. It was nevertheless predicted that dipolar interactions could alter such a mode, provided that the atomic sample has a size larger than a natural wavelength set by the DDI~\cite{Lepoutre2018csm}.

\subsubsection{Out-of-equilibrium spin dynamics at constant magnetisation}\label{subsec:spindyn_bulk}

Strong dipolar effects have been observed in the spin dynamics of magnetic atoms even on time scales short compared to the dipolar relaxation processes. Studies where performed on $S=3$ $^{52}$Cr atoms, initially polarised in the lowest energy spin state $m_s=-3$, and after homogeneously quenching a BEC into a spin-excited state. This was first performed using an engineered tensor light shift to promote all atoms into a well-defined Zeeman state $m_s=-2$~\cite{depaz2016psd}. In a latter experiment, the spin excitation was performed by simply rotating the spins using a rf pulse~\cite{lepoutre2017sma}.

In both cases, out-of-equilibrium spin dynamics was monitored by means of a Stern-Gerlach procedure to measure the population of atoms in the different Zeeman states as a function of time. It was found that spin dynamics was driven by an interplay between spin-dependent contact interactions and DDI. Spin-dependent contact interaction  played the dominant role in the spin dynamics because they are relatively strong in the case of $^{52}$Cr atoms. 

In the case where the atomic spins are tilted compared to the magnetic field by an rf pulse, it was nevertheless demonstrated that initial spin dynamics were entirely triggered by the DDI. This is a consequence of the SU(2)-symmetric nature of contact interactions, which cannot trigger dynamics after a simple rotation of the atomic spin initially in a stretched (ferromagnetic) state. Therefore,  the onset of spin dynamics after rotation of the spins seen in Ref.~\cite{lepoutre2017sma} is a purely dipolar effect. Finally, in the specific case where the spins were tilted by $\pi/2$ compared to the magnetic field, it was found that the spin dynamics vanishes. The mean-field theory provides a natural explanation for this phenomenon, as the torque associated with the inhomogeneous magnetic field carried by the atoms vanishes when the magnetisation of each atom vanishes~\cite{Kawaguchi2007csd}. 

In the $\pi/2$ rotation case, spin-dynamics could be recovered by applying a magnetic field gradient providing the necessary spin-orbit coupling to trigger dynamics. It was then discovered that spin dynamics develops while preserving the local spin length of the condensate~\cite{lepoutre2017sma}. This protection of the initial ferromagnetic character of the gas was attributed to an energy gap provided by sufficiently large  spin-dependent contact interactions. This dynamical protection of ferromagnetism is the reason why trapped magnon excitations could be observed in the Cr experiments and described using the hydrodynamic equation of a ferrofluid, indicating that the Cr BEC behaves like a genuine ferrofluid; see Section \ref{subsec:magnons} and \cite{Lepoutre2018csm}.  

It should be noted that the protection of ferromagnetism in the $^{52}$Cr condensates is tied to the relatively small strength of the DDI compared to spin-dependent contact interactions; it is likely that similar experiments performed (for example with Ln atoms) in the regime where dipolar interactions overwhelm spin-dependent contact interactions may lead to qualitatively different behaviour, since the DDI does not preserve the spin length. As we will see in Sec.~\ref{sec:lattice}, working in optical lattices is another way to investigate a purely dipolar spin system. In this case,  the collective spin length is indeed reduced during dynamics~\cite{gabardos2020rot}.   

\subsubsection{Effective spin-1/2 mixture of dipolar fermions}
\label{subsec:fermispin}
Beyond the case of BEC described above, recent work has been dedicated to the study of spinor systems of dipolar fermions~\cite{Burdick2015fso,Burdick2016lls,Baier2018roa}.
This system is of great interest for the study of how the DDI affects superfluid pairing and the celebrated crossover from delocalised Cooper pairs in the BCS regime to a BEC of molecules~\cite{Bloch2008qca,Baranov2012cmt}. By choosing specific spin mixtures, spinor gases of constant magnetisation can be studied over long time scales, even for highly magnetic species of Er and Dy. This is due to the quantum statistical suppression of dipolar relaxation~\cite{Burdick2015fso}; see also Sec.\,\ref{fsuppress}. This feature has been recently used to realise a long-lived spin-1/2 spin-orbit-coupled degenerate dipolar Fermi gas~\cite{Burdick2016lls}. This experiment will be discussed in Sec.~\ref{SecSOC}. 

A spin-1/2 degenerate Fermi mixture of strongly magnetic atoms $^{167}$Er was created~\cite{Baier2018roa}, and the collisional behaviour of this mixture experimentally investigated.  This experiment used a deep 3D optical lattice as a tool during the preparaiton of the mixtures, which serves to inhibit collisional losses induced by large magnetic field sweeps and the crossing of numerous FRs: A quantum-degenerate spin-polarised sample is first loaded into a deep 3D optical lattice at low magnetic field, such that double occupancy are precluded and fermions are spatial separated. The magnetic field can there be ramped up to a value where quadratic Zeeman shifts are significant, allowing spin state preparation using  using rf sweeps, and later be ramped down to the low-field region, where  the physics of the spin mixture is studied. To do so, the 3D lattice is slowly ramped down and the Fermi mixture loaded back into a 3D trap. As already observed for the spin-polarized case -- see  Sec.\,\ref{subsec:fesbachsec}) -- the collisional properties of the spin mixtures as a function of the bias magnetic field amplitude, $B$, is marked by a large number of intra- and inter-spin FRs. Baier \textit{et al.}~\cite{Baier2018roa} mapped the FR spectra for $B$ in the $[0,2]$G range via loss spectroscopy. Many narrow and overlapping features are obserrved. A comparatively broad and isolated interspin FR is also identified. Baier \textit{et al.} used this resonance to reach the strongly-interacting regime. Here, a large collisional stability of the balanced spin-mixture at $T/T_F\approx 0.3$ is observed, in particular on the repulsive side of the FR. This paves the way toward study of BEC-BCS physics in such systems.  

\subsection{Spinor physics with free magnetisation}
\label{freemag}
Besides the studies at constant magnetisation described above, the DDI introduces fundamental new features due to the presence of magnetisation-changing collisions; see Eq.~\eqref{relaxationterms}. In this context, the linear Zeeman effect can impact out-of-equilibrium physics and phase diagrams. In particular, magnetisation-changing collisions can free the total magnetisation of a gas, so that the system now becomes sensitive to the linear Zeeman effect. Different regimes may be reached depending on the ratio of the Larmor precession energy  to  other energy scales. A rather complex phenomenology unfolds, depending on the Larmor frequency $g_J \mu_B B / h $ compared to the following energy scales; note that typical values are indicated on the right column for the specific case of $^{52}$Cr atoms and $g_J$ is the Land\'e factor:

\vspace{.5cm}

\begin{tabular}{|l|l|} \hline
Physical process & Energy scale \\ \hline
Trap depth & $U_0/h = 500$ kHz \\
Thermal excitations & $k_B T/h = 2.5$ kHz \\
Fermi energy & $\epsilon_F/h = 2.5$ kHz \\ 
Spin-dependent  & $\Gamma_{\rm exc}/h =  250$ Hz  \\
interactions &  \\
Trap frequency & $\omega_L / 2 \pi = 250$ Hz  \\
DDI & $V_{\rm dd}/h = 25$ Hz  \\ \hline
\end{tabular}

$\Gamma_{\rm exc}$ was defined in Eq.~\eqref{Gamma_exchange}.

 \vspace{.5cm}

\begin{itemize}

\item $g_J \mu_B B > U_0$. Dipolar relaxation leads to (spin-sensitive) losses. 

\item $U_0 > g_J \mu_B B$. Dipolar relaxation does not lead to losses, but introduces heating. If $ g_J \mu_B B  \gg k_B T$, most of the atoms are polarised in the lowest energy state at equilibrium. 

\item $g_J \mu_B B \approx k_B T \gg \Gamma_{\rm exc}$. The thermal gas is spontaneously depolarised, populating a few Zeeman sublevels at equilibrium. For bosons, Bose-stimulation insures that the BEC remains polarised in the lowest energy state. However, for Fermi gases, the DFG may be depolarised if  $g_J \mu_B B < \epsilon_F$.

\item $g_J \mu_B B < \Gamma_{\rm exc}$. A demagnetised spinor BEC may be produced provided spin-dependent contact interactions do not favour a polarised BEC in the lowest energy state. 

\item When $g_J \mu_B B \approx \hbar \omega_L$, the  released energy due to dipolar relaxation corresponds to one unit of excitation in the trap. This is favourable for measuring the change in angular momentum associated with dipolar relaxation; this is a quantum gas analogue of the Einstein-de-Haas effect.

\item Reaching $g_J \mu_B B \leq V_{dd}$ would be very interesting as the dipolar field then overcomes the ambient magnetic field. The (inhomogeneous) dipolar field itself sets the lowest-energy spin-textures. 

\end{itemize}

\subsubsection{Thermodynamics of a Bose gas with free magnetisation}
\label{subsec:thermo}

Thermodynamics of a Bose gas with a spin degree of freedom is derived by including the single-particle magnetic energy in the Bose occupation factor. For non-interacting particles, 
\begin{equation}
    f_{k,m_s}=\frac{1}{\exp  \left[ \left(\epsilon_{k,m_s}- \mu\right)/ k_{\rm B} T\right] -1},
\end{equation}
where $\mu$ is the chemical potential, and $\epsilon_{k,m_s}$ is the single-particle energy of the (trapped) states labeled by the index $k$, in the Zeeman state $m_s$. When magnetisation is free (case of dipolar particles) $\epsilon_{k,m_s}$ includes the linear Zeeman effect, whereas when magnetisation is fixed (as for example of Rb and Na atoms), the linear Zeeman effect is gauged out and $\epsilon_{k,m_s}$  includes only the quadratic Zeeman effect.

\begin{figure}[t]
\centering
\includegraphics[width= 8 cm]{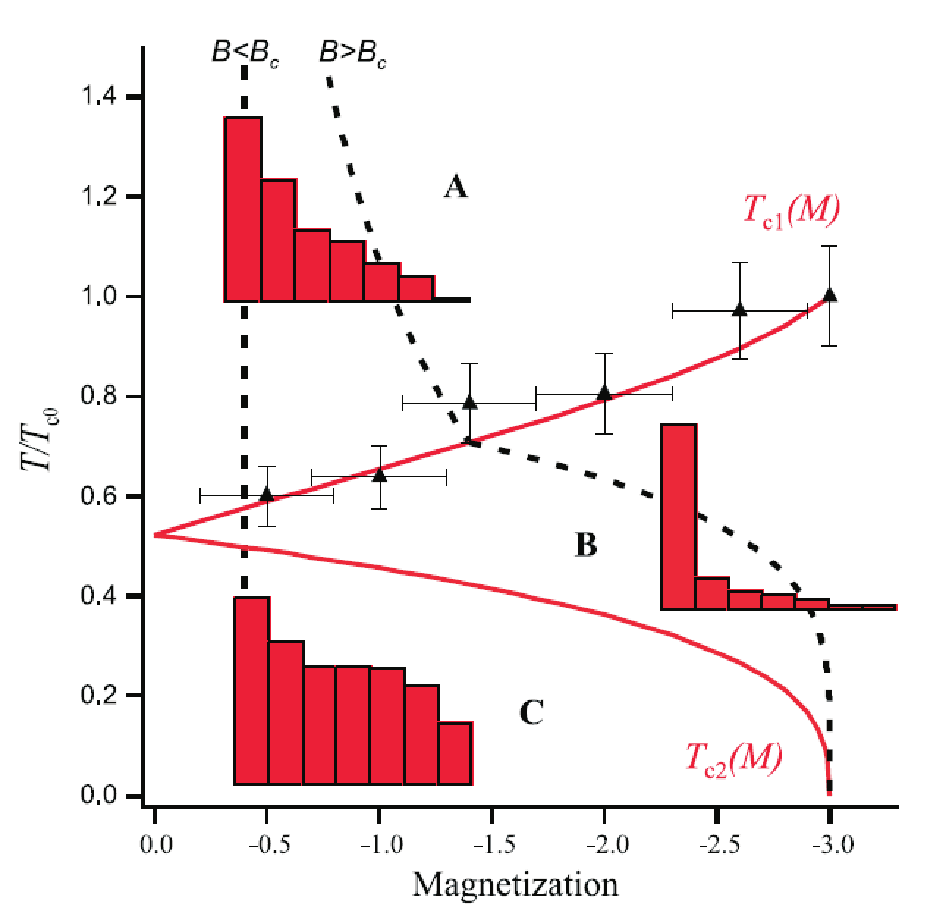}
\caption{Measured and predicted  phase diagram of a spin 3 Cr BEC. Three phases are observed: (A) thermal gas; (B) polarised BEC in stretched state; (C) spin-full (depolarised) BEC. The histograms show typical experimental population distributions. Black triangles are the measured $T_1$. Solid lines shows the predictions of $T_{1,2}$ in the non-interacting case. Dashed lines show typical evolution predicted by theories with no interactions when the temperature is lowered at a given magnetic field $B$, for two different regimes depending on whether magnetisation is fixed (left curve) or free (right curve). Adapted from~\cite{Pasquiou2012toa}.
}
\label{phasediagram}
\end{figure}

For the generic case of a $F=1$ spinor Bose gas interacting solely via contact interactions (and thus at constant magnetisation and without the quadratic Zeeman effect), the phase diagram has been worked out by Isoshima~\textit{et al.}~\cite{Isoshima2000dpt} as a function of the temperature $T$ and the magnetisation $M$. Two phase transitions were predicted. The first transition, at a critical temperature $T_1(M)$ separates the normal phase, and a phase where a condensate forms in the (most populated) stretched state. Below a second critical temperature $T_2(M)$, all other spin components condense simultaneously (see Fig.~\ref{phasediagram}).
 
First investigations of spinor physics with free magnetisation were performed using cold Cr atoms in an optical dipole trap. It was observed that the spin degrees of freedom equilibrated to the gas mechanical temperature~\cite{Pasquiou2012toa}. At high temperature, the population of the different Zeeman states followed a distribution close to the Boltzmann distribution. The spin temperature was equal to the mechanical temperature, and thermal equilibrium between spin and orbital degrees of freedom was ascribed to magnetisation-changing dipolar collisions. Below the critical temperature for BEC, a spontaneous accumulation into the Zeeman state of lowest energy was observed. This led to a bimodal spin distribution, arising from BEC and the spontaneous accumulation of the atoms in the lowest-energy spin state. 

By varying the temperature and the magnetic field, it was possible to map out the 
corresponding critical line $T_1(M)$ (see Fig.~\ref{phasediagram}). However, the second predicted phase transition 
at $T_2(M)$ could not be studied. Indeed, when magnetisation is free, Bose stimulation towards the single-particle lowest-energy state insures that the BEC is only produced in this polarised state~\cite{Pasquiou2012toa}. The second transition was recently observed in the case of sodium atoms with negligible DDI by Frapolli \textit{et al.}~\cite{Frapolli2017sbe}. The authors also studied the impact of spin-dependent interactions on the double-condensation scenario. 

In a later study, it was possible to reach a multicomponent phase with Cr atoms by applying rapid, forced evaporative cooling to a depolarised gas, thus performing the experiment fast compared to the time scale for magnetisation-changing collisions. However, the spinor BECs were very small due an interplay between Bose condensation and spin dynamics~\cite{Naylor2016cbb}. This result pointed out the difficulty of fully thermalising the spin degrees of freedom.  This is a prominent effect to be taken into account for very large spin systems such as Dy and Er where all spin states must be saturated for a stable multicomponent BEC to be produced. 

The connection between magnetic order and superfluid order and the BEC transition is an intriguing question that remains largely unexplored. While these orders are intrinsically connected due to Bose stimulation, it was predicted that strong spin-dependent interactions induce spin ordering at a finite $T$ above the BEC transition~\cite{natu2011pfa}. In low dimensions, the connection between magnetic order and superfluidity promises to be especially interesting. In particular, in the 2D polarised case, superfluidity does not arise from condensation but by the pairing of vortices with opposite circulation through the Berezinskii-Kosterlitz-Thouless mechanism.  This gives rise to a topological order~\cite{berezinskii1972dol,kosterlitz1973omp,Hadzibabic2006bkt,Hadzibabic2006bkt,hadzibabic2009tdb}. The case of a depolarised gas promises new scenarios for superfluidity, due to the existence of topological excitations such as half-quantum vortices that include the spin degree of freedom~\cite{Mukerjee2006tda}.

We note that the case of fermions will drastically differ from the scenario described above due to the presence of the additional energy scale provided by the Fermi energy. One consequence is that spontaneous demagnetisation of a noninteracting Fermi sea is expected provided the Larmor energy is smaller than the Fermi energy (Pauli paramagnetism). How interactions modify this picture is an open question and connected to Stoner instability.

\subsubsection{Spontaneous demagnetisation of a BEC}

One of the fascinating avenues of research for dipolar spinor physics is the investigation of the zero-temperature phase diagram at low magnetic fields when the Larmor energy is comparable to spin-dependent interactions. For the experiments performed with alkalis, spin-dependent contact interactions are extremely small compared to the linear Zeeman effect. In addition, the gas magnetisation remains constant for all practical purposes because contact interactions are isotropic, and anisotropic DDIs between alkali atoms are small. Consequently, the true ground state of these system, with free magnetisation at extremely low magnetic fields, has never been experimentally investigated.

Quantum gases made of strongly magnetic atoms, on the other hand, offer the new possibility of free magnetisation introduced by the DDI. Moreover, spin-dependent contact interactions provide an energy scale that corresponds to a magnetic field in the few 100-$\mu$G range for Cr atoms (compared to 10 $\mu$G for alkalis). It therefore becomes possible to investigate phase transitions driven by the competition between the Zeeman effect and  spin-dependent interactions. For example, the phase diagram of Cr atoms has been calculated (ignoring the DDI) by~\cite{Santos2006scb, Diener2006csc}, and is expected to display a number of transitions separating ferromagnetic, cyclic, polar, and biaxial nematic phases as a function of the magnetic field. The phase diagram of Ln atoms is unknown to the best of our knowledge, though some have speculated~\cite{Lian2012sfn}. We note that the case of Lns may be quite different, because contact interactions themselves contain an anisotropy associated with the large electronic angular momentum in the ground state. 

\begin{figure}[t]
\centering
\includegraphics[width= 8 cm]{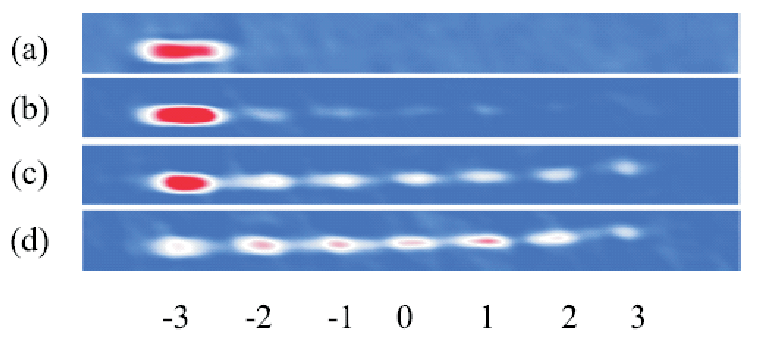}
\caption{Absorption images of a Cr BEC held in a magnetic field of (a) 1mG, (b) 0.5 mG, (c) 0.25 mG, and (d) 0 mG $\pm 100\,\mu$G. These were obtained after Stern-Gerlach separation of the seven Zeeman components. The BEC spontaneously depolarises when the field is lowered below a critical value. Figure adapted from~\cite{Pasquiou2011sdo}.}
\label{demag}
\end{figure}

The exploration of the spinor phase diagram at very low magnetic field was started with a Cr BEC in the $B\sim 100\,\mu$G regime~\cite{Pasquiou2011sdo}. While the BEC remains polarised in the lowest energy single-particle state for large enough magnetic fields, it was then shown that a dipolar BEC spontaneously depolarises when the Zeeman energy is quenched  below a (density-dependent) critical value (see Fig.~\ref{demag}). It was furthermore shown that, while the DDI was presumably too small to significantly modify the observed phases, the dynamics of depolarisation  was driven by dipolar interactions. 

A number of open questions remain after the first results described in \cite{Pasquiou2011sdo}. One interesting question concerns the possibility of characterising the excitations that may be produced when the phase transition is crossed. As magnetisation is dynamically modified, the sample should start rotating, in the spirit of the Einstein--de-Haas effect. Such rotation could not be observed. However, simulations indicate that the threshold for demagnetisation could originate in the resonant dynamics associated with this exchange between spin and orbital momentum~\cite{Swislocki2014rdo}. 

Furthermore, the impact of the DDI on the phases at low field also remains unexplored. Note that it would be especially interesting to study the regime where the Zeeman energy is smaller than the mean-field energy associated with the dipolar field, which could provide a scenario for symmetry breaking and modify the possible quantum phases~\cite{yi2004qpd}. The requirements in terms of magnetic field, below 50 $\mu$G, are demanding, yet within reach of state-of-the-art experiments~\cite{Farolfi2019dac}.

\subsubsection{New cooling methods}

\paragraph{Demagnetisation cooling}\label{demagcooling}

The coupling between the spin degree of freedom and the orbital degree of freedom due to the DDI also presents the possibility of exchanging Zeeman energy and mechanical energy for cooling purposes. This strategy, first suggested in Ref.~\cite{Hensler2005dco}, is based on the depolarisation of a gas of trapped atoms. Similar to adiabatic demagnetisation cooling, the coupling between the internal spin reservoir of the gas and the external kinetic reservoir via dipolar relaxation reduces the temperature of the gas. Although a single dipolar relaxation event per atom is insufficient to significantly cool the sample, it was suggested that optical pumping  can bring  the atoms back into the initial state, cool the spin reservoir, and begin a repeat of the cooling process.  

Demagnetisation cooling of a gas of ultracold $^{52}$Cr atoms was demonstrated soon after by Fattori et al.~\cite{Fattori2006dco}. Demagnetisation was driven by inelastic dipolar collisions, and optical pumping was used to magnetise the system and drive continuous demagnetisation cooling. An increase of the phase space density by one order--of--magnitude was demonstrated, with nearly no atom loss. In Ref.~\cite{Volchkov2014edc}, demagnetisation cooling of a Cr gas was further studied in a deep optical dipole trap, which allowed the exploration of  a large temperature range and could access high densities up to 5.10$^{19}$ m$^{-3}$. An increase of two orders in phase space density was shown, up to 10$^{-2}$. For Cr atoms, inelastic collisions between one ground state atom and one atom optically excited by the repumping laser was shown to be the main process limiting the increase in phase space density. 

Nevertheless, demagnetisation cooling offers a realistic potential for reaching degeneracy by optical cooling only. In particular, dysprosium or erbium atoms are good candidates, due to smaller recoil energy (larger mass), and the existence of narrow lines for optical repumping. In addition, the use of spin-changing collisions for cooling purposes may even be used for non-dipolar particles using spin-exchange contact interactions~\cite{Hamilton2014caz}. 

\paragraph{Purification of a BEC by spin-filtering}

Another closely related cooling scheme, which is efficient below quantum degeneracy, has been demonstrated by Naylor \textit{et al.}~\cite{Naylor2015coa}. This scheme also relies on redistributing population between different spin states, with free magnetisation. The key idea is that only non-condensed atoms may populate spin-excited states since Bose thermodynamics enforce a fully polarised condensate; see Section \ref{subsec:thermo}. Therefore, expelling spin-excited atoms from the trap provides a way to engineer losses specific to non-condensed atoms, thus cooling and purifying the condensate.

The scheme, experimentally demonstrated using $^{52}$Cr atoms, starts with a partially condensed Bose gas polarised in the lowest energy spin state. Demagnetisation of the thermal component is triggered by lowering the magnetic field, such that the Larmor energy is comparable to the thermal energy. Then, the spin-excited thermal components produced by magnetisation-changing collisions are filtered out by a magnetic field gradient. It was found that when the initial BEC fraction is high enough, this scheme ends up with a polarised BEC in the lowest energy state with an increased BEC fraction. This provides a thermodynamic cycle  that, in principle, decreases the entropy by a factor of up to $(2s+1)^{3/4}$ (where $s$ is the atomic spin), and could also be in principle repeated. The obtained reduction in entropy was typically a factor of 2 for one cycle in the Cr experiment.

In the experiment, technical limitations arose from the difficulty to control the magnetic field at very low values. This limitation is directly related to the fact that dipolar gases are sensitive to the linear Zeeman effect. It was suggested in Ref.~\cite{Naylor2015coa} that the purification of a BEC by spin-filtering could be extended to non-dipolar species by using spin-dependent contact interactions and spin-exchange interactions. Both the dipolar and the non-dipolar cooling schemes were theoretically explored in~\cite{swislocki2021sdc}. A related cooling scheme was demonstrated with Rb atoms~\cite{Olf2014tac}.

Note that, despite their highly magnetic character, such cooling techniques could be difficult to apply to Lns, because the Feshbach spectrum of these atoms might be too dense at low field, below typically 10 mG. Up to now, these techniques haven not yet been reported using Ln species.

\begin{figure}[t]
\centering
\includegraphics[width= 8 cm]{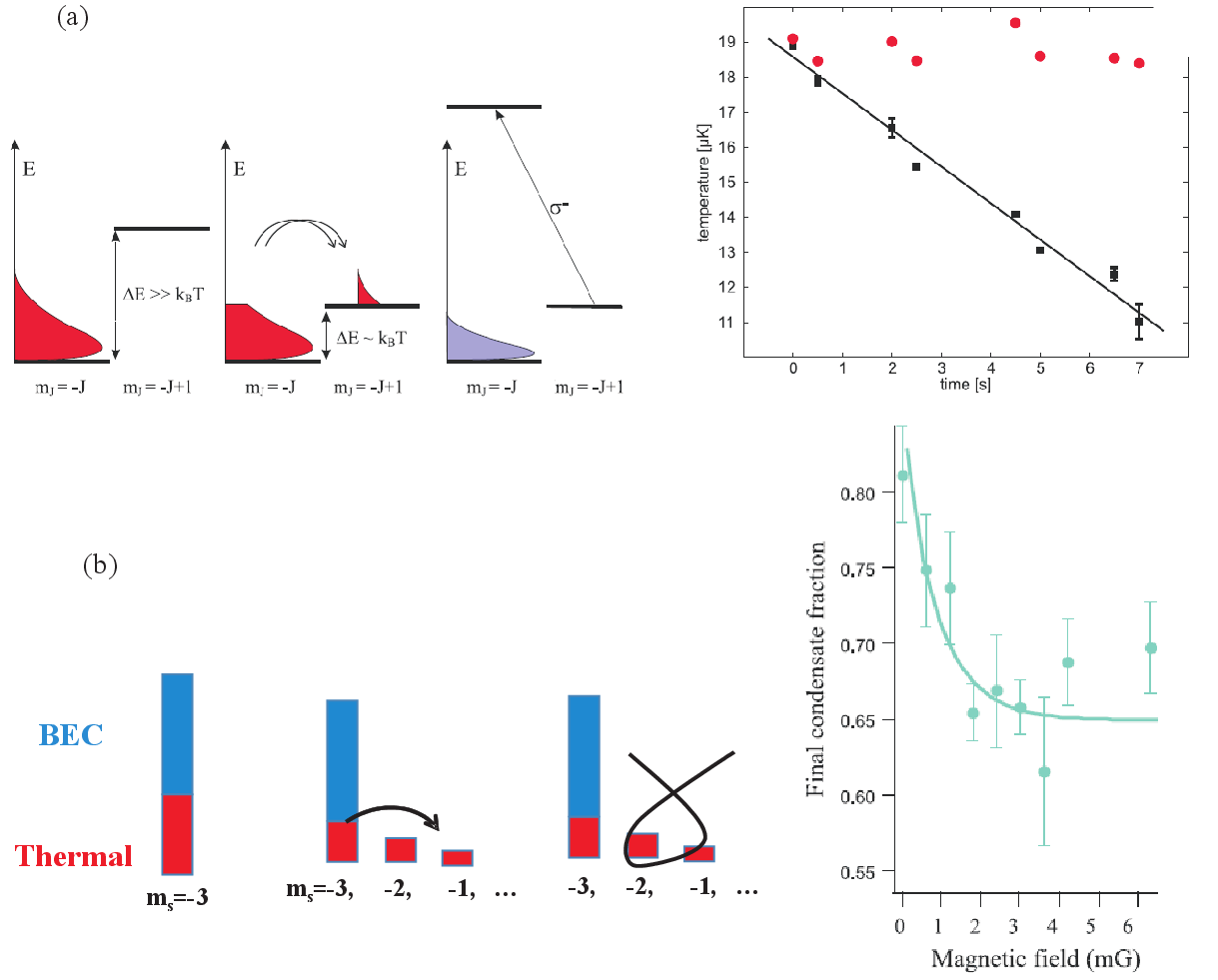}
\caption{Comparison of demagnetisation cooling and spin filtering. (a) Demagnetisation cooling. Spin-changing collisions transfer kinetic energy of the most energetic atoms to magnetic energy, allowing a net cooling of the remaining atoms. Optical pumping can bring back spin-excited atoms to the initial spin state, enabling continuous cooling with no atom loss (graph). (b) Spin filtering. In presence of a condensed Bose gas, spin-changing collisions  affect only thermal atoms that can then be selectively removed. This removal of entropy results in the cooling of the remaining gas and an increase of the condensed fraction (graph).}
\label{cooling}
\end{figure}
 
Both approaches described here are ways to use the spin degrees of freedom to efficiently store and remove entropy from a gas. We point out that ideas using the spin degrees of freedom have already been developed in the context of lattice-based large-spin Fermi gases~\cite{Taie2012asm}, mixtures of Bose gases~\cite{Medley2011sgd}, and Bose and Fermi gases~\cite{Vincent2018dco}. Alternatively, it has also been suggested that the spin degrees of freedom, and/or spin-changing collisions, may be employed for thermometry purposes down to extremely low temperatures.  Such temperatures are typically impossible to measure in BECs with usual thermometric techniques~\cite{Olf2014tac,Mele-Messeguer2013tsf2}. 



\subsection{Light-induced spin-dependent Hamiltonians in dipolar gases}\label{sec:lightspincoupling}

Apart from their large DDI, the large spin and rich electronic properties of highly magnetic atoms open interesting directions to explore spin-light coupling; see Secs.~\ref{subsec:Lnspectrum},\ref{subsec:atomlightLn}. 
The wide variety of transition linewidths allow spin manipulation with reduced heating and the strong vector and tensor parts of the atomic polarisability allow the strong coupling of the electron's angular momentum to light; as detailed in Sec.\,\ref{subsec:atomlightLn}.  These higher-rank polarisabilities  are intrinsically stronger in magnetic atoms due to their larger $L$ and $S$ values. In addition, in the case of Lns, $LS$ coupling occurs in the electronic ground state itself. This is in contrast to the case of alkali atoms for which spin-dependent coupling to light occurs only through the mediation of the excited electronic levels.  This yields a different scaling of the vector and tensor polarisabilities with the detuning of the laser to the atomic transition, $\Delta_a$: scaling as $\propto \Delta_a^{-1}$ and $\propto\Delta_a^{-2}$ in Lns and alkali-metals, respectively~\cite{Geremia2006tpa,Cui2013sgf}. This makes Lns promising candidates with which to implement spin-dependent Hamiltonian tailored by light. Indeed, even if such synthetic spin-dependent Hamiltonian have been realised with alkali-atoms, they have been subjected to strong heating associated with incoherent light scattering.  As a result, some of the most interesting regimes could not be reached with long-lived gases.

\subsubsection{Synthetic spin-orbit coupling in gases of magnetic atoms}\label{SecSOC}

\myparagraph{Synthetic spin-orbit coupling and artificial gauge fields.}

The coupling of a particle's momentum and spin underlies many important phenomena in quantum systems, and in particular in electronic, solid-state systems~\cite{Soumyanarayanan2016epi}.  For example, topologically nontrivial materials often require the coupling of the electron spin to momentum~\cite{Hasan2010cti,Qi2011tia}. Here, SOC typically arises from the movement of electrons through the crystal electric field.  The effect of a magnetic field on charged particles also yields intriguing phenomena, the most famous of them being the quantum Hall effect~\cite{Klitzing1980nmf}. This effect has its roots in the special form of the Lorentz force, which, at a quantum level, yields a kinetic term of the form:
  \be\label{eq:SOCHk}
\hat{H}_k = \frac{[\hat{{\bf p}}-\hat{{\bf \mathcal{A}}}(\hat{\bf r})]^2}{2m}, 
\ee
where $\hat{{\bf p}}$ is the canonical momentum and $\hat{{\bf \mathcal{A}}}(\hat{\bf r})$ is the vector potential or gauge potential. 
Even more exotic behaviour, such as the fractional quantum Hall states~\cite{Tsui1982tdm}, Laughlin liquids~\cite{Laughlin1983loe}, and topological superconductors harbouring Majorana excitations~\cite{Alicea2012ndi} can arise if
$\hat{{\bf \mathcal{A}}}(\hat{\bf r})$ is not just a classical vector (i.e., three complex numbers), but a set of three operators $\hat{A}_x(\hat{\bf r})$, $\hat{A}_y(\hat{\bf r})$, and $\hat{A}_z(\hat{\bf r})$ that do not commute. This yields a non-Abelian field.

Interestingly, ultracold atoms enable the combination of SOC and gauge potentials by the engineering of a realisation of Eq.~\eqref{eq:SOCHk}. This mimics the effect of a Lorentz force on neutral matter while, additionally, the set of operators $\hat{A}_i$ connect to the atomic spin operators. If realised in high dimensions (i.e., greater than one), non-Abelian gauge fields may then be synthesised. This technique relies on laser-field dressing~\cite{Dalibard2016itt}. In the following, we will review this SOC protocol and discuss its extension to magnetic atoms. We note that several other strategies, not connected to SOC, have been developed over time to realise artificial gauge fields on neutral atoms, e.g., by engineering trap rotations~\cite{Fetter2009rtb}. 

As introduced in Sec.~\ref{subsec:processes}, and reviewed in detail in the above Sec.~\ref{freemag}, the DDI provides \textit{intrinsic} SOC through the dipolar relaxation terms of Eq.~\eqref{relaxationterms}. However, this is not generally the sort of SOC that results in the physics of artificial gauge fields mentioned above. It gives rise to an interaction term that does not conform to that of Eq.~\eqref{eq:SOCHk}. (See Ref.~\cite{Peter2013ddf} for a method that does.)  
In contrast, light-induced SOC schemes applied to ultracold quantum gases do realise a Hamiltonian term of the form of Eq.~\eqref{eq:SOCHk}. The basic idea can be formulated as follows: Light fields are applied to couple Zeeman sublevels via a two-photon Raman transition. Recoil momentum is transferred as the optically coupled spin flips. The adiabatic evolution of these dressed states as the atom moves in the Raman field creates a synthetic gauge field~\cite{Dum1996gsi}.  This field provides SOC when these states form a (near) degenerate manifold.  The general form of the SOC Hamiltonian under the 1D Raman coupling along the direction $\hat{x}$ of two Zeeman ground states is  
\be\label{SOCHam}
\hat{H}({\bf p}) = \frac{[{\bf p}-\hbar{\bf k}\hat{\sigma}_z]^2}{2m} +  \frac{\hbar\Omega}{2}\hat{\sigma}_x + \frac{\hbar\Delta}{2}\hat{\sigma}_z, 
\ee
where $\hat{\sigma}_{x,y,z}$ are the Pauli spin matrices. $\Omega$ and $\Delta$  are the  two-photon Raman  Rabi frequency and detuning, respectively. The form of this coupling is the sum of Rashba ($k_z\sigma_z+k_y\sigma_y$) and Dresselhaus ($k_z\sigma_z-k_y\sigma_y$) SOC with equal weights.  The resulting dispersion relation is of the form of a double well centred at $k=0$, as is typical of SOC systems. That is, the momentum of spin up atoms is oriented in the opposite direction from that of spin down atoms. In the 1D case described by Eq.~\eqref{SOCHam},  $\hat{A}_z = \hbar{\bf k}\hat{\sigma}_z$ and $\hat{A}_{x,y}=0$, which does not describe a non-Abelian coupling because the single component of $\hat{A}$ commutes with itself. Generalisations to 2D and 3D with a truly non-Abelian gauge potential have been proposed~\cite{Dalibard2016itt} and realised in the case of 2D SOC~\cite{Wu2016rot,Huang2016ero}.
 
\myparagraph{Spin-orbit-coupling in quantum gases: limitations and prospects}
Spin-orbit coupling in BECs~\cite{Lin2011soc,Galitski2013soc} and  DFGs~\cite{Cheuk2012sis,Wang2012soc,Huang2016ero} of alkali-metals have been achieved. However, heating due to spontaneous emission is severe. This arises because Raman coupling is not very efficient compared to spontaneous light emission due to the generally weaker tensor versus scalar polarizability. The optimal ratio of Raman coupling to spontaneous emission occurs at a Raman-laser detuning approximately equal to the fine-structure splitting~\cite{Spielman2009pra}: alkali's small fine structure splitting implies a large spontaneous emission rate if sufficiently strong coupling strengths are to be achieved. 
The heating from spontaneous emission leads to loss of quantum degeneracy and atomic population and the short lifetimes severely hamper the  study of quantum many-body phenomena.  

We note that lattices of fermionic alkaline-earth atoms~\cite{Kolkowitz2017soc,Wall2016clsso,Mancini2015ooc} have also been used in SOC experiments.   However, optical lattice confinement and inelastic collisions~\cite{Yamaguchi2008ici,Traverso2009iae,Uetake2012sdc} among atoms limit the future ability to explore a wide variety of many-body phenomena.

By contrast, Ref,~\cite{Cui2013sgf} suggested that fermionic open-shell Ln atoms like Dy and Er might better serve due to their large orbital angular momentum and narrow-line transitions. Spontaneous emission can be eliminated while still producing large Raman coupling even without lattice confinement or narrow lines because their ground-state orbital angular momentum is $L>0$.  This is due to the fact that the vector and tensor polarizabilities that factor into the Raman coupling scale as the inverse atomic detuning $\Delta_a^{-1}$ of the Raman lasers from the atomic transitions in these systems, as opposed to the faster  $\Delta_a^{-2}$ scaling in alkali-metals once $\Delta_a$ exceeds the fine-structure splitting~\cite{Deutsch1998qsc,Geremia2006tpa}.   Thus, in open-shell Lns, one can always choose a detuning that provides a large Raman coupling $\Omega$ while minimising heating from incoherent scattering~\cite{Cui2013sgf}. The realisation of long-lived SOC Fermi gases would open new avenues to experimentally study topological matter not easily realisable in the solid state~\cite{Galitski2013soc,Zhang2014fgw,Goldman2014lig,Zhai2015dgs,Dalibard2016itt}. For example,  one could create and study topological superfluids and exotic quantum liquids~\cite{Galitski2013soc,Sato2009nat,Jiang2011mfi,Zhu2011pna,Alicea2012ndi,Ruhman2015tsi,Nascimbene2013rod,Celi2014sgf} in a well-controlled manner.  

\myparagraph{Spin-orbit-coupling in assemblies of magnetic atoms}

Burdick~\textit{et al.}~\cite{Burdick2016lls} first reported the realisation of SOC in degenerate Fermi gases of Ln atoms using Dy. The SOC was induced using Raman light near the 741-nm transition. Its 1.8-kHz width reduces spontaneous emission rates below the background lifetime of the gas with only modest $\sim$GHz detuning, which is large compared to the hyperfine splittings, but smaller than the fine-structure; see Refs.~\cite{Lu2011soa,Kao2017ado}. The lifetime of SOC gases was then limited, not by  spontaneous emission, but rather by dipolar relaxation.

\begin{figure}[t!]
 \includegraphics[width=0.95\columnwidth]{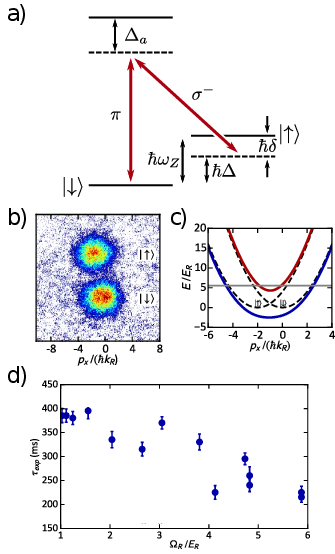}
  \caption{Spin-orbit-coupling of fermionic $^{161}$Dy. (a)  Coupling scheme for $^{161}$Dy at a magnetic field high enough that   the lowest two Zeeman states are isolated by the quadratic Zeeman shift. $\delta$ is the two-photon Raman detuning and $\omega_Z$ is the Zeeman splitting.  (b)  Momentum distribution after  removing the coupling and separating the spin states with a magnetic field gradient.  (c) Quasimomentum dispersion curve for the spin-orbit-coupled cloud in (b). The solid horizontal line indicates the Fermi energy.    (d)  Experimental lifetimes for a gas of $1\times10^{4}$ $^{161}$Dy atoms with $T/T_{F}=0.4$.   Adapted from Ref.~\cite{Burdick2016lls}.}
  \label{socfig}
\end{figure}

Due to the suppression of inelastic dipolar scattering (see Sec.~\ref{fsuppress} for details), relatively long-lived SOC Fermi gases could be realised while working at a bias field of a few tens of G. Such magnetic fields were required,  not only because the suppression effect scales as $\sqrt{B}$, but also because the isolation of an effective pseudospin $1/2$ requires a large the quadratic Zeeman shift obtainable only at large $B$. However, as discussed in Sec.~\ref{subsec:fesbachsec}, the extreme density of FRs in fermionic Lns, increasing with $B$, complicates the situation, and Ref.~\cite{Burdick2016lls} even reported the overlapping of resonances at these fields.  Nevertheless, a field region  at 33.846(5)~G was found that allowed fermionic spin mixtures to live enough to allow spin-orbit coupled $^{161}$Dy gases to be created with a lifetime of $\sim$400~ms at a Raman coupling strength of $\hbar\Omega = 1E_R$; see Fig.~\ref{socfig}(b). At this field, quadratic Zeeman shifts ensures that the Raman fields couple only the $m_F = -21/2$ and $m_F = -19/2$ states, providing maximum fermionic suppression of dipolar relaxation; see Fig.~\ref{socfig}(a). Note that by contrast, the lifetimes of SOC $^{40}$K and $^{6}$Li  were lower by $\sim$10 and 100, respectively~\cite{Cheuk2012sis,Wang2012soc,Huang2016ero}.  Moreover, the SOC $^{161}$Dy lifetime is similar to that of free-space bosonic SOC alkali gases~\cite{Lin2011soc} and $\sim$10$\times$ longer than that achieved in a bosonic lattice system~\cite{Li2016soc}. 

Reference~\cite{Burdick2016lls} also reported that Raman-coupled bosonic Dy has a  short lifetime of less than 10 ms at low $B$ field. This is just as short as $^{6}$Li, demonstrating the importance of fermionic statistics in preventing fast relaxation. Finally, the effect of dipolar interactions on Rabi oscillations was observable, highlighting the interacting dipolar character of the fermionic SOC system.

\myparagraph{Dipolar relaxation of Raman-dressed spins}

The physics of dipolar relaxation of  Raman-dressed spins differs from the scenario of equilibrium gases described in Sec.~\ref{Drelaxationsec}.   While one might suppose that atoms colliding in dressed spins are identical---the spins are just rotated on the Bloch sphere---and should therefore enjoy the full suppression of identical fermions (see Sec.~\ref{fsuppress}), this is not the case for SOC fermions.  Since spin is locked to momentum, atoms colliding with different momenta along the 1D SOC axis no longer behave as identical particles (they have different spin states)~\cite{Burdick2016lls}. 

Although dipolar relaxation along the 1D-coupling axis  proceeds unhindered, not all is lost: dipolar relaxation remains suppressed for collisions with momenta aligned primarily along axes transverse to the SOC coupling.  A simulation of these effects shows that in terms of relaxation rate, the gas neither behaves as identical fermions enjoying full dipolar relaxation suppression, nor as distinguishable particles, but roughly halfway in between~\cite{Burdick2016lls}.  

Dipolar relaxation remains a nuisance when coupling more than two spin states together, since such spin states are not protected. In Ref.~\cite{Burdick2015fso}, the lifetime of SOC Dy gases at low fields, around 1 G, were observed to be of around 10~ms, for bosons as for fermions.  Thus, SOC in dipolar gases are limited to high fields, setting limitations on the realisation of certain exotic many-body states~\cite{Deng2012soc,Gopalakrishnan2013qqo,Cui2013sgf,Lian2014sms}. 

\myparagraph{Realisation of a synthetic quantum Hall system using magnetic atoms}

Reference~\cite{Chalopin2020pce} reported on the realisation of the equivalent of a quantum Hall system, i.e., that dictated by Eq.~\eqref{eq:SOCHk}, with $\hat{A}_x(\hat{\bf r})=eB\hat{y}$, $\hat{A}_y(\hat{\bf r})=0$, using Ln atoms. 
This achievement relied on SOC in one spatial dimension ($x$), using light near the 626-nm transition, and the coupling all the spin states of bosonic Dy, where  the large electronic spin $J=8$ of Dy was interpreted as a synthetic dimension ($y$), i.e., $ \hat{J}_z \leftrightarrow \hat{y}$. The SOC Hamiltonian, similar to \eqref{SOCHam}, can then be interpreted as the Hall equivalent formulated above~\cite{Celi2014sgf}. Similar work has been performed using non-magnetic atoms, yet bulk properties remained elusive due to the narrowness of the synthetic dimension used~\cite{Stuhl2015ves,Mancini2015ooc,Kolkowitz2017soc,An2017doc}

In this work, Chalopin \textit{et al.}~demonstrated that their system shows distinct sectors with respectively bulk and edge behaviours in the ground band. This demonstrates the relevance of large spin Lns systems to realise quantum many-body systems with non-trivial topology. The observed behaviour in each sector of the ground band are reminiscent of Landau-level physics: In the bulk, dynamics is suppressed due to the flatness of the band. In contrast, the edge modes exhibit a ballistic dispersion, yet with a motion allowed in one direction only, i.e., showing chirality. The authors also characterised the elementary excitations above the ground band and directly observe cyclotrons orbits, in the bulk, and skipping orbits at the edges. The cyclotron gap is roughly constant over the bulk sector and the corresponding frequency is large ($\approx 40$KHz) compared to both the decay rate (see above) and the temperature scale. Finally, the Hall response of the ground band was measured, by applying a potential difference in the synthetic dimension. The response shows a characteristically nearly quantised behaviour, with a Chern number of nearly one in bulk and a vanishing mobility in the edge sectors.  This is suggests the topological protection of the edge states.  While the limited lifetime of the bosonic system due to dipolar relaxation may hamper the exploration of some many-body physics, this results helps to establish Ln SOC systems as promising platforms for the realisation of synthetic topological quantum many-body states.   

\subsubsection{Synthetic spin-dependent interactions in gases of magnetic atoms and generation of entangled quantum states}\label{SecSC}

In recent works using ultracold bosonic gases of $^{162}$Dy and $^{164}$Dy, a spin-dependent Hamiltonian was generated using  laser light close to the 626-nm atomic line~\cite{Chalopin2018qes,Evrard2019ems}. The laser field has a linear polarisation along $x$ while the  magnetic field points along $z$. The resulting atom-light interaction yields an effective coupling term $\hbar\omega \hat{J}_x^2$, with $\hat{J}$ being the total atomic angular momentum. This scheme, similar to that previously investigated with a large ensemble of room temperature Cs atoms~\cite{Fernholz2008sso}, realises a so-called \textit{one-axis twisting} Hamiltonian~\cite{Kitagawa1993sss}, which has also been investigated in many other systems~\cite{Leroux2010ode,Bohnet2016qsd,Molmer1999meo,Hosten2016qpm}. 
Typical coupling strengths $\omega$ are on the MHz scale and can be made to greatly exceed the kHz-scale Larmor precession frequency as well as the dipolar relaxation rate. 

Highly non-classical spin states could be generated under the effect of this coupling term~\cite{Kitagawa1993sss}. Reference~\cite{Chalopin2018qes} experimentally demonstrates the creation of Schr\"{o}dinger cat states resulting from the coherent superposition of the two stretched spin states $|\pm J\rangle$. In contrast to ensembles of $s=1/2$ particles, for which entanglement can only be generated between different atoms of the ensemble, here the non-classical states arise because of the entanglement between electrons within each individual atom~\cite{Fernholz2008sso}. This possibility is a key  feature provided by large-spin systems. These states appear in the time evolution of the atomic spin (at the single particle level) after a quarter of coupling period, $T=2\pi/\omega$. Furthermore, several revivals of the cat state spaced by $T/2$ as well as a repolarisation into the Fock states $|J\rangle$ and $|- J\rangle$ at intermediate times (half integer multiples of $T$) occur during the time evolution.
Coherence is seen to decay under the effect of classical $B$-field noise, to which the cat state is more sensitive than a classical (coherent) state. Nevertheless, the cat state's lifetime reaches up $60\mu{\rm s}$. 
Focusing on a similar dynamics but shorter time scales (set by $\tau=1/\left(\sqrt{2J}\omega\right)$), Ref.~\cite{Evrard2019ems} probes the formation of non-Gaussian ``oversqueezed'' states.

%% file: Section7/7_lattice_reorder.tex
\section{Magnetic atoms in strongly confined geometries: low-dimensional gases and lattice systems}
\label{sec:lattice}

Up to now, we have discussed bulk dipolar effects, i.e., that which happens in assemblies of atoms free to move and collide in space, both in the fully polarised case (Secs. \ref{Sec:RepulsiveGases} and \ref{Sec:DCQD}) and considering additionally the spin degree of freedom (see Sec.~\ref{spinorsection}). A complementary trend has developed in the community, loading atoms in strongly confined geometries, either using tight anisotropic traps or periodic potentials made by light standing waves (so called optical lattices)~\cite{Jaksch1998cba,Bloch2005uqg,Bloch2008mbp, Bloch2012qsw,Lewenstein2007uag}. This has proven to be a new platform to achieve the strongly interacting regime and investigate interesting open questions in quantum many-body physics~\cite{Bloch2005uqg,Bloch2008qca,Bloch2008mbp,Esslinger2010fhp,Lewenstein2007uag,Bloch2012qsw,Gross2017qsw}. 
In this section, we discuss the new physics arising from loading highly magnetic atoms in strongly confined geometries, focusing on the experimental achievements to date. In a first part we discuss the case of dipolar gases effectively constrained to a one-dimensional (1D) space, using a two dimensional optical lattice. In this case, the atomic motion remain free in the direction transverse to the lattice, realising an array of 1D gases. In a second part (Sec.~\ref{subsec:spinless_lattice}), we discuss the new physics brought by the long-range and anisotropic dipolar interactions in a spin-polarised sample in  3D periodic potentials. In this case, the atomic motion occurs in direction in which the external potential is a lattice, realising Hubbard-like models. In the third part (Sec.~\ref{subsec:spinlattice}), we still consider 3D periodic potential but additional allow for an internal spin degree of freedom, relating to the very broad topic of quantum magnetism.

\subsection{One-dimensional dipolar gases}\label{Sec1D}

By using tight traps in one or two directions of space, the motion of ultracold atomic gases is constrained, reducing  the dimensionality of the system. Such systems have been  studied in contact-interacting gases~\cite{Giamarchi2003qpi,hadzibabic2009tdb}. In this section, we focus on 1D gases of magnetic atoms and describe the new behaviour resulting from strong DDIs.

The physics of interacting quantum particles in dimensions higher than 1D is often reducible to an effective description of non-interacting quasiparticles composed of the original particles ``dressed'' by their interactions.  The excitations of such a system are nearly single-particle-like.  That is, as we have seen in Sec.~\ref{subsec:bose}, MF and Bogoliubov treatments are often adequate for bosons, while high-dimensional fermionic systems are accurately described by the celebrated MF Fermi liquid theory of Landau~\cite{Nozieres1997toi,Landau1957tto}; see Sec.~\ref{subsec:fermi}. 

Many-body quantum physics in 1D is far more strange.  Particles, being constrained to move on a line, cannot avoid each other, and so all excitations are collective in nature.  Moreover, quantum fluctuations play a large role in determining how the system may organise; e.g., strong fluctuations prevent the establishment of long-range order~\cite{Mermin1966aof}. Fermi liquid theory is no longer applicable in 1D. Instead, Tomonaga-Luttinger liquid theory captures the low-energy physics of both interacting fermions and bosons~\cite{Giamarchi2003qpi}. For example, it describes the strange phenomenon of the spin-charge separation of excitations:  excitations fractionalise into spinons that carry spin but no charge, and holons that carry charge without spin~\cite{Giamarchi2003qpi}.  

The role of quantum statistics also changes in 1D. Because particles can no longer exchange their positions without undergoing a collision, strongly interacting fermions behave like bosons and vice-versa, at least for some local observables like the density distribution.  Indeed, the Lieb-Liniger model for contact interacting bosons,
 \begin{equation}\label{LLM}
 H_{LLM} = -\sum_\text{atoms}\frac{\hbar^2}{2m}\frac{\partial^2}{\partial x^2} + \sum_\text{pairs}g_\text{1D}\delta(x),
  \end{equation}
describes how the physics of bosons map onto that of free fermions in the limit of infinitely strong interactions ($g_\text{1D} \rightarrow \infty$)~\cite{Cazalilla2011odb}.  The coupling strength $g_\text{1D}$ is approximately related to the 3D scattering length $\as$ through  $g_\text{1D}= 2\hbar^2a/ma_\perp^2$, where   $a_\perp = \sqrt{\hbar/m\omega_\perp}$  and $\hbar\omega_\perp$ is the excitation energy of the tightly confined directions perpendicular to the 1D axis~\cite{Olshanii1998asi}.  For the system to be considered within the 1D regime, $\hbar\omega_\perp$ should be much larger than the chemical potential of the gas $\mu$.

The bosonic wavefunction describing the system in the $g_\text{1D} \rightarrow \infty$ limit, referred to as a Tonks-Girardeau (TG) gas, is equal to the absolute value of a fermionic wavefunction~\cite{Yurovsky2008cca,Cazalilla2011odb}.  The `fermionised'  bosons repel each other strongly enough that their two-body correlation function $g^{(2)}(0)$ vanishes just like that of identical fermions---repulsion in 1D mimics the effect of the Pauli exclusion principle.  The TG regime is reached when the Lieb-Liniger parameter, 
\begin{equation}\label{TGparam}
\gamma\equiv mg_\text{1D} / n_\text{1D}\hbar^2, 
\end{equation}
is much greater than one; $n_\text{1D}$ is the 1D density.  Intuitively, $\gamma$  compares the mean interparticle distance $1/n_\text{1D}$ to the length scale  $\hbar^2/mg_\text{1D}$ over which the repulsive interaction `bends' the wavefunction. In this high-$\gamma$ regime, the density-density correlation $g^{(2)}(x=0)\rightarrow 0$, i.e., the quasiparticles antibunch, as expected from their fermion-like character.  

There exist even more exotic states, such as the super-Tonks-Girardeau (sTG) gas. These highly excited states are characterised by stronger-than-ideal Fermi gas correlations. Such gases can be accessed by quenching $g_\text{1D}$ from $+\infty$ to $-\infty$, as has been observed in Cs and  Dy~\cite{Astrakharchik2004qod,Astrakharchik2005btt,Batchelor2005eft,Haller2009roa,Chen2010tfa,Kao2020tpo}; see Sec.~\ref{subsec:sTG}.

Many aspects of 1D fermionic systems have been explored in the condensed matter setting~\cite{Giamarchi2012set}, and bosonic 1D gases have been realised and studied both near and below the TG limit using ultracold atomic gases confined in arrays of 1D optical potentials using 2D optical lattices or in magnetostatic potentials from atom chips~\cite{Paredes2004tgg,Kinoshita2004ooa,Yurovsky2008cca,Hofferberth2008pqa,Cazalilla2011odb,Langen2015uao,DAlessio2015fqc,Langen2016pau}.   Several characteristic properties have been observed, including antibunching~\cite{Tolra2004oor,Kinoshita2005lpc}; unusual transport~\cite{Fertig2005sit}; quantum integrability (i.e., the existence of an extensive number of integrals of motion)~\cite{Kinoshita2006aqn}; long-lived metastability of a sTG gas~\cite{Haller2009roa}; pinning quantum phase transitions~\cite{Haller2010pqp};  unusual excitation spectra~\cite{Fabbri2015dsf,Meinert2015pte};  velocity of sound~\cite{Yang2017qca}; and rapidity (i.e., quasiparticle momentum) distributions~\cite{Wilson2020ood}. One-dimensional Fermi gases have also been created with excitation spectra consistent with Luttinger liquid theory~\cite{Pagano2014aod,Yang2018mot}.

Dipolar interactions enrich the physics accessible with 1D gases. Theoretical results have primarily focused  on purely dipolar quantum gases, in which the only interaction comes from the DDI~\cite{Astrakharchik2008stg,Arkhipov2005gsp,Citro2007eol,Pedri2008ceo,Deuretzbacher2010gsp,Deuretzbacher2013egs,Girardeau2012stg}. Various effects of the DDI have been predicted, in particular including increased correlations of akin to sTG gases, or even crystallisation in the ground state.

\subsubsection{The effective quasi-1D DDI and realisation of 1D dipolar gases}

\begin{figure}[t!] \vspace{-6mm}
\includegraphics[width=0.95\columnwidth]{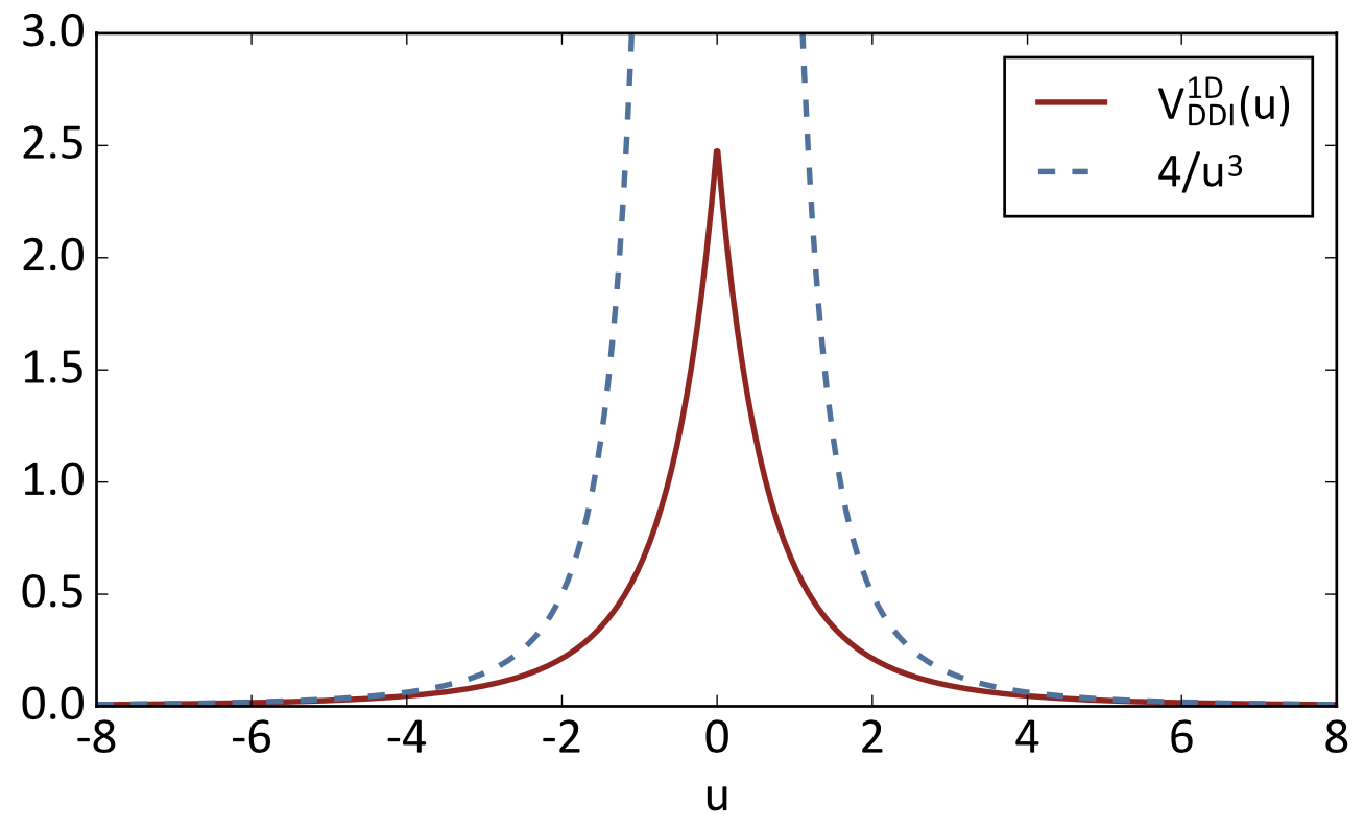}\vspace{-3mm}
\vspace{-3mm}
\caption{Red line is the 1D DDI, while dashed line is the 3D DDI.  The 1D DDI approaches the 3D curve at large normalised distance $u=x/a_\perp$.   Unlike the 3D divergence at short range, the 1D DDI assumes a finite value. Figure reproduced from Ref.~\cite{Tang2018tni}.}
\label{1DDDI}
\end{figure}

In a quasi-1D geometry, the external motion of the atoms is effectively frozen in two directions when $\hbar\omega_\perp \gg k_{\rm B}, \mu$. The effective form of the DDI in quasi-1D geometries can be derived from the 3D expression, Eq.~\eqref{vdd}, by integrating out the transverse degrees of freedom, as shown in Refs.~\cite{Sinha2007cdg,Deuretzbacher2010gsp,Deuretzbacher2013egs}. Under a single-mode approximation, this yields 
 \begin{equation}
       U^\text{1D}(u) = V(\theta)\left[ {V}^\text{1D}(u) - \frac{8}{3} \delta (u) \right],
   \label{ddi_1d} 
   \end{equation}
where
 \begin{equation}
  V(\theta) = \frac{\mu_0 \mu^2}{4\pi} \frac{1-3\cos^2{\theta}}{4 a_\perp^3},
   \end{equation}
   and
   \begin{equation}\label{vtilda}
       {V}^\text{1D}(u) = -2|u| + \sqrt{2\pi} (1+u^2) e^{u^2/2} \text{erfc}(|u|/\sqrt{2}).
   \end{equation}
Here, $u=x/a_\perp$, and  $\text{erfc}(u)$ is the complementary error function.    There are two contributions to the short-range part of the 1D DDI.  The first is the $\delta$-function term.  It comes from the point limit of an extended dipole~\cite{Griffiths1998hsi} and has an opposite sign to ${V}^\text{1D}(u)$.  The second contribution is an effective delta-function term that arises from the fact that ${V}^\text{1D}(u)$ becomes more sharply peaked as  $a_\perp$ shrinks~\cite{Deuretzbacher2010gsp,Deuretzbacher2013egs}.  At distances $|x|\gg a_\perp$, ${V}^\text{1D}(u) \rightarrow 4/|u|^3$.  This $x^{-3}$ long-range potential is just like the DDI in 3D, but with diminished magnitude at ranges on the order of $a_\perp$. Within a distance set by $a_\perp$, the DDI ceases to seem 1D to atoms that approach near each other. The underlying 3D spatial nature is manifest and a small attractive (repulsive) contribution to the DDI emerges from the  part of their wavefunctions that extend transversely by $a_\perp$, even if the long-range  interaction along the 1D axis is  repulsive (attractive)~\cite{Deuretzbacher2010gsp,Deuretzbacher2013egs}.

The DDI in 1D thus has  both long and short-range components~\cite{Astrakharchik2008stg,Deuretzbacher2010gsp,Deuretzbacher2013egs}.  The chemical potential remains intensive in 1D, which is indicative of a short-range interacting system. However, like a long-range interacting system, there is no asymptotic phase shift (scattering length) that can be defined for two-body collisions in 1D~\cite{Astrakharchik2008stg}.  Away from collisional resonances, the short-range part of the DDI can add to the van der Waals contact pseudopotential  to yield a total hard-core contact interaction strength $g_\text{1D}$, as considered in Refs.~\cite{Tang2017tni,Palo2020vba}.  In addition to the FR-tunability of the van der Waals contact interaction (discussed in previous sections), the 1D DDI provides wide tunability in the properties of 1D many-body systems because both the short and long-range parts of the DDI in 1D can be set to a positive or negative sign, or made to completely vanish at the magic angle  $\theta_m = 54.7^\circ$.

Dipolar 1D gases have been created by loading BECs of either $^{52}$Cr or $^{162}$Dy into 2D optical lattices creating arrays of 1D tubes~\cite{Pasquiou2011sra,Tang2017tni}.  In this geometry, the DDI affects different aspects of the interactions: 1) the dependence of $g_\text{1D}(\theta)$ on the short-range component of the 1D DDI; 2) the intratube long-range DDI; and 3) the mutual long-range DDI of atoms in nearby tubes.  The intertube DDI does not change the dimensionality of the system, but rather is a multichannel effect that couples different flavors of 1D quasiparticles~\cite{Tang2017tni}; i.e., coupled 1D systems remain 1D because the phase space degrees of freedom are still 1D when there is no transport between tubes.   

The first experimental work with 1D dipolar gases used Cr and explored magnetisation-changing collisions and their suppression in 2D optical lattices~\cite{Pasquiou2011sra}; see Fig.~\ref{relaxation1D} and Sec.~\ref{subsec:inel_confin} for more details. The first experiments  with strongly interacting gases entering the fermionised $\gamma > 1$ regime used Dy with $\sim$50 Dy atoms per tube~\cite{Tang2017tni}.  

\subsubsection{Thermalization in a dipolar quantum Newton's cradle}\label{subsec:cradle}

Tang~\textit{et al.}~\cite{Tang2017tni} investigated the thermalization of a near-integrable quantum system.  The Lieb-Liniger model is an example of a quantum integrable Hamiltonian, in which there exists an extensive number of conserved quantities, resulting in regular, non-chaotic, non-thermalizing dynamics for any value of $g_\text{1D}$~\cite{Yurovsky2008cca,Cazalilla2011odb}.  This is due to the 1D geometrical constraint imposed on interactions: Two-body interactions permit non-diffractive scattering between nearest neighbours only, leaving the set of incoming momenta  unchanged with respect to the set of outgoing momenta~\cite{Lamacraft2013dso}. As a result, the scattering matrix obeys the Yang-Baxter equation, a sufficient condition for integrability~\cite{Yang1967ser,Baxter1972pfo,Cazalilla2011odb,Lamacraft2013dso},  since the binary collisions are unable to alter the distribution functions. Therefore, an empirical `smoking gun' for integrability would be the persistence of an out-of-equilibrium momentum distribution beyond the intrinsic dynamical time scale.  

The group of David Weiss at Penn State University observed this persistent non-equilibrium momentum distribution in 1D gases of contact-interacting Rb atoms, thereby experimentally establishing the integrability of the Lieb-Liniger model.  Their experiment relied on quantum quenches using what they called a `quantum Newton's cradle'~\cite{Kinoshita2006aqn}:  Atoms in 1D traps were set in motion using a Bragg diffraction pulse as a system quench. The atoms oscillated in counterpropagating packets and collided twice each period under the strong-coupling condition of $\gamma \gg 1$.  This is akin to the desktop Newton's cradle toy, except that instead of the metal spheres reflecting upon each collision, the Rb atoms could also pass through one another as a manifestation of the quantum nature of the system. Rather than rapidly come to a steady state (i.e., a stationary Gaussian momentum distribution), the packets were observed to oscillate many times (far longer than what one would expect in a 3D  gas) before the onset of heating from spontaneous emission. Thus, the Weiss group established that a strongly interacting integrable quantum system could be studied in the laboratory.  
Although both longitudinal confinement and transverse (virtual) motional excitation break integrability, these detrimental effects are suppressed in the $\gamma \gg 1$ regime~\cite{Mazets2008boi,Mazets2010tia,Tan2010roa}.

\begin{figure}[t!] \vspace{-6mm}
\includegraphics[width=0.99\columnwidth]{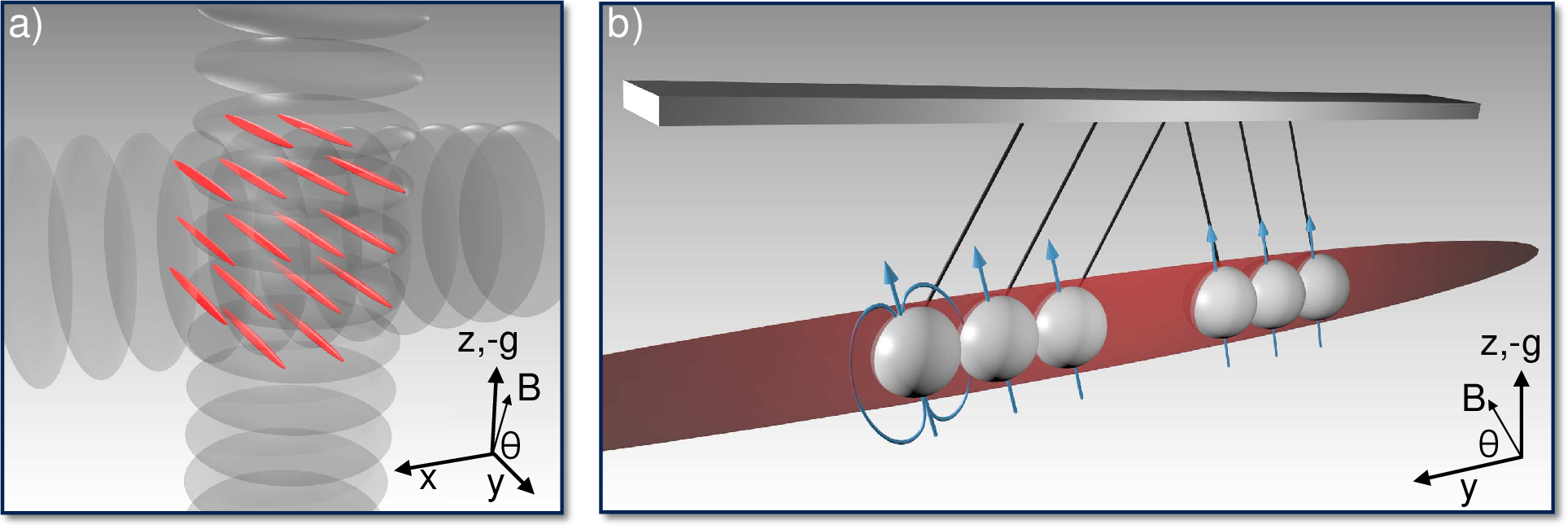}
\vspace{-3mm}
\caption{ a) An array of 1D tubes for Dy atoms is created by a 2D optical lattice. b) Cartoon of dipolar quantum Newton's cradle.  Dipolar Dy atoms are made to collide within the 1D tubes (only one in the is array shown).  The dipolar interaction is controlled by an external magnetic field making an angle $\theta$ with respect to the 1D axis along $\hat{y}$.  This provides a knob with which to control this integrability-breaking interaction.}
\label{DQNC}
\end{figure}

Tang~\textit{et al.}~\cite{Tang2017tni} used the magnetic DDI in a 1D Dy gas as the controllable interaction with which to break integrability in a dipolar version of the quantum Newton's cradle experiment. It is known from classical physics that magnetic spheres cause the motion of a toy Newton's cradle to be chaotic due to the long-range interaction; the long-range DDI should likewise break integrability in the quantum systems, since it allows for ``diffractive'' collisions among atoms. Indeed, the DDI allows for ``diffractive'' collisions among atoms: i.e., in addition to two colliding atoms swapping their momentum, as in the case of contact collisions (which are non-diffractive), their momentum may also be imparted to a third particle. This violates the Yang-Baxter condition, and hence breaks integrability. Since the DDI strength falls off slowly in space, the probability for a diffractive interaction is not suppressed by the usual need for three particles to be at the same place at once. Furthermore, by simply controlling $\theta$, the strength of the integrability-breaking perturbation can be tuned.  Figure~\ref{DQNC} depicts the dipolar quantum Newton's cradle.

The dipolar quantum Newton's cradle opens new avenues to explore how quantum thermalization arises upon the introduction of a perturbation that lifts integrability.  For example, theoretical consensus is lacking regarding whether relaxation involves two distinct timescales or three for strongly interacting quantum near-integrable systems (i.e., first prethermalization, then a prethermal plateau of a near steady-state prethermal state, and finally a decay to the thermal state)~\cite{Eckstein2009taa,Kollar2011gge,Nessi2014qqa,Babadi2015ffe,Bertini2015pat,Bertini2016tal}.  More generally, thermalisation of a near-integrable system is an old question with a celebrated answer in the realm of classical physics~\cite{Kolmogorov1954otc}.  In the 1950's, Kolmogorov, Arnold, and Moser developed what became known as KAM theory, establishing that chaos, and hence thermalization, sets in as a smooth crossover as the strength of a nonlinear, integrability-breaking perturbation is increased~\cite{dumas2014tks}.  In the quantum realm, it has  long  been wondered whether any meaningful analogue to KAM theory  exists, and a general theory for quantum thermalization in near-integrable systems is lacking despite much work~\cite{Burkov2007ddi,Moeckel2008iqi,Rigol2009bot,Rigol2009qqa,Eckstein2009taa,Kollar2011gge,Marcuzzi2013pia,Rigol2016fai,Nessi2015glb,Babadi2015ffe,Bertini2015pri,Bertini2015pat,Brandino2015goa,Bertini2016tal,Langen2016pau,Biebl2017tri}.

In their experiment to investigate this issue, Tang~\textit{et al.}~\cite{Tang2017tni} observed relaxation and thermalization dynamics in the time evolution of the momentum distribution of the kicked 1D gas. The distance to a thermal distribution is quantified and is observed to undergo \textit{two} distinct exponential decay regimes. The first evolution is a prethermalization (dephasing) decay, and the second consists of the relaxation from the prethermal state towards a Gaussian thermal distribution determined by the Gibbs ensemble.

Tang~\textit{et al.} found that the thermalization rate in this second step obeys a simple scaling expression that depends only on $\theta$ without any free parameters.  The expression accounts for the DDI perturbation using Fermi's golden rule formula, multiplied by the cross section for short-range interactions to occur. This describes the dominant integrability-breaking diffractive interaction term: that of two atoms scattering via the contact interaction while simultaneously interacting with a third atom at long range via the DDI. It is remarkable that such a simple description appropriately describes the thermalization dynamics of such a strongly interacting system near an integrable point.

The experiment was supported by an exact diagonalization calculation of a two-rung XXZ quantum magnetism model~\cite{Sachdev2011qpt} with long-range hopping serving as the perturbation. Similar two-stage thermalization dynamics was observed despite the microscopic dissimilarity of the systems. 

Future work could attempt to understand how generally applicable is this simple, Fermi golden rule-like expression for a wider variety of quantum quench experiments. One could also explore how more sophisticated descriptions based on, e.g., adaptations of generalised hydrodynamics, may better describe the dipolar 1D near-integrable system.   More generally, this work sets the stage for a wide array of inquiries into the physics of strongly interacting quantum systems near integrability. In the next section, we describe one such investigation, that of quantum many-body prethermal `scar' states, enabled by the realisation of dipole-stabilised \textit{excited} 1D quantum gases.

\subsubsection{Dipolar stabilisation of super-Tonks-Girardeau gases}\label{subsec:sTG}

As mentioned above, the TG state is one in which 1D bosons with a divergently repulsive coupling strength behave like fermions.  The two-body wavefunction vanishes when the atomic positions coincide, and so possesses exclusion correlations as if it were an ideal Fermi gas. A quantum quench of the TG gas prepares an eigenstate of the attractive Lieb-Liniger model, the so-called super-TG gas mentioned above in Sec.~\ref{Sec1D}.  The TG and sTG state wavefunctions and energies are smoothly connected as a function of $g_\text{1D}$, so while the effective model exhibits a discontinuity, the system remains adiabatic throughout the parameter quench. The result is a highly excited sTG state wherein the 1D bosons that attractively interact at short range behave as if they were ground-state fermions \textit{repulsively} interacting at \textit{long} range. That is to say, the bosons develop even stronger exclusion correlations than the ideal Fermi gas. The two-body wavefunction node inherited from the quench from the TG state extends into a pair of nodes separated by the length $a_\text{1D}$, effectively introducing a rigid exclusion zone similar to the classical hard rod model.  Moreover, its excited-state many-body wavefunction exhibits many more nodes than the TG wavefunction, which enhances the stiffness of the sTG gas~\cite{Astrakharchik2005btt,Tempfli2008eoa}. 

Such metastable attractive gases typically collapse into bound states (cf.~the `Bose-nova' implosion of strongly attractive BECs in 3D~\cite{Roberts2001cco}).  By contrast, however, the strong antibunching correlations in the sTG gas wavefunction prevent atoms from approaching each other close enough to bind into cluster-like states. The sTG gas is expected to never collapse, despite the multitude of bound states of lower energy.  The deeper reason for this surprising metastability lies in the aforementioned integrability of the Lieb-Liniger model, for any $g_{1D}$ value.  That is, the sTG state is also a solution to the Bethe ansatz equations of the Lieb-Liniger model~\cite{Chen2010tfa}. 

Experimentally, however, the sTG gas \textit{does} collapse for attractive interactions weaker than those in the unitary $g_\text{1D}\rightarrow -\infty$ regime, as first observed in the nondipolar Cs system~\cite{Haller2009roa}.  While the system remains effectively integrable in the unitary regime, since the interaction term dominates all others in the perturbed Lieb-Liniger Hamiltonian, the system can collapse when the interactions become sufficiently weak outside the unitary regime. That is, the breaking of the integrability of the Lieb-Liniger model becomes manifest once the system is no longer unitary near the resonance.  Experimental imperfections such as the longitudinal harmonic trap and the virtual excitation of the higher motional bands of the transverse 2D optical lattice forming the array of 1D traps all break integrability, if only weakly~\cite{Mazets2008boi,Mazets2010tia,Tan2010roa,Kao2020tpo}.  These yield nonzero matrix elements coupling the initial state and molecular/cluster-like bound states in the near-integrable system that enable the collapse of the sTG state at  finite negative $g_\text{1D}$ coupling strength.

This may be understood in analogy to the classical 1D gas of hard rods of length $a_\text{1D}$. The rod length $a_\text{1D}$ extends an exclusion zone that prevents rods from overlapping.  When more rods $N$ are stuffed into a box of length L than can fit end-to-end (i.e., when $Na_\text{1D}>L$), then the system ``collapses'' by kinking the rods out of 1D alignment.  In the quantum system, $a_\text{1D}$ is the 1D scattering length definable through $g_\text{1D} = 2\hbar^2a/ma_\perp^2 = -2\hbar^2/ma_\text{1D}$.  Both $g_\text{1D}$ and $a_\text{1D}$ are tunable using a confinement induced resonance (CIR)~\cite{Olshanii1998asi}.  These arise in quasi-1D traps due to the open-channel-like role of higher-energy transverse motional states of the trap and were first observed in Cs~\cite{Haller2010cir}. The CIR provides tunability through the dependence of $a$ on $B$-field near a 3D FR:
\be
g_\text{1D}(B)=\frac{2\hbar^2 a(B)}{ma_\perp^2}\frac{1}{1-C a(B)/a_\perp}, 
\ee
where $a_\perp$ is the transverse oscillator length and $C\approx 1.46$~\cite{Olshanii1998asi}.  

\begin{figure}[t!] \vspace{-6mm}
\includegraphics[width=0.99\columnwidth]{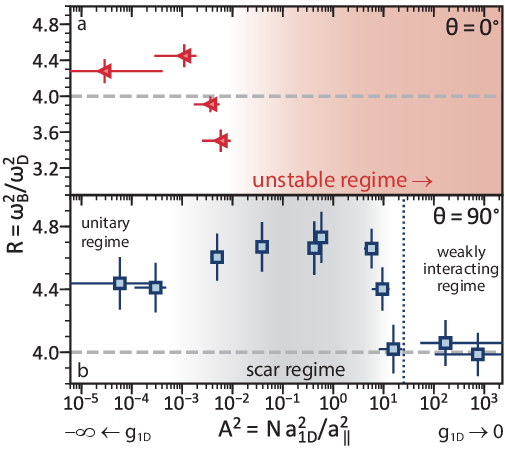}\vspace{-3mm}
\vspace{-3mm}
\caption{Comparison of excited state stiffness $R$ versus coupling parameter $A^2\propto g_\text{1D}^{-2}$ showing a) increased instability due to an attractive DDI in contrast to b) the complete stabilisation of the excited gas due to the repulsive DDI.  $A$ is a normalised form of the coupling constant~\cite{Astrakharchik2005lda},  $a_\parallel$ is the longitudinal oscillator length in the 1D trap, and $N$ is the average number of atoms per tube.  The regime of collapse is indicated in panel (a), while the unitary, scar, and weakly interacting regimes are roughly sketched in panel (b). Figure adapted from Ref.~\cite{Kao2020tpo}.}
\label{stabilization}
\end{figure}

Kao~\textit{et al.}~\cite{Kao2020tpo} explored the effect of DDIs on the stability of the sTG gas.  An array of 1D Dy gases were quenched into the sTG state under various angles $\theta$ with respect to the 1D axis.  CIRs were found near a broad 3D FRs in $^{162}$Dy, allowing $g_\text{1D}$ to be tuned from $0^+\rightarrow {+}\infty\rightarrow{-}\infty\rightarrow 0^-$ by increasing the magnetic field. Note that unlike theory predictions~\cite{Sinha2007cdg,Giannakeas2013dci,Shi2014odc,Guan2014qod}, molecular binding energy measurements of the CIRs employed in Ref.~\cite{Kao2020tpo} showed that they did not exhibit a dependence on $\theta$, neither in the repulsive ($\theta = 90^\circ$), attractive ($\theta = 0^\circ$), nor non-dipolar $\theta = \theta_m$ configurations. The state of the system was revealed through gas stiffness measurements obtained from the observations of the square of the ratio of the breathing--to--dipole oscillation frequencies.  This quantity $R$ is greater than four in an sTG gas, but dives toward zero when a gas collapses into bound states due to diverging compressibility. Figure~\ref{stabilization} contrasts the repulsive and attractive DDI cases.

Intuitively, one might expect that a repulsive DDI would inhibit the sTG gas from collapsing, by adding a repulsive energy barrier between atoms, at least until the contact interaction becomes too weak. Conversely, an attractive DDI might cause the sTG to collapse at a more strongly negative $g_\text{1D}$ than in a nondipolar gas.  However, this intuitive picture is confused by the fact that the DDI breaks integrability, see Sec.\,\ref{subsec:cradle},  leading to collapse through eigenstate mixing with cluster states. 

Reference~\cite{Kao2020tpo} indeed found that the repulsive DDI stabilises the excited gas, and surprisingly, does so  regardless of how close to zero the negative $g_\text{1D}$ is tuned. That is, the DDI does not simply expand the region of sTG gas stability by a small margin, but prevents collapse for \textit{all} attractive contact interactions.  By contrast, an attractive DDI hastened the collapse, as expected, while collapse in the nondipolar case occurred at a coupling strength similar to that found in Cs. 

The reason for the dramatic influence of DDI on the stability of the excited states, despite being too weak to affect the phase diagram of the ground state, remains unclear, pending future studies. This experimental discovery nevertheless provides  access to a brand-new near-integrable system where prethermal scar-like states can be explored. We note that stable gases with $R<4$ were also observed, suggesting that gas-like few-body cluster states had been formed as predicted in Ref.~\cite{Batchelor2005eft}.  In the next section, we discuss how the stabilisation enabled a novel state-preparation scheme that exploits a quantum holonomy inherent to the Lieb-Liniger Hamiltonian.   

\subsubsection{Quantum holonomy, topological pumping, and strongly correlated prethermal states}

Systems with a quantum holonomy possess the property that eigenstates can change after a cyclic, adiabatic variation of Hamiltonian parameters: while the Hamiltonian stays the same, the resulting eigenstate differs~\cite{Bohm2003tgp}. The Lieb-Liniger Hamiltonian harbours one such exotic quantum holonomy~\cite{Yonezawa2013qhi}.  It is realised by cycling $g_\text{1D}$ through $0^+\rightarrow {+}\infty\rightarrow{-}\infty\rightarrow 0^-$, resulting in \textit{different} eigenstates of higher energy upon the return to the \textit{same} $g_{1D}$ value. 

Such holonomies provide a \textit{topological} means with which to pump a system into higher energy states. A simple, single-particle example of an energy pump arises in the case of a 1D infinite square well:  Periodically imposing a delta function potential between the infinite barriers pumps the ground state wavefunction to higher energy eigenstates~\cite{Kasumie2016aeo,Tanaka2020gan}. Topological pumping \textit{in space} has been known at least since Archimedes discovered the use of a screw to move water up an incline. Rotating Archimedes' screw by $2\pi$ returns it to the same configuration, but the water within the screw advances by one screw site. Similarly, the Thouless charge pump translates an electron one lattice site due to topological properties of the system~\cite{Thouless1983qop}, inducing quantised transport in an insulator; such spatial quantum topological pumps have been demonstrated using quantum gases in special optical lattices~\cite{Lohse2015atq,Nakajima2016ttp}.  

\begin{figure}[t!] \vspace{-6mm}
\includegraphics[width=0.99\columnwidth]{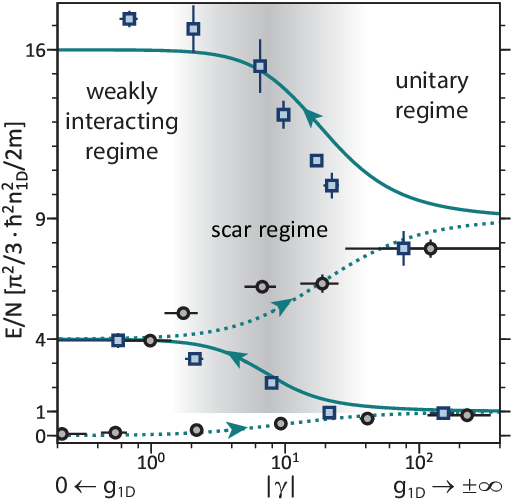}\vspace{-3mm}
\vspace{-3mm}
\caption{The hierarchy of energy eigenstates accessible via topological pumping using the quantum holonomy point at $g_\text{1D} = \pm \infty$.  Two complete quantum holonomy cycles are shown verses  energy per particle $E/N$.  The Bethe ansatz solutions for the  repulsive (attractive) LL model are shown in dotted (solid) curves, along with the associated data on the repulsive (attractive) branches in black circles (blue squares).  The intermediate coupling region associated with scars is shaded (starting above the ground-state repulsive branch).  To the left and right are the weakly interacting and  effectively integrable unitary regimes, respectively.  Figure adapted from Ref.~\cite{Kao2020tpo}.}
\label{hierarchy}
\end{figure}

The many-body quantum holonomy of the Lieb-Liniger Hamiltonian was realised in Ref.~\cite{Kao2020tpo} using Dy 1D gases by scanning $B$ through a sequence of CIRs. This enabled the first topological pumping of a quantum many-body gas up a hierarchy of extensively higher energy eigenstates because the gas was stabilised by the repulsive DDI; see Sec.\,\ref{subsec:sTG}.  Figure~\ref{hierarchy} shows this hierarchical ladder.

In the very strongly interacting $g_\text{1D} \rightarrow \pm\infty$ regime of the energy eigenstate spectrum, the system is effectively integrable because the large contact interaction overwhelms all integrability-breaking terms in Hamiltonian. 
By contrast, in the opposite, weakly interacting regime, the system is effectively single-body (mean-field).  It is thus only in the intermediate-coupling regime where the topological pump prepares a hierarchy of exotic, ergodicity-avoiding prethermal states in a system that remains many-body and correlated but also effectively not integrable because of non-negligible contributions from the trap and the DDI. Surprisingly, despite the lack of perfect integrability, its energy density follows the solutions to the Bethe ansatz equations throughout the coupling regimes, within experimental uncertainty. 

These particular excited states resemble the atypical, nonergodic quantum many-body scar states that fail to immediately thermalize~\cite{Serbyn2020qmb}. Their name derives from the regular patterns (scars) traced by special wavefunctions through the otherwise ergodic phase-space of single-particle chaotic quantum systems~\cite{Heller1984bse}.  Until very recently, it was believed that long-lived far-from-equilibrium states of quantum many-body system only exist in integrable and many-body localised systems~\cite{Vasseur2016nqd,Abanin2019cmb}. The discovery of ``quantum many-body scars" in a Rydberg-atom lattice~\cite{Bernien2017pmb,Turner2018web} showed that such far-from-equilibrium states can be long-lived outside these limits, and thus potentially serve as long-lived quantum memories~\cite{Serbyn2020qmb}. 

The novel scars exhibited by the dipolar-stabilised excited 1D gas are the first  observed in a continuous, rather than lattice-based system. The fact that scars might form near integrability was first pointed out in Refs.~\cite{Kollar2011gge,Khemani2019soi}, and their observation in the 1D dipolar system makes this connection experimentally explicit.  Measurements of stiffness and energy--per--atom versus time show that the states do not heat; rather, they persist far longer than the collective-oscillation time scale in a prethermal state~\cite{Kao2020tpo}:  entropy rather than energy increases while the gas weakly thermalizes. 

The dipolar-stabilised 1D gas results mark an advance in our understanding of the exciting and highly active topic of quantum nonequilibrium many-body dynamics: not only do they provide a new way of creating a class of interesting nonequilibrium states that are relatively unexplored from both theoretical and experimental points of view, but they also draw a rich new correspondence between quantum scars, topological pumping, and near-integrability. In addition to preserving quantum many-body correlations, these scars may play a role in the quantum simulation of exotic fermionic many-body systems, since Feynman's `no-node' theorem does not apply to the wavefunctions of excited bosonic states~\cite{Feynman2018sma,Wu2011ube}.

\subsection{Spinless dipoles in lattices.}
\label{subsec:spinless_lattice}
Even in absence of a spin degree of freedom, because of its long-range and anisotropic character, the DDI drastically affects the  physics of lattice-confined atomic assemblies,  both its static properties and in its dynamics. In the first part, we describe assemblies of spin-polarised dipolar particles, loaded in three-dimension optical lattice potentials, and the status of their experimental achievements with highly magnetic atoms.  

\subsubsection{Extended Hubbard Hamiltonian and consequences}

Assemblies of spinless interacting particles, confined in deep lattices (tight-binding regime), are typically well described via Hubbard models of the form:
\begin{equation}
\label{Hubbard1}
H=-\sum_{\bs i\neq \bs j} t_{\bs i, \bs j}{\hat a}_{\bs i}^{\dagger}{\hat a}_{\bs j}+ \frac{1}{2}\sum_{\bs i,\bs j}U_{\bs i,\bs j,\bs k,\bs l}{\hat a}_{\bs j}^{\dagger}{\hat a}_{\bs l}^{\dagger}{\hat a}_{\bs k}{\hat a}_{\bs i}
\end{equation}
Here, ${\hat a}_{\bs i}$ is the (Wannier) destruction operator for a particle at site $\bs i$, and we note the number operator ${\hat n}_{\bs i}={\hat a}_{\bs i}^{\dagger}{\hat a}_{\bs i}$. We first keep the discussion general, and the operators ${\hat a}^{\dagger}_{\bs i}$, ${\hat a}_{\bs j}$ abide either the canonical commutation or anti-commutation relations for bosonic or fermionic particles, respectively.   In ultracold atomic systems, the Hubbard model is derived by writing the Hamiltonian of the particles moving in the periodic external potential of the lattice in the basis of the associated Wannier wavefunctions $\{w_{\bs i}\}_{\bs i \text{, lattice sites}}$ in the lowest band and restricting the motion to this lowest band~\cite{Jaksch1998cba,Hofstetter2002hts,Bloch2005uqg,Bloch2008qca,Bloch2012qsw,Lewenstein2007uag}. The $t_{i,j}$ coefficients correspond to the energies from  the single-particle kinetic and external (lattice) potential terms for a particle initially at site $i$ and moving to site $j$.  They are thus denoted hopping or tunnelling coefficients. Due to the weak overlap between Wannier functions at different sites in the tight-binding regime, hopping is generally restricted to the nearest-neighbour hopping, $\bs j= \bs i+{\vecu}_{\alpha}$, ${\vecu}_{\alpha}$ denoting the lattice unit cell vector in direction $\alpha={x,y,z}$. The $U_{i,j,k,l}$ describe interparticle interaction energies for two particles located at sites $i$ and $k$ initially and, after the interaction, at sites $j$ and $l$. Explicitly, those coefficients write $U_{i,j,k,l}= \int  {\rm d}r{\rm d}r' w_{\bs j}^*(\bs r) w_{\bs l}^*({\bs r'})U({\bs r}-{\bs r'})w_{\bs k}({\bs r'})w_{\bs i}(\bs r)$ with $U({\bs r})$, the interparticle interaction potential. Because interaction potentials usually decay with distance, the largest interaction contribution for bosonic assembly  is the on-site interaction ($i=j=k=l$) yielding a term $U {\hat n}_{\bs i}({\hat n}_{\bs i}-1)$, with $U=U_{i,i,i,i}$. We note that the on-site term cancel for fermions due to the Pauli exclusion principle and the single-band assumption, precluding multiple occupancy at one lattice site.  Off-site interactions terms may also arise. In deep lattices (i.e., additionally accounting for the weak overlap between Wannier functions), the largest contribution of off-site interactions are those involving no particle motion, i.e.,  $i=j \neq k =l$ between two particles located at cite $i$ and $k$ respectively, yielding the coefficients $V_{i,k}=V{i,i,k,k}$. 

In the case of contact-interacting assemblies, off-site interactions vanish with the Wannier functions overlap. Therefore, the standard Hubbard model for bosonic spin-polarised contact-interacting samples in a cubic lattice (spacing $d$) has only two parameters: the tunnelling rate $t=t_{i,i+{\vecu}_{\alpha}}$, independent on the bond direction $\alpha$, and the onsite interaction $U$~\cite{Jaksch1998cba}. This yields the so-called Bose-Hubbard model:
\begin{equation}
\label{HubbardBose}
H=- t\sum_{\langle \bs i, \bs j\rangle} {\hat a}_{\bs i}^{\dagger}{\hat a}_{\bs j}+ \frac{U}{2}\sum_{\bs i}{\hat n}_{\bs i}({\hat n}_{\bs i}-1),
\end{equation}
with $\langle i,j\rangle$ denoting nearest neighbours sites. 
For the spinless fermionic case, in absence of long-range interactions,  the standard Hubbard model only yields tunnelling terms, which intimately interplay with Pauli exclusion principle. We note that, in the contact-interacting fermionic case, two spin components are usually considered, which also yields a two-parameter (tunnelling/on-site interaction) model called the Fermi-Hubbard model~\cite{Hofstetter2002hts}. This is:
\begin{equation}
\label{HubbardFermi}
H=- t\sum_{\langle \bs i, \bs j\rangle,\sigma} {\hat a}_{\bs i,\sigma}^{\dagger}{\hat a}_{\bs j,\sigma}+ \frac{U}{2}\sum_{\bs i, \sigma \neq \sigma'}{\hat n}_{\bs i,\sigma}{\hat n}_{\bs i,\sigma'},
\end{equation}
with ${\hat a}_{\bs i,\sigma}$ (${\hat n}_{\bs i,\sigma}$) being the fermionic destruction (number) operator for particle of spin $\sigma$ at site $\bs i$. This expression matches the original model heuristically introduced by Hubbard in the context of condensed matter systems, where strongly correlated electrons move in the ionic crystal. In this context, Hubbard models have been powerful in predicting or explaining quantum phases of matter~\cite{Hubbard1963eci,Hubbard1978gws,Ashcroft1976ssp}. 

Atomic systems offer direct and convenient control of the Hubbard model parameters. In particular, for contact-interacting particles (spinless bosons or spin-$1/2$ fermions) on a cubic lattice created by standing-wave of light  (e.g., retroreflection of a laser beam)~\cite{Bloch2005uqg,Bloch2008mbp}, $t$ and $U$ intrinsically depend on the light intensity. The light intensity indeed sets the depth of the periodic lattice potential, $\Vlatt$. Consequently, it controls both the external potential in the Hamiltonian (coming in the expression of $t$) and the extension of the Wannier wavefunction, at the basis of the Hubbard derivation. Defining $s=\Vlatt/E_{\rm r}$ to be the ratio of the lattice depth to the recoil energy, $E_{\rm r}=\hbar^2 k_{\rm L}/2m$  ($m$ is the atomic mass and $k_{\rm L}=2\pi/d$ the reciprocal lattice constant),  $t$ is found to vary as $t \propto s^{3/4}\exp\left(-2s^{1/2}\right)$ and $U$ as $s^{3/4}$ at large lattice depths~\cite{Bloch2008mbp}. Therefore, the ratio $t/U \propto \exp\left(-2s^{1/2}\right)$ decreases for increasing lattice depths. Furthermore, $U$ is proportional to the scattering length and can be additionally controlled thanks to Feshbach resonances~\cite{Chin2010fri}.
Relying on these control knots, ultracold-gas experiments have  revealed and explored the spectacular effects related to the dynamics and thermodynamics of Hubbard models. For instance, the transitions from superfluid or metallic phases to strongly interacting insulating ones, called Mott phases, arising from the competition between $U$ and $t$, have been observed both in bosonic and in fermionic systems~\cite{Greiner2002qpt,Joerdens2008ami,Schneider2008mai,Mark2011pmo,Bloch2005uqg,Bloch2008qca,Esslinger2010fhp,Bloch2012qsw,Gross2017qsw}.

In the reminder of this section, we only consider spinless atoms. In the case of highly magnetic atoms, the long-range and ansiotropic DDI gives rise to additional interaction terms in the Hubbard model, following the general form of Eq.~\eqref{Hubbard1}. Additionally,  the DDI introduces  dependencies in the coefficient values related to the orientation of the dipoles defined by the angles $(\theta, \phi)$ compared both to the on-site Wannier function anisotropy or to the lattice geometry~\cite{Mazzarella2006ebh,Trefzger2011udg,Baranov2012cmt,Dutta2015nsh}. For simplicity, we first consider the case of dipolar bosons. In the tight binding regime, the relevant extended Bose-Hubbard model can be written as~\cite{Dutta2015nsh}
\begin{eqnarray}
\nonumber
&&H=-\sum_{\langle \bs i, \bs j\rangle} t_{\bs i, \bs j}{\hat a}_{\bs i}^{\dagger}{\hat a}_{\bs j}
+ \frac{U_{\rm c} + U_{\rm dd}(\theta,\phi)}{2}\sum_{\bs i}{\hat n}_{\bs i}({\hat n}_{\bs i}-1) \\
\nonumber
&&+ \frac{1}{2}\sum_{\langle \bs i, \bs j\rangle / \vecu_\alpha =\bs i-\bs j}V_{\alpha}(\theta,\phi){\hat n}_{\bs i}{\hat n}_{\bs j}
\\
\label{HubbardDDI}
&&+ \frac{1}{2}\sum_{\langle \bs i, \bs j\rangle / \vecu_\alpha =\bs i-\bs j}\Delta t_{\bs i,\bs j}{\hat a}_{\bs j}^{\dagger}{\hat a}_{\bs i}({\hat n}_{\bs j}+{\hat n}_{\bs i}-1),
\end{eqnarray}
which includes the terms with contributions up to first order in the Wannier wavefunction overlap. The dominant effect of the DDI is the on-site dipolar-interaction term $U_{\rm dd}(\theta,\phi)$, which adds to the contact-interaction one, $U_{\rm c}$, and which depends on the Wannier function anisotropy compared to the dipole orientation  due to the DDI anisotropy. For an isotropic Wannier function, $U_{\rm dd}$ cancels.  Because of the long-range character of the DDI, off-site terms also contribute. These are:  (i) a nearest-neighbour interaction (NNI) $V_{\alpha}$ between two atoms localised on neighbouring sites $i$ and $i+{\vecu}_{\alpha}$; (ii) a density-induced tunnelling (DIT) term of strength $\Delta t_{\alpha}$, arising from the interaction between one atom localised at site $i$ or $i+{\vecu}_{\alpha}$ and one delocalized between $i$ and the neighbouring site $i+{\vecu}_{\alpha}$.   Remarkably, at zero order in the wavefunction overlap, the NNI does not cancel, but becomes independent of the lattice depth, its strength is approximately given by $V_{\alpha}=U_{\rm dd}(d \vecu_\alpha )$ with $U_{\rm dd}$ given in Eq.~\eqref{Udd}. In contrast, the DIT only exists at first order in the Wannier overlap between sites. Additionally, due to the DDI anisotropy, $V_{\alpha}$ depends on the dipole orientation, mainly compared to the direction $\vecu_{\alpha}$ of the lattice bond on which the interaction occurs. $\Delta t_{\alpha}$ also depend on $(\theta,\phi)$ both via the bond direction and the Wannier wave function anisotropy. 

The competition between the additional terms appearing in Eq.~\eqref{HubbardDDI} (i.e., $V_{\alpha}$ and $\Delta t_{\alpha}$) and the conventional ones (i.e., $U$ and $t$) are expected to yield new physics.  Such extended Hubbard models have been extensively studied theoretically, in various lattice geometries and dimensionalities, for various parameters strengths and lattice filling $n$, as well as for different particle statistics (fermions or bosons)~\cite{Dutta2015nsh}. Rich phase diagrams, including unconventional quantum phases supporting strong and exotic correlations, have been predicted.
The most studied case is the one of bosons in a 1D or 2D square lattice~\cite{Sengupta2005svp,Batrouni2006spi,DallaTorre2006hoi,Mitra2009hwc,Danshita2009sos,CapogrossoSansone2010qpo,Rossini2012pdo,Batrouni2013csa,Batrouni2014cpp,Kawaki2017pdo,Suthar2020ssp,Kraus2020spi,Kraus2022qpo}. Here, \textit{charge-density waves}, i.e.,~insulating phases with modulated density, have been predicted for strong enough NNI, $V_{\alpha}$~\cite{Sengupta2005svp,Mitra2009hwc,Danshita2009sos,Dalmonte2011hai,Rossini2012pdo,Kawaki2017pdo,Kraus2022qpo}. 
Depending on the filling and on the dipole orientation, star, stripe or chequerboard spatial patterns have been predicted~\cite{CapogrossoSansone2010qpo}. The transition from superfluid to density-wave states can occur directly without an intermediate \textit{``supersolid"} (or \textit{``lattice-supersolid"}) phase~\cite{Mitra2009hwc,Danshita2009sos,Rossini2012pdo,Kawaki2017pdo,Kraus2022qpo}, i.e.,~a superfluid with a spatial density modulation different from the lattice itself~\cite{Goral2002qpo}. Yet, under some conditions (typically low filling $n<0.5$ and finite $U>t$), a lattice-supersolid phase has also been predicted in simulations~\cite{Sengupta2005svp,Batrouni2006spi,Batrouni2013csa,Kawaki2017pdo,Suthar2020ssp,Kraus2022qpo}. It is in particular expected to be stabilised by doping (adding holes or particles) of the density-wave state away from the rational fillings. At incommensurate fillings and large enough $V_\alpha$, phase separation into a pure solid and a homogeneous superfluid has also been predicted to occur~\cite{Sengupta2005svp}.  Phase separation is favoured by large $U$, yet it is destabilised, contrarily to the supersolid, by considering the next order of extension of the DDI, and may ultimately disappear when one consider the full range of the DDI.
In 1D, an Haldane-like phase has also raised high interest~\cite{DallaTorre2006hoi,Batrouni2013csa,Batrouni2014cpp,Kraus2022qpo,Kawaki2017pdo}. This insulating phase occurs at unit filling for $U\sim V>t$; it does not break the translation symmetry but stands out by its special correlations. Compared to the Mott insulator phase, the particle-hole fluctuations indeed appear with an alternating order. These unconventional phases have critical temperatures typically scaling with $V_\alpha/k_{\rm B}$.

For spinless fermions, a similar expression Eq.~\eqref{HubbardDDI} can also be derived, yet because of the Pauli exclusion principle, the second and fourth terms of the sum in Eq.~\eqref{HubbardDDI} cancels, only retaining the tunnelling and NNI terms. Theoretically, this case is also of interest, yet turns out to be more complicated to treat than the boson one~\cite{Mikelsons2011dwp,Bhongale2012bos, Gadsbolle2012dfi,Gadsbolle2012htd,Zeng2014soa,CamachoGuardian2016spo,Baranov2012cmt,Dutta2015nsh,vanLoon2016idl}. Few studies have shown that various kinds of charge-density-wave and supersolid phases also forms in dipolar fermions lattice systems in 2D~\cite{Mikelsons2011dwp,Bhongale2012bos, Gadsbolle2012dfi,Gadsbolle2012htd,CamachoGuardian2016spo} and 3D~\cite{Zeng2014soa}. Additionally, $p$-wave superfluidity may arise. 
The interplay between lattice and dipole-induced Fermi surface deformation has also been predicted to yield a topological phase transition, of a Lifshiftz type~\cite{vanLoon2016idl}. Of even larger interest has been a case recovering a situation closer to the bosonic model, with (effective) spin-1/2 fermions as for the conventional Hubbard model; see Eq.~\eqref{HubbardFermi}. This particular case will not be discussed further here, interested readers can consult e.g. refs.~\cite{Dutta2015nsh,Baranov2012cmt}.

\subsubsection{Experimental status}
\begin{figure}[t]
\centering
\includegraphics[width= 8 cm]{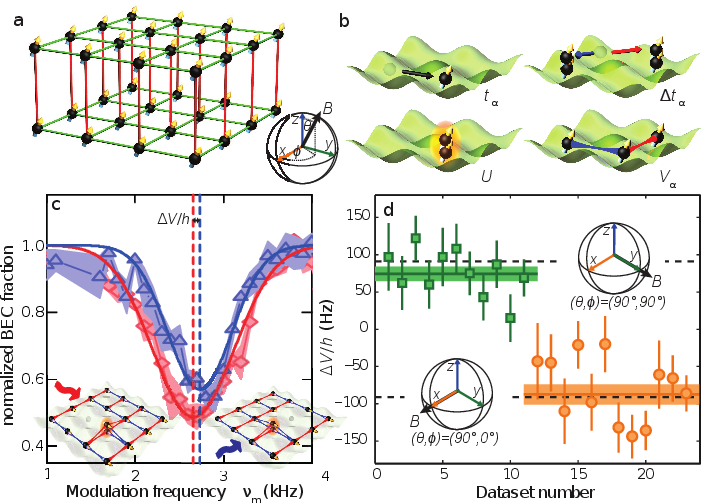}
\caption{Measurement of the extended Bose-Hubbard model from a spin-polarised $^{168}$Er BEC loaded in a 3D anisotropic lattice~\cite{Baier2016ebh}. (a) illustrates the lattice  and atomic dipoles geometry. (b) illustrates the different term of the Hamiltonian. (c-d) shows the measurement revealing the NNI term via its effect on the particle hole excitation energy. (c) exemplifies the principle of the differential measurement, comparing the resonance position from the modulation in two different excitation directions using the same overall configuration. The shift between $y$ (red) and $x$ excitation is attributed to the loss of one distinct NNI unit $V_{y}$ or $V_{x}$ as illustrated in the insets. (d) summarise all the differential measurement performed with the dipole oriented along $y$ (green) or along $x$ (orange) and the comparison to the theory prediction (dotted line). Adapted from~\cite{Baier2016ebh}}
\label{fig:eBH}
\end{figure}
Up to now, one experiment has reported on dipolar effects in quantum gases of spin-polarised highly-magnetic atoms confined in a deep lattice potential~\cite{Baier2016ebh}. It is based on  $^{168}$Er BEC loaded in a 3D optical lattice of parallelepipedic unit cell with spacings $(266,266,532)$\,nm, see Fig.\ref{fig:eBH}(a). This work demonstrates the relevance of the extended Hubbard model (Eq.~\eqref{HubbardDDI}) for magnetic bosons, see Fig.\ref{fig:eBH}(b). By varying both the lattice depths independently in each spatial direction and the dipole orientation, Baier \textit{et al.}~\cite{Baier2016ebh} reveal the impact of the DDI in the lattice system --- that is on the excitation spectrum, in particular affecting the gap of the Mott-insulator phase, and on the phase diagram itself, in particular shifting the superfluid-to-Mott-insulator transition.  Baier \textit{et al.}~\cite{Baier2016ebh} additionally quantified the various terms of the extended Hubbard model (Eq.~\eqref{HubbardDDI}). They measured the dipolar on-site interaction contribution , $U_{\rm dd}$, by varying the Wannier wavefunction anisotropy and the dipole orientation. They also observed the two leading long-range interaction terms, i.e. the NNI and DIT. 
The strength of the NNI was isolated and quantified thanks to a differential measurement of the particle-hole excitation gap, comparing two excitation directions in an otherwise identical system, see Fig.\ref{fig:eBH}(c). Based on Eq.~\eqref{HubbardDDI}, the excitation gap depends on on the excitation direction compared to the dipole orientation by the contribution of $-V_{\alpha}(\theta,\phi)$.  The differential measurement shows that $V_{\alpha}$ is tunable from $-h$ 30\,Hz to $h$ 60\,Hz, in the $266$-nm-spacing directions, in agreement theoretical expectations, see Fig.\ref{fig:eBH}(d). The DIT was evidenced by the observed shift of the superfluid-to-Mott-insulator transition, which cannot be entirely explained by a theory accounting for the effect of the on-site DDI but is when accounting for DIT additionally
~\footnote{We note that DIT is also relevant for contact-interacting bosons in a lattice. Yet it occurs at a higher order expansion in the Wannier overlap than in the DDI case, as the interaction potential is fully localised at one position. Its impact was previously successfully demonstrated in this context, see refs~\cite{Luehmann2012moa,juergensen2014ood}}.

The predicted exotic phases of the extended Hubbard model have not yet been observable in the experiment. 
We note that the strength of the NNI, being mainly set by  the simple magnetic moment $\mu$ and by the lattice spacing, cannot be easily increased in the experiment. Its relative weakness in systems made of magnetic atoms impose stringent restrictions on the temperature below which the phases of interest would become observable as well as on the time needed for their spontaneous emergence or their (adiabatic) preparation. Such regimes are on the edge of the experimental possibilities to-date. To mitigate these constraints, interesting approaches consist in designing optical lattices with small (sub-wavelength) spacings, thus increasing the NNI coefficient ~\cite{Nascimbene2015dol,Lacki2016nds,Jendrzejewski2016swo,Wang2018dso,Lacki2019spo,Tsui2020roa,Ge2020dso}. Finally, preparation and detection schemes at the single-atom level~\cite{Sherson2010sar,Bakr2009aqg,Gross2017qsw} are a direction of broad interest which would ease the observation of exotic phases as well as many others effects of interests,  e.g.\,the influence of the DDI on the correlations or the dynamics of impurities. 
Beyond the case of spin-polarised bosons in rectangular lattices, very important prospects relate to the case of the distinct fermionic statistics, as well as to the change of the lattice geometry and the investigation of frustration or disorder effects.

\subsection{Exploration of spin lattice models and quantum magnetism}
\label{subsec:spinlattice}
\subsubsection{Introduction}

Beyond the spin-polarised case, magnetic atoms in optical lattices make it possible to explore quantum magnetism. For this reason, these are exciting system that may realise an analogue quantum simulator of canonical open problems associated with  quantum many-body physics~\cite{Bloch2008mbp,Bloch2012qsw,Lewenstein2007uag,Dutta2015nsh,Gross2017qsw,Baranov2012cmt}. 

In typical experiments with ultracold atoms interacting at short range, a spin degree of freedom can be included by, e.g., populating different Zeeman sublevels in the ground-state hyperfine manifold, see also Sec.~\ref{subsec:large_spin}. In this case, the system is well described by a \textit{spinfull} Hubbard model: Extending the case of Eqs.\,\eqref{HubbardBose},\eqref{HubbardFermi}, it takes the general form:
\begin{equation}
\label{HubbardSpin}
H=- t\sum_{\langle \bs i, \bs j\rangle,m} {\hat a}_{\bs i,m}^{\dagger}{\hat a}_{\bs j,m}+ \frac{1}{2}\sum_{\bs i, k,l,m,n}U_{k,l,m,n}{\hat a}_{\bs i, l}^{\dagger}{\hat a}_{\bs i, n}^{\dagger}{\hat a}_{\bs i,m}{\hat a}_{\bs i,k},
\end{equation}
where ${\hat a}_{\bs i,m}$ is the destruction operator of an atom of spin $m$ at site $\bs i$ and $U_{k,l,m,n}$ is the on-site spin-dependent interaction that one can deduce from the $a_{S}$ scattering lengths; see also Secs.\,\ref{subsec:fesbachsec} and \ref{sec:scatt}. Here, we assume a spin and direction independent tunnelling rate $t$. If $t$ remains weak compared to the on-site interactions, 
tunnelling processes between two adjacent occupied lattice sites are energetically forbidden and they contribute  as only a second-order virtual process. In such a process, an atom tunnels to an occupied site where spin-dependent onsite interaction may lead to a change of the individual spin of the atoms before the atom tunnels back to its original (unoccupied) site.
In the case where there is one particle per lattice site, this results in an effective lattice model with spin-spin nearest-neighbour interaction. In the case of effective spin-1/2 systems (or larger spin with SU($N$)-symmetric interactions), with interpsin on-site interaction strength $U$, the system is well described by the Heisenberg Hamiltonian,
\begin{equation}
H_{\rm ex}=\frac{J_{\rm ex}}{2} \sum_{\langle \bs i,\bs j \rangle}\left[ \hat{S}_{\bs i}^z.\hat{S}_{\bs j}^z+\frac{1}{2}\left(\hat{S}_{\bs i}^+.\hat{S}_{\bs j}^-+\hat{S}_{\bs i}^-.\hat{S}_{\bs j}^+\right) \right],
\label{heisenbergham}
\end{equation}
with $J_{\rm ex} = \pm 4 t^2/U$
~\cite{Kuklov2003cso,Duan2003cse,GarciaRipoll2003sdf}, where 
$+$ holds for fermions, and $-$ for bosons. $\hat{S}_{i}^{(-,+,z)}$ are the spin operators  on site $i$ and defined as $\hat{S}_{i}^{(-,+,z)}=\sum_{m,m'}S_{m,m'}^{(-,+,z)} \hat{a} _{i,m}^\dagger \hat{a} _{i,m'} $, with $\hat{a} _{i,m}$ the Wannier operator annihilating a particle on site $i$ in the internal state $m$ and $S_{m,m'}^{(-,+,z)}$ the usual spin matrices elements $\langle m|\hat{S}^{(-,+,z)}|m'\rangle$.  In the case of a spin $1/2$ system, they are the Pauli matrices. This effective interaction, known as super-exchange, is the direct analogue of the exchange interaction between strongly interacting electrons, which arises from the interplay between tunnelling and Coulomb blockade~\cite{Ashcroft1976ssp}. For the goal of gaining a deeper understanding of condensed matter phenomena, physics arising from the super-exchange interaction in atomic assemblies has been widely studied, both in bosonic~\cite{Trotzky2008tro,Nascimbene2012ero,Fukuhara2013moo,Hild2014ffe,Brown2015tds} and in fermionic~\cite{Greif2013srq, Grief2015fad,Hart2015ooa, Drewes2017aci, Parsons2016srm, Boll2016sad, Cheuk2016oos} assemblies; see also Ref.~\cite{Gross2017qsw} for a review. In the case discussed here of atoms interacting at short range, the resulting Heisenberg Hamiltonian is limited to isotropic (XXX) interactions between nearest neighbours, which conserve the total longitudinal magnetisation.

In the case of particles interacting via the DDI, the situation is qualitatively different. This is because the DDI introduces a direct coupling between spins which, in contrast to super-exchange interactions, is long-ranged (not limited to nearest-neighbour interactions), and anisotropic (following the intrinsic anisotropy of the dipolar forces). In addition, as a direct consequence of the anisotropy, the total longitudinal magnetisation is not conserved. The associated two-body interaction potential is given in Eq.~\eqref{ddispinform} and leads, by projection on the Wannier basis (see also Sec.~\ref{subsec:spinless_lattice}), to on- and direct off-site interactions terms comprising similar elastic, exchange and relaxation processes in the lattice Hamiltonian.

When tunnelling is absent, as is the case in very deep lattices, it becomes possible to study spin lattice models due to the mere DDI, i.e., without the addition of super-exchange effects. In practice, for magnetic atoms, the DDI can noticeably exceed super-exchange interactions even when $t$ is not negligible, which eases the requirements for investigating equilibrium versus out-of-equilibrium quantum dipolar magnetism. In addition, compared to the situation governed by a Heisenberg Hamiltonian, it is expected that these dipolar interacting spin ensembles will display a number of exotic quantum magnetic behaviours~
\cite{Peter2012abo}. 

\subsubsection{Free magnetisation and XYZ spin models} 

Perhaps the most striking differences between the spin-lattice models of the super-exchange Hamiltonian (Eq.~\eqref{heisenbergham}) and the DD lattice Hamiltonian in Eq.~\eqref{ddispinform} arise from the presence of magnetisation-changing collisions. Through the conservation of total angular momentum, these magnetisation-changing collisions introduce an intrinsic nonlinear coupling between the spin degrees of freedom and the orbital degrees of freedom, see also Sec.~\ref{spinorsection}. In the context of lattice systems, this coupling can result in an effective XYZ Hamiltonian
\begin{equation}
H_{XYZ}=
 \sum_{i,j}J_{i,j}\left(\alpha \hat{S}_{i}^x.\hat{S}_{j}^x+\beta\hat{S}_{i}^y.\hat{S}_{j}^y+\gamma\hat{S}_{i}^z.\hat{S}_{j}^z\right),
\label{XYZham}
\end{equation}
where $\alpha$, $\beta$, and $\gamma$ differ. Note that in Eq.~\eqref{XYZham}, the coupling rate $J_{i,j}$ may also depend on the intersite bond, and the sum is not restricted to nearest neighbours. Since $\alpha \hat{S}_{i}^x.\hat{S}_{j}^x+\beta\hat{S}_{i}^y.\hat{S}_{j}^y = \frac{\alpha+\beta}{4} \left( \hat{S}_{i}^+.\hat{S}_{j}^- + \hat{S}_{i}^-.\hat{S}_{j}^+   \right) + \frac{\alpha-\beta}{4} \left( \hat{S}_{i}^+.\hat{S}_{j}^+ + \hat{S}_{i}^-.\hat{S}_{j}^-  \right)$, the fact that $\alpha$ and $\beta$ differ is directly associated with the DDI couplings that do not conserve magnetisation. 

This Hamiltonian is associated with novel quantum phases presenting non-trivial topologies~\cite{Syzranov2014sdi, Wall2015ruq, Peter2015tpw, Glaetzle2015dfq}. In these proposals, the topological states arise due to the spin-orbit coupling associated with the DDI, and the associated circulation.  
As for the bulk case of dipolar spinor physics (see Sec.~\ref{freemag}), the actual use of the spin-orbit term to achieve coherent coupling between discrete single-particle states with different external (orbit) and internal (spin) properties, may require extremely fine tuning of the experimental parameters such as the magnetic field with a precision better than the 10 $\mu$G level. Similar to the case of the control of dipolar relaxation, described in \ref{subsec:inel_confin}, the presence of a lattice and its associated band gap might relax these conditions.

\subsubsection{Fixed magnetisation and the secular NMR Hamiltonian (XXZ model)}

The experiments have first concentrated on the regime where magnetisation-changing collisions may be ignored. Such a situation may be realised naturally when particles never share the same lattice site, as dipolar relaxation is a localised phenomenon when the magnetic field is sufficiently high, see Sec.~\ref{range_inel}. This applies when $R_{\rm dr}(B)\ll d$, where $d$ denotes the lattice spacing, see Eq.~\eqref{rdr}. It yields $B\gg \frac{2}{\pi^2}\frac{E_r}{g_F\mu_B}$, where $E_r$ is the lattice recoil energy, typically of a few kHz, while $g_F\mu_B$ is of a few kHz/mG (see Sec.~\ref{freemag}) and off-site dipolar relaxation is practically suppressed for $B\gtrsim 10\,$mG. Additionally, as described in Sec.~\ref{subsec:inel_confin}, a very strong reduction of magnetisation-changing collisions can also be obtained in optical lattice, when the energy released in a dipolar relaxation event does not match the energy for band excitation. This applies for both on-site and off-site processes and as long as the magnetic field is low enough so that the Zeeman splitting is smaller than the lattice band gaps \cite{depaz2013rdo}.  Typically the lattice depth $\Vlatt$ is a few tens to hundreds of $E_r$ and the above suppression holds for small $B$ up to few tens of mG.

The effective Hamiltonian that describes 
 interacting dipoles in a lattice in absence of dipolar relaxation is known from the nuclear magnetic resonance community as the secular dipolar Hamiltonian~\cite{depaz2013nqm}, which reads:    
\begin{eqnarray}
H_{\rm sec}=\frac{1}{2} \sum_{i,j} \frac{\mu_0 (g_F \mu_B)^2}{4 \pi r_{i,j}^3} (1-3\frac{z_{i,j}^2}{r_{i,j}^2}) \times \nonumber \\
  \left[S_i^z.S_j^z - \frac{1}{4}\left(S_i^+.S_j^-+S_i^-.S_j^+\right)\right],
  \label{XXZham}
\end{eqnarray}
where $r_{i,j}$ and $z_{i,j}$ are the distance between sites $i$ and $j$ and its projection along the quantization axis $z$, respectively. Compared to the Heisenberg Hamiltonian of Eq.~\eqref{heisenbergham} introduced above, the main differences are (i) the long-range and anisotropic nature of the coupling 
$J_{i,j}\propto \frac{1}{r_{i,j}^3} (1-3\frac{z_{i,j}^2}{r_{i,j}^2})$ not restricted to nearest neighbours, and (ii) the fact that the magnitude of the exchange term $\left(S_i^+.S_j^-+S_i^-.S_j^+\right)$ is different, changed by a factor $-1/2$. This seemingly slight modification in fact breaks the $SO(3)$ rotational symmetry of the Hamiltonian and can have important consequences both for the magnetic phases in the ground state~\cite{Peter2012abo}, and for out-of-equilibrium properties~\cite{hazzard2013ffe}. $H_{\rm sec}$ in fact corresponds to an XXZ Heisenberg model ($\alpha=\beta$ compared to the general XYZ model of Eq.~\eqref{XYZham}).

\subsubsection{Large spin magnetism}~\label{subsec:large_spin_lattice}

The Hamiltonians $H_{\rm XYZ}$ and $H_{\rm sec}$ introduced above show that magnetic atoms can become a useful platform to investigate spin lattice models for frozen particles. As previously discussed in the bulk case in Sec.~\ref{subsec:large_spin}, the large DDI from magnetic atoms is the consequence of a large total (spin plus angular) orbital momentum $s>1/2$, so that these lattice models can typically be investigated  for rather large spins. One of the interesting practical consequences is that the large spin of atoms provide new measurement protocols because spin dynamics can be monitored by measuring the evolution of the population of the different Zeeman states directly, contrary to the case of $s=1/2$ particles. 

Moreover,  the existence of a large spin also introduces novel physics compared to that associated with the pure dipolar-based models, $H_{\rm XYZ}$ or $H_{\rm sec}$, as well as with $H_{\rm ex}$ for $s=1/2$-particles. For example, spin-dependent contact interactions, directly related to the existence of spin-dependent scattering lengths (as discussed in Sec.~\ref{subsec:spin_int}), also need to be taken into account. As for the bulk case, the description of the interaction and the different ground state phases becomes increasingly complicated with the spin length, even without dipolar interactions. Reference~\cite{Imambekov2003sei} studies the specific case of spin-1 atoms with antiferromagnetic interactions ($a_2 > a_0$) and shows a phase diagram similar to that obtained with spinless bosons, but with a polar superfluid phase and singlet to nematic phase transitions now present inside the insulating lobes. This spin-1 phase diagram was first experimentally investigated with sodium atoms \cite{Jiang2016fos}, with negligible dipolar interactions. 

Studying the interplay between spin-dependent contact interactions and dipolar interactions in a lattice setting is an extremely appealing prospect, in particular at large lattice depth where spin-exchange processes within a lattice site  will result in a spin-dependent super-exchange interaction between neighbouring sites. New spin-dependent terms, thereafter denoted as $H_{\rm sd}$, thus arise in the nearest-neighbour interaction in addition to $H_{\rm ex}$ and $H_{\rm sec}$. This results in a very complicated Hamiltonian which has now become experimentally available.

\subsubsection{Magnetism as a function of lattice depth}

In the presence of tunnelling, super-exchange processes $H_{\rm ex}$ and $H_{\rm sd}$ and dipolar interactions $H_{\rm sec}$ must all be taken into account. Interestingly, while the onsite interaction $U$ (possibly spin-dependent) increases with lattice depth, the tunnelling $t$, and $H_{\rm ex}$ and $H_{\rm sd}$ decrease, while $H_{\rm sec}$ is more or less independent of lattice depth. By varying the lattice depth, it is thus possible to study the interplay between these different mechanisms. When the lattice depth is weak and tunnelling is allowed, magnetism is driven by the interplay between long-range dipolar interactions and short range physics. When the lattice is very deep and tunnelling if frozen, the arrangement of the atoms in the lattice is typically very regular due to the Mott insulator transition~\cite{Jaksch1998cba,Greiner2002qpt, Joerdens2008ami, Schneider2008mai,Bloch2008qca,Bloch2012qsw,Gross2017qsw} (see also discussion in Sec.~\ref{subsec:spinless_lattice}), $H_{\rm ex}$ and $H_{\rm sd}$ vanish, and one can then revisit spin lattice models associated with the secular Hamiltonian. 
 
Varying the lattice depth thus offers a unique and exotic situation to study the cross-over between quantum magnetism associated with Heisenberg-like Hamiltonian, the t-J model at intermediate lattice depths, where quantum magnetism and transport compete, and, at low lattice depth, spinor physics. The intermediate regime is especially enticing, as it represents a challenge for a realistic theoretical simulation. While most of the research on magnetic atoms has up to now focused on Bose systems~\cite{depaz2013nqm, Lepoutre2018eoo}, dipolar Fermi gases are also available~\cite{Patscheider2019cde} and should be fascinating systems with which to study quantum magnetism, as tunnelling and spin-dynamics are expected to be strongly coupled and the atoms are under the strong influence of the Pauli exclusion principle.  

\subsubsection{Other experimental systems}

Similar spin models can also be studied using either heteronuclear molecules \cite{Yan2013ood,Hazzard2014mbd,Bohn2017cmp,Moses2017nff} or Rydberg atoms with electric dipoles \cite{Barredo2015cet,Barredo2016aab,Endres2016aba,Bernien2017pmb}. In these cases, the reduced collisional lifetime forbids working in the regime where tunnelling is significant. Important experimental results have  recently been obtained for these systems exploring the physics of spin-lattice models for an assembly of effective spin-$1/2$ particles pinned in a periodic potential. One of the attractive features of these systems is that the relative strength of the exchange and Ising terms of the interaction can be experimentally controlled.

Furthermore, ions have become a prominent platform to study spin-lattice models~\cite{Blatt2012qsw,Zhang2017ooa,Friis2018ooe}. In these ensembles of crystallised ions, an effective spin-spin interaction is mediated between the ions using phonons. These systems possess the unique possibility to vary the range of the effective interaction potential between the particles.

While these other systems are extremely promising and offer complementary paradigms of quantum magnetism, a few characteristics distinguish the prospects offered by highly magnetic atoms to date:
\begin{itemize}
    \item Magnetic atoms uniquely realise large-spin systems, beyond the effective spin-$1/2$ case (see also Sec.~\ref{subsec:large_spin_lattice}). 
    \item Up to now, systems of ions, Rydberg atoms and molecules remain limited in the size of the sample under studies or in their densities. While ultracold atoms offer many-body systems of several thousands or tens of thousands of atoms with lattice filling factors on the order of unity, Rydberg and ion assemblies are restricted to a few tens to hundred particles. This size limitation is particularly problematic for the case of long-range interacting systems, for which border and finite-size effects are important. Despite the tremendous progress made on molecular systems, the fillings achieved remain  below 0.5~\cite{Moses2015coa}.  
    \item Magnetic atoms primarily offer a competition between spin-spin interactions and  tunnelling, as both terms can be allowed and tuned on the same scale. This enables the exploration of quantum magnetism when transport competes with spin-spin interactions, which is perhaps the most relevant regime from the point of view of quantum simulation.
\end{itemize}

\subsubsection{Experimental status with magnetic atoms}

The study of dipolar lattice systems has started with two complementary experiments performed with ultracold KRb molecules~\cite{Yan2013ood} and a BEC of Cr atoms~\cite{depaz2013nqm}. For the experiments with KRb, an effective spin 1/2 system is encoded in two rotational states of the vibrational ground state. NMR-like experiments have been performed which revealed the impact of inter-site pair-wise dipolar interactions on the decay of the contrast of a molecular Ramsey interferometer. This experiment was performed in a rather dilute environment with a filling factor of typically 5 to 10 percent, and tunnelling was absent to prevent inelastic collisions when two molecules physically meet~\cite{Yan2013ood}. In a latter study, the experimental results were compared to a new theoretical model based on a cluster expansion technique, and an extremely good agreement was found~\cite{Hazzard2014mbd}.

In the Cr experiment~\cite{depaz2013nqm}, a BEC was loaded in a 3D optical lattice, which, at large lattice depth, led to the production of a Mott insulating state with a core with double occupancy and a shell with unit filling. Spin dynamics was studied after spin excitation in a well-defined state. The experiments revealed a non-equilibrium spinor dynamics resulting from inter-site Heisenberg-like spin-spin interactions provided by non-local DDI. While mean-field theories could not reproduce the experimental data, a model based on exact diagonalization techniques on a plaquette provided a good agreement with the experiment for short times. This showed the many-body character of spin dynamics, and the importance to take quantum correlations into account. For doubly occupied site, a complex spin dynamics was observed, involving both short-range interactions and intersite DDI. This experiment therefore showed the potential of lattice gases made of strongly magnetic atoms for the study of quantum magnetism of high-spin systems.

\begin{figure}[t]
\centering
\includegraphics[width= 8 cm]{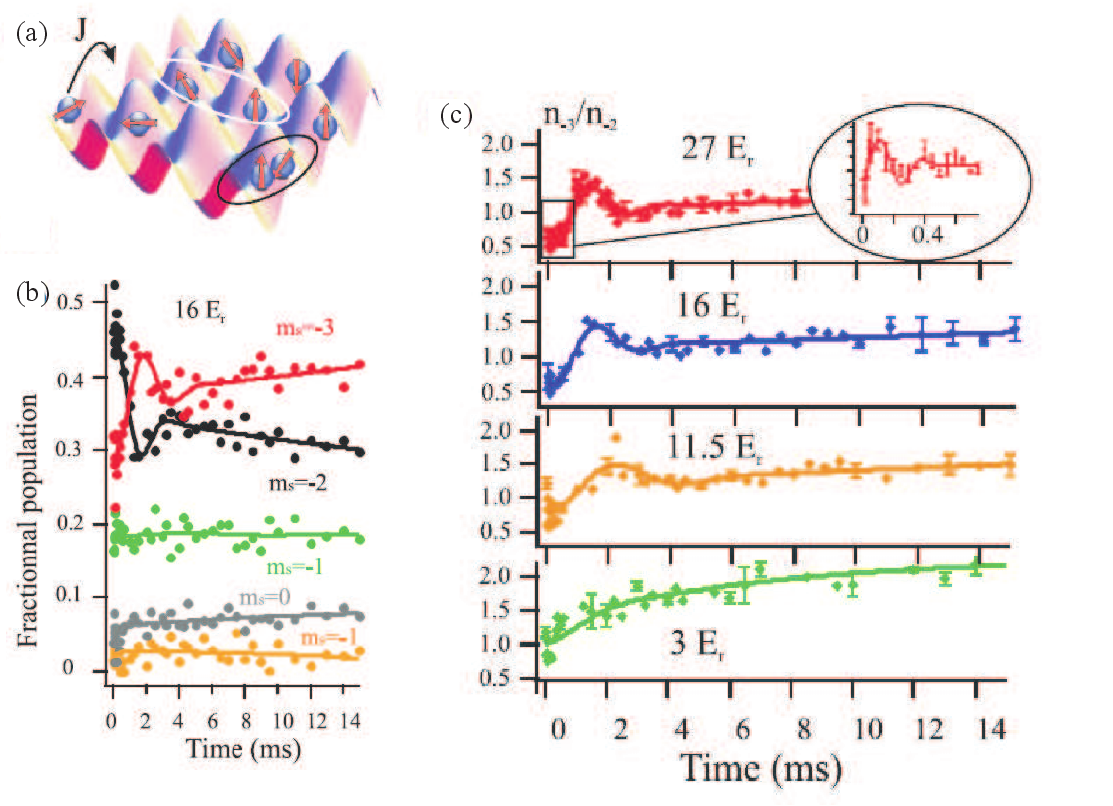}
\caption{Spin dynamics of chromium atoms as a function of lattice depth, from \cite{depaz2016psd}: (a) Simple representation of the system close to the Mott-to-superfluid transition. Atoms interact both due to intersite (white ellipse) and on-site (black ellipse) interactions. (b) Measurement of the spin components ($m_s=-3$ to $m_s=1$) as a function of time for a lattice depth of 16 Er. (d) Time evolution the ratio between populations in $m_s$ states -3 and -2, for four different lattice depths (27Er, 16Er, 11.5Er, and 3Er, from top to bottom). Lines are guides for the eye resulting from fits.}
\label{latticedynamics}
\end{figure}

In a another set of experiments, the group in Villetaneuse also studied the impact of the lattice depth on spin dynamics~\cite{depaz2016psd, Fersterer2019doa}. The experiment unveiled a smooth crossover from a complex oscillatory behaviour to an exponential behaviour of the spin populations throughout the Mott-to-superfluid transition, as shown in Fig.~\ref{latticedynamics}. The experiment provided data in the intermediate regime between superfluid and Mott insulating, where dipolar interactions, contact interactions, and superexchange mechanisms compete. In this strongly correlated regime, spin dynamics and transport are coupled, which constitutes a challenge for theoretical models of quantum magnetism. As exact modelling of the experimental dynamics exceeds the capabilities of classical computation, experimental results in \cite{Fersterer2019doa} were compared to approximate models.  The comparison between experimental data and theory  demonstrated that the dynamics at low lattice depth is qualitatively reproduced by mean-field calculations based on the Gutzwiller ansatz. In this regime, it was found that transport and contact interactions both play an essential role in spin dynamics; on the contrary, only a beyond-mean-field theory could account for the dynamics at large lattice depths, which is then mostly driven by dipolar interactions \cite{Fersterer2019doa}. 

The potential offered by magnetic atoms to study out-of-equilibrium quantum magnetism was recently further demonstrated by two works that realised clean instances of the above-described XXZ Heisenberg model~\cite{Lepoutre2018eoo,Patscheider2019cde}. The experimental systems were prepared in order to form large unit-filled arrays of magnetic atoms in deep lattices. An out-of-equilibrium dynamics, occurring under the pure effect of intersite DDI, was initiated by preparing the atomic assembly in a given spin state. In the first study~\cite{Lepoutre2018eoo}, the first shell from a Mott insulator made of bosonic Cr was isolated and a spin excitation was performed by tilting the spins with respect to the magnetic field orientation (quantisation axis), yielding a coherent spin state. In the second study~\cite{Patscheider2019cde}, a band insulator of fermionic Er was realised (ensuring single filling from Pauli exclusion principle) and the spin excitation was performed by fully transferring the population from the lowest $m=-F$-state to an excited $m$-state, thus forming a spin Fock state.  In both studies, the dynamics was observed directly on the evolution of the $m$-states populations. The effect of quantum correlations on the dynamics was demonstrated by discrepancies in the comparison with mean-field theories. 

The good agreement with simulations based on a generalisation of the discrete truncated Wigner approximation~\cite{Schachenmayer2015mbq} to the case of large spins, additionally proves the role of quantum correlation and shows that the spin dynamics leads to a growth of entanglement. In Ref.~\cite{Lepoutre2018eoo}, this was characterised by a calculation of the Renyi entropy. Here, the role quantum correlations was observed to increase when the initial tilting angle with respect to the magnetic field approaches $\pi / 2$. Experimentally, the isolated ensemble of atoms approaches an effective thermal equilibrium, in the spirit of the Eigenstate Thermalisation Hypothesis \cite{Deutsch1991qsm,Rigol2008tei,Kaufman2016qtt}, in which the growth of entanglement conveys a thermal character to local observables. 

By performing a Ramsey experiment and analysing the contrast of the interferometer as a function of time for the ensemble of unit-filled Cr atoms, it was also observed that, during spin dynamics, the collective spin length of the atoms decayed under the effect of dipole-dipole interactions, as was theoretically expected \cite{hazzard2013ffe}. Interestingly, for a pure homogeneous system, the dynamical reduction of the collective spin length is a purely beyond mean-field effect. The results in \cite{gabardos2020rot} show that the decay of the spin length was slower than expected by a spin model of frozen particles, a finding that could not be previously deduced from the measurements on population dynamics. This illustrates how measurements of spin coherences provide valuable and complementary information on quantum many-body systems, and shows that further experiments as well as new observables are needed to fully characterise the growth of correlations in this system.

In Ref.~\cite{Patscheider2019cde}, the rate of the spin dynamics was additionally tuned thanks to the two distinct knobs available to control the single-spin-state energies. 
The level spacing could thus be tuned to be equal, providing a resonant condition for the spin-exchange process of $H_{\rm sec}$. The two control knobs are the light-shift induced by off-resonant laser beams (e.g., the one forming the optical lattice potential, see also Sec.~\ref{subsec:atomlightLn}) and the Zeeman effect induced by the bias magnetic field. While both effects also exist for bosonic Cr, the missing point lies in the existence of quadratic Zeeman shifts (induced by the hyperfine structure of the fermions) which can be tuned to compensate the quadratic light shifts in the case of fermionic Er. This exquisite control enables a thorough investigation of the secular Hamiltonian. The large quadratic light shifts can be additionally used for a fast (and potentially local) control of the spin dynamics.  
Reference~\cite{Patscheider2019cde} additionally studies the effect of the excited $m$-state on the rate of the initial spin dynamics, deducing a universal scaling, independent on the detail of the initial preparation and probes the effect of tilting the quantization axis compared to the lattice geometries, providing first steps towards more generic lattice models. 

These first experiments reveal that dipolar lattice gases provide a new arena with which to study an exotic quantum magnetism of large spin systems, driven by the competition of long-range and short-range interactions and of tunnelling. From the theoretical point of view, the full description of such a system in the presence of tunnelling (thus being analogous to the the t-J model of magnetism) is already a challenge. From the experimental point of view, a number of challenges remain ahead. Here, we only point a few of the challenges and perspectives for magnetic atoms:

\begin{itemize}
    \item {\bf Impact of quantum statistics:} Up to now, experiments were only performed with bosons {or with fermions in the frozen regime}. The study of fermions in presence of tunnelling would be extremely interesting, with an expected interplay between spin and motional  dynamics, and the Pauli exclusion principle.
    \item {\bf Revealing the expected growth of entanglement in large spin systems:} The use of entanglement witnesses is a possibility, although preliminary studies show that the extension of the existing witnesses to large spins is not straightforward nor favourable. Bipartite entanglement could also be revealed by in situ measurements of spin fluctuations. 
    \item {\bf Magnetic phases:} Up to now, only out-of-equilibrium experiments have been performed. The magnetic ground state close to zero temperature remains out of reach of our current technology, limited by a rather large entropy. However, the analysis of the equilibrium state reached after spin dynamics has occurred is also an interesting avenue for future research, which is readily accessible. For example, the nature of the equilibrium state, the presence of quantum correlations within it, and how it differs from a thermal state, are open questions.  Likewise, the study of the nature of this quasi-equilibrium state as a function of the lattice filling factor, and as a function of entropy, is also a very interesting question related to many-body localisation.
\end{itemize}

%% file: Section10/10_conclusions.tex
\section{Perspectives}

The present time is highly exciting for research based on magnetic atoms.  Many substantial advances are being made at an impressively rapid pace. 
New experimental apparatus are coming online soon and promise to bring even more excitement. 
State-of-the-art techniques, recently implemented in other ultracold-atom experiments, will soon become available in lanthanide experiments.   These promising to shed new light on the physics that has been revealed within the few last years.  For example, single-atom  `quantum gas' microscopes are well-poised to reveal additional details regarding the strong correlations of the many-body states that have been observed both in bosonic and fermionic dipolar quantum gases as well as open the way to extend investigations of their far-from-equilibrium dynamics. Control on the single-atom level will also provide novel understanding, at the few-body level, of scattering phenomena in these complex lanthanide systems. Complete magnetic shielding will enable the exploration of the zero-field regime of spin physics. Exploiting the rich optical transition spectrum of the lanthanides will also provide novel capabilities for controlling spins and interactions more exquisitely. The prospect of realising topological quantum states is one of many central aims. 

Highly versatile trap geometries will also play a role and provide new prospects for exploring the interplay between the anisotropic and long-range character of the DDI and the geometry of the system.  Other ingredients, such as disorder and frustration could also be added to the systems.  The mixture of different magnetic and non-magnetic species will reveal new physics, and in particular,  could provide access to dipolar physics with an internal degree of freedom and distinct dipolar moments but without dipolar relaxation. Mass imbalance effects could also be explored. This opens new avenues toward exotic few-body as well as many-body phenomena, including Efimov states, $p$-wave superfluidity, and other exotic states of matter. One cannot capture all the possible new directions enabled by highly magnetic atoms, but undoubtedly, brand-new physics will arrive in the next years, and we hope for new surprises to match those this review has presented.

\subsection*{Acknowledgements}

All the authors would like to warmly thank their countless collaborators over the years, both from the experiment and for the theory sides. They were all crucial actors to the discoveries summarised in this review. Additionally, we are grateful to Steven Lepoutre, Fabian Boettcher, Wil Kao, and Kuan-Yu Li for their careful reading of the manuscript. 
L.C., I.F.B., F.F., and T.P.  acknowledge the support from the international collaborative network on dipolar quantum gases funded by the DFG and FWF - FOR 2247/PI2790. F.F., B.L.T., and T.P. acknowledge support through the QuantERA grant MAQS. 
L.C. acknowledges support by the Deutsche Forschungsgemeinschaft (DFG, German Research Foundation) under
Germany’s Excellence Strategy EXC 2181/1-390900948 (the Heidelberg STRUCTURES Excellence Cluster) and Project-ID No. 273811115 (SFB1225 ISOQUANT). 
B.L.T acknowledges support from CNRS, Conseil Régional d’Ile-de-France under SIRTEQ Agency, and Agence Nationale de la Recherche (Project
No. ANR-18-CE47-0004).
B.L.L. acknowledges support from the NSF and AFOSR. T.P acknowledges support from the European Research Council (ERC)  (Grant agreement No. 101019739).